\pdfoutput=1
\RequirePackage{silence}
\ErrorsOff*
\documentclass[twoside,onecolumn,11pt,a4paper,openright]{book}
\usepackage{anuthesis6}
\usepackage{everyshi}
\usepackage{graphicx}
\usepackage{ifthen}
\usepackage{calc}
\usepackage{pdfpages}
\usepackage{fancyhdr}
\usepackage{amsmath}
\usepackage{amsfonts}
\usepackage{amssymb}
\usepackage{amstext}
\usepackage{array}
\usepackage{xspace}
\usepackage{xkeyval}
\usepackage{upgreek}
\usepackage{mathrsfs}
\usepackage[hidelinks]{hyperref}
\usepackage{cite}
\usepackage{url}
\usepackage[figure]{hypcap}
\usepackage{float}
\usepackage[CaptionAfterwards]{fltpage}
\usepackage{nomencl}

\makenomenclature

\usepackage[hyperpageref]{backref}
\renewcommand*{\backref}[1]{}
\renewcommand*{\backrefalt}[4]{%
\ifcase #1 %
[Not cited]%
\or
[Cited p.~#2]%
\else
[Cited pp.~#2]%
\fi
}

\let\origdoublepage\cleardoublepage
\newcommand{\clearemptydoublepage}{%
  \clearpage
  {\pagestyle{empty}\origdoublepage}%
}

\let\cleardoublepage\clearemptydoublepage

\newcommand{\be}{\begin{equation}}
\newcommand{\ee}{\end{equation}}
\newcommand{\beqn}{\begin{eqnarray}}
\newcommand{\eeqn}{\end{eqnarray}}

	\newcommand{\chapquote}[2] {\begin{quote}
			\emph{#1}\vspace{-0.5cm}	\begin{flushright}$\sim$ #2 \end{flushright}
	\end{quote}}
	
\newcommand{\dg}{^{\circ}}
\newcommand{\blankpage}{
\newpage
\thispagestyle{empty}
\mbox{}
\newpage
}

{ \end{list} }

{ \end{list} }

\begin{document}
 
\includepdf[pages={1-14}]{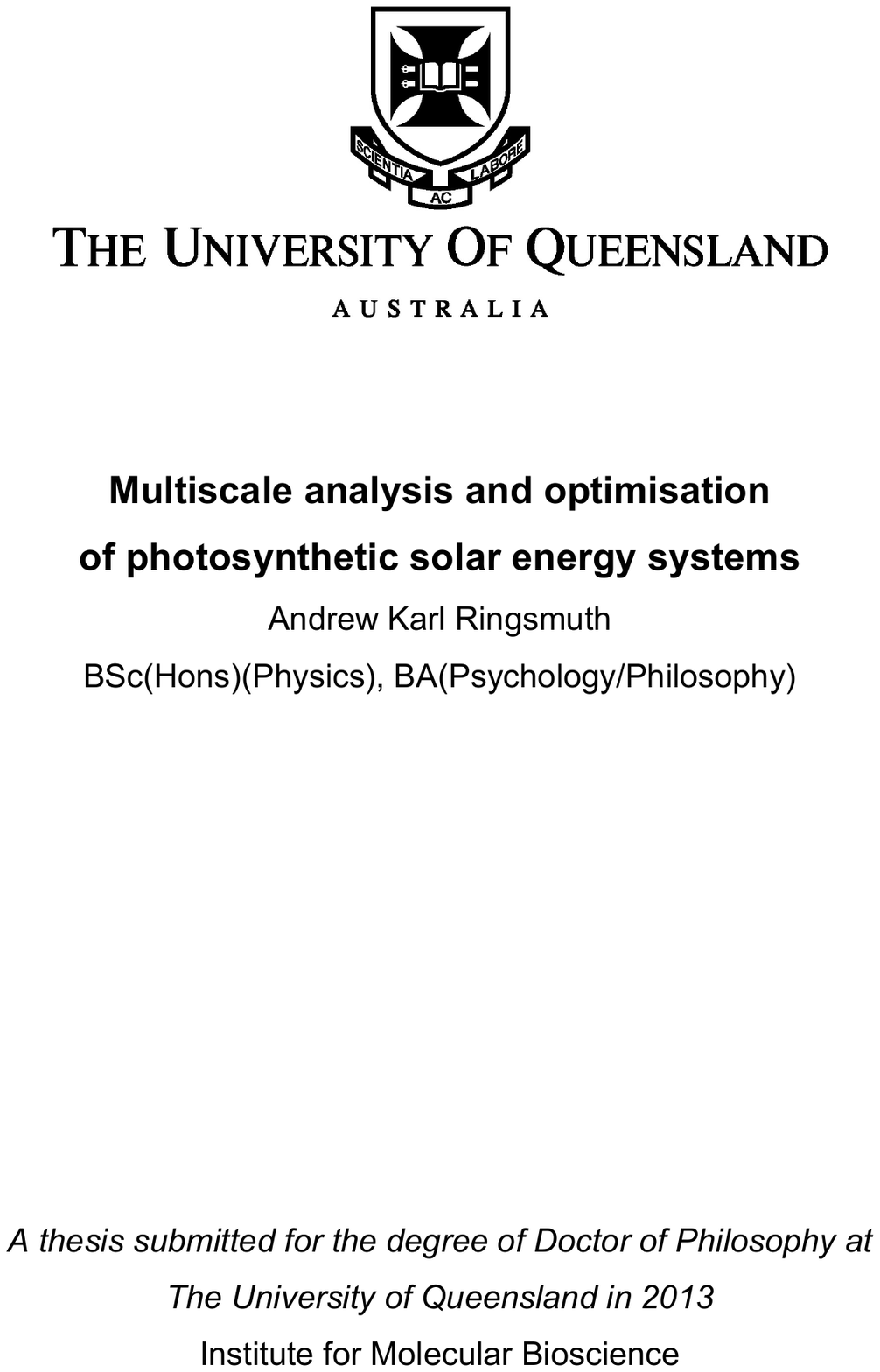}

\setlength{\parindent}{20pt}
\setlength{\parskip}{2ex plus 0.8ex minus 0.2ex}

\bibliographystyle{myunsrt}

\pagenumbering{roman} 

\setcounter{page}{13} 
\chapter*{Dedication}
\addcontentsline{toc}{chapter}{Dedication}

To my parents, Helmuth and Patricia Ringsmuth, \\
who have given me more than was given to them \\
and more than any son could have reasonably asked for. 
\\\\
And in memory of Matthias Rist (03.02.1982 -- 28.06.2009), \\
who shared a childhood with me in an uncommon corner of the Earth\\
and worked to protect its natural beauty after we were grown.

\tableofcontents
\cleardoublepage
\thispagestyle{empty}
\addcontentsline{toc}{chapter}{List of figures}
\listoffigures \markboth{List of figures}{}
\cleardoublepage
\addcontentsline{toc}{chapter}{List of tables}
\listoftables \markboth{List of tables}{}
\cleardoublepage
\addcontentsline{toc}{chapter}{List of abbreviations}\markboth{List of abbreviations}{}
\printnomenclature 

\thispagestyle{empty}
\chapter[Photosynthesis in global energy systems: realities and prospects]{Photosynthesis in global energy systems: realities and prospects} \label{chp:global}
\pagenumbering{arabic} 
\setcounter{page}{1}  
\chapquote{Our bodies are stardust; our lives are sunlight.}{O. Morton \cite{morton2007}}

\section{Motivation: Earth in one day}\label{sec:earthday}

Four-and-a-half billion years ago, the early universe had cooled enough for the Solar System and Earth to form \cite{bowring1995}. Three hundred million years later, the planet's surface was covered with oceans \cite{nutman2006, wilde2001}. Though the origins of life remain mysterious, fossil cells date from 3.5 billion years ago at least \cite{schopf2002}. Levels of free oxygen rose suddenly around 2.4 billion years ago, this `Great Oxidation Event' being attributed to the concerted activity of photosynthetic cyanobacteria \cite{sessions2009, hohmann2011}, and the life-protecting ozone layer formed shortly thereafter. If Earth's history until the present were rescaled to fit within one of its days, life on land would have appeared around 9:30pm \cite{shear1991}. Humans emerged just eight seconds before midnight \cite{hershkovitz2010} and drilled the planet's first commercial, fossil oil well one one-thousandth of one second before midnight \cite{yergin1991, smil2010(2)}. Within the remaining moment, the global human population has grown hyperbolically \cite{nielsen2005}, by a factor of roughly seven \cite{uscensus}.

Humanity's current role in the Earth system is historically unique. For the first time, a vertebrate species has become a geophysical force, modifying on a global scale the long-evolved natural systems that support its existence \cite{barnosky2012, rockstrom2009}. Compared with the history of evolution on Earth, an individual human life is brief. Yet it is estimated that the next human generation is likely to witness extinction of some 20--30\% of known species if the world economy continues with `business as usual' \cite{ipcc2007, stern2006}, with up to $30\%$ of all mammal, bird and amphibian species to be threatened with extinction before the end of this century \cite{rockstrom2009}. This rate of species loss is estimated be 100 to 1000 times higher than what could be considered natural in the absence of human influences \cite{rockstrom2009} and there is evidence that humanity is forcing the biosphere towards a `critical transition', which would abruptly override gradual trends to produce a global-scale state shift with unanticipated biotic effects \cite{barnosky2012}.

Concurrent with these disruptions in natural systems have been unprecedented advancements in human technology, including sustained exponential growth in information processing power \cite{dreslinski2010, kurzweil2006}, and progress in engineering biological and artificial systems at near-atomic scales \cite{niemeyer2006, mirkin2007, karkare2008}. An opportunity exists to use these technologies judiciously in addressing urgent ecological challenges, targeting systems across the spectrum of scales from global Earth systems and the global economy to the nanoscale physics and chemistry of energy, information-processing and pollution-mitigating technologies. 
 
A natural phenomenon that spans these scales -- a process fundamentally important to life on Earth -- is photosynthesis. Photosynthesis, which stores energy from sunlight in chemical bonds within reduced forms of carbon, provides almost all of the energy driving the biosphere and ultimately, through fossil fuels, also the human economy. It is nature's solar energy conversion technology, engineered by natural selection over three billion years \cite{hohmann2011,leslie2009}. Due to growing concerns over the unsustainability of fossil-fuelled energy systems, in recent years there has been increasing interest in the potential for photosynthesis to help meet humanity's energy needs sustainably, on a global scale, through harnessing higher plants \cite{haberl2010, larkum2010, mcconnell2010, field2008, barber2009, hambourger2008}, microalgae \cite{stephenson2011, singh2011, day2012, larkum2011, stephens2010, brennan2010, larkum2010, mata2010, pienkos2009, beer2009, barber2009, hambourger2008, kruse2005}, and artificial photosynthetic systems \cite{magnuson2012, gust2012, tachibana2012, blankenship2011, jiang2010, mcconnell2010, gray2009, barber2009, hambourger2008, lewis2006}. This potential is the foundation for the work presented in this thesis.

\section{Photosynthetic systems, scales and sciences}

A system is a collection of interacting components that carries out a function or purpose \cite{meadows2008}. Accordingly, the following definition may be formulated:
\begin{quote}
    A photosynthetic system is a collection of interacting components that carries out the process of photosynthesis.
\end{quote}
Importantly, the photosynthetic performance\footnote{Various performance measures may be chosen. Two common measures are energy conversion efficiency and energy conversion rate (`productivity').} of the whole system generally differs from the sum of its components' performances when each acting in isolation, by virtue of interactions between components (e.g. The photosynthetic productivity [section \ref{subsec:PP}] of two leaves in direct sunlight differs from the productivity of the same two leaves when one shades the other [chapter \ref{chp:multianalysis}]). This phenomenon is known as emergence \cite{chalmers2006} and is a hallmark of a complex system. The above, very general definition is more than a theoretical curiosity; later chapters demonstrate its practical utility in detail. Here, it is helpful in outlining the approach and scope of this thesis. 

Intuitively, the term, `photosynthetic system' may evoke ideas of a plant or the physicochemical processes inside its cells (section \ref{sec:photosynthesis}) and indeed both of these systems satisfy the above definition. However, a forest satisfies the definition equally well, as does the entire Earth. While the definition is scale-invariant, what differs between these cases is the scale of the photosynthetic system, the available choices of constituent subsystems, the mechanisms of their interactions, and the scientific disciplines relevant for describing each.  

According to Blankenship \cite{blankenship2002}, 
\begin{quote}
`\textit{Photosynthesis is perhaps the best possible example of a scientific field that is intrinsically interdisciplinary... (spanning) time scales from the cosmic to the unimaginably fast, from the origin of Earth 4.5 billion years ago to molecular processes that take less than a picosecond. This is a range of nearly thirty orders of magnitude.}' 
\end{quote}
Spanning this range is a hierarchy of scientific disciplines, each associated with its own level(s) of organisation in the world. These are taken in the reductionist view to be reducible, one to the next, in something like the following order: geoscience, environmental science, ecology, physiology, cell biology, molecular biology, chemistry, many-particle physics, elementary particle physics. However, the success of this reductionism does not imply that constructionism from smaller to larger scales will be equally successful; at each larger level of organisation, new properties emerge that cannot be predicted from the laws describing smaller scales alone \cite{anderson1972}. Consequently, the literature on photosynthesis is distributed across diverse scientific disciplines concerned with different scales of organisation. Comprehensive analysis of photosynthetic systems in the context of global energy systems must face the need to integrate analyses across the hierarchy of disciplines, ranging from global-scale to nanoscale. This presents a challenge to the high level of specialisation endemic in modern science, and an opportunity to break new ground between fields not traditionally thought of as related. 

\subsection{Scope of thesis}
This thesis asks whether and how photosynthetic systems can be optimised as solar energy harvesting technologies able to sustainably power the human economy on a globally significant scale. The analysis traverses the scalar and disciplinary hierarchies described above, beginning with the overall Earth system in this introductory chapter and focussing on subsystems at smaller scales in later chapters. Principles from systems theory are used to combine scale-focussed studies utilising tools from the discipline(s) relevant at each scale: geoscience, environmental science, ecology and economics at the global scale, biology, chemistry and classical physics at intermediate scales, and structural biology, physical chemistry and quantum physics at the nanoscale. 

The analyses at larger scales provide context and constraints for analyses of smaller-scale subsystems, helping to define focus questions and criteria for optimal subsystem functioning. Reciprocally, the smaller-scale analyses help to determine whether the criteria for optimal (sub)system functioning at larger scales can be met. For example, the geophysical, ecological and economic limitations of global energy systems identified in this introductory chapter dictate functional requirements of photosynthetic energy systems later analysed at smaller scales. In turn, the properties of those photosynthetic systems, determined in smaller-scale analyses, constrain whether they can meet the needs of economies and ecosystems at larger scales. Overall the goal is to provide a framework for analysing photosynthetic energy systems using an integrated, multiscale approach, as well as to answer specific technical questions in the context of that framework. It is hoped that this will inform practical engineering efforts aimed at providing solutions to the economic and ecological problems that arise from the provision of energy to human societies.

\section{Photosynthesis}\label{sec:photosynthesis}
\chapquote{Life is woven out of air by light.}{J. Moleschott \cite{perkowitz1996}}
A photosynthetic system of any scale must carry out photosynthesis. The process first appeared in nature prior to the Great Oxidation Event of 2.4 billion years ago, though the various lines of geological and biological evidence have yet to converge on an exact time or mechanism of emergence. There is, however, little doubt that it was oxygenic photosynthesis by cyanobacteria that first oxidised mineral deposits on Earth's surface and, thereafter, its atmosphere. This radically altered the planet's chemistry and enabled oxygenic life \cite{sessions2009, morton2007, hohmann2011}.
 
Photosynthesis stores energy from electromagnetic radiation by using it to break and create chemical bonds against the chemical equilibrium. Naturally occurring in higher plants, eukaryotic algae and a few genera of bacteria, it is the process by which almost all carbon fixation and atmospheric oxygen evolution on Earth have occurred. It is a solar-powered heat engine which builds relatively low-entropy organic products such as carbohydrates and lipids out of relatively high-entropy environmental carbon dioxide (CO$_2$)\nomenclature{CO$_2$}{Carbon dioxide} and water, while oxygen (O$_2$)\nomenclature{O$_2$}{Molecular oxygen} is evolved as a waste product \cite{smil2008, jennings2005}. 
	
\begin{figure}
\includegraphics[angle=0,width=1\textwidth]{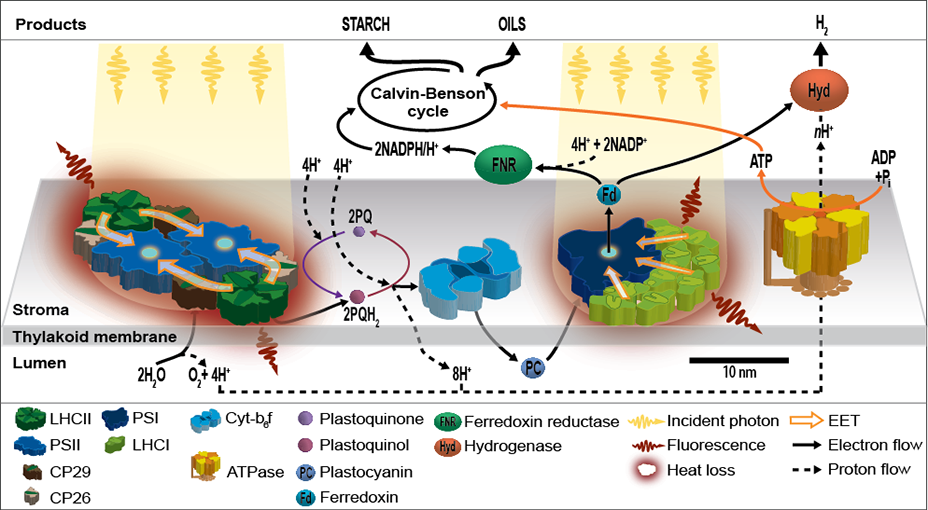}
\caption[Photosynthesis in green algae and higher plants]{\textbf{Photosynthesis in green algae and higher plants.} See text for explanation, and figure \ref{fig:xraystruct} for high-resolution structures of membrane protein complexes. Photon wavelengths shown are reduced by 2--3 orders of magnitude for practicality.}
\label{fig:zscheme}
\end{figure}
	
	Traditionally, photosynthesis is partitioned into two stages (fig. \ref{fig:zscheme}). First are the so-called light reactions, in which pigment-binding protein complexes absorb light and transfer its energy to photochemical reaction centres (RCs\nomenclature{RC}{Reaction centre}) where chemical charge-separation generates energised electrons and protons. These are processed through a series of enzymatic steps to create reducing equivalents that are chemical precursors for the second stage of photosynthesis. The second stage is known as the Calvin-Benson cycle, carbon reactions or dark reactions, because it proceeds without direct addition of radiant energy. Although the two stages include a complex web of photochemical and biochemical reactions \cite{taiz2006}, the following general equation summarises the overall chemistry: 
		\beqn
		\text{Carbon dioxide}+\text{Water}&\xrightarrow{\text{Light~energy}}&\text{Carbohydrate}+\text{Oxygen}\nonumber\\
	 	2n~\text{CO}_2+2n~\text{H}_2\text{O}&\xrightarrow{\text{Photons}}&2\text{(CH}_2\text{O)}_n+2n~\text{O}_2. \nonumber
   	\eeqn
	Physically, the process is analogous to a photovoltaic cell (light reactions) coupled to an electrochemical cell (dark reactions), which together use photons to drive fuel synthesis.	As the detailed biophysical analyses in this thesis are concerned mainly with the light reactions, these form the primary focus of the ensuing summaries, with the dark reactions described somewhat superficially, in the interest of completeness. 
	
		\subsection{The light reactions: from photons to reducing equivalents}\label{subsec:lightreact}
		In green algae and higher plants, incident light is absorbed by photosystem `core' complexes I and II (PSI\nomenclature{PSI}{Photosystem I} and PSII\nomenclature{PSII}{Photosystem II}) and their coupled `antenna' light-harvesting complexes (LHCs\nomenclature{LHC}{Light-harvesting complex}), embedded in the `thylakoid' lipid membrane within the chloroplast (fig. \ref{fig:zscheme}). PSI and PSII, together with their LHCs, form the PSI-LHCI and PSII-LHCII supercomplexes, in which the primary role of the LHCs is light absorption and excitation energy transfer (EET\nomenclature{EET}{Excitation energy transfer}) to RCs within PSI and PSII. There the energy drives the light reactions, which are linked to form the so-called photosynthetic electron transport chain \cite{rochaix2011}. 
		
		The first step is catalysed by PSII-LHCII \cite{renger2008}. Light is predominantly absorbed by the major LHCII proteins \cite{barros2009} (Lhcb 1, 2 and 3 in higher plants and Lhcbm 1--9 in green algae), which bind chlorophyll \textit{a}, chlorophyll \textit{b} and carotenoid chromophores. This creates electronic excitations which are transferred nonradiatively\footnote{Chapter \ref{chp:qeet} describes the mechanisms of nonradiative excitation energy transfer in detail.} through the minor LHCII proteins (CP29, CP26 and CP24, the latter being absent from algae)\cite{ginsberg2011, debianchi2008, tokutsu2012} to PSII (comprising more than 20 subunits, including CP47, CP43, D1, D2, Cyt--$b559$, PsbO, PsbP and PsbQ \cite{hankamer2005}) to drive the splitting of water molecules, yielding protons, electrons and oxygen (fig. \ref{fig:zscheme}). The electrons are passed \emph{via} plastoquinone (PQ)\nomenclature{PQ}{Plastoquinone}, cytochrome $b_6f$ (Cyt--$b_6f$)\nomenclature{Cyt--$b_6f$}{Cytochrome $b_6f$} and plastocyanin (PC\nomenclature{PC}{Plastocyanin}), on to PSI where the associated LHCI proteins (Lhca 1--4 in higher plants, and additionally Lhca 5--9 in green algae) transfer electronic excitations to the photochemically active PsaA and PsaB subunits of PSI (evolutionarily related -- homologous -- to the CP47, CP43, D1 and D2 proteins of PSII \cite{dekker2005, barber1999}). Finally, electrons are passed from PSI to ferredoxin (Fd), where they are ultimately used in the production of nicotinamide adenine dinucleotide phosphate (NADPH)\nomenclature{NADPH}{Nicotinamide adenine dinucleotide phosphate}, a reaction catalyzed by the ferredoxin-NADP+ oxidoreductase. `Linear' flow of electrons through the electron transport chain (cf. cyclic electron flow \cite{iwai2010(1)}) is accompanied by simultaneous release of protons into the thylakoid lumen by PSII, and the PQ/plastoquinol (PQH$_2$)\nomenclature{PQH$_2$}{Plastoquinol} cycle and Cyt--$b_6f$. This results in the buildup of a proton gradient across the thylakoid membrane, which drives subsequent adenosine triphosphate (ATP)\nomenclature{ATP}{Adenosine triphosphate} production \emph{via} the proton-pumping ATP synthase (ATPase)\nomenclature{ATPase}{ATP synthase}.
		
		\subsection{The dark reactions: primary production of biomass and fuels}
		The NADPH and ATP produced in the light reactions are used in the dark reactions and other biochemical pathways to produce sugars, starch, oils and other bio-molecules \cite{taiz2006}. These energy-dense products of cellular metabolism -- photosynthates -- collectively form biomass (or `phytomass'), which can itself be used as a fuel or further refined into biofuels (section \ref{sec:solfuels}). Additionally, some photosynthetic micro-organisms such as the model green alga \textit{Chlamydomonas reinhardtii} \nomenclature{\textit{C. reinhardtii}}{\textit{Chlamydomonas reinhardtii}} possess the ability to recombine protons and electrons extracted from water (or starch) into molecular hydrogen -- a process driven by an anaerobically inducible hydrogenase (Hyd\nomenclature{Hyd}{Hydrogenase}) -- thus driving the direct production of an immediately usable biofuel \cite{amaro2012, melis2000}. Similarly, other organisms are under development to directly synthesise hydrocarbons (alkanes), providing fungible replacements for mineral gasoline, diesel and kerosene (jet fuel) (section \ref{sec:solfuels}).	

\subsection{Photosynthesis in global energy systems}\label{subsec:phoglo}     
The remainder of this chapter considers Earth as coupled photosynthetic and `metabolic' systems. The coupling is taken to comprise two components: 1) a short-time-scale component through which current photosynthetic biota supply photosynthates\footnote{A photosynthate is any compound that is a product of photosynthesis.} to meet the metabolic needs of natural ecosystems and a small fraction of the human economy; 2) A long-time-scale component through which fossil fuels -- stored, geochemical products of ancient photosynthates -- meet the majority of the current human economy's metabolic needs. The central question considered is whether component 1 can be engineered to supplant a large enough fraction of component 2 to sustain the human economy while also continuing to sustain natural ecosystems. Geophysical, ecological and economic constraints to this possibility are quantified. 

To begin, Earth's solar energy resource is described since this provides the ultimate limit to energy available through photosynthesis. Next is an assessment of the total energy currently provided to the biosphere through primary photosynthates and this is compared with human energy consumption. Fossil fuels are then introduced, their role in modern civilisation discussed, and limits to its sustainability assessed. In the final section, potential for a global transition from fossil to solar fuels is considered, and promising solar fuel production technolgies -- photosynthetic energy systems -- introduced and compared. Key pathways are identified for innovating these technologies towards economic scalability and this provides the impetus for detailed biophysical studies presented in later chapters. Salient numbers from the text are compared in table \ref{tab:globalenergetics}.

\begin{table*}
\centering
\includegraphics[angle=0,width=0.97\textwidth]{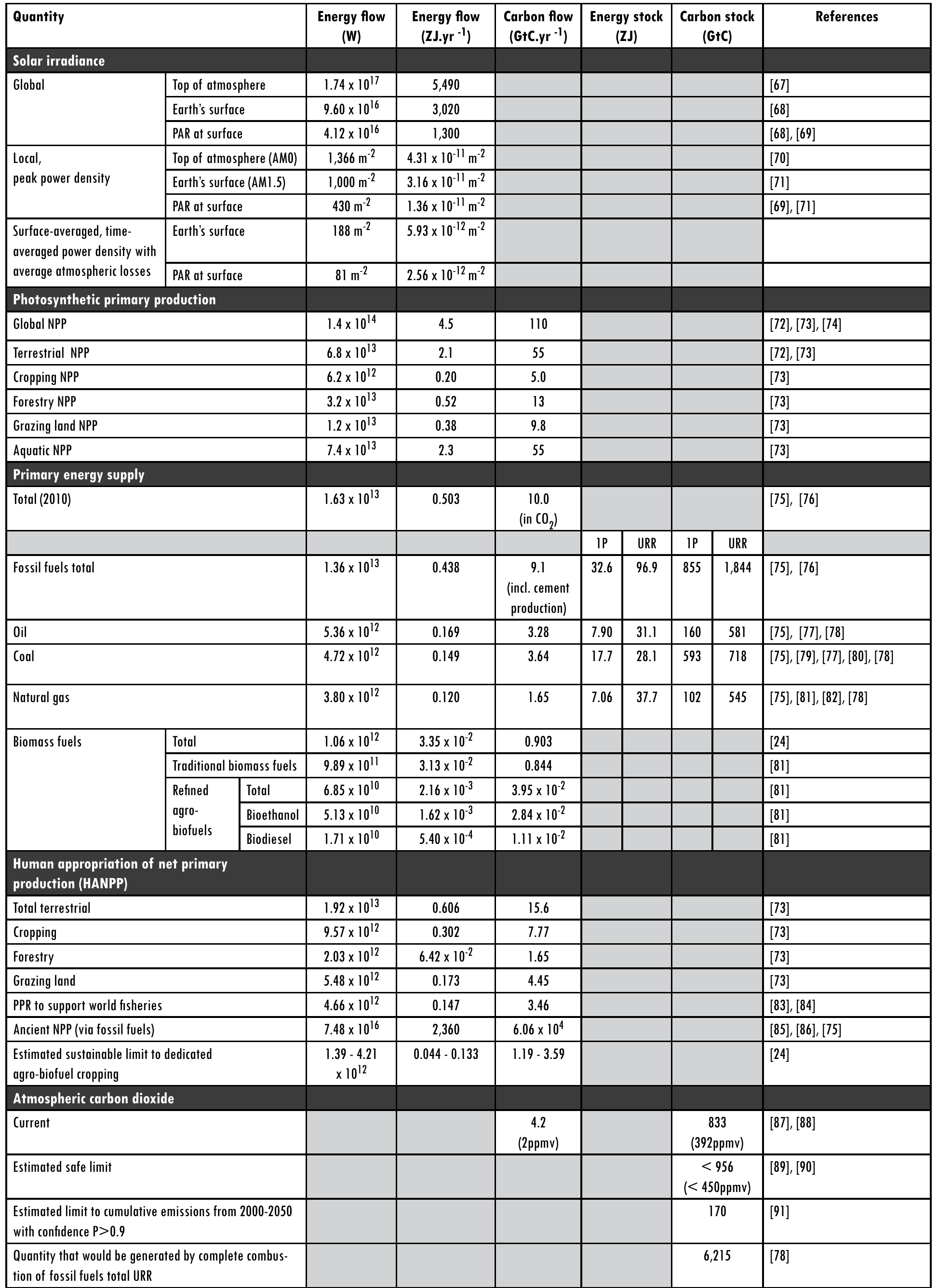}
\caption[Summary of global, photosynthesis-related energetics]{\textbf{Summary of global, photosynthesis-related energetics}. See List of Abbreviations and text for more information. \cite{nasa2012(2), nasa2012(1), alados1996, nasa2012, usnrel10,  running2012, haberl2007, taucher2011, bp2011, peters2011, lequere2009, ward2012, mohr2009, mohr2010, iea2010, haberl2010, iea2009, chassot2010, pauly1995, dukes2003, iea2000, co2now, lal2008, anderson2011, lenton2011, zickfeld2009}}
\label{tab:globalenergetics}
\end{table*} 

\section{Earth's solar energetics}\label{sec:earthsolar}

	\subsection{Solar radiation}
	
	\begin{figure}
\centering
\includegraphics[angle=0,width=0.9\textwidth]{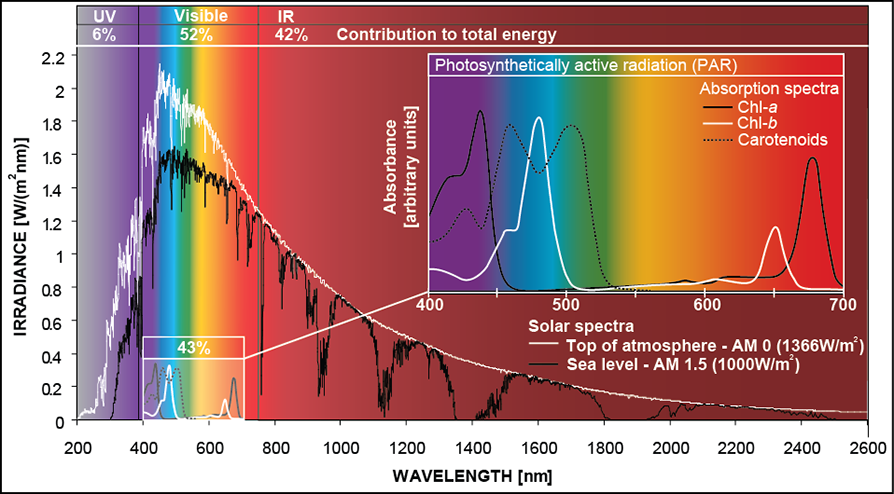}
\caption[Terrestrial solar irradiance and photosynthetic absorption spectra] {\textbf{Terrestrial solar irradiance and photosynthetic absorption spectra.} AM0 and AM1.5 standard solar spectra are shown (see text). Atmospheric absorption bands are visible in the AM1.5 spectrum. Inset shows \textit{in vivo} absorption spectra for pigments from higher plants and green algae. Inset adapted from \cite{govindjee2004}. Overall figure adapted from \cite{kruse2005, govindjee2004}.}
\label{fig:solarspectra}
\end{figure}

	The weight of the Sun's $1.991\times10^{31}$ kg mass generates sufficient pressure and temperature within the innermost quarter of its radius (its `core') that hydrogen (H) is fused into helium (He) \emph{via} the proton-proton chain \cite{phillips1992}, generating a total power output (luminosity) of $3.846\times10^{26}$ W \cite{nasa2012(2)}. The solar emission spectrum (fig. \ref{fig:solarspectra}) is well approximated by a blackbody at a temperature of \mbox{$\sim5,780$ K}, with an emission peak in the green at $\sim500$ nm \cite{smil2008, usnrel10}. Most solar energy (52\%) occupies the visible spectrum (wavelengths 390--750 nm) and a large fraction (42\%) is in the infrared (IR)\nomenclature{IR}{Infrared} (0.7--300 $\upmu$m). Most of the remaining 6\% is accounted for by the ultraviolet (UV)\nomenclature{UV}{Ultraviolet} spectrum (10--390 nm) \cite{usnrel10}. The spectrum conventionally defined as photosynthetically active radiation (PAR)\nomenclature{PAR}{Photosynthetically active radiation} is 400--700 nm \cite{alados1996} (fig. \ref{fig:solarspectra} inset), corresponding closely but not exactly to the visible spectrum. 

Assuming negligible attenuation between the Sun and Earth's atmosphere, the time-averaged solar irradiance at the top of the atmosphere (the `solar constant') is simply the quotient of the Sun's luminosity to the surface area of the sphere with radius equal to Earth's average distance from the Sun (149.6 Gm, 1 astronomical unit \cite{standish}), or 1,367.5 W.m$^{-2}$. Time-averaged satellite measurements \cite{nasa2012} give a value of $\sim1,366$ W.m$^{-2}$, revealing attenuation of only $\sim0.1\%$. As the mean radius of Earth is 6,371 km, the total power incident on its cross-sectional area (before atmospheric interference) can be calculated to be $1.74\times10^{17}$ W (174 PW). Over a year this delivers 5,490 ZJ of energy. In 2010 this represented $\sim11,000$ times the total primary energy\footnote{`Primary energy' refers to energy sources harvested from nature by humans, prior to any human-induced conversions. Examples include raw fossil and nuclear fuels, solar radiation, biomass sources and wind.} supply (TPES\nomenclature{TPES}{Total primary energy supply}) of the global economy, which was $0.503$ ZJ \cite{bp2011}.

	\subsection{Earth's radiation fluxes}\label{subsec:radfluxes}
	Earth must re-radiate into space as much energy as it absorbs, to maintain thermal equilibrium. However, on average the outgoing radiation is red-shifted; the peak in Earth's emission spectrum is at 10 $\upmu$m, far into the IR, corresponding to an average surface temperature of 288 K \cite{smil2008}. Accordingly, the outgoing radiation is higher in entropy. Together with the radiation energy balance, this entropy imbalance accounts for Earth's persistent low-entropy state, essential for life. As Boltzmann described it in 1886, `\textit{There exists between the Sun and the Earth a colossal difference in temperature... The energy of the Sun may, before reaching the temperature of the Earth, assume improbable transition forms. It thus becomes possible to utilize the temperature drop between the Sun and the Earth for performing work, as is the case with the temperature drop between steam and water.}' \cite{boltzmann1886}
	
	\paragraph{Transmission to surface}
	The solar spectrum incident at the upper atmosphere is known as the Air Mass 0 (AM0\nomenclature{AM0}{Air Mass 0 (reference solar spectrum)}) spectrum (fig. \ref{fig:solarspectra} main, white curve), since it is not subject to atmospheric interference \cite{usnrel10}. Upon entering the atmosphere, radiation may follow a large range of possible pathways to eventual re-emission into space and some of these allow its energy to enter the biosphere. However, approximately 26\% of solar irradiance is reflected back into space by the atmosphere and clouds \cite{nasa2012(1)}, leaving 4,060 ZJ.yr$^{-1}$ available to systems on Earth and within its atmosphere. The atmosphere and clouds absorb a further 19\% of the total incident energy, leaving 3,020 ZJ.yr$^{-1}$ (96 PW) available for absorption at Earth's surface \cite{nasa2012(1)}. Approximately 80\% (2,420 ZJ.yr$^{-1}$ or 77 PW) is incident on  the oceans (which cover 70.8\% of the planet's surface \cite{stephens2010}), and 20\% (600~ZJ.yr$^{-1}$ or 19 PW) on land (29.2\% of the surface) \cite{smil2008}. Currently, Earth's surface overall reflects around 4\% of the total energy incident at the upper atmosphere \cite{nasa2012(1)}, though this may be reduced through, for example, reforestation or large-scale deployment of other solar energy harvesting systems.
	
	\paragraph{Surface insolation}
	Solar irradiance at Earth's surface, termed insolation, varies over both location and time. Radiation falling within the planet's cross-sectional area is projected onto its approximately-spherical surface, reducing the average irradiance by a factor of four to 342 W.m$^{-2}$ even without atmospheric interference \cite{smil2008}. Including the effects of the atmosphere, average insolation is reduced to 188 W.m$^{-2}$ (16.2 MJ.m$^{-2}$.day$^{-1}$). Instantaneous, local insolation varies widely from this average, however. A commonly used standard for midday insolation is the Air Mass 1.5 (AM1.5)\nomenclature{AM1.5}{Air Mass 1.5 (reference solar spectrum)} reference solar spectrum (fig. \ref{fig:solarspectra} main, black curve), which corresponds to spectral filtering through 1.5 cloudless atmospheric depths, delivering 1,000 W.m$^{-2}$ of irradiance \cite{usnrel10}. At the opposite extreme, night time irradiance does not exceed $2.4\times10^{-4}$ W.m$^{-2}$ under a full moon on a clear night \cite{scholes2011}. Local, time-averaged daily insolation depends on geographical location and weather, and typically ranges from 12--405 W.m$^{-2}$ (1--35 MJ.m$^{-2}$.day$^{-1}$) \cite{bom2012}. Annual, local averages in Australia fall between 69--278 W.m$^{-2}$ (6--24 MJ.m$^{-2}$.day$^{-1}$ or 2.2--8.8 GJ.m$^{-2}$.yr$^{-1}$) \cite{bom2012}. 
	
	These time-averaged, local insolation values are an order of magnitude lower than typical power production densities of thermal power plants such as coal-fired or nuclear\footnote{Considering the land area of the entire thermal power plant facility (rather than just the core), but not including the land area used for fuel mining. Including the latter generally does not tip the balance in favour of solar power plants \cite{smil2008}.} \cite{smil2008}. Solar energy conversion through photosynthesis, photovoltaic panels or solar-thermal systems widens the disparity by a further 1-2 orders of magnitude. These comparably low power densities, as well as the intermittency of solar power, currently challenge the economic competitiveness of solar energy systems despite the abundant total supply of solar radiation across Earth's surface \cite{smil2008}. 
	
	\paragraph{Greenhouse effect}
	Were Earth to interact with radiation simply as a blackbody, its average surface temperature would depend only on its orbital radius and albedo (reflectivity), and would be 255~K \cite{common2005,smil2008}. The disparity between this and the actual average surface temperature of 288~K is accounted for by the greenhouse effect, in which atmospheric gases known as `greenhouse' gases (GHGs)\nomenclature{GHG}{Green house gas}, most notably water vapour (H$_2$O)\nomenclature{H$_2$O}{Water}, CO$_2$ and methane (CH$_4$)\nomenclature{CH$_4$}{Methane}, that are relatively transparent to the short-wavelength radiation incident from the Sun absorb the IR radiation emitted by Earth's surface. The absorbed energy is re-radiated isotropically, $\sim50\%$ back towards Earth's surface. Resulting from this `radiative forcing', the lower atmosphere is warmer and the upper layers cooler than they otherwise would be, while Earth's overall radiation balance is maintained \cite{raval1989}. 

	Atmospheric GHG concentrations are affected by many processes including photosynthetic carbon fixation, human emissions of CO$_2$ and CH$_4$ from `production' (mining, drilling) and combustion of fossil fuels (sections \ref{sec:fossil} and \ref{sec:limfoss}), as well as through land-use changes such as deforestation (section \ref{sec:fossilciv}). Increases in GHG concentrations strengthen the greenhouse effect, causing global warming \cite{houghton2009, ipcc2007}. To quantify each gas' contribution to warming (its radiative forcing), its concentration and `global warming potential' (GWP)\nomenclature{GWP}{Global warming potential} must both be considered (as well as complex feedback effects not discussed here). A gas' GWP incorporates its optical properties and atmospheric lifetime, and is a relative measure of the heat trapped per unit mass of gas over a set time interval compared with CO$_2$ \cite{houghton2009, ipcc2007}. For example, the 100-year GWP of CH$_4$ is $\sim25$ \cite{houghton2009}, making it a significantly more potent GHG than CO$_2$ (GWP 1) on a mass basis. However, average atmospheric concentration of CH$_4$ is far lower than that of CO$_2$, making it a lesser contributor to global warming overall \cite{ipcc2007}. Global warming and its consequences are explored in section \ref{subsec:climate}.	
	 
	\subsection{Photosynthetic primary production}\label{subsec:PP}
	Primary production is the synthesis of new biomass from inorganic precursors. Though this may be achieved through either photosynthesis or chemosynthesis, the latter occurs on drastically smaller scales and in largely inaccessible areas such as the deep oceans. Accordingly, primary production and photosynthetic primary production are often referred to interchangably. The \textit{rate} of primary production (usually annualised) -- primary productivity -- depends on both environmental and organismal parameters. This section reviews estimates of annual global primary production and compares its magnitude with current human energy consumption.
	
	Primary production is ultimately limited by PAR insolation. Approximately 43\% of the 1,000 W.m$^{-2}$ insolation under AM1.5 conditions -- 430 W.m$^{-2}$ -- is photosynthetically active \cite{usnrel10} because it falls within the spectral range absorbed by the chlorophyll and carotenoid chromophores of the photosynthetic protein complexes. Since global irradiance at Earth's surface annually delivers 3,020 ZJ.yr$^{-1}$ (section \ref{subsec:radfluxes}), a first estimate for the energy globally available to drive photosynthesis is $0.43\times3,020$ ZJ.yr$^{-1}=1,300$ ZJ.yr$^{-1}$. However, actual global production is limited by geographic variations in insolation, climate, species composition, availabilities of water and nutrients, and inefficiencies within the photosynthetic machinery itself (chapter \ref{chp:multianalysis}). 
	
	There are three different conventions for expressing primary production: assimilated carbon (C)\nomenclature{C}{Carbon}, dry phytomass and stored energy\footnote{Approximate conversion factors: 1 kg dry phytomass $\equiv$ 0.45 kgC $\equiv$ 17.5 MJ \cite{jenkins1998, smil2008}. This compares with 1 kg dry \textit{woody} phytomass $\equiv$ 0.5 kgC $\equiv$ 19 MJ \cite{erb2009, smil2008, vanloo2008, jenkins1998}.}. One metric is gross primary production (GPP)\nomenclature{GPP}{Gross primary production}, which is the total amount of the chosen measure (carbon, phytomass or energy) fixed after losses due to photorespiration during the dark reactions of photosynthesis. GPP is diminished by autotrophic respiration to give net primary production (NPP\nomenclature{NPP}{Net primary production}), which is the most commonly used measure of primary production. NPP represents the resource of photosynthates available to other organisms. \cite{haberl2007, vitousek1986} 
	
	\paragraph{Global net primary production}	
	NPP is difficult to measure on geographical scales. Estimates are increasingly satellite-based and synthesise data from more traditional land-based sources. A recent analysis \cite{running2012} of nearly 30 years of satellite and weather data \cite{zhao2010, nemani2003} found that global terrestrial NPP has been remarkably stable over that period, with less than 2\% annual variation. Mean global NPP was estimated at 53.6 GtC.yr$^{-1}$ (gigatonnes of carbon, annually), in reasonable agreement with another recent study \cite{haberl2007} that employed more land-based methods, combining vegetation modelling, agricultural and forestry statistics, and geographical information systems data on land use to obtain an estimate of 59.22 GtC.yr$^{-1}$. Accordingly, annual, terrestrial NPP is here estimated to be 2.1 ZJ (55 GtC). 
	
	Aquatic NPP is even more difficult to estimate. Recent studies place it within a similar range, $55\pm5$ GtC.yr$^{-1}$ \cite{taucher2011}. This equates to annual energy storage of $2.3\pm0.2$ ZJ.yr$^{-1}$, assuming aquatic biomass energy density\footnote{For energy density the conventional `heat of combustion' or `higher heating value' (HHV\nomenclature{HHV}{Higher heating value}) is used here. This is the energy released as heat when a compound undergoes complete combustion with oxygen under standard ambient temperature and pressure (298.15 K, 101.325 kPa), which provides a theoretical upper limit to the energy available from a fuel in practical applications. The unit used here for HHV is MJ.kg$^{-1}$, rather than MJ.L$^{-1}$ or MJ.mol$^{-1}$, to aid in comparing gaseous, liquid and solid fuels of various compositions.} and carbon content of 20 MJ.kg$^{-1}$ \cite{illman2000} and  0.47 kgC.kg$^{-1}$ \cite{geider2002}\nomenclature{kgC}{Kilograms of carbon} respectively, as for typical green microalgae (table \ref{tab:fuelproperties} for a comparison of biomass types). Annual, global NPP is therefore estimated at 4.4 ZJ (110 GtC), which is larger than the human economy's TPES in 2010 by a factor of nine \cite{bp2011}. Nonetheless, based on these estimates, photosynthetic biota currently store only 0.3\% of the $\sim1,300$ ZJ of PAR annually incident at Earth's surface, or 0.1\% of the 3,020 ZJ of annual, global insolation.  
	
	\paragraph{Productivity densities and fuel production}	
Mean areal productivity densities of terrestrial and aquatic NPP across the Earth's surface respectively are estimated at 0.45 W.m$^{-2}$ of ice-free land and 0.13 W.m$^{-2}$ of ocean \cite{smil2008}. However, there is variation with geography and ecosystem type. Mean productivity density of tropical forests is estimated at 1.3 W.m$^{-2}$, while temperate and boreal forests respectively average 0.5 W.m$^{-2}$ and 0.2 W.m$^{-2}$ \cite{smil2008}. 

Agriculture\footnote{Cropping and grazing but not forestry.}, which now appropriates 38\% of Earth's ice-free land area \cite{running2012, foley2011} (section \ref{subsec:microal}), often delivers lower NPP than the natural ecosystems replaced but concentrates growth in biomass components valued by humans \cite{running2012}. Haberl \textit{et al} \cite{haberl2010} estimate that global cropland accounted for 0.20 ZJ of NPP in the year 2000, while forestry and grazing land respectively accounted for 0.52 ZJ and 0.38 ZJ of NPP. Crop productivities differ substantially between species, conditions and cultivation methods but average global productivity density is estimated at 0.4 W.m$^{-2}$ \cite{smil2008}. The highest annual-average productivity density of any vegetation is 5.0 W.m$^{-2}$, for natural stands of the grass, \textit{Echinochloa polystachya} on the Amazon floodplain \cite{somerville2010}. Sugarcane has also approached this rate \cite{smil2008} and is currently a major agro-biofuel crop, used as a feedstock for bioethanol production \cite{iea2009(1)}; its high productivity densities provide a benchmark for future photosynthetic energy systems.       
	
	  NPP constrains metabolism of photosynthates within natural ecosystems and the human economy. Humans appropriate photosynthates for many uses beyond their indispensible role as food (section \ref{subsec:currentnpp}). Traditional biomass fuels\footnote{The International Energy Agency defines \cite{iea2010} `traditional biomass' as `biomass consumption in the residential sector in developing countries and refers to the often unsustainable use of wood, charcoal, agricultural residues and animal dung for cooking and heating.' Victor and Victor \cite{victor2002} tabulate the fractions of primary energy consumption accounted for by these types of biomass fuel in 16 developing countries. The values used for `traditional biomass fuel' in this thesis were calculated using the mean fuel fractions across these countries (76.8\% wood, 19.7\% crop residues, 3.5\% dung), and also assuming that crop residues and dung both have the energy and carbon content of `dry phytomass' as detailed above.} have been an essential driver of economic activity in preindustrial societies throughout history and remain so today in the undeveloped world. However, in 2008 such fuels supplied only 6.1\% of TPES ($3.1\times10^{-2}$ ZJ) \cite{iea2010}. The overwhelming majority of primary energy in the modern, industrialised world is supplied by a large repository of geochemically processed, ancient photosynthates: fossil fuels \cite{iea2010}. 
	  
\section{Fossil fuels}\label{sec:fossil}
\chapquote{You will be astonished when I tell you what this curious play of carbon amounts to.}{M. Faraday, 1861 \cite{faraday2011}}

When Earth formed 4.5 billion years ago, the Sun was $\sim50\%$ less luminous than at present \cite{guinan2002}. Assuming for simplicity linear growth in solar luminosity over Earth's history, the total solar energy incident upon its surface has been approximately $1\times10^{13}$ ZJ. Most of this energy was re-radiated to space following a brief chain of interactions with the planet's matter, including in many cases photosynthetic primary production followed by re-oxidation of photosynthates. However, under favourable conditions during some geological periods, photosynthates entered into geological processes that initiated long-term storage of solar energy in large quantities. Over periods of time ($10^7$--$10^8$ years) that are large when compared with human history, slow heat and pressure transformations of accumulated biomass, both terrestrial and aquatic, formed the fossil fuels that have powered human civilisation's frenetic expansion since the industrial revolution, and which in 2010 accounted for $\sim87\%$ of TPES \cite{bp2011}. 

\begin{table*}
\centering
\includegraphics[angle=0,width=1\textwidth]{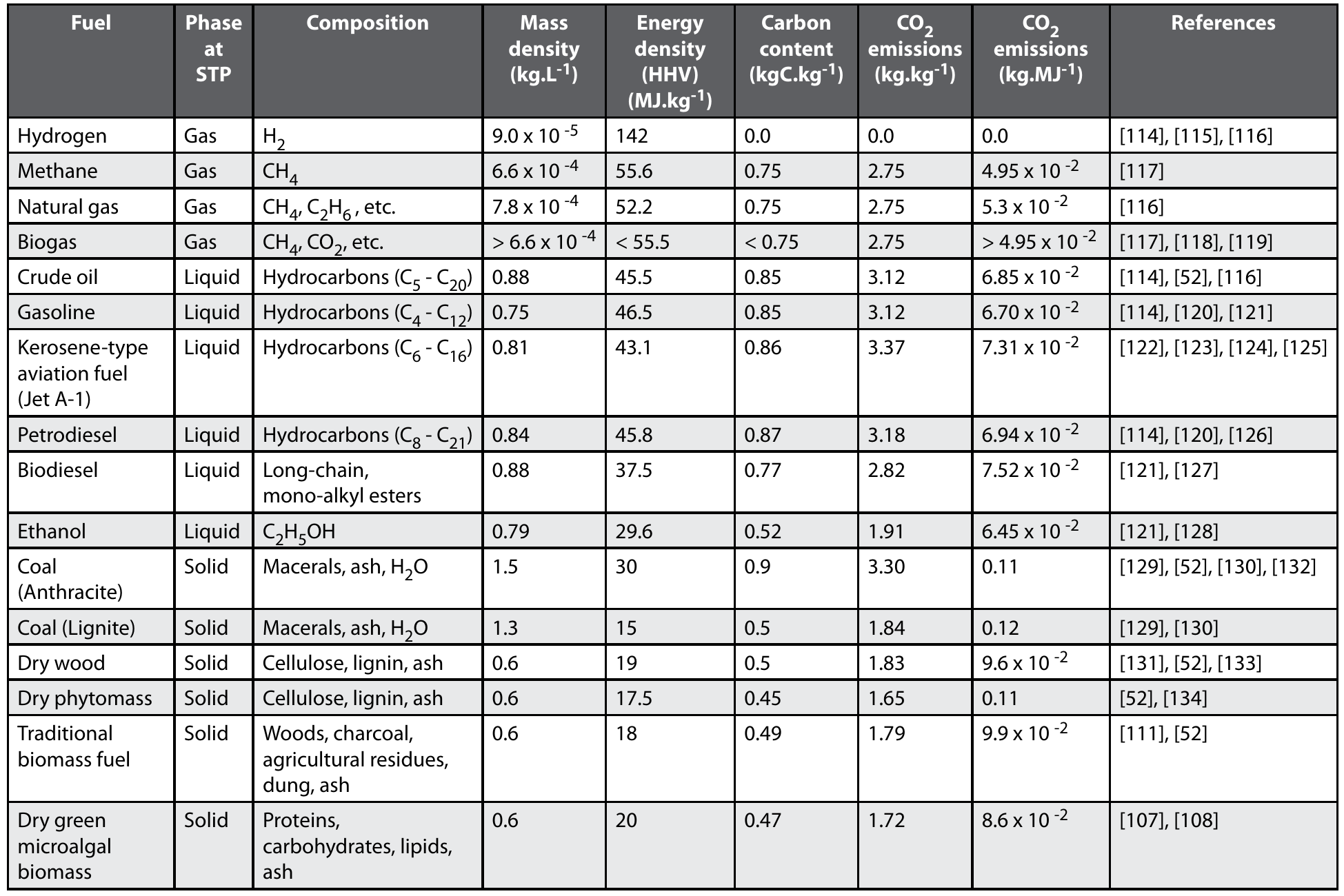}
\caption[Fuel properties at standard ambient temperature and pressure]{\textbf{Fuel properties at standard ambient temperature and pressure (STP)}. Hydrogen, methane and ethanol are pure compounds with invariant properties at STP. All other fuels listed are variable mixtures of compounds and their stated properties should therefore be taken as representing a range. HHV refers to `higher heating value', defined in section \ref{subsec:PP}. \cite{supple2007, usdoe2001, oakridge2011, cuellar2008, klass1998, walsh1988, smil2008, afdc2011, coronado2009, airbp2008, nojoumi2009, starik2008, edwards2003, epa2011, biodieselboard2011, shapouri2003, lett2004, keeling1973, henry2010, perry1984, lamlom2003, witkowski1991, victor2002, illman2000, geider2002}}
\label{tab:fuelproperties}
\end{table*} 

\begin{figure}
\centering
\includegraphics[angle=0,width=0.65\textwidth]{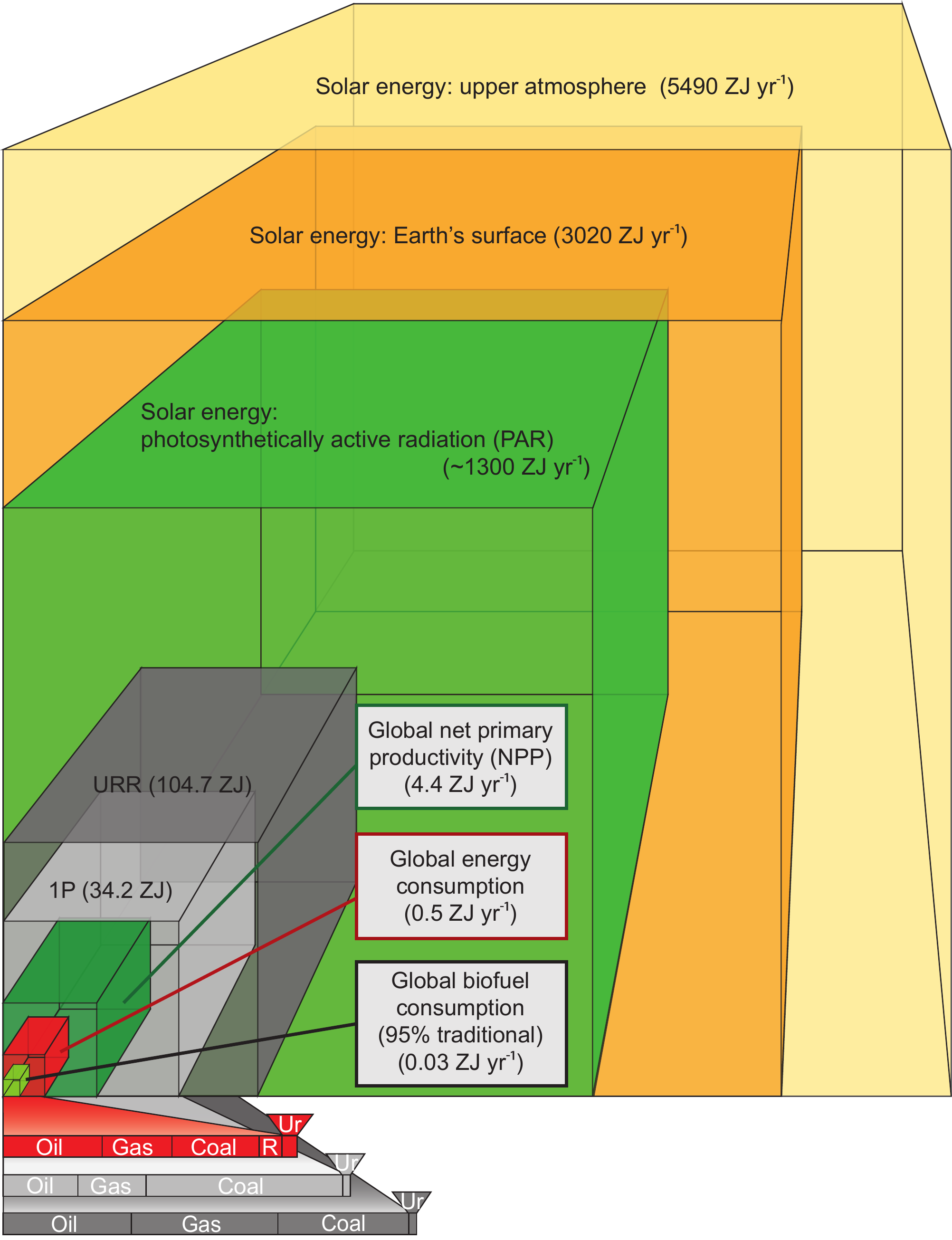}
\caption[Global energy resources and consumption]{\textbf{Global energy resources and consumption.} Cubic volumes correspond to magnitudes of labelled quantities. Horizonal bars show decompositions of global energy consumption (red), ultimately recoverable resources (URR -- dark grey) and 1P reserves (1P -- light grey) into fractions by primary energy source. `Ur' denotes uranium, and `R' denotes all sources of renewable energy combined. URR and 1P are defined in the text. Uranium reserves (1.59 ZJ) and resources (5.04 ZJ conventional; 2.79 ZJ unconventional) are given by \cite{rogner2012}.}
\label{fig:energycubes}
\end{figure}

The rate at which humanity is now oxidising fossil fuels constitutes a biogeophysical event of global significance \cite{ipcc2007}, effectively reversing within centuries the carbon fixation carried out by the planet's collective photosynthetic biota over geological time scales. It is useful here to review fossil fuels and their role in the global economy because they set the standard for potential replacements such as solar fuels\footnote{`Solar fuels' here refers to all fuels generated from current photosynthates, whether produced by living species (biofuels) or artificial photosynthetic systems.}. Table \ref{tab:fuelproperties} compares physicochemical properties of various fuels, and figure \ref{fig:energycubes} compares current global energy consumption with available resources. 

	\subsection{Coals}
		Coals are sedimentary rocks made up of partially decomposed organic matter, inorganic minerals and water. They are composed primarily of heterogeneous organic compounds -- macerals -- derived from woody phytomass \cite{lett2004, smil2008}. Coal compositions span a wide continuum that is conventionally divided into four classes\footnote{This is the United States classification system. The European system has 15 classes.} by gravimetric energy and carbon contents (in decreasing order): anthracite, bituminous, sub-bituminous and lignite. The purest anthracites contain less than 5\% water and 95\% C, while the poorest lignites have up to 65\% water and only 15\% C. Ash is the collective term for the incombustible minerals, which by mass range from negligible to 40\% \cite{lett2004, smil2008}.  
				
		In 2010 humans harvested 0.149 ZJ of energy, or 29.6\% of TPES, from coal \cite{bp2011}. According to the International Energy Agency (IEA\nomenclature{IEA}{International Energy Agency}) \cite{iea2010}, in 2008 two thirds of coal consumption supplied electricity generation, of which coal was by far the largest source at 41\% of generation, nearly twice the second largest source, natural gas, at 21\%. One fifth of coal consumption was in the industrial sector and the remainder mostly in the building and agriculture sectors, as well as coal-to-liquids and coal gasification fuel transformations. Coal contributed 40\% of fossil fuel CO$_2$ emissions in 2008 \cite{lequere2009}; assuming the same fraction in 2010, coal contributed 13.4 GtCO$_2$ (3.64 GtC) in that year (section\ref{subsec:climate}).  

 		Coal is thought to be the most abundant fossil fuel, with 17.7~ZJ of proven (`1P') reserves\footnote{Proven reserves are generally taken to be those quantities that geological and engineering information indicates with reasonable certainty can be recovered from known deposits under existing economic and operating conditions \cite{bp2011}. `1P' reserves are considered to be marketable with a probability of 90\% using only existing infrastructure \cite{owen2010, stephens2010}.} in 2010 \cite{bp2011}. Worldwide ultimately recoverable resources\footnote{`Ultimately recoverable resources' are the sum of 1P reserves and the increasingly less certain and more costly 2P, 3P and 1C reserves \cite{stephens2010}.} (URR\nomenclature{URR}{Ultimately recoverable resources}) of coal are estimated\footnote{Despite the diversity of coal types, 29.3 GJ.t$^{-1}$ is an accepted conversion factor for a `tonne of coal equivalent' \cite{mohr2010, schilling1987} and the carbon content of this unit is 0.746 $\pm$ 2\% by mass \cite{cdiac2013} (taken here to be 0.75\% by mass for simplicity).} at 28.1 ZJ \cite{ward2012, mohr2010}. 
		 
	\subsection{Crude oils (Petroleum)}
		Crude oils are liquid mixtures of hydrocarbons with various structures (cycloalkanes, alkanes and arenes) and chain lengths (generally C$_5$--C$_{20}$). The chemical composition of a given crude oil is particular to a given deposit, with its own unique geological history. Crude oils are formed from aquatic biomass such as algae, rather than woody biomass \cite{deffeyes2005, smil2008}.
		
		Global oil consumption in 2010 was 0.169~ZJ, accounting for 33.6\% of TPES and constituting the single largest source of primary energy \cite{bp2011}. In 2008, 53\% of oil consumed was used in the transport sector, which was powered 96\% by oil-based fuels. Industry, building and agriculture accounted for the remaining consumption \cite{iea2010}. Oil contributed 36\% of fossil fuel CO$_2$ emissions in 2008 \cite{lequere2009}; assuming the same fraction in 2010, oil combustion generated $\sim12.0$~GtCO$_2$ in that year (section\ref{subsec:climate}). 

		In 2010, proven global reserves of oil, conventional and unconventional, totalled 7.90~ZJ \cite{bp2011}. Ultimately recoverable resources of conventional and unconventional oil are respectively estimated at 10.0 ZJ and 21.1 ZJ \cite{ward2012, mohr2010}.

	\subsection{Natural Gases}
		Predominantly mixtures of the three smallest alkanes, CH$_4$ (73--95\%), ethane (3--13\%) and propane (0.1--1.3\%), natural gases can also contain butane, pentane and trace amounts of larger alkanes. Contaminants are present as hydrogen sulphide, nitrogen, helium and water vapour \cite{smil2008}, lowering the energy density, which is typically $\sim52.2$~MJ.kg$^{-1}$ \cite{oakridge2011}, compared with 55.6~MJ.kg$^{-1}$ for pure CH$_4$ \cite{cuellar2008}. Natural gases are formed by the same processes that form crude oil, though gas genesis requires temperatures and pressures sufficiently high to crack longer hydrocarbons into short-chain hydrocarbons ($<C_5$). 
		
		In 2010, the world consumed 0.120~ZJ of energy from natural gas, or 23\% of TPES \cite{bp2011}. Electricity generation provided the highest sectoral demand for gas in 2008, at 39\% \cite{iea2010}. The second largest gas-consuming sector was residential-and-commercial heating, followed by industry \cite{iea2010gas}. In the same year, natural gas combustion emitted $\sim6.1$~GtCO$_2$, 16.5\% of global emissions.  
		
		A use of natural gas noteworthy in the context of bioenergy is ammonia production \emph{via} the Haber-Bosch process. This ammonia is mainly used to produce nitrogenous fertilisers upon which modern agricultural yields (and therefore, agro-biofuel production) depend. Natural and biological processes for nitrogen fixation from abundant-but-inert atmospheric dinitrogen into reactive, bioavailable compounds are few and limited in yield. Availability of reactive nitrogen compounds is almost always the productivity-limiting factor in intensive agricultures \cite{smil2001}. Synthetic fixation of nitrogen into ammonia has enabled a `green revolution', dramatically increasing yields over prior agricultures. However, presently this process depends on CH$_4$ from natural gas which, through steam reformation, yields hydrogen for subsequent reaction with atmospheric nitrogen. Currently this consumes $\sim3-5\%$ of global natural gas production and $\sim1\%$ of TPES \cite{smith2002}, and the $>100$ million tonnes of fertilisers it generates sustains approximately two fifths of the world's population \cite{smil2001}. Natural gas could be replaced in this role by other hydrogen sources such as electrolytic or photosynthetic water-splitting, but these are not yet cost-competitive. This fossil-fuel dependence constrains the ability of large-scale agro-biofuel production to replace fossil fuels.  
				
		Proven reserves of natural gas in 2010 were 7.06~ZJ \cite{bp2011}. Ultimately recoverable resources of conventional and unconventional natural gas are respectively estimated at 15.6 ZJ and 22.1 ZJ \cite{ward2012, mohr2010}.
		
\section{Photosynthates in fossil-fuelled civilisation}\label{sec:fossilciv}	
	
At present the human economy is powered almost exclusively by products of photosynthesis. In assessing the sustainability of this energy system, one may compare economic metabolism of photosynthates with Earth's primary productivity. This comparison has two components. First, estimating human appropriation of \textit{current} net primary production (HANPP)\nomenclature{HANPP}{Human appropriation of net primary production} indicates the extent to which that appropriation could be increased to compensate for fossil fuel depletion. Second, estimating human appropriation of \textit{ancient} NPP \emph{via} fossil fuels aids in understanding the value provided by fossil fuels and also how much current NPP would be needed in order to replace them.   
	  
	  \subsection[Human appropriation of current net primary production (HANPP)]{Human appropriation of current net primary production}\label{subsec:currentnpp}	  
\begin{figure}
\centering
\includegraphics[angle=0,width=0.55\textwidth]{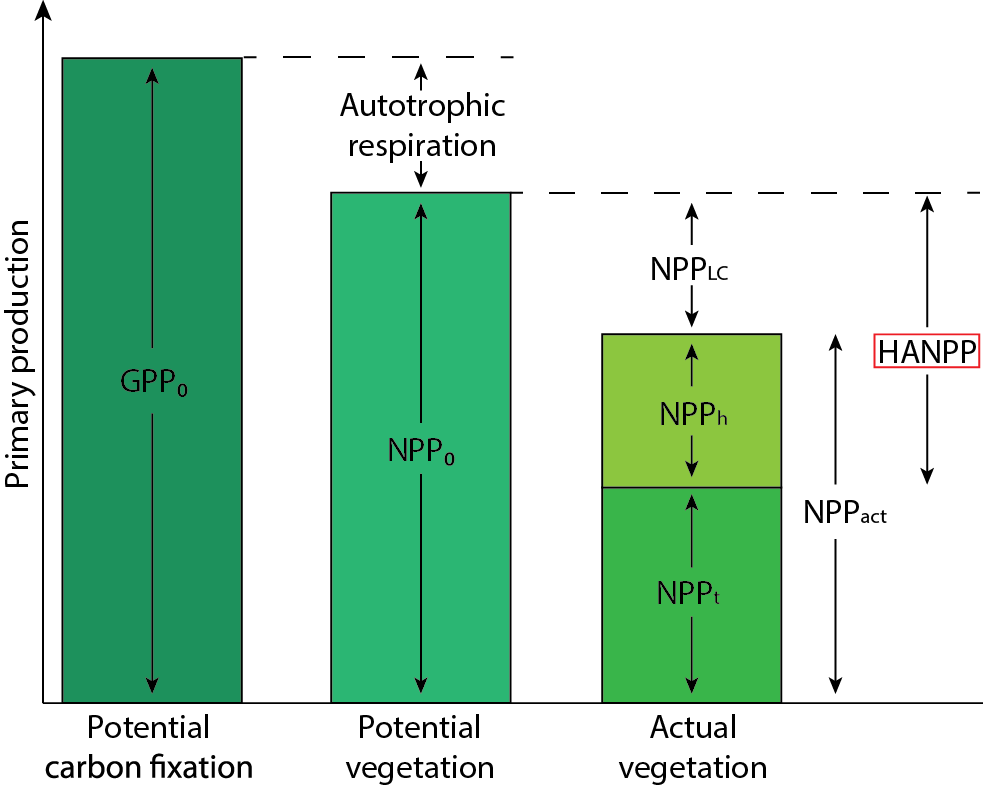} 
\caption[Human appropriation of net primary production (HANPP)] {Graphical representation of the standard definition of human appropriation of net primary production (HANPP). Terms are defined in the main text. Figure adapted from \cite{erb2009}.}
\label{fig:hanpp}
\end{figure}

Whittaker and Likens gave an early estimate of the total biomass consumed by humans \cite{whittaker1973, erb2009}. They accounted for only the harvest of food and wood used directly by humans, concluding that this appropriated 3\% of Earth's annual NPP. Various authors \cite{vitousek1986, wright1990, rojstaczer2001, imhoff2004, haberl2007} have since further developed the study of HANPP, broadening its scope and refining its definition. It is now conventionally defined (fig. \ref{fig:hanpp}) \cite{erb2009} to include two interdependent processes: 1) land-use changes that modify the NPP of vegetation (denoted $\Delta NPP_{LC}$), compared with the \textit{potential} undisturbed vegetation that would persist in the absence of human interference ($NPP_0$), by the formula $\Delta NPP_{LC}=NPP_{0}-NPP_{act}$ (here $NPP_{act}$ is the actual NPP of the human-modified land\footnote{Estimates of global NPP discussed in section \ref{subsec:PP} constitute global $NPP_{act}$.}); and 2) extraction or destruction of a share of NPP for human purposes ($\Delta NPP_{h}$), such as through biomass harvest or livestock grazing. HANPP is the sum of the two components ($HANPP=\Delta NPP_{h}+\Delta NPP_{LC}$) and indicates land use intensity, explicitly linking natural with socioeconomic processes. Depending on the precise definitions used for components 1 and 2, a third component, human-induced destruction of NPP without purpose (e.g. by human-induced fires), may also be included. Importantly, the conventional definition of HANPP accounts for the fraction of NPP which in strongly human-controlled ecosystems such as plantation forests or grazing pastures is \textit{not} appropriated by humans ($NPP_t=NPP_0-HANPP$)\cite{erb2009}. 

\paragraph{Human appropriation of terrestrial net primary production} Haberl \textit{et al} \cite{haberl2007} provide the most recent and rigorous estimate of global terrestrial HANPP, employing data for the year 2000 from vegetation modeling, agricultural and forestry statistics, and geographical information systems data on land use, land cover and soil degradation. These authors estimate terrestrial HANPP at 0.606 ZJ or 28.3\% of $NPP_{act}$ (global terrestrial NPP). Of this HANPP, harvest ($NPP_h$) contributed 53\%, land conversion ($NPP_{LC}$) 40\%, and human-induced fires 7\%. 

Contributions of cropping and forestry to HANPP are of interest in considering large-scale agro-biofuel production. Respectively, they contributed 49.8\% and 10.6\% of terrestrial HANPP in the year 2000. Grazing land contributed an additional 28.5\% of HANPP. 

\paragraph{Human appropriation of aquatic net primary production} Published estimates of HANPP in aquatic ecosystems are fewer and less sophisticated than those for terrestrial ecosystems, focussing on $\Delta NPP_{h}$ through fish harvest and largely neglecting other impacts (the `aquatic equivalent' of $\Delta NPP_{LC}$). Vitousek \textit{et al} \cite{vitousek1986} estimated HANPP in aquatic ecosystems at 2.2\%. 

Pauly and Christensen \cite{pauly1995} introduced the metric of `primary production required' (PPR)\nomenclature{PPR}{Primary production required} to sustain the world's fish harvest and gave a detailed estimate based on multiple trophic models of various aquatic ecosystem types. Results suggested that 8\% of global aquatic primary production was required to support the harvest and discarded bycatch of world fisheries, averaged over the years 1988--1991. Only 2\% of NPP was needed to support fisheries in open ocean systems but 24--35\% was required in fresh water systems \cite{pauly1995}. In a more recent study, Chassot \textit{et al} \cite{chassot2010} apply the same metric with more sophisticated analysis to find an annual, global PPR value equivalent to 6.3\% of global, aquatic NPP, averaged over the years 2000--2004. 

Different metrics indicate strong human impacts on aquatic ecosystems beyond fish harvest throughout the world's oceans \cite{halpern2008}. This suggests that a comprehensive assessment of aquatic HANPP, including a measure of ecosystemic degradation `equivalent to' $\Delta NPP_{LC}$, may yield a significantly larger estimate. 

\subsection{Human appropriation of ancient net primary production}\label{subsec:ancientnpp}
Dukes \cite{dukes2003} uses published data on energy conversion efficiencies of the numerous biological, geochemical and industrial steps required for formation and extraction of coal, oil and gas to estimate the ancient NPP that supplies the fossil fuels consumed by the modern human economy. It is found that formation and extraction of coal are together $\sim9\%$ efficient, while formation and extraction of oil and gas are respectively $\sim0.009\%$ and $0.008\%$ efficient. Using these efficiencies together with data on fossil fuel properties and consumption, Dukes concludes that the fossil fuels globally consumed in 1997 (0.318~ZJ \cite{iea2000}) were derived from 1710~ZJ of ancient biomass energy. Scaling this result for fossil fuel consumption in 2010 (0.438~ZJ \cite{bp2011}) gives 2360~ZJ of ancient biomass energy. This equates to 550 years of Earth's current annual NPP, demonstrating the extraordinary value provided by fossil fuels, notwithstanding the low efficiencies of their genesis.
 
\subsection[Limits to sustainable HANPP, and biofuel production]{Limits to sustainable HANPP, and biofuel production}\label{subsec:sushanpp}
Stating HANPP as a percentage of $NPP_0$ quantifies the fraction of total potential biospheric energy flows diverted by humans. If robust, biodiverse ecosystems are to be sustained there must be a limit to sustainable HANPP, since other species also need NPP. Biofuels -- NPP-based fuels -- constitute a fraction of HANPP so their quantity is constrained by the limit to sustainable HANPP together with the magnitudes of other HANPP fractions: food, materials and other ecosystem services. 

Bishop \textit{et al} \cite{bishop2010} estimate the limit to sustainable HANPP, assessing the maximum level that would allow for land conversion and climate change small enough to sustain ecosystem services essential to humanity. The results assert that in 2005, humanity ought to have appropriated at least 59\% less than the actual HANPP estimated from the data of Haberl \textit{et al} \cite{haberl2007}. Bishop \textit{et al} use an HANPP metric different from the conventional definition\footnote{used by Haberl \textit{et al} \cite{haberl2007}. Bishop \textit{et al} instead base their study on the `high estimate' definition of Vitousek \textit{et al} \cite{vitousek1986}) and use the data of Haberl \textit{et al} \cite{haberl2007}.} (section \ref{subsec:currentnpp}), so the resulting estimated limit to sustainable HANPP cannot be compared directly with the HANPP level found by Haberl \textit{et al} \cite{haberl2007} in section \ref{subsec:currentnpp}. However, given that the two estimates arise from the same data, the finding by Bishop \textit{et al}, that humanity has significantly exceeded the sustainable limit of HANPP, strongly suggests that energy technologies requiring increased HANPP cannot sustainably meet future human energy demand. This is consistent with other global sustainability assessments based on different but related metrics such as the `ecological footprint', which have also found the global economy to have significantly overshot the regenerative capacity of some of Earth's life-support systems \cite{wackernagel2002, rockstrom2009}. 

\paragraph{Sustainable HANPP for biofuel production in terrestrial ecosystems}
Based on a review of recent literature, Haberl \textit{et al} \cite{haberl2010} estimate that the global technical potential of agro-bioenergy\footnote{Bioenergy production based on traditional agricultural and forestry systems.} in the year 2050, considering sustainability criteria, is in the range\footnote{Based on the authors' assumption that dry biomass is 50\% carbon by mass and has an energy density of 18.5~MJ.kg$^{-1}$.} of 0.160--0.270~ZJ.yr$^{-1}$. The authors estimate that 0.044--0.133~ZJ.yr$^{-1}$ could come from dedicated bioenergy crops, with the remainder to be contributed by currently unused residues and through efficiency gains in biomass utilisation (more efficient HANPP). The estimates for total agro-bioenergy potential equate to 31--53\% of \textit{current} global energy demand, and the estimates for dedicated bioenergy crops, 9--26\%. These latter estimates also correspond to 8--23\% of terrestrial HANPP in 2000, leaving a substantial fraction of HANPP available for other uses within the sustainable limit advocated by Bishop \textit{et al} \cite{bishop2010}. Nonetheless, with global energy demand projected to grow at least threefold this century \cite{kruse2005}, these findings further suggest that agro-bioenergy can sustainably supply at most a small fraction of future energy demand. The interested reader is directed to the detailed analysis of agro-bioenergy's global potential by Giampietro and Mayumi \cite{giampietro2009}.

\paragraph{Sustainable HANPP in aquatic ecosystems}
Although the methods used for studying terrestrial HANPP have yet to be adapted for aquatic ecosystems, studies using other metrics raise concerns over aquatic NPP reductions due to human-induced biodiversity loss. For example, Worm \textit{et al} \cite{worm2006} predict the possible collapse of 100\% of global fisheries by the year 2048 if existing trends in aquatic biodiversity loss continue. Chassot \textit{et al} \cite{chassot2010} provide strong evidence for the basic claim that aquatic NPP ultimately constrains fisheries catches. However, the effect of fisheries collapse on overall aquatic NPP is unclear because it is difficult to predict the complex effects across trophic levels resulting from changes in predator populations at higher levels \cite{bruno2005}. 

The current lack of comprehensive measures for HANPP in aquatic ecosystems is of limited consequence in discussing the potential to replace fossil fuels with fuels derived from current NPP because almost all proposals for producing the latter involve land-based production systems. However, Wang \cite{wang2010} gives an ambitious proposal for large-scale domestication of open ocean for bio-energy production using aquatic micro-organisms. In light of available evidence for strong pre-existing human interference in aquatic ecosystems, such a proposal suggests that comprehensive study of aquatic HANPP may be increasingly important.       
	
\section{Limits to fossil-fuelled civilisation}\label{sec:limfoss}
\chapquote{One can't have something for nothing.}{A.L. Huxley \cite{huxley2008}}
\vspace{-0.88cm}
\chapquote{Nearly always people believe willingly that which they wish.}{J. Caesar\cite{caesar2011}}

This section describes the dynamics of fossil fuel depletion and anthropogenic climate change, reviewing forecasts of their trajectories and consequences for global ecosystems and the human economy. Of particular interest are the time scales on which these consequences of fossil fuel combustion are predicted to constrain economic activity. Since these time scales also constrain options for transitioning to new energy systems, they indicate the speed with which these systems must be developed and scaled up.  

\subsection{Fossil fuel supply}\label{subsec:fossilsupply}
Fuel reserve estimates (table \ref{tab:globalenergetics}) are commonly stated together with reserves-to-production (R/P)\nomenclature{R/P}{Reserves-to-production} ratios, which compare proven reserves with current production rate as a measure of reserves longevity \cite{bp2011}. However, estimates of reserves and resources are frequently revised due to fuel price changes, improvements in exploration and production technologies, and definitional changes\footnote{For example, there may be disagreement on classifications of conventional and unconventional resources \cite{verbruggen2010, sorrell2010(1)}.} \cite{hughes2011, verbruggen2010}. Production rates can depend on many factors including geophysical constraints, demand, consumption efficiency improvements and sociopolitical dynamics \cite{hall2012, hughes2011, verbruggen2010, heinberg2005}. Even aside from these variables, the R/P ratio is a crude measure of fuel security because depletion-related economic consequences are likely to arise long before reserves are actually depleted \cite{murray2012, lutz2012, hall2012}. 

Debate over the relative importances of the above factors to energy policy is divided between two camps: `technological cornucopians' (`optimists') argue for medium-to-long-term supply security, due primarily to anticipated technological improvements. Conversely, `peak oilers' (`pessimists') assert that resource depletion trajectories depend strongly on geophysical factors and only weakly on technological improvements and other `economic' factors. Some concepts in this section may apply broadly to other nonrenewable resources \cite{heinberg2010, hubbert1956} but, in fitting with the literature concerning energy constraints on the economy, the focus is on crude oil.

\subsubsection{Peak oil}
`Peak oil' describes the time at which the production rate of an oil field or producing region peaks before going into decline. Production rate usually follows an approximately bell-shaped curve over time, the `peak' (sometimes plateau) of which is reached after approximately half of the recoverable reserves have been produced \cite{meng2008}. This was first empirically modelled by Hubbert in 1956, using a `logistic' (sigmoid) function for cumulative production of an oil field, parameterised by the field's time of discovery, production history and an estimate of its URR \cite{hubbert1956, hughes2011, hall2012}. Hubbert also assumed that production would peak when half of the resource had been depleted \cite{hubbert1956, hughes2011}. Notably, although Hubbert's model does not depend directly on economic variables but rather, only geophysical variables, it predicted a production peak date for the United States (US) in close agreement with the peak eventually observed in 1970 \cite{hughes2011, hall2012}. Similar production peaks have since been observed in more than 60 (of 95) other oil-producing nations \cite{hall2012, meng2008} and Hubbert's methods have been refined and applied to other oil-producing regions and other resources, proving accurate for Indonesian gas production and United Kingdom (UK)\nomenclature{UK}{United Kingdom} coal production \cite{nel2009}, for example. 

There has been wide interest in Hubbertian forecasts for the peaking of worldwide oil production because oil is the largest source of primary energy (section \ref{sec:fossil}). Predictions of a global oil peak ultimately depend on the assumptions that all oil fields eventually peak and global production equates to a sum over individual fields \cite{hughes2011}. Conversely, peak-oil critics point to the theory's heavy reliance on uncertain URR estimates and other limitations of the Hubbert model \cite{hughes2011, meng2008}. Hughes and Rudolph \cite{hughes2011} collate 40 published forecasts for global peak oil. Thirty-one predict a peak between 2000 and 2020 while nine predict a peak as late as the 2030s or argue that production will grow indefinitely. 

\paragraph{Conventional oil} 
Petroleum extracted from wells drilled in the traditional manner is known as conventional oil. Recent, prominent commentaries observe that conventional oil production has not increased globally for non-OPEC \nomenclature{OPEC}{Organisation of the Petroleum Exporting Countries} nations (outside the Organisation of Petroleum Exporting Countries) since 2004 \cite{kerr2011}, nor for the whole world since 2005 \cite{murray2012, hall2012, fantazzini2011}. This has been despite generally high (though volatile) oil prices and contrasts with a history of strong correlation between prices and production \cite{murray2012}. Moreover, production at existing fields is now declining at rates of 4.5--6.7\%.yr$^{-1}$; only by adding new fields is global production holding steady \cite{murray2012, kerr2011}, yet oil discovery peaked in 1963 and has fallen short of production since 1980 \cite{sorrell2010}. Murray and King argue \cite{murray2012} that since 2005 the oil market has undergone a `phase transition' to a new state in which global production is inelastic to demand, causing increased price volatility. This suggests that production may be geophysically limited; global peak oil may have occurred already, though at this stage it is too soon for certainty \cite{murray2012, kerr2011}. 

\paragraph{Unconventional oil} 
Critics of peak-oil theory predict that declining production at conventional fields will be more than offset by adding new fields yet to be discovered (e.g. deepwater) plus natural gas liquids (NGLs)\nomenclature{NGL}{Natural gas liquid} and `unconventional' sources of oil such as `heavy' oil, `tight' (shale) oil and tar sands, as well as coal-to-liquid (CTL\nomenclature{CTL}{Coal-to-liquid}) and gas-to-liquid (GTL\nomenclature{GTL}{Gas-to-liquid}) conversions. In 2010, all unconventional oil sources together supplied only $3\%$ of the global market \cite{smil2010}. However, a recent boom in shale gas and shale oil production rates in the US \cite{kerr2010} has led to a prediction that these will respectively grow three- and sixfold from 2011 levels by 2030 \cite{bp2013}. The IEA recently predicted \cite{iea2012} that a `\textit{surge in unconventional supplies, mainly from light tight oil in the United States, and oil sands in Canada, natural gas liquids, and a jump in deepwater production in Brazil, (will push) non-OPEC production up (by 8\% compared with 2011) after 2015... until the mid 2020s}.' It also predicted a net increase in global oil production until at least 2035, driven entirely by unconventional oil \cite{iea2012}. 

Some analysts describe these forecasts as overly optimistic \cite{hughes2013, murray2012, kerr2011}. Hughes \cite{hughes2013} suggests that industry-standard productivity models of shale oil and gas fields systematically underestimate their decline rates which, at $>$40\%.yr$^{-1}$ are in fact very rapid compared with 4.5--6.7\%.yr$^{-1}$ for conventional fields \cite{murray2012}. Consequently, suggests Hughes \cite{hughes2013}, the high drilling rates required to compensate for such rapid decline are likely to hasten production peaks compared with industry predictions \cite{hughes2013}. More time is needed for data collection to settle these uncertainties. Importantly however, aside from questions of production rate, shale oil and gas extraction is more expensive, resource intensive and environmentally damaging than conventional oil and gas drilling (on land) \cite{hughes2013}. This is also generally true of other unconventional oil production methods as well as deepwater drilling for conventional oil \cite{hughes2013, murray2012, kerr2011, verbruggen2010}. Moreover, these costs are likely to increase because the economic incentives are to exploit the highest-quality, lowest-cost resources first \cite{verbruggen2010}. As exploration and production become more difficult, the energy return on (energy) investment (EROI)\footnote{EROI is defined as the ratio of gross fuel energy extracted to cumulative primary energy required, directly and indirectly, to deliver the fuel to society in a useful form \cite{cleveland1984}. The analyst may choose the `useful form' of interest; it may be the raw fuel immediately after extraction (e.g. crude oil `at the wellhead') or a derivative after subsequent conversions and/or transportation (e.g. refined gasoline distant from the wellhead). For a summary of other subtleties in net-energy analysis, see \cite{cleveland2011}.} declines, driving up the price of oil and oil-dependent products and services \cite{king2011}. 

An EROI of 1 may represent a strict limit to the economic viability of fuel production\footnote{A possible exception is a process which converts a less economically useful form of energy to a more useful form; such a process may remain economical even with EROI$<$1.}, though it has been argued \cite{king2011} that a higher value between 1--10 may be required at a minimum, for the primary energy supply of a modern industrial society to sustain its complexity. The EROI of conventional oil itself has declined from 100 in 1930 to $\lesssim$20 currently (`at the wellhead', before transportation and/or conversions) \cite{fantazzini2011, cleveland2011}. This value of $\lesssim$20 nonetheless remains high compared with equivalent EROI estimates for shale oil and tar sands (after extraction, before transportation and conversion), and CTL; these are respectively $\sim$2 \cite{cleveland2011}, 2--4 \cite{murphy2010} and $\sim$2 \cite{kreider2007}. This limits the capacity of unconventional oil to economically substitute for conventional oil and mitigate the effects of its peaking \cite{hall2012, hughes2011}. 

The rising costs of oil production have significant economic implications. History suggests that recessions typically occur when oil expenditures exceed $\sim5.5\%$ of gross domestic product (GDP\nomenclature{GDP}{Gross domestic product}) \cite{hall2012}. Of the 11 recessions in the US since the Second World War, ten, including the most recent `Global Financial Crisis' (GFC)\nomenclature{GFC}{Global financial crisis (of 2008--)} beginning in 2008, were preceded by spikes in oil prices \cite{murray2012, hall2012}. Naturally, the limitation of EROI applies to all energy technologies, and is discussed for solar fuel production technologies in section \ref{sec:solfuelsys}.  
 
Verbruggen and Al Marchohi report carbon intensities for unconventional oil sources ranging from 10-80\% higher than conventional oil \cite{verbruggen2010}. This is closely related to the low EROI values for these fuels since exploration and production are powered mainly by fossil fuels. Unconventional oils are therefore prone to deepening the already serious problem of anthropogenic climate change (section \ref{subsec:climate}) and are also vulnerable to the constraints of carbon pricing, compounding their costliness \cite{hughes2011}.

Although the date of global peak oil remains uncertain, the evidence reviewed here strongly suggests that scalable substitutes for conventional oil are now required, and the high financial and environmental costs associated with unconventional oil brings into question its capacity to fill this role sustainably. According to Hughes and Rudolph \cite{hughes2011},
\begin{quote}
`\textit{It is no longer a case of `if' or `when' there will be a plateau and eventual peak in world oil production, it is now a question of how jurisdictions can prepare for a world with less oil, one in which energy security and climate policy play a dominant role.}'
\end{quote} 

	\subsection{Anthropogenic climate change}\label{subsec:climate}
The global climate system comprises atmosphere, ocean, land, ice and biosphere \cite{houghton2009}. These components interact through biogeochemical cycles including the carbon cycle, in which photosynthetic NPP (section \ref{subsec:PP}) is a major driver \cite{lal2008}. Human interference in the carbon cycle may now be of sufficient scale to cause lasting, global changes in the climate, chiefly through global warming due to the greenhouse effect (section \ref{subsec:radfluxes}) \cite{ipcc2007, houghton2009, smil2008}. This section assesses the scale of human influence on atmospheric CO$_2$ concentration, compares this with global photosynthetic NPP and describes the scientific consensus on its likely consequences for the climate. Recent literature assessments of mitigation efforts required to avoid dangerous climate change are also reported.

\paragraph{Interference in the carbon cycle} Carbon dioxide is the GHG most significant to climate change, with a radiative forcing\footnote{The radiative forcing accounts for the GWP of a GHG and also its atmospheric concentration. The numbers given here were current when reference \cite{ipcc2007} was published and have since changed in proportion to small changes in GHG concentrations.} of 1.66~W.m$^{-2}$, compared with 0.48~W.m$^{-2}$ and 0.16~W.m$^{-2}$ respectively for CH$_4$ and nitrous oxide (N$_2$O), the 2nd and 3rd most significant GHGs \cite{ipcc2007}. At the time of writing, atmospheric CO$_2$ is concentrated at 392 parts per million by volume (ppmv) \cite{co2now}, which equates\footnote{Conversion factor \cite{ballantyne2012}: 2.124~GtC.ppmv$^{-1}$.} to a carbon pool of 833~GtC. This is significantly above preindustrial concentrations, which for the past $4.2\times10^5$ years oscillated over $1\times10^5$-year cycles by $\sim$100ppmv, between $\sim$180--280ppmv \cite{falkowski2000}. Since the industrial revolution, cumulative carbon emissions have totalled $\sim$450~GtC from fossil fuel combustion (280~GtC) and land use changes (150--200~GtC) \cite{smil2008}, with emission rates rising over time with economic activity (e.g. 70\% increase in GHG emissions between 1970--2004 \cite{ipcc2007}). 
  
The largest net GHG flow into the atmosphere is currently CO$_2$ from fossil fuel combustion, at a rate of $\sim$3.5~GtC.yr$^{-1}$ \cite{lal2008, co2now}. In 2010, fossil fuel combustion and cement production together added $\sim$9.1~GtC to the atmosphere \cite{peters2011}. Combined with emissions from land-use change (0.9~GtC), this put total anthropogenic carbon emissions for 2010 at $\sim$10.0~GtC \cite{peters2011}. Total CO$_2$-equivalent (CO$_2$e\nomenclature{CO$_2$e}{Carbon dioxide-equivalent}) emissions of all GHGs in 2010 were $\sim$48~GtCO$_2$e \cite{rogelj2011}.
 
 Global photosynthetic NPP annually fixes more than ten times the CO$_2$ emitted by the human economy, though $\sim$96\% returns to the atmosphere through heterotrophic respiration on the same time scale \cite{lal2008}. Net carbon stored through NPP and soil production from photosynthates, after heterotrophic respiration and soil erosion, is estimated at $\sim$4~GtC.yr$^{-1}$. Together with net oceanic CO$_2$ absorption of $\sim$2.3~GtC.yr$^{-1}$, this accounts for the discrepancy between gross CO$_2$ emission rate and the rise in atmospheric concentration \cite{lal2008}. 
  
\paragraph{Climate change} Ongoing changes observed in Earth's climate and predictions of their dynamics over coming decades are complex subjects, under intensive study and heated discussion. Nonetheless, a high level of scientific consensus supports the hypothesis that human interference in the carbon cycle is affecting\footnote{Climate change is defined in the United Nations Framework Convention on Climate Change (UNFCCC)\nomenclature{UNFCCC}{United Nations Framework Convention on Climate Change} as `\textit{... a change of climate that is attributed directly or indirectly to human activity that alters the composition of the global atmosphere and that is in addition to natural climate variability observed over comparable time periods}' \cite{ipcc2007}.} the climate, and its strength among commentators varies proportionally with level and relevance of scientific expertise \cite{anderegg2010}. At the time of writing, the Intergovernmental Panel on Climate Change's `IPCC Fourth Assessment Report: Climate Change 2007', summarised in \cite{ipcc2007}\nomenclature{IPCC}{Intergovernmental Panel on Climate Change}, remains the accepted consensus on observed and predicted climate change dynamics. Salient among its assertions are \cite{ipcc2007}: 
 \begin{itemize}
\item Warming of the climate system is now unequivocal, evidenced by increases in global average air and ocean temperatures, widespread melting of snow and ice, and rising global average sea level.
\item Many natural systems are being affected by regional climate changes, particularly temperature increases. The effects include melting glaciers/permafrost, altered hydrology in snow-fed rivers and lakes, poleward migration of plant and animal species in terrestrial systems, shifts in ranges and abundances of marine and freshwater species, and ocean acidification due to increased carbon uptake from the atmosphere. 
\item Predicted future effects of climate change include (among others, such as increases in effects already observed) more frequent extreme weather events, accelerated species loss, reduced (increased in some regions) agricultural productivities and increased human migration.
\end{itemize}

Aside from these continuous trends, there is a threat of `large-scale discontinuities' in the complex climate system: rapid shifts causing events such as irreversible loss of major ice sheets, reorganizations of oceanic or atmospheric circulation patterns and abrupt shifts in critical ecosystems \cite{lenton2011}. Such discontinuties present perhaps the greatest risk posed by climate change. Originally established by the UNFCCC, the conventionally accepted limit to `safe' average global warming above the average temperature before the industrial revolution is $2 ~\dg$C \cite{lenton2011, anderson2011, parry2008}. However, more recent studies have shown that large-scale climate discontinuities could occur before warming reaches $2 ~\dg$C, so this target is likely inadequate to ensure safe climate stabilisation and may be better described as `the threshold between 'dangerous' and 'extremely dangerous' climate change' \cite{anderson2011, lenton2011}. Policy recommendations aiming to limit warming to $2 ~\dg$C should therefore be seen as minimal measures.

Numerous analysts have estimated GHG emission limits consistent with limiting average global warming to $2 ~\dg$C. Zickfeld \textit{et al} \cite{zickfeld2009} present a coupled climate-carbon cycle model, finding that to stabilise mean global warming at $2 ~\dg$C with a probability of at least 0.66 ($P(\langle\Delta T\rangle\leq2~\dg\text{C})\geq0.66$), cumulative CO$_2$ emissions from 2000--2500 must not exceed a median estimate of 590~GtC (range: 200--950 GtC). Furthermore, to ensure $P(\langle\Delta T\rangle\leq2~\dg\text{C})\geq0.9$, median total allowable CO$_2$ emissions are estimated at 170~GtC (range: -220--700 GtC). This equates to only $\sim38\%$ of human emissions since the industrial revolution. This is also small compared with the carbon content of remaining fossil fuels (section \ref{sec:fossil}), being equal to 20\% of 1P reserves ($\sim863$~GtC) and 9\% of URR ($\sim$1844~GtC). Confidently avoiding dangerous climate change will require the remaining fossil fuel resources to be left largely unproduced, unless effective carbon capture and storage methods can be rapidly developed and scaled up \cite{lackner2012, lenton2010}(section \ref{sec:transition}). Rogelj \textit{et al} \cite{rogelj2011} reanalyse a large set of published emission scenarios from integrated (techno-enviro-economic) assessment models in a risk-based climate modelling framework. In scenarios in which $P(\langle\Delta T\rangle\leq2~\dg\text{C})\geq0.66$, GHG emissions peaked between 2010--2020 and fell to a median level 8.3\% below the 2010 rate by 2020. Together with the findings of Zickfeld \textit{et al}, this indicates that deep cuts in global GHG emissions are urgently required. 

Evidence of fossil fuel supply limitations reviewed in section \ref{subsec:fossilsupply} mandates a rapid transition away from fossil-fuel dependence, though the precise time scale required remains uncertain. Analyses of anthropogenic climate change, however, suggest that less than one decade remains in which to deploy mitigation efforts to reduce GHG emissions below 2010 levels. This time window is brief compared with previous global energy transitions (section \ref{subsec:energytrans}), indicating the urgency with which a transition away from the fossil fuel-based energy paradigm to low-carbon energy systems is needed.

\section{Transitioning to solar fuels}\label{sec:transition}

 This section examines the potential for a global-scale transition to sustainable energy systems with the rapidity needed to avert dangerous climate change and mitigate the damaging effects of fossil fuel supply limitations. 
 
 \subsection{The importance of fuels}\label{subsec:impfuels}
	\chapquote{Because sunlight is diffuse and intermittent, substantial use of solar energy to meet humanity's needs will probably require energy storage in dense, transportable media \emph{via} chemical bonds.}{D. Gust \textit{et al.} \cite{gust2009}}
	
 According to the IEA, in 2008, $\sim17\%$ of global final energy consumption was supplied by electricity, $\sim3\%$ by heat and the remaining $\sim80\%$ directly by fuels \cite{iea2009}. Applying these sectoral fractions together with TPES for 2010 (section \ref{sec:earthsolar}) and assuming efficiencies of $38\%$ for electricity generation \cite{bp2011}, and $93\%$ for electricity transmission and distribution \cite{eia2012}, gives an estimate\footnote{Additional assumptions: 1) 0.95 is taken as a representative value for transportation and distribution efficiency of fuels, based on \cite{huang2011}; 2) transmission and distribution efficiency of heat is taken to be 0.9, based on \cite{nve2009}.} that $\sim36\%$ and $\sim2\%$ of global primary energy respectively are currently used to supply final electricity and heat consumption; the remaining $\sim62\%$ of primary energy supply, therefore, meets demand directly with chemical fuels. 
 
 Electricity is the most versatile energy carrier and worldwide the share of fossil fuels converted to it has increased by an order of magnitude since 1900 \cite{smil2008}. This is projected to continue over coming decades, increasing electricity consumption from 17\% to 22\% of global final energy consumption by 2030 \cite{iea2010}. Nonetheless, there are vast industries, particularly aviation and long-distance road transportation, in which it is currently thought that electricity cannot foreseeably replace chemical fuels, due to the relatively low energy densities of existing charge-storage systems \cite{warshay2011, hedenus2010, rye2010}. Current batteries have energy densities two orders of magnitude lower than liquid fuels \cite{fischer2009}. 
 
 Shorter-range road transportation is a market more amenable to electrification \cite{fischer2009}, though electric vehicles (EVs)\nomenclature{EV}{Electric vehicle} at present remain a negligibly small fraction of the roughly billion-strong on-road fleet. Recent techno-economic models \cite{higgins2012, nemry2010} suggest that market penetration of pure (battery-only) EVs in developed economies is unlikely to exceed 10\% by 2030. Including plug-in hybrid-electric vehicles, which rely on liquid fuels for long-range travel, total market penetration is not predicted to exceed 30\% by 2030 \cite{higgins2012, nemry2010}. Moreover, existing high-efficiency, gasoline-only and gasoline-hybrid vehicles emit less CO$_2$ compared with coal-fired electricity-powered EVs; in order to mitigate carbon pollution, adoption of EVs would need to be coordinated with decarbonisation of the consumer electricity supply \cite{smil2010}. The latter process is projected to take place over several decades \cite{battaglini2009, verbruggen2009, smil2010}. An alternative, which may help to accelerate a large-scale transition from fossil fuels, is to supply the economy with nonpolluting fuels compatible with existing (or minimally modified) fuel distribution systems and fuel-powered technologies. This approach can be complementary to decarbonisation of electricity generation.
 
 Existing renewable energy conversion technologies that generate electricity provide no direct carbon sequestration capabilities. Conversely, renewable fuels (section \ref{sec:solfuels}) synthesised using atmospheric CO$_2$ have the added advantage that their production cycles can theoretically be made carbon-negative, sequestering more carbon in terrestrial stores than combustion of the fuels re-emits. For example, components of biomass not used for biofuel production can be pyrolysed\footnote{Made to undergo thermochemical decomposition at elevated temperatures without the participation of oxygen.} into so-called biochar, which can be added to soil or buried for long-term sequestration \cite{lehmann2007}. In the case of microalgal biomass, as much as 55\% of the carbon can be recovered as biochar \cite{sayre2010}.
 \subsection{Pathways to globally sustainable solar fuel production}\label{subsec:possusfuels}
 Current HANPP, from both ancient and current sources, is thought to be unsustainable (section \ref{subsec:sushanpp}). Nonetheless, it may be possible to sustainably increase reliance on HANPP from current sources to supply the human economy with low-carbon energy on scales adequate to assist in a global transition away from fossil fuels. This will likely require either efficiency gains in socioeconomic utilisation of biomass ($NPP_h$ -- see figure \ref{fig:hanpp}) \cite{haberl2010}, an increase in Earth's photosynthetic productivity beyond that of current biota ($NPP_{act}$), or a combination of the two.
 
The need for sustainability restricts increases in $NPP_{act}$ through expanded agriculture (expanded $NPP_{LC}$ and $NPP_h$, and correspondingly reduced $NPP_t$) and/or further intensification of cropping and forestry\footnote{Scale is important to consider here; smaller-scale agro-bioenergy systems may be sustainable in some localities. However, a global-scale increase in HANPP from agricultural intensification for fuel production is not a sustainable option (section\ref{subsec:sushanpp}).} (increased $NPP_h$, with corresponding reduction in $NPP_t$) \cite{bishop2010, haberl2010, giampietro2009, smil2010}. The remaining option is to increase $NPP_0$ by deploying bioengineered and/or artificial photosynthetic systems on otherwise-unproductive or low-productivity lands, or on productive lands in ways that do not degrade natural ecosystems. 
 \subsection{Time scales of global energy transitions}\label{subsec:energytrans}
 According to Smil \cite{smil2010}, `\textit{An energy transition encompasses the time that elapses between the introduction of a new primary energy source (coal, oil, nuclear electricity, wind captured by large turbines) and its rise to claiming a substantial share of the overall market.}'\footnote{Smil acknowledges that `substantial share' is arbitrary and explains why he argues for at least 15\%. Further, Smil claims that for an energy source to be an absolute leader, it must contribute more than 50\% of TPES. \cite{smil2010}} The large scale of global fossil fuel consumption and urgency with which a transition away from it is required raise the question of how rapidly such a transition could be orchestrated.
  
\begin{figure}
\centering
\includegraphics[angle=0,width=0.5\textwidth]{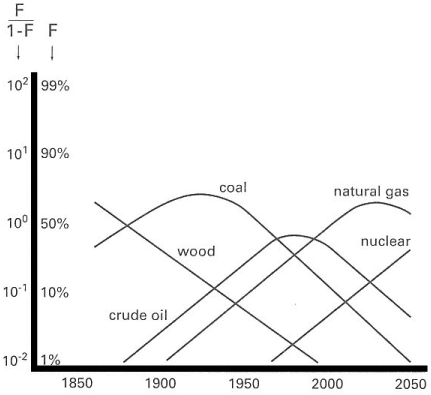}
\caption[Time scales of global energy transitions]{\textbf{Time scales of global energy transitions}, predicted by Marchetti \cite{marchetti1977}. $F$ is market share of primary commercial energy consumption. Figure reproduced from \cite{smil2008}.}
\label{fig:marchetti} 
\end{figure}

  Previous global energy transitions lend insight. Coal replaced biomass in 1885 as the energy source having the largest share of TPES, and oil replaced coal by 1960 \cite{smil2008, smil2010(2)}. In 1977, Marchetti \cite{marchetti1977} found that individual energy sources' market shares ($F$) approximated Gaussian curves over time (presented over a logarithmic scale in figure \ref{fig:marchetti}). For more than a century (1850--1970), successive transitions appeared to follow a predictable pattern, robust to major economic perturbations such as wars, booms and depressions \cite{marchetti1977, smil2008}. This suggested a characteristic time scale; each transition required almost one hundred years for the emerging energy source to rise from 1\% to 50\% market share \cite{smil2010(2)}. Applying his model to more recently adopted energy sources, natural gas and nuclear electricity, Marchetti projected similar, near-century transition times. In reality, these transitions have been slower than predicted; since 1970, different sources' market shares have stabilised and therefore diverged from Marchetti's model, with coal and oil retaining large shares for longer than predicted. Smil writes \cite{smil2010(2)}, `\textit{There is only one thing that all large-scale energy transitions have in common: Because of the requisite technical and infrastructural imperatives and because of numerous (and often entirely unforeseen) social and economic implications (limits, feedbacks, adjustments), [such transitions] are inherently protracted affairs. Usually they take decades to accomplish, and the greater the degree of reliance on a particular energy source ..., the longer [the substitution] will take.}'
  
 The evidence reviewed in section \ref{sec:limfoss} indicated that a single decade remains in which to reduce CO$_2$ emissions by replacing heavily entrenched \cite{unruh2000}, high-quality\footnote{Cleveland \textit{et al} \cite{cleveland2000} define energy quality as the relative economic usefulness per heat equivalent unit of different fuels and electricity.} fossil fuels with inescapably lower-quality, renewable energy sources. The historical precedents for slow dynamics in large-scale energy transitions provide a sobering check on the potential for humanity to avert dangerous climate change or the consequences of fossil fuel supply limitations. The detuning between these timescales indicates the need for an unprecedented level of sociopolitical will and cooperation in deploying serious efforts to accelerate, as much as possible, the transition away from fossil fuels.
	 	\section{Solar fuels}\label{sec:solfuels}
	 	A fuel is a reduced substance that can be oxidised, usually with oxygen, to emit practically useful energy \cite{gust2009}. A given fuel's chemical synthesis may be amenable to different methods and/or driven by different power sources. Currently, the only industrial-scale fuel synthesis operations energised renewably are biofuel systems, powered by sunlight. In addition to traditional biomass fuels (section \ref{subsec:PP}), primary photosynthates can be refined into high-energy-density biofuels. 
	 	
	 	In 2009, 0.00216~ZJ of bioethanol and biodiesel were produced globally (0.43\% of TPES) \cite{iea2010}. Presently these two fuels constitute almost the entire global biofuel sector \cite{iea2010}, though demand for others is growing, especially biogas \cite{iea2011}. This section canvasses properties of the most significant biofuels: ethanol, biodiesel and biogas. Biologically produced hydrogen and hydrocarbons, which are not yet commercial but show promise, are also introduced. Established production methods for each fuel, from biological feedstocks, are summarised (and the production of feedstocks is described in section \ref{sec:solfuelsys}). Excepting ethanol and hydrogen, the fuels discussed are not pure compounds but variable mixtures of similar compounds falling within a range. Physicochemical properties of different fuels are given in table \ref{tab:fuelproperties}.  		
	 	
		\paragraph{Ethanol}
			Fuel ethanol is either pure anhydrous ethanol or it may contain a small fraction (4--5$\%$) of water, depending on production methods and intended use. It is a volatile, flammable, colourless liquid. 
			The stages of large-scale ethanol production are: microbial sugar fermentation, distillation and dehydration. The last stage may be omitted if complete purity is not required; purification by distillation alone limits purity to 95--96$\%$, due to the formation of an azeotrope. The resulting solution can be used as a fuel alone but is immiscible in gasoline and so cannot be used in gasoline blends \cite{madsonandmonceaux2003}.
			In 2009, 0.00162~ZJ of fuel ethanol were produced worldwide \cite{iea2010}. 

		\paragraph{Biodiesel}
			Biodiesel is a variable mixture of alkyl (methyl, propyl or ethyl) esters of fatty acids. It is a volatile, flammable, transparent, yellow liquid \cite{mousdale2008}. Traditionally, biodiesel is produced by transesterification of biological oil feedstock such as vegetable or animal oil. This is done by reacting an alcohol (usually methanol or ethanol) with the feedstock, in a basic medium \cite{pahl2005}. Biodiesel can also be produced by the multi-stage biomass-to-liquid (BTL\nomenclature{BTL}{Biomass-to-liquid}) process in which biomass from a wide variety of possible sources is gasified\footnote{Reacted with a controlled amount of oxygen and/or steam at high temperature to produce syngas, a mixture of carbon monoxide and hydrogen.} and the gas then polymerised through Fischer-Tropsch synthesis to produce diesel-range hydrocarbons (table \ref{tab:fuelproperties}). Whereas traditional biodiesel production uses only part of the available biomass (oil), BTL allows all of the biomass to be used, thus using NPP more efficiently. BTL technology is not yet mature however, and is under continuing development \cite{stocker2008}. In 2009, 0.0054~ZJ of biodiesel were produced worldwide \cite{iea2010}. 
							
		\paragraph{Biogas}
			Biogas is produced by fermentation or anaerobic digestion of organic materials such as biomass, manure, sewage and municipal waste \cite{klass1998}. Its composition varies depending on production process but is predominantly CH$_4$ (40--75\%, typically $\sim65\%$) and CO$_2$ (25--60\%), with trace amounts of other gases such as water vapour, hydrogen and hydrogen sulphide \cite{klass1998, walsh1988}. The useful energy content of biogas is contained almost exclusively in its CH$_4$, which under standard conditions has an energy density of 55.6~MJ.kg$^{-1}$ \cite{cuellar2008}. The energy density of biogas scales linearly with CH$_4$ content; for a typical mole fraction of 65$\%$ it is $\sim$30~MJ.kg$^{-1}$. Carbon emissions per volume of biogas combustion are effectively independent of composition since the CH$_4$ fraction is converted to CO$_2$ and the initial CO$_2$ fraction is unchanged \cite{cuellar2008}. However, due to the scaling of energy density with composition, CO$_2$ emissions per megajoule are a function of composition. For 65mol$\%$ CH$_4$ biogas, CO$_2$ emissions per megajoule are equivalent to typical biodiesel (table \ref{tab:fuelproperties}). Notably, the GWP of biogas is reduced through combustion, since the 100-year GWP of CO$_2$ is approximately 25 times weaker than that of CH$_4$. This provides an additional argument for combusting biogas that would otherwise be vented to the atmosphere. 	
		
		\paragraph{Hydrogen}	
		  `Biohydrogen' is hydrogen produced by living organisms. It is a highly flammable, colourless, odourless gas under standard conditions, with the highest gravimetric energy density of any chemical fuel. Its combustion emits no CO$_2$ but rather, only water. Biological systems provide various methods for hydrogen production, including indirect water photolysis, direct water photolysis, photofermentation, and dark fermentation. Cyanobacteria evolve hydrogen through indirect water photolysis by first photosynthesising carbohydrates and then fermenting them to produce H$_2$ and CO$_2$ \cite{levin2004}. Other species, photoheterotrophs, have the ability to photoferment carbohydrates provided by autotrophic species \cite{levin2004}. These species are of interest for the conversion of waste organic compounds into hydrogen. In dark fermentation, anaerobic bacteria grown in the dark on carbohydrate-rich substrates ferment the carbohydrates to produce a biogas mixture which is predominantly H$_2$ and CO$_2$ but which may also contain trace amounts of CH$_4$, carbon monoxide and/or hydrogen sulphide \cite{levin2004}. Green microalgae such as the model species \textit{C. reinhardtii} can produce hydrogen by direct water photolysis under sulphur-deprivation conditions \cite{melis2000}.		  
							
\begin{figure}
\centering
\includegraphics[angle=0,width=1\textwidth]{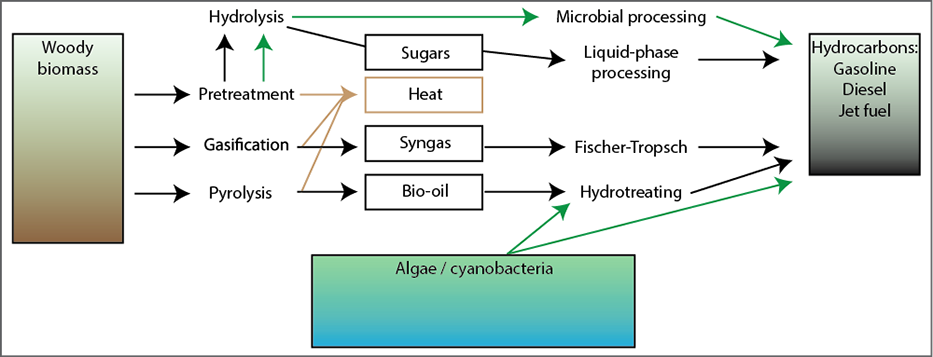}
\caption[Biohydrocarbon production pathways]{\textbf{Biohydrocarbon production pathways}. Cellulosic biomass (e.g. wood) and lipids (e.g. from microalgae) can be converted to biohydrocarbons. Conversion pathways may combine biological (green arrows), thermal (brown) and catalytic (black) processes. Direct biosynthesis production (e.g. in cyanobacteria) is also under development. Figure adapted from \cite{regalbuto2009}.}
\label{fig:biohydrocarbons}
\end{figure}

		\paragraph{Hydrocarbons}
			`Biohydrocarbons' are hydrocarbons produced by living organisms or chemically from biological feedstock. They offer a directly fungible substitute for fossil hydrocarbon fuels. In principle, biohydrocarbon fuel mixtures can be produced that exactly mimic petroleum-derived gasoline, diesel and aviation fuel \cite{regalbuto2009}. Multiple kinds of aviation fuel are in use worldwide; a typical example is kerosene-type BP Jet A-1. Figure \ref{tab:fuelproperties} compares the properties of gasoline, petrodiesel and aviation fuel. Biohydrocarbons can be produced by chemically processing biological feedstocks including cellulosic biomass, bio-oils and sugars. Regalbuto \cite{regalbuto2009} describes these pathways, which are summarised in figure \ref{fig:biohydrocarbons}. Production may also occur through direct biosynthesis, such as recently characterised for alkanes in cyanobacteria \cite{schirmer2010}, although such direct production remains in the early stages of research. 
			
	 	\section{Solar fuel production systems}\label{sec:solfuelsys}
	 		\chapquote{... current options to harness and store (solar) energy are too expensive to be implemented on a large scale. Hence the objective to science is to develop new materials, reactions and processes to enable solar energy to be sufficiently inexpensive to penetrate global energy markets.}{D.G. Nocera \cite{nocera2009}}
	 			 
	 In generic terms, photosynthetic fuel production requires \cite{gust2009}: 1) an energy source such as sunlight; 2) a material that can be oxidised to emit electrons; and 3) a material that can be reduced by those electrons to produce a fuel or feedstock for chemical processing into fuel. These elements are combined in a wide variety of morphologies across biological species and are also being engineered into artificial photosynthetic systems. Fuel production through biological photosynthesis in general constitutes three steps. First, energy absorbed from radiation generates electrochemical excitations/redox equivalents. Next, the light-induced redox potential is used to catalyse water oxidation, generating protons, electrons stored as reducing equivalents, and oxygen. Finally, another catalytic system uses the reducing equivalents to produce high-energy, low-entropy chemicals such as carbohydrates, lipids, hydrocarbons or molecular hydrogen, which can be chemically processed into practical fuels \cite{gust2009} (the particular realisation of these steps found in higher plants and green algae is explained in section \ref{sec:photosynthesis}). This section introduces and compares solar fuel production systems based on mass culturing of microalgae and artificial photosynthetic systems, as these both offer promising alternatives to unscalable agro-biofuel production systems (section \ref{subsec:sushanpp}). 
 	 
 	  \subsection{Microalgal cultivation systems}\label{subsec:microal}
 	  
 	  \chapquote{On the arid lands there will spring up industrial colonies without smoke and without smokestacks; forests of glass tubes will extend over the plains and glass buildings will rise everywhere; inside of these will take place the photochemical processes that hitherto have been the guarded secret of the plants, but that will have been mastered by human industry which will know how to make them bear even more abundant fruit than nature, for nature is not in a hurry and mankind is.}{G. Ciamician, 1912 \cite{gust2012}}
 	  
\begin{figure}
\centering
\includegraphics[angle=0,width=0.9\textwidth]{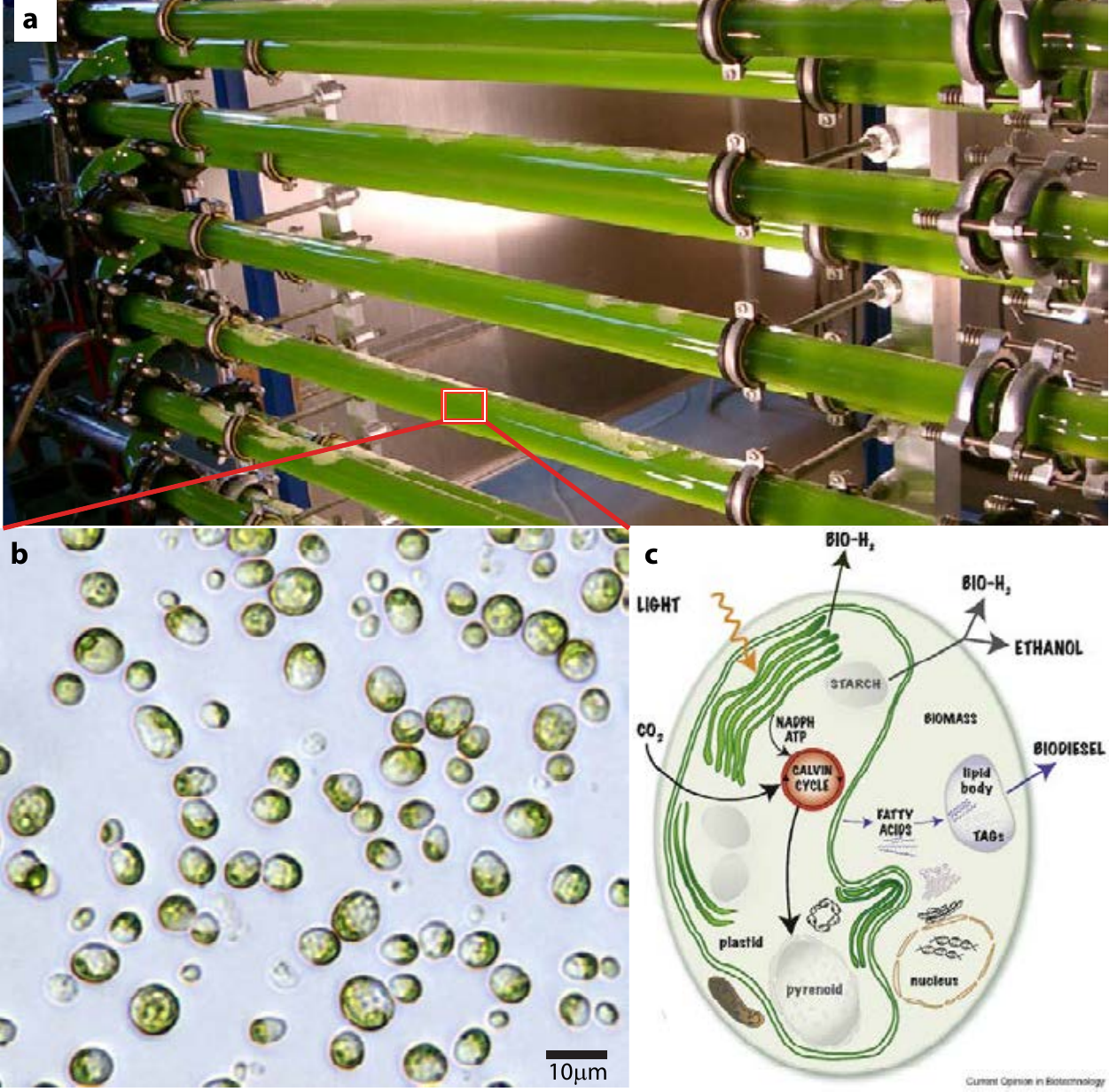}
\caption[Microalgae and photobioreactor]{\textbf{Microalgae and photobioreactor}. \textbf{a}, Laboratory-scale photobioreactor cultivating green microalgae. Image reproduced from \cite{solarbiofuels2013}. \textbf{b}, Unknown species, local to Brisbane, Australia. Image courtesy of G. Jakob, used with permission. \textbf{c}, Schematic microalgal anatomy and fuel synthesis pathways. Image reproduced from \cite{beer2009}}.
\label{fig:microalgae}
\end{figure}
	  
Microalgae (fig. \ref{fig:microalgae}) are unicellular eurkaryotes, typically $1-10~\mu m$ in extent, that are ubiquitous across Earth's ecosystems \cite{larkum2011}. They are generally photoautotrophic, with nutritional requirements for only PAR, H$_2$O, CO$_2$, nitrogen, phosphorous and comparably small amounts of inorganic nutrients. Possessing great metabolic plasticity, microalgae can synthesise a diverse range of chemical compounds despite their simple nutritional requirements, and are adapted to survive under a broad range of environmental stresses (e.g. heat, cold, drought, salinity, photo-oxidation, anaerobiosis, osmotic pressure and UV radiation), giving them access to habitats inhospitable to higher plants \cite{amaro2012, larkum2011}. Microalgae and other plankton are currently responsible for more than 90$\%$ of global oceanic NPP \cite{chassot2010}, or 45$\%$ (2.0~ZJ) of total global NPP. Microalgal cultivation systems are either open ponds or closed photobioreactors \cite{brennan2010, posten2009}.

There is now growing interest in the potential for microalgae to economically produce fuels and other products, due to beneficial properties described in recent reviews: \cite{stephenson2011, singh2011, day2012, larkum2011, stephens2010, brennan2010, larkum2010, mata2010, pienkos2009, beer2009, barber2009, hambourger2008, kruse2005}. They offer higher productivity densities than higher plants and, unlike higher plant cropping, algaculture does not require arable land. It can also be coupled to multiple waste streams (wastewater, saline water [including seawater], CO$_2$ emissions), combining waste treatment (wastewater treatment, desalination, carbon sequestration) with value-adding processes (production of fuel, food, high-value-products (HVPs\nomenclature{HVP}{High value product}) such as pharmaceuticals, and fresh water). The flexible metabolism of microalgae can be tuned to produce the full spectrum of solar fuels/biofuels, either directly or through biomass used as feedstock for chemical processing. Moreover, complete genome sequences are available for some species, including the well-studied model species, \textit{C. reinhardtii}, facilitating optimisation through genetic engineering. 

Microalgal cultivation systems have the potential to offer increased $NPP_h$ simultaneously with other economic and environmental services. However, they remain at an early stage of development, with research and commercialisation milestones remaining before economic scalability can be achieved. This section considers some important constraints on the environmental and economic viability of scaling up microalgal cultivation systems to a globally significant level, and identifies key innovation pathways.  

\paragraph{Productivity densities}
Literature surveys \cite{weyer2010, griffiths2009} of daily biomass productivity densities in microalgal cell cultures grown in outdoor open ponds\footnote{Open pond cultivation systems are a more mature technology than photobioreactor systems, and are accordingly better represented in the literature at present.} under nutrient-replete conditions find a range\footnote{Assuming biomass energy density of 20~MJ.kg$^{-1}$.} of 2.3--11.1~W.m$^{-2}$ (daily average). Weyer \textit{et al} \cite{weyer2010} estimate 7.6--9.7~W.m$^{-2}$ as a `best case' range, representing an optimistic target for production, based on realistic efficiencies and geographical locations with favourable conditions. Stephens \textit{et al} \cite{stephens2010} use 4.6~W.m$^{-2}$ as conservatively representative of current open-pond systems, and 10.4~W.m$^{-2}$ as achievable with current photobioreactors. These figures compare favourably with the global average for agricultural crops, $\sim$0.4~W.m$^{-2}$ and even the conservative productivity estimate for open ponds approaches the highest recorded cropping rate of 5.0~W.m$^{-2}$ (section \ref{subsec:PP}). 

Assuming average insolation of 20~MJ.m$^{-2}$.day$^{-1}$, which is realistic for a well-selected cultivation site \cite{weyer2010, stephens2010}, average productivity densities of 4.6~W.m$^{-2}$ and 10.4~W.m$^{-2}$ respectively equate to average photosynthetic efficiencies of 2$\%$ and 4.5$\%$. This is well within what is thought to be the theoretical efficiency limit of $\sim12\%$ for glucose production\footnote{Biohydrogen production can in principle achieve higher efficiencies. See chapter \ref{chp:multianalysis}.} (chapter \ref{chp:multianalysis}), leaving room for engineering improvements. 

\paragraph{Land constraints}
Assuming insolation of 20~MJ.m$^{-2}$.day$^{-1}$ and 2$\%$ photosynthetic conversion efficiency, an area of $3.44\times$10$^6$~km$^{2}$ would be required to displace global TPES in 2010 (0.503~ZJ). This equates to 2.3$\%$ of total land, 3.4$\%$ of non-agricultural land, or 31.3$\%$ of estimated marginal land \cite{stephens2012}, suggesting that global scale-up of microalgal energy systems is not strictly land-limited. However, a comprehensive analysis needs to account for geographical availabilities of essential resources beyond light (CO$_2$, H$_2$O, nutrients, funds), as well as for the detailed resource requirements and outputs of microalgal cultivation systems at the local scale. These are studied in life-cycle analyses.

\paragraph{Life-cycle analysis and costing: Energy, water, nutrient and financial balances}
Life-cycle assessment (LCA\nomenclature{LCA}{Life-cycle assessment}) evaluates environmental sustainability through comprehensive accounting of all energy and material inputs and outputs associated with a particular product or process over all stages of its life cycle: extraction of raw materials, manufacturing, transport, use and recycling or disposition \cite{colosi2012}. Life-cycle costing (LCC\nomenclature{LCC}{Life-cycle costing}) assesses economic sustainability through similarly comprehensive financial accounting. Whereas `first-order' such analyses consider only the energy (financial) production and consumption flows of system operation, `second-order' analyses also include energy (cost) embedded in materials used for system construction \cite{beal2012(1)}. Numerous LCA (in some cases, also LCC) analyses for microalgal cultivation systems have been published in recent years. The earlier such studies are critically reviewed by Colosi \textit{et al} \cite{colosi2012} and Grierson and Strezov \cite{grierson2012}. It is found that these LCA studies use diverse metrics, demonstrate limited methdological consistency, and many do not include uncertainty analyses. Consequently, their findings are difficult to evaluate and compare.

A more recent study by Beal \textit{et al} \cite{beal2012} proposes a standard methodology for comprehensive evaluation of microalgal biofuel production, using metrics for EROI (first- and second-order), financial return on investment (FROI\nomenclature{FROI}{Financial return on investment}), water intensity, nutrient requirements and CO$_2$ requirements. The authors apply this methodology to both an experimental lab-scale system (accounting for lab-scale artifacts), and a theoretical `Highly Productive' case representing an optimised commercial-scale system. They find that the EROI and FROI are less than 1 in both cases. Consistent with multiple previous studies reviewed in \cite{colosi2012, grierson2012}, these findings indicate that with existing production and processing methods, microalgal systems are neither energetically nor financially competitive with conventional or unconventional fossil fuels. Furthermore, Beal \textit{et al} find that both production cases considered have much higher water intensities than both fossil fuel production and non-irrigated biofuels from conventional (non-microalgal) feedstocks, and only the Highly Productive case (which assumes very efficient water use) is found to have water intensity competitive with irrigated agro-biofuel crops. However, the water intensity metrics used do not account for water quality, and microalgal systems have the advantage that they are not reliant on pure freshwater availability; the availability of saline, brackish or wastewater streams at a cultivation site may significantly reduce the `effective' water intensity of a microalgal system, improving its competitiveness. 

The authors note that the experimental case is also rendered infeasible at an economically significant scale of deployment by its demands for CO$_2$, nitrogen and electricity. The Highly Productive case presents markedly lower demands for these resources, which are concluded to be `more manageable, but still large' compared with availabilities at the national scale in the US.

Despite the infeasibility of the microalgal systems studied by Beal \textit{et al}, similarly to other studies the authors emphasise that feasible systems may be attainable through innovation. The following innovation pathways are suggested \cite{beal2012}:
\begin{quote}
\textit{`1. using waste and recycled nutrients (e.g., waste water and animal waste);\\ 2. using waste heat and flue-gas from industrial plants, carbon in wastewater, or developing energy-efficient means of using atmospheric CO$_2$;\\ 3. developing ultra-productive algal strains (e.g., genetically modified organisms);\\ 4. minimizing pumping;\\ 5. establishing energy-efficient water treatment and recycling methods;\\ 6. employing energy-efficient harvesting methods, such as chemical flocculation, and\\ 7. avoiding separation \emph{via} distillation.'}
\end{quote} 

Overall, these suggestions indicate the importance of an `ecological', holistic approach to system design: utilising environmental waste streams as resources and optimising system performance to maximise competitiveness under local conditions. 

\subsection{The importance of optimising light-harvesting efficiency} 
Consistent with suggestion 3 above by Beal \textit{et al} \cite{beal2012}, many other authors have emphasised the importance of optimising light-harvesting efficiency, which is a fundamental determinant of photosynthetic productivity \cite{day2012, ort2011, stephenson2011, singh2011, larkum2011, wijffels2010, posten2009, melis2009, beer2009, mitra2008, johnson2008, hankamer2007}. Biological photosynthetic systems have evolved not for maximal productivity but rather for survival and reproduction under the conditions of their native habitats. Most higher plants and green algae have evolved genetic strategies to assemble large light-harvesting antennas comprising arrangements of LHCI and LHCII light-harvesting complexes respectively surrounding the PSI and PSII core complexes (which contain the photochemical reaction centres [RCs]) (section \ref{subsec:lightreact})\cite{hankamer2007}. This gives the organism a competitive advantage under light-limited conditions, in which irradiance is considered to be the only factor limiting productivity. 
	
	However, a high LHC-to-RC ratio is disadvantageous under light-supersaturated conditions in which the antennas absorb energy from radiation faster than the resulting electronic excitations can be used by the RCs. So-called `nonphotochemical quenching' (NPQ\nomenclature{NPQ}{Nonphotochemical quenching}) (or `photoprotective') mechanisms have evolved to deal with this kinetic imbalance primarily by wasting surplus excitations as heat \cite{ruban2012}. Melis \textit{et al} \cite{melis2009, melis1998} estimate that in a typical high-light environment a microalgal mass culture or dense plant foliage can over-absorb and dissipate through NPQ approximately 60$\%$ of the daily irradiance, with cells at the directly illuminated surface wasting over 80$\%$ of absorbed irradiance through NPQ (section \ref{sec:procpart}). Re-optimising the complex light-harvesting machinery of photosynthesis for maximal productivity in different light environments is therefore a critically important innovation pathway to increasing the EROI of microalgal solar fuel production systems. This challenge of system-optimal light harvesting, which designers of artificial photosynthetic systems also face, forms the basis of the detailed biophysical analyses presented in this thesis.  
	
\subsection{Artificial photosynthetic systems}\label{subsec:artps}
Many different kinds of systems have been described under the term `artificial photosynthetic system' (APS\nomenclature{APS}{Artificial photosynthetic system}). According to Gust \textit{et al} \cite{gust2012}, these include `\textit{photovoltaic cells based on inorganic semiconductors for electricity production, dye-sensitized solar cells, photovoltaics based on organic semiconductors, systems for fuel production based on these types of devices, a large variety of devices for direct photochemical conversion of light excitation energy to a fuel using organic or inorganic molecular photocatalysts, natural organisms or their components interfaced to synthetic materials, and combinations of these approaches.}' Here the term is taken to refer to any predominantly abiological system that directly photocatalyses chemical fuel production without intermediate electricity generation. 

While some systems fitting this definition, such as concentrated solar thermal collectors coupled to thermochemical fuel production cycles \cite{magnuson2012}, do not closely mirror natural photosynthesis, other systems are more directly biomimetic. Arguably the closest are nano-scale systems variously comprising large organic molecules, metals, semiconductors, nano-structured materials, or combinations of these. These systems are often isomorphic to natural photosynthetic systems, having light-harvesting antennas coupled to photochemical reaction centre complexes that energise catalysts for the splitting of water and reduction of protons and/or carbon dioxide \cite{magnuson2012, gust2012, tachibana2012}. 

One of the strongest motivations behind developing APSs is that are thought able to ultimately exceed natural photosynthetic efficiencies because no energy is required to cover the metabolic cost of living \cite{blankenship2011, magnuson2012}. In natural photosynthetic organisms, this loss is typically on the order of 50\% of the chemical energy stored by photosynthesis (section \ref{sec:procpart}). It has also recently been argued \cite{blankenship2011} that systems using photovoltaic cells to power electrochemical cells for fuel production have a significant efficiency advantage over natural and bioengineered photosynthetic systems. However, the efficiency advantages claimed for both types of artificial system are only first-order. In a comprehensive assessment of economic and environmental feasibility these must be considered within a second-order LCA and LCC. Indeed, the energy required for the production and maintenance of an artificial system may be compared to the energy required by a natural system to maintain cell metabolism for survival and reproduction. APSs and high-efficiency photovoltaics both currently depend on scarce and expensive materials, and energy-intensive construction \cite{magnuson2012, peter2011}. Conversely, photosynthetic organisms self-assemble from abundant materials at ambient temperatures and the abiological components of existing microalgal cultivation systems comprise relatively simple, abundant materials \cite{posten2009}. 

Leveraging natural photosynthesis only indirectly, through biomimicry, the development of APSs is at an earlier stage than that of microalgal photosynthetic systems. No APS has yet been proven feasible outside of the laboratory \cite{magnuson2012, barber2009} and accordingly, no fair comparison can yet be made between comprehensive LCAs and LCCs for APSs and microalgal systems. Moreover, most APS development presently focusses on systems that produce molecular hydrogen, a fuel which has limited utility in today's carbon fuel-powered economy but which is seen as promising for the longer-term \cite{magnuson2012, tachibana2012}. In principle, carbon-fixing APS systems can also be developed but this is seen as a more difficult and distant goal \cite{magnuson2012, barber2009}. 

With many fundamental technical hurdles still to be cleared, the present state of APS development strongly suggests that these systems are unlikely to be viable on globally significant scales within the narrow window of time available for meaningful climate-change and peak-oil mitigation efforts. Nonetheless, APSs are a promising solar fuel production technology for the longer term and their development can benefit from ongoing advances in understanding biological photosynthetic systems such as microalgal cultivation systems, which are more rapidly approaching economic scalability.
	\section{Summary and outlook}\label{sec:chp1sum}
	Photosynthetic systems are intrinsically multiscale. Comprehensive analysis of photosynthetic systems in the context of global energy systems must face the challenge of integrating analyses across the hierarchy of scientific disciplines, ranging from global-scale to nanoscale. The geophysical, ecological and economic limitations of global energy systems identified in this chapter constrain the functional requirements for photosynthetic energy systems at smaller scales.
	
Solar energy is available at Earth's surface far in excess of global economic demand but is incident at power densities low compared with incumbent thermal power generating systems (e.g. coal, gas, nuclear), which challenges its economic utility. Earth's current biota photosynthetically store only $\sim0.1\%$ of the solar energy annually available, though the resulting NPP is still nine times larger than the human economy's TPES. This NPP also fixes ten times more atmospheric carbon than total annual anthropogenic CO$_2$ emissions but currently, $\sim96\%$ returns to the atmosphere on the same time scale without entering long-term storage. 

Over geological time scales, photosynthesis by ancient biota and geochemical processing of ancient NPP has created fossil fuels, which currently supply $\sim87\%$ of TPES. This provides hundreds of years' worth of ancient global NPP per year of modern consumption (largely because fossil-fuel genesis and extraction are together highly inefficient). Simultaneously, the economy appropriates $\sim25\%$ of current NPP, both directly through biomass harvest and indirectly through land conversion, and this level of HANPP is estimated to be significantly higher than the sustainable limit. Recent analyses strongly suggest that when sustainability criteria are considered, agro-biofuels can be expected to supply only a small fraction of future global energy demand. 

Crude oil is the single largest source of primary energy for the economy and worldwide conventional crude oil production rate may have already peaked (or plateaued). Some analysts are optimistic that unconventional oil will adequately offset declines in conventional oil. However, the high financial, energetic and environmental costs associated with unconventional oil bring into serious question its capacity to fill this role sustainably. Human interference in the Earth's carbon cycle, chiefly through emissions of CO$_2$ from fossil fuel combustion, is very likely to be disrupting the climate through global warming. An urgent transition to low-carbon energy systems is required, leaving the majority of remaining fossil fuel resources unproduced, if dangerous climate change is to be confidently averted.

Carbon-based, chemical fuels currently supply the majority of total final energy consumption, and the ongoing transition to electricity is gradual compared with the speed required for climate-change and peak-oil mitigation. Moreover, some economic sectors such as aviation and long-distance road transportation will likely depend on chemical fuels for the foreseeable future. Developing and deploying fungible, sustainable fuels is therefore critically important. Time scales of previous global energy transitions indicate that an unprecedented level of sociopolitical will and cooperation is required to accelerate the transition away from fossil fuels as much as possible. 

Sustainable, increased reliance on HANPP from current sources, to supply low-carbon fuels to the economy on globally significant scales, may be possible. This is likely to require an increase in Earth's total potential net primary productivity (NPP$_0$) through deployment of bioengineered and/or artificial photosynthetic systems on otherwise unproductive or low-productivity lands. 

A variety of solar fuels are already in use and two, bioethanol and biodiesel, are being produced on industrial scales, based on (unscalable) agricultural feedstocks. Solar fuels at earlier stages of development include biohydrogen and biohydrocarbons, the latter being directly fungible for petroleum-based fuels. Microalgal cultivation systems are rapidly emerging as solar fuel production systems with the potential to offer increased NPP$_0$ simultaneously with other economic and environmental services. However, comprehensive assessments indicate that significant technical innovation is still required to achieve economic and environmental feasibility. Artificial photosynthetic systems are seen as promising technologies for the longer-term and stand to benefit in the shorter-term from ongoing advances in understanding biological photosynthetic systems. 

A key challenge in developing high-productivity photosynthetic energy systems is optimising light-harvesting efficiency under a range of light environments including high-light conditions well suited to solar fuel production. This challenge is addressed by the chapters to follow in this thesis. In chapter \ref{chp:multianalysis} a multiscale framework for analysing light-harvesting energetics in photosynthetic energy systems is formulated based on complex systems theory. This reveals the importance of modelling interdependences between material composition, structure and energetics at each scale of organisation in such a system. Chapters \ref{chp:qeet} and \ref{chp:structure} present quantitative studies exploring these interdependences at the micro- and nano-scales of the thylakoid membrane. Chapter \ref{chp:qeet} analyses how the mechanisms of excitation energy transfer (EET) depend on structure in a chromophore network with the generic structural and energetic features of a thylakoid membrane. The focus is on how this dependence changes with length scale. Renormalisation theory is used to assess the largest length scale up to which quantum dynamical effects may mediate EET. Chapter \ref{chp:structure} then asks how the multiscale structure of the thylakoid membrane depends on its protein composition. The focus is on structural changes observed in so-called `antenna-mutant' strains of microalgae, in which the size and composition of the light-harvesting antennas have been genetically modified from wild-type species. Experimentally determined structures are presented for protein complexes and supercomplexes, and these are discussed in the context of a broader program integrating experimental and theoretical modelling to elucidate composition-structure relationships in the thylakoid. Finally, chapter \ref{chp:multiopt} extends the multiscale \\\textit{analysis} framework from chapter \ref{chp:multianalysis} to a framework for integrated, multiscale system \textit{optimisation}; this provides a conceptual basis for future work. Together, these investigations provide a framework for analysing light harvesting in photosynthetic energy systems using an integrated, multiscale approach, answer specific technical questions in the context of that framework, and open new avenues for future research. This constitutes progress toward the urgent goal of developing photosynthetic energy systems able to sustainably power the human economy on a globally significant scale.

\chapter[Multiscale systems analysis of light harvesting in photosynthetic energy systems]{Multiscale systems analysis of light harvesting in photosynthetic energy systems} \label{chp:multianalysis}
\chapquote{No one understands all the relationships that allow a tree to do what it does. That lack of knowledge is not surprising. It's easier to learn about a system's elements than about its interconnections.}{D.H. Meadows \cite{meadows2008}}


\section{Introduction}\label{sec:multiintro}
A key challenge in developing high-productivity photosynthetic energy systems is optimising light-harvesting efficiency under a range of light conditions. In particular, high-irradiance conditions are well suited to solar fuel production but saturate the light-harvesting kinetics in biological photosynthetic systems such as microalgae, which induces energy wastage through nonphotochemical quenching (NPQ) mechanisms \cite{ruban2012}. Simultaneously, $\sim$57\% of the energy incident in solar radiation falls outside the photosynthetically active spectrum and cannot be utilised by wild-type organisms (fig. \ref{fig:solarspectra}); this is another major source of inefficiency. Finally, uneven light distribution in cultivation systems means that some parts of the system experience excess light while others are light-limited. This seemingly paradoxical problem of simultaneously absorbing too much and too little radiation demands that photosynthetic energy systems be engineered to better utilise incident radiation. This requires optimal distribution through the system, not only of light but also of other resources involved in converting and storing the energy absorbed from light, such as CO$_2$, H$_2$O and nutrients, as well as effective disposal of wastes from photosynthesis such as O$_2$ and heat. 

Efforts to engineer microalgae and cultivation systems for efficient fuel production have so far typically aimed at improving the efficiency of some chosen part(s) of a specific model system. The implicit assumption is that subsystem improvements sum linearly to give global system improvements, so \textit{any} advance in \textit{any} subsystem will improve the system overall. However, this neglects the phenomenon of emergence, whereby the global properties of a system can differ from the sum of its components' properties when acting in isolation, by virtue of component interactions \cite{chalmers2006, anderson1972}. This means that it is possible for subsystem improvements to make no difference to, or actually impair global system performance if they do not form part of a holistic design strategy. For example, engineering chloroplasts to expand their light-absorption spectrum \cite{chen2011, gundlach2009} would increase their light-absorption efficiency under solar irradiation. However, under high-irradiance conditions where the photosynthetic kinetics are already saturated, increased light absorption may serve only to increase subsequent losses through NPQ, resulting in no net efficiency gain unless the downstream kinetics were also sped up commensurately. While sophisticated techniques are now available for engineering the components of photosynthetic systems at all scales from molecular to macroscopic, no methodology yet exists for integrated, whole-system design. Recent economic and environmental feasibility assessments at both local and global levels indicate that such a systems analysis methodology is needed (section \ref{sec:solfuelsys}) \cite{stephens2012, beal2012, malcata2011, stephens2010(1)}. 

This chapter introduces so-called `hierarchy theory', a multiscale, hierarchical systems analysis framework based on complex systems theory (section \ref{sec:hierarchy}), which has not previously been applied to the study of photosynthetic energy systems. This is proposed as a system-design tool complementary to the linear-process framework traditionally employed for analysing photosynthetic energetics (section \ref{sec:procpart}), on which previous engineering efforts have largely been based. Microalgal cultivation systems are the focus, though reference is also made to higher-plant systems throughout; the latter have been more extensively studied and show many structural and functional similarities to microalgae at intracellular scales. They also show multiscale, hierarchical organisation over larger scales, which offers insights into strategies for cooperative, multicellular photosynthesis. These natural systems provide a template from which the multiscale framework is abstracted and then mapped onto a generic microalgal system. However, it is emphasised that photosynthetic energy systems have different system-scale objectives and constraints from higher plants (efficient, cost-effective energy conversion rather than survival and reproduction in native ecosystems), so an optimal system configuration should not necessarily be expected to resemble a plant in detail. Rather, the goal of the plant-inspired multiscale approach is a generalised framework for rigorously linking global system-scale objectives and constraints with engineerable parameters at different scales in a coordinated way that is conducive to whole-system optimisation. In principle the framework developed may also be adapted for artificial photosynthetic systems; design of these systems has so far focussed mainly on microscopic components (section \ref{subsec:artps}) but, similarly to microalgae, commercialisation will require these components to co-operate within complex, macroscopic systems constrained by environmental and economic conditions. 

This chapter introduces hierarchy theory \textit{conceptually}, as a basis for system analysis. Chapter \ref{chp:qeet} then applies the theory quantitatively in a detailed study of excitation energy transfer dynamics across a range of scales in the thylakoid membrane. Chapter \ref{chp:structure} subsequently presents a hierarchical, multiscale study of thylakoid structure. Finally, chapter \ref{chp:multiopt} extends the hierarchical \textit{analysis} framework for integrated, hierarchical system \textit{optimisation}, as a speculative basis for future work.

\section{Linear-process analysis of photosynthetic systems}\label{sec:procpart}
\begin{figure*}
\centering
\includegraphics[angle=0,width=1\textwidth]{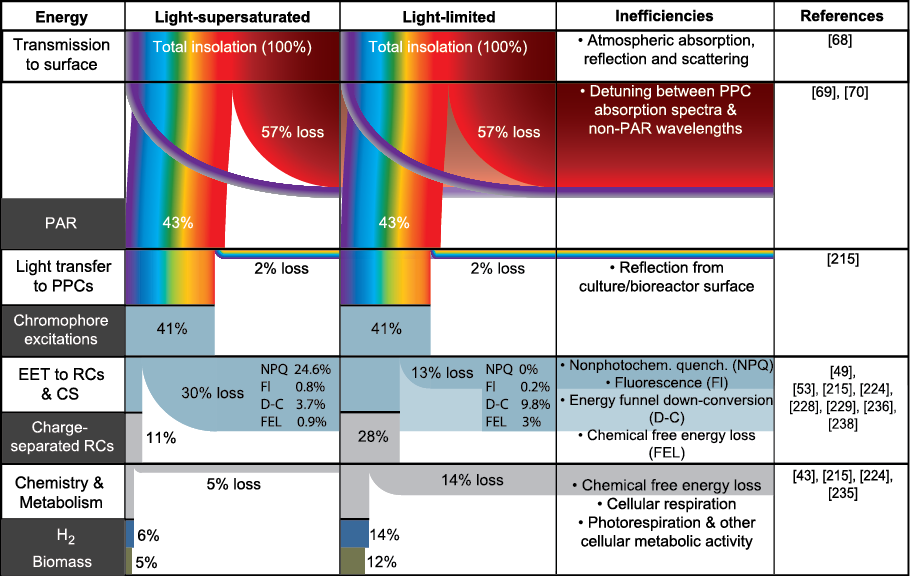}
\caption[Linear-process analysis of photosynthetic energetics in green microalgae]{\textbf{Linear-process analysis of photosynthetic energetics in wild-type green microalgae under light-limited and light-supersaturated conditions.} See text for explanation. Abbreviations: PAR -- photosynthetically active radiation; PPC -- pigment-protein complex; EET -- excitation energy transfer; RC -- reaction centre; CS -- charge separation; NPQ -- Nonphotochemical quenching; Fl -- fluorescence; D-C -- down-conversion; FEL -- free energy loss. \cite{nasa2012(1), nasa2012, alados1996, weyer2010, henriques2009, blankenship2002, jennings1993, melis2009, weyer2010, fan2009, ruban2012, melis1998, blankenship2011, zhu2010}}
\label{fig:procpart}
\end{figure*}

In a standard analysis of photosynthetic energetics the system is partitioned stepwise into subsystems that each complete one of the energy transfer and/or conversion steps spanning the free-energy gradient from incident light to biochemical products \cite{zhu2010, weyer2010, melis2009} (fig. \ref{fig:procpart}). Each process subsystem suffers inefficiencies, dissipating useful energy carriers and/or converting them into waste heat for subsequent dissipation. It is typically assumed that these inefficiencies add linearly to yield the net system inefficiency in light-to-chemical energy conversion \cite{zhu2010, weyer2010, melis2009} (fig. \ref{fig:procpart}). To illustrate the utility of the hierarchical approach developed in later sections, this section provides a careful linear-process analysis of photosynthetic energetics in a system cultivating wild-type green microalgae. Two light regimes are considered, based on current literature: light-limited conditions in which irradiance level limits the photosynthetic kinetics, and light-supersaturated conditions in which the irradiance level exceeds that required to saturate the photosynthetic kinetics. The proportion of incident light energy remaining after each process step is shown in figure \ref{fig:procpart}. 

\paragraph{Photosynthetically active radiation }
Forty-three percent of the energy in standard AM1.5 solar spectral irradiance is in photosynthetically active radiation (PAR) (section \ref{sec:earthsolar}) and is therefore taken as the maximum available for harvesting by natural microalgae. 

\paragraph{Light transfer to pigment-protein complexes} 
PAR incident on a microalgal cultivation system must be transferred to the cells and pigment-protein complexes (PPCs\nomenclature{PPC}{Pigment-protein complex}) within them. The first step is transmission into the aqueous culture. Weyer \textit{et al} \cite{weyer2010} calculate that for an open-pond system, 5$\%$ of PAR or 2$\%$ of total incident radiation is reflected from the culture surface over a day, leaving 41$\%$ of the total incident energy available for absorption within the culture. Assuming limited absorption and scattering by components other than the PPCs, almost all PAR entering the culture can be absorbed by the PPCs \cite{weyer2010}. In a photobioreactor however, light may encounter multiple culture-container-air interfaces, affecting reflection losses. This depends on the specific photobioreactor design so, for simplicity, reflection losses are here assumed for the `base case' of an open-pond system. 

\paragraph{Excitation energy transfer to photochemical reaction centres} 
Electronic excitations resulting from light absorption by the PPCs are transferred non-radiatively through the chromophore network of LHCI, LHCII and minor antenna proteins to reaction centres (RCs) in the PSI and PSII core complexes (section \ref{sec:photosynthesis} and chapter \ref{chp:qeet}). There the energy is used by P680 (in PSII) and P700 (in PSI) to drive primary charge separation reactions, generating the electron and proton flows of the photosynthetic light reactions (section \ref{subsec:lightreact}). Under both light-limited and light-supersaturated conditions, some energy loss to other excitation decay pathways such as fluorescence and internal conversion is unavoidable (section \ref{subsec:relax}). Henriques reports \cite{henriques2009} that a minimum of $\sim$0.5$\%$ of absorbed light energy is wasted as fluorescence under light-limited conditions, and a maximum of $\sim$5$\%$ under light-supersaturated conditions. Each charge-separation event requires an input of 1.80 eV (the energy contained in a 680 nm, red photon) in the case of PSII and 1.75 eV (700nm, red photon) for PSI \cite{blankenship2002}. Additional energy absorbed from shorter-wavelength photons is wasted as heat through internal conversion during excitation energy transfer (EET) (figs. \ref{fig:zscheme} and \ref{fig:enaqt}) \cite{blankenship2002, jennings1993}. This effective down-conversion of, for example, blue photons to red photons during EET through the `energy funnel' landscape of the antenna (fig. \ref{fig:enaqt}a [top]) is a source of \textit{energy} inefficiency even though the \textit{quantum} efficiency of EET approaches unity under light-limited conditions \cite{fan2009, melis2004}. The maximum photon-utilization efficiency can be estimated as the ratio of the average energy used for charge separation (1.78 eV, red) to the average absorbed photon energy (2.34 eV, green), giving 76$\%$ \cite{melis2009, weyer2010}. Further, thermodynamically dictated losses occur during charge separation itself \cite{kern2007}, leaving on average 1.6 eV stored per primary charge separation event. This reduces the estimated maximum photon-utilization efficiency to 68$\%$ and further small losses are incurred by occasional charge recombination in the RC \cite{fan2009}. Overall therefore, photon utilization losses account for $\sim$13$\%$ of the incident solar energy under light-limited conditions (fig. \ref{fig:procpart}). 

Under light-supersaturated conditions the light-harvesting complexes (LHCs) absorb energy from radiation faster than the RCs can utilise the resulting excitations (the RCs are `closed'). To avoid photoinhibition (oxidative damage at the RCs \cite{ruban2012}), excess excitations are dissipated as heat through NPQ mechanisms \cite{ruban2012}. Melis \textit{et al} \cite{melis2009, melis1998} estimate that in a typical high-light environment a microalgal mass culture or dense plant foliage can over-absorb and dissipate through NPQ approximately 60$\%$ of the daily irradiance, with cells at the directly illuminated surface wasting over 80$\%$ of absorbed irradiance through NPQ. Daily, this is equivalent to $\sim$25$\%$ of total solar energy lost \emph{via} NPQ (fig. \ref{fig:procpart}). The precise amount of wastage through various non-photochemical relaxation pathways (section \ref{subsec:relax}) depends on species, acclimation state, irradiance and spectral quality of incident light, temperature, and system macrostructure. Accordingly, a broad range of estimates exists in the literature for absorbed photon utilization efficiency under physiological, light-supersaturated conditions; Weyer \textit{et al} \cite{weyer2010} state a range from 10--30$\%$. In figure \ref{fig:procpart}, 27$\%$ (based on daily NPQ losses of 60$\%$ of absorbed light) has been chosen as representative.

\paragraph{Chemistry and metabolism}
The balanced equation of CO$_2$ fixation \emph{via} the Calvin-Benson cycle (section \ref{sec:photosynthesis}) is \cite{stryer1995}: 
	\beqn
		6~\text{CO}_2 + 18~\text{ATP} + 12~\text{NADPH} + 12~\text{H}_2\text{O} \xrightarrow~\text{C}_6\text{H}_{12}\text{O}_6 + 18~\text{ADP} + 18~\text{P}_\text{i} + 12~\text{NADP}^+ + 6~\text{H}^+.\nonumber
   	\eeqn
Three ATP and two NADPH molecules are required to fix one molecule of CO$_2$ into a hexose sugar such as glucose. The reduction of NADP$^+$ is a two-electron process. To transfer two electrons to NADP$^+$ \emph{via} linear electron transport requires four photons (two to transfer two electrons from water through PSII and two more to transfer them through PSI). Ideally, eight photons would therefore be required to produce the two NADPH molecules needed to fix one molecule of CO$_2$ into one hexose sugar molecule. The three ATP molecules are produced simultaneously. Fixation of one molecule of CO$_2$ into a hexose sugar would therefore require eight photons. The measured number of photons per O$_2$ evolved or CO$_2$ fixed is $\sim$9.5 \cite{melis2009}. Thus, 84$\%$ of the energy used to drive charge separation in PSII and PSI is stored as carbohydrates \cite{melis2009}. 

Weyer \textit{et al} \cite{weyer2010} quote literature values ranging from 11--89\% for the fraction of energy captured by photosynthesis that algal cells require for respiration and housekeeping. Following Weyer \textit{et al} \cite{weyer2010}, 50\% is chosen as a representative value (fig. \ref{fig:procpart}). This effectively raises the number of photons required per CO$_2$ fixed to $\sim$14. 

\paragraph{Hydrogen}
The H$_2$-producing hydrogenase (HydA) is tightly coupled to the photosynthetic electron transport chain (fig. \ref{fig:zscheme}). This reduces energy losses associated with the more extensive biochemical pathways involved in the synthesis of carbohydrates and lipids, resulting in higher theoretical light-to-biofuel conversion efficiencies. The metabolic load of maintaining the cell is of course still required. The theoretical upper limit of efficiency in microalgal H$_2$ production is estimated to be 12--14$\%$ \cite{melis2009}.

\paragraph{Carbohydrate and oil-rich biomass}
Due to the different energy requirements of differing biochemical pathways, the number of photons required to store one mole of carbon in a given molecule varies. The more energy-dense the final biomass product, the higher the number of photons required to produce it. Thus, while low-lipid biomass typically needs $\sim$14 photons, $\sim$20 photons are typically needed to achieve high lipid content. Overall light-to-biomass production efficiencies are calculated to be $\sim$12$\%$ and $\sim$5$\%$ respectively for the light-limited and light-supersaturated conditions shown in figure \ref{fig:procpart} \cite{blankenship2011, zhu2010, weyer2010, melis2009}. 

\paragraph{Key efficiency limitations }
Figure \ref{fig:procpart} shows that the greatest inefficiencies in the photosynthetic process are nonabsorption of incident wavelengths falling outside the PAR spectrum and dissipation of unusable electronic excitations through NPQ. Under light-limited conditions, nonabsorption of non-PAR is the dominant source of inefficiency. Under light-supersaturated conditions, NPQ dominates.

\subsection{Limitations of linear-process analysis as a design tool for photosynthetic energy systems}
The challenge of designing maximally efficient light harvesting is complicated by the fact that light-supersaturated and light-limited conditions coexist in real systems in a typical high-light environment. Light-supersaturated conditions exist near to the illuminated surface(s), and light-limited conditions in shaded regions. The analysis above demonstrates the importance of matching PAR irradiance level to the level that saturates the photosynthetic kinetics; in both light regimes a mismatch between these rates underlies the dominant source of inefficiency. Maximally efficient light harvesting would be achieved if every photosynthetically active system component received PAR in balance with its maximum-achievable productivity, determined only by the thermodynamic limits of its internal machinery (assuming that other inputs such as CO$_2$ and H$_2$O are not rate-limiting) \cite{terashima1995}.  

Accordingly, strategies for improving system efficiency have largely focussed on achieving more equitable irradiance distribution throughout the system. However, the standard linear-process analysis gives no account of how irradiance distribution depends simultaneously on parameters over a range of scales in the system (e.g. optical properties of pigments, proteins, cells, photobioreactor components and the overall photobioreactor), nor for how those parameters depend on each other. Rather, the highly simplifed account of light transfer to PPCs given above in section \ref{sec:procpart}, which assumes a particular, simple system geometry and refers to optical properties at only one scale in the system, is typical of accounts found in the literature \cite{weyer2010, zhu2010, melis2009}. Hence, strategies to improve light distribution have tended to focus at a single scale (or a relatively small subset of the scales of organisation present in the system), tuning compositional and structural parameters such as LHC antenna protein composition \cite{ort2011, murphy2011, mussgnug2007} (chapter \ref{chp:structure}), cell culture density \cite{dillschneider2013, posten2009, ugwu2008}, and photobioreactor structure and surface properties \cite{dillschneider2013, posten2009, ugwu2008, gordon2002}. 

Moreover, engineering efforts to date have typically begun from a model algal species and/or photobioreactor design and improved system performance incrementally by tuning only a small number of parameters. Due to this strong basis on existing systems and focus at particular scales, there exists no general theory for optimal system design which rigorously links global system-scale objectives and constraints (e.g. maximally efficient light harvesting at minimal financial cost) with engineerable parameters at different scales in a coordinated way (e.g. LHC antenna composition, cell density, photobioreactor structure and surface properties). Ideally, such a theory would allow the designer to start from a minimal number of fixed parameters and let system configuration emerge organically from an optimisation algorithm. The result may be beyond what is reachable based on intuition, starting from an existing model. Obtaining such a theory is a major challenge and it is proposed here that analysing photosynthetic systems using complex systems theory may offer progress.

\subsection{Complexity in photosynthetic systems}\label{subsec:comppho}
\paragraph{Multiscale complexity}
Photosynthetic light harvesting involves mechanisms of energy transfer and conversion spanning length scales from individual chromophores (pigment molecules) ($10^{-9}-10^{-10}$ m) to an overall photobioreactor array or plant stand ($>10^1$ m), and time scales from energy-transfer processes at the molecular level ($10^{-14}$ s) to annual climatic cycles ($10^7$ s). At progressively larger scales of organisation in the system, properties emerge from interactions between components and these properties are often not predictable only from properties of the smaller-scale components \cite{korn2005, anderson1972} (e.g. the shape of a leaf cannot be predicted only from the shapes of its constituent cells because the cells can be arranged in many ways [section \ref{sec:hierarchy}]). As a result, the process(es) within each process subsystem (e.g. `light transfer to the PPCs') can scale nonlinearly such that different models are needed at different scales (section \ref{subsec:suggestedhierarchy}) \cite{jarvis1995, li2004}. For example, so-called `big-leaf' models of leaf-canopy photosynthesis, which treat an overall canopy as a scaled-up single leaf, are found to predict canopy productivity inaccurately \cite{depury1997}. This is due to the difference between time-averaged and instantaneous distributions of absorbed irradiance at the canopy scale, caused by penetration of sunflecks and the range of leaf angles in canopies \cite{depury1997}. That is, although the leaves within a canopy may be considered analogous to chloroplasts within a leaf (fig. \ref{fig:multiscale}b,c), the emergence of different properties at the canopy scale causes irradiance to be distributed differently between components (leaves/chloroplasts) at each of the two scales, so productivity models do not scale linearly between them. Different radiative transfer models again are needed at intra-chloroplast scales, where subwavelength structures are significant \cite{margalit2010, meszena1994, paillotin1993} (section \ref{subsec:suggestedhierarchy}). 

Since processes spanning larger scales must also span smaller scales (e.g. radiation interacting with a photobioreactor also must interact with at least one of its structural components [fig. \ref{fig:multiscale}a,b]), and smaller-scale processes must exist against a background of larger-scale processes (each component operates within the light environment transmitted to it by the overall photobioreactor), altering parameters at a particular scale can have complex effects on the energetics over multiple scales (e.g. applying an anti-reflective coating to the surface of an entire photobioreactor may increase the photosynthetic efficiency of individual components in light-limited regions while simultaneously decreasing the overall system efficiency because the gains in light-limited regions are more-than compensated by increased NPQ losses in light-supersaturated regions). Therefore, parameters at different scales should be optimised collaboratively. For example, optical properties of the overall photobioreactor should ideally be optimised for total absorption of PAR \textit{over the scale of the photobioreactor} while also distributing irradiance equitably among its components; optical properties of each component should simultaneously be optimised to absorb irradiance in balance with what it can utilise \textit{while also helping to distribute excess irradiance to other components}. These optimisations must be performed collaboratively because the optical properties of the overall photobioreactor depend on the optical properties of its components. 

\paragraph{Multiprocess complexity}
In addition to complexities between scales, parameters at a given scale affect multiple energetic processes at that scale (as well as at other scales) simultaneously. For example, LHCs are highly multifunctional, serving roles not only in transfer and absorption of light over multiple scales \cite{pilon2011, terashima1995, meszena1994, richter2007} but also in EET \cite{renger2009, lambrev2011}, NPQ \cite{ruban2012}, heat transfer \cite{kochubey2010, park2009, gulbinas2006, lervik2010} and electron transport \cite{kirchhoff2008}. Therefore, optimising parameters such as light-harvesting antenna composition \textit{only} for light transfer through the cell culture, can also affect other energetic processes, generating complex effects on system performance (e.g. the full effects of engineering light-harvesting antenna composition are not yet known [chapter \ref{chp:structure}]). Parameters at each scale therefore should be optimised for multiple energetic processes simultaneously, such that a compromise between the needs of different processes is achieved.  
 
 According to Li \textit{et al} \cite{li2004}, complex systems have the following features:
\begin{itemize}
\item consist of many subsystems interacting with one another;
\item show hierarchical, multiscale structure;  
\item are always dynamic, stabilised by exchanging energy, matter and information with their environment;
\item are governed by at least two dominant mechanisms which compromise with each other.
\end{itemize}

As explained above, photosynthetic systems show all of these features. Overall, a multiscale, multiprocess design framework is needed, which rigorously balances reductionism with holism, linking system-scale objectives and constraints to engineerable parameters within subsystems across the spectrum of smaller scales \cite{ge2007, li2004, coppens2012}. This poses a considerable theoretical challenge, demanding a balance between analytical rigor and practicality.

\section[Multiscale, hierarchical analysis of photosynthetic systems]{Multiscale, hierarchical analysis of photosynthetic systems}\label{sec:hierarchy} 

Multiscale, hierarchical network structure is a common feature of complex systems \cite{ahn2010, clauset2008, sales2007, li2004} and multiscale analytical methods have arisen from studies of diverse systems within the social \cite{galam2008}, biological \cite{wilkinson2009, woods2008, lischke2007, wu2002}, chemical \cite{salciccioli2011, ge2007, li2004, li2003}, physical \cite{albert2002, aoki2000, chen1996, wilson1975} and informational sciences \cite{moses2008, abry2002}. Previous studies of natural photosynthetic systems have accounted (sometimes implicitly) for their multiscale complexity to various levels of sophistication. A common approach has been to use a mechanistic light transfer model at larger scales (such as the radiative transfer equation or analytical solution thereof, such as the Beer-Lambert law or two-flux approximation) across a cell culture or leaf canopy, coupled to an empirical model of photosynthetic productivity at cellular scales \cite{murphy2011, packer2011, molina1994, kim2001, anisimov1997}. Such models have successfully described experiments linking environmental parameters, such as irradiance, CO$_2$ concentration and temperature, with biomass productivity in terms of parameters such as cell density and pigment composition. Explicitly multiscale studies of microalgal systems have bridged phenomenological models of processes at well-separated time scales (such as EET, biomass production, and photoinhibition\footnote{Oxidative damage to the PSII reaction centre, causing reduced productivity \cite{ruban2012}.}), aiming to inform control algorithms for biomass production under varying environmental conditions \cite{fernandez2012, hartmann2012, papacek2010, celikovsky2010}. All studies of these types neglect mechanical details underlying their coarse-grained phenomenological and/or empirical models and so are limited in the guidance that they can provide to increasingly sophisticated engineering capabilities. Consequently, more detailed, mechanistic models are needed in order to resolve causal relations between system parameters and performance across scales. 

Ecological studies of higher-plant systems have pursued this level of mechanistic, multiscale analysis. It has been observed that such systems comprise hierarchies of recursively nested structural and functional components, spanning scales from chloroplast to landscape (fig. \ref{fig:multiscale}a--c, `Higher plants' column). While some authors have described energetics at different scales largely independently \cite{smith2004, terashima1995}, others have quantified scaling trends in system parameters and dynamics, yielding insights into complex ecological systems \cite{lischke2007, wu2002, vangardingen1997, jarvis1995, leuning1995, kull1995, harley1995, baldocchi1995}. One example is the failure of big-leaf models to accurately describe leaf-canopy photosynthesis, described in section \ref{subsec:comppho}. Such findings have implications for microalgal photobioreactor design but there have so far been no comparable multiscale analyses accounting for hierarchical structure in microalgal cultivation systems (comparing for example irradiance distribution between structural components of the photobioreactor with irradiance distribution between cells within the cell culture [fig. \ref{fig:multiscale}b,c]). Nor have there been, to the author's knowledge, any such studies in higher plants which have extended their scaling hierarchies to intra-chloroplast scales. 

Moreover, previous scaling analyses have focussed on \textit{describing} natural systems rather than \textit{engineering} optimal systems. The latter goal permits more freedom in modelling because system configuration can emerge organically from the optimisation algorithm. This is also arguably simpler than attempting to characterise the full complexity of natural systems, which are optimised not merely for efficient solar energy storage but rather for the more complex problem of survival and reproduction in their native ecosystems. Therefore, rather than viewing natural photosynthetic systems as templates for detailed biomimicry in engineered systems, a more fruitful approach may be to view them as sources of inspiration for  general principles relating parameters and dynamics within and between different scales. These principles may then provide a basis for general, theoretical models of photosynthetic systems, which can be optimised across the scalar hierarchy under system-scale objectives and constraints that may differ from those of naturally evolved systems. Computing power, modelling and optimisation algorithms for complex systems, as well as knowledge of structure and energetics in photosynthesis, have all progressed in recent years such that mechanistic, multiscale-optimal design of photosynthetic energy systems \textit{in silico} is becoming a realistic prospect. Considerable technical development will be required towards this end, particularly in balancing explanatory power with computational tractability. Hierarchy theory, which constitutes a simple and intuitive formulation of multiscale, hierarchical analysis, is now presented as a possible theoretical foundation for such development. The presentation is qualitative; the reader interested in the mathematical basis of hierarchy theory is directed to references \cite{auger2003}, \cite{lischke2007}, and others cited below. Chapter \ref{chp:qeet} presents a mathematically formulated hierarchical model based on renormalisation theory, which may be thought of as a particular formulation of hierarchy theory (see below).

\subsection{Hierarchy theory}\label{subsec:hierarchy}

\begin{figure*}
\centering
\includegraphics[angle=0,width=1\textwidth]{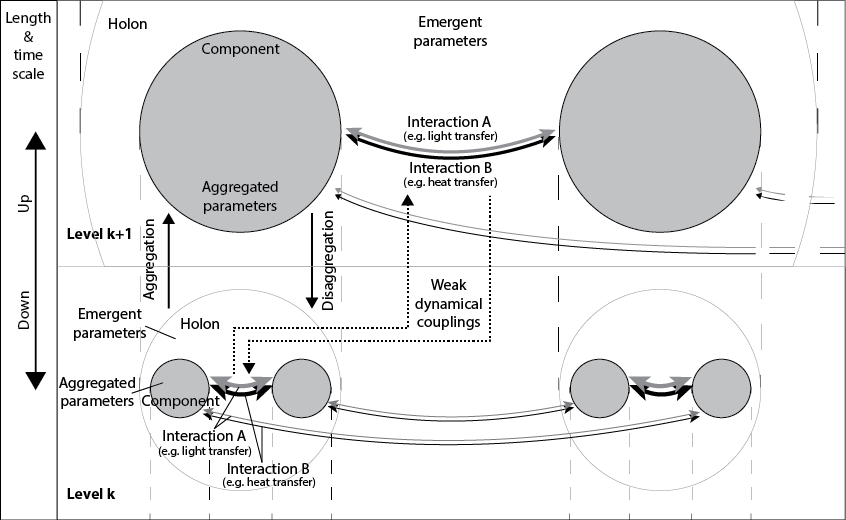}
\caption[Schematic summary of hierarchy theory]{\textbf{Schematic summary of hierarchy theory.} A complex dynamical system is partitioned into a hierarchy of recursively nested subsystems with well-separated characteristic length and time scales to simplify system analysis. At each level in the hierarchy an encompassing subsystem or `holon' contains `component' subsystems, which are simultaneously holons at the next, downscale level. Each subsystem represents a repeated functional unit that interacts \emph{via} strong and weak couplings with other subsystems at the same and other levels respectively. See text for detailed explanation. Figure adapted from \cite{lischke2007}.}
\label{fig:hierarchy}
\end{figure*}

Hierarchy theory arose within general systems theory and can be applied to any system showing differences in scale between levels of organisation \cite{lischke2007, auger2003, salthe1991, ahl1996}. The theory partitions a complex dynamical system into a hierarchy of recursively nested subsystems with well-separated characteristic length and time scales in order to simplify system analysis. The multiscale analysis of the hierarchy is constructed by studying internal properties and dynamics of each scale subsystem independently, as well as couplings between scale subsystems. 

\paragraph{Nomenclature}
Nomenclature used for different scale subsystems and their interactions varies in the literature. Here, \textit{upscale} and \textit{downscale} subsystems are referred to generically, where the scales need not, though may, be adjacent. In a pair of scale-adjacent subsystems, the convention is adopted here of calling the upscale subsystem the \textit{holon} and the downscale, nested subsystem a \textit{component} (fig. \ref{fig:hierarchy} and fig. \ref{fig:generic}a). Subsystem partitions (theoretical boundaries chosen by the analyst) divide a holon at level $k+1$ into components at level $k+1$, which are simultaneously holons at level $k$. The partitions at level $k+1$ are located between clusters of strongly interacting components at level $k$ such that intra- and inter-cluster pairwise component interaction strengths are scale-separated. This means that components interact relatively strongly within a holon and relatively weakly between holons. A holon at any scale may also contain subsystems of other types at the component scale (nested within the holon but not within the components). In a well-formed hierarchy, holons and components are well separated at all scales.

\paragraph{Dynamical couplings between scales}
Couplings between holon dynamics and component dynamics are generally weak due to their separated characteristic time scales; a scale-adjacent pair is a `slow-fast' system. When scales are well separated, upscale dynamics are effectively stationary compared with downscale dynamics, which in turn are comparatively so rapid that only equilibrium values (or time-averaged values, where the dynamics are non-equilibrium; i.e. driven) of downscale dynamical variables are significant to upscale subsystems. For example, light transmission through a 10 \textup{$\mu$m}-thick cell (fig. \ref{fig:multiscale}c,d) spans roughly one ten-thousandth of the time needed for transmission through a 10 \textup{cm}-thick `phytoelement' (photobioreactor tube or panel element; figure \ref{fig:multiscale}b). Accordingly, intracellular optical processes generally equilibrate (e.g. through absorption by chromophores) before any `noticable' (to the cell) changes take place in the incident light field due to phytoelement-scale dynamics in the cell distribution.

Consequently, a well-formed hierarchy is said to be `near-decomposable' and formal simplifications can be made at low explanatory cost in modelling couplings between scales. For example, perturbation theory can be used to approximate upscale and downscale dynamical variables as constants, which act as boundary conditions for focal-scale dynamics \cite{auger2003, salthe1991}. More sophisticated methods have been developed to deal with cases where `scale-adjacent' subsystems are not properly scale separated and regular perturbation theory is inadequate. Relatively simple models may be analytically tractable using singular pertubation theory and/or renormalisation group theory \cite{cronin1999, chen1996}, and recourse can be made to various numerical methods for more complex models \cite{keil2012, salciccioli2011, jaworski2011, king2010, chen2010, steinhauser2007, vlachos2005, weinan2004, brandt2002}.

\paragraph{Transforming models between scales}
If a dynamical model has not been established for a subsystem at a given scale, it is often possible to derive an approximate model through `upscaling' and/or `downscaling' transformations of models and parameters from other scales. For in-depth discussion of formal up/downscaling techniques see \cite{lischke2007} and \cite{jarvis1995}. In systems which have previously been modelled at different scales, so-called `heuristic upscaling' can be used to link the established models between scales \cite{lischke2007}. In such cases, interscale parameter transformations must be established. 

\paragraph{Transforming parameters between scales} 
Parameter upscaling transformations may be called constitutive relations or aggregations. An aggregation maps a parameter distribution over components (e.g. chlorophyll content of individual cells within a leaf) to a single parameter of the holon (e.g. total leaf chlorophyll content). The aggregation can take different forms, though it is often a simple sum (as in the chlorophyll example) or average over subsystems \cite{lischke2007, chen2010, jarvis1995}. Parameter aggregation links upscale and downscale models consistently, at the cost of some explanatory power at the larger scale due to information loss intrinsic to the aggregation. Its form should therefore be chosen to balance simplification with retention of explanatory power according to the goals of a particular model \cite{lischke2007}. 

The opposite of aggregation is referred to here as parameter disaggregation: the mapping of a single, upscale parameter (e.g. leaf chlorophyll content) to a distribution of downscale parameters (e.g. cellular chlorophyll content). In general there are multiple disaggregations that will invert a given aggregation; different distributions of cellular chlorophyll content will sum to the same total leaf chlorophyll content. This can allow some freedom of choice, though in some cases this is reduced by the need for consistency with parameters at yet smaller scales. 

\paragraph{Parameter emergence} 
In addition to parameters resulting from aggregations and/or disaggregations of down/upscale parameters, a dynamical model at a given scale can also depend on parameters which emerge uniquely at that scale. For example, while total leaf chlorophyll content can be aggregated over cellular chlorophyll contents, leaf shape cannot be aggregated over cell shapes because the cells can be arranged in many ways. Leaf shape is an emergent parameter at the leaf scale, resulting from intercellular interactions. 

\paragraph{Explanatory power}
In a well-formed hierarchical system, the comparatively simple and less computationally expensive analysis provided by hierarchy theory can predict system behaviour with accuracy approaching that of a fully detailed all-in-one analysis \cite{dewit2011, xiong2010, lischke2007, auger2003}. Significant divergence between these models may indicate poorly chosen partitioning of the hierarchy such that subsystems are not well separated, and/or poor choices of subsystem dynamical models or parameter aggregations/disaggregations between scales. A multiscale, hierarchical model also has the advantage that it can be compared against empirical models at multiple scales simultaneously. This can be done to validate the hierarchical model and/or to compare empirical data between scales \cite{lischke2007}.

\begin{FPfigure}
\hspace*{0.34cm}
\includegraphics[angle=0,width=0.94\textwidth]{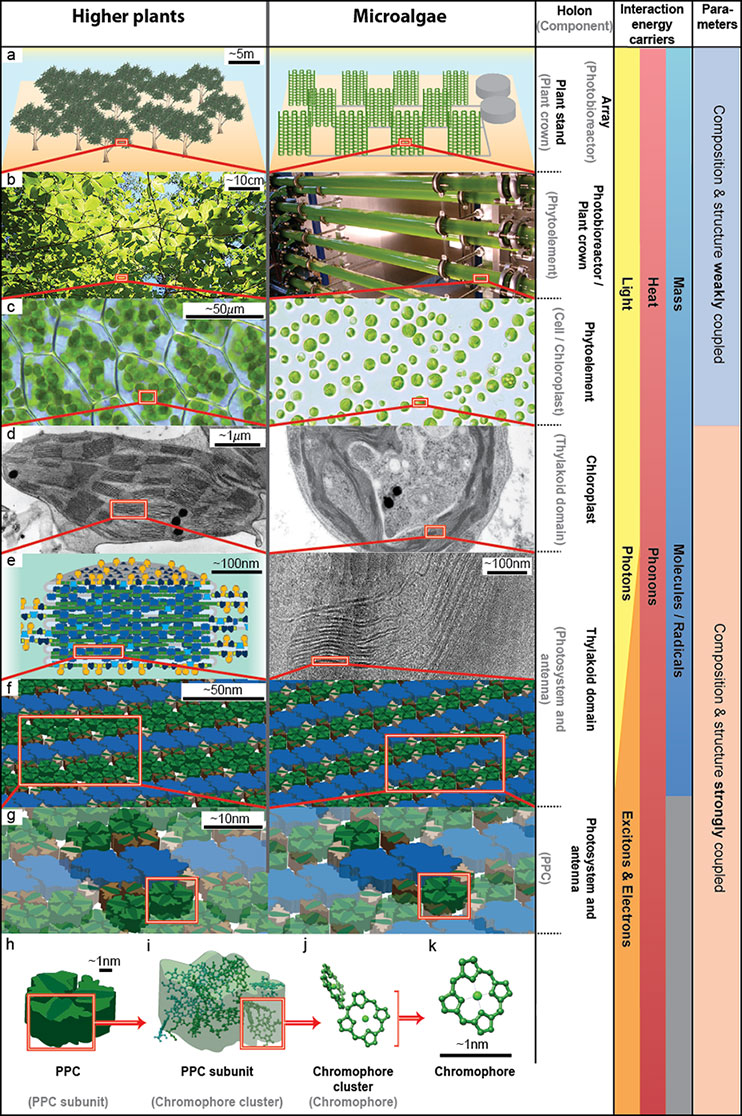}
\caption[Multiscale, hierarchical partitioning of photosynthetic systems]{\textbf{Multiscale, hierarchical partitioning of higher plant and microalgal cultivation systems.} Each row shows components within the holon, energy carriers that mediate transfer between components, and whether the composition and structure of the system are strongly or weakly interdependent at that scale. The particular structures shown at each scale a--g and j each represent a range of possible structures. \textbf{a}, (left) Schematic plant stand; (right) schematic photobioreactor array. \textbf{b}, (left) Higher-plant phytoelements; (right) photobioreactor phytoelements. Imaged reproduced from \cite{solarbiofuels2013}. \textbf{c}, (left) Higher-plant chloroplasts within cells (Image $\copyright$ Kristian Peters 2006. Reproduced pursuant to Creative Commons licence.); (right) microalgal chloroplasts within cells (image courtesy of G. Jakob, used with permission). \textbf{d}, (left) Transmission electron micrograph of a sugarcane chloroplast. Densely stacked grana (dark) regions of the thylakoid membrane are visible (image courtesy of R. Birch, used with permission); (right) transmission electron micrograph of cell from the green microalga large-antenna mutant strain \textit{stm3}. Densely stacked pseudograna (dark) regions are visible in the chloroplast (image courtesy of E. Knauth, used with permission). \textbf{e}, (left) Schematic higher-plant `thylakoid domain', comprising granum and surrounding stroma lamellae; (right) transmission electron micrograph of a typical `thylakoid domain' from \textit{stm3}, comprising a pseudogranum and adjacent stroma lamellae (image courtesy of E. Knauth, used with permission). \textbf{f}, Schematic semicrystalline arrays of PSII(blue)-LHCII(green, brown) supercomplexes within the thylakoid. \textbf{g}, (left) C$_2$S$_2$M$_2$-type PSII-LHCII supercomplex comprising reaction centre core dimer and peripheral LHCs forming a light-harvesting antenna; (right) C$_2$S$_2$-type PSII-LHCII supercomplex comprising reaction centre core dimer and peripheral LHCs forming a light-harvesting antenna functional domain. Nomenclature for different PSII-LHCII supercomplex types is explained in section \ref{subsec:spathy}. Different PSII-LHCII types with different antenna sizes can coexist in a membrane, and the connectivity between supercomplexes can also vary depending on ultrastructural conformation of the membrane. \textbf{h}, Trimeric LHCII: one type of pigment-protein complex (PPC). \textbf{i}, Monomeric subunit of LHCII, with chlorophyll crystal structure overlaid. This complex binds 14 chlorophyll (8 chlorophyll-\textit{a}, 6 chlorophyll-\textit{b}) and 4 carotenoid chromophores. \textbf{j}, Strongly coupled dimer (\textit{a}613-\textit{a}614) of chlorophyll-\textit{a} chromophores. \textbf{k}, Chlorophyll-\textit{a} chromophore.}
\label{fig:multiscale}
\end{FPfigure}

\subsection{A suggested hierarchical partitioning for a generic green-microalgal cultivation system}\label{subsec:suggestedhierarchy}

Figure \ref{fig:multiscale} shows a multiscale, hierarchical system partitioning, chosen heuristically for a generic microalgal system cultivating green microalgae. The heuristic used is that each component (e.g. cell -- fig. \ref{fig:multiscale}c) within a holon (phytoelement -- fig. \ref{fig:multiscale}b) is a repeated functional unit that exchanges energy through multiple mechanisms (light transfer, heat transfer, mass [CO$_2$, H$_2$O, O$_2$] transfer) with other components, both inside the holon and in neighbouring holons. In accordance with hierarchy theory, the strengths of intra- and inter-holon interactions are assumed to be scale-separated, and the parameters affecting them at each scale are proposed to include both emergent parameters and parameters aggregated from parameters of downscale components. 

`Equivalent' levels of organization in a higher-plant system are shown in parallel, though the latter system arguably has extra levels of organisation (e.g. multiple chloroplasts per cell, multiple levels of clustering in the plant crown) that have been neglected here for consistency with the microalgal system. Identifying similarities between microalgal and higher-plant scale subsystems allows the more extensive literature describing the latter to be leveraged. It is stressed that the partitioning chosen here is heuristic only; a precisely defined partitioning would require component interaction strengths to be quantified throughout the system for each of their various interactions, which would be a nontrivial undertaking. However, in chapter \ref{chp:qeet} this is done for the electrodynamic interactions that underly EET across a hierarchy of pigment-binding components within the thylakoid membrane: chromophores (figure \ref{fig:multiscale}k), chromophore clusters within PPC subunits (j), PPC subunits (i), PPCs (h), and aggregates and supercomplexes of PPCs (g,f). Beyond this collection of scale subsystems within an overall photosynthetic system, recently developed algorithms for quantifying the hierarchical structures of complex networks \cite{ahn2010, clauset2008, sales2007}, and for choosing optimal partitioning for system design optimisation \cite{allison2009}, may be helpful in future work for improving the heuristic partitioning suggested in the present chapter. 

Each row in figure \ref{fig:multiscale} shows components within the holon, energy carriers that mediate transfer between components, and whether the composition and structure of the system are strongly or weakly interdependent at that scale. At larger scales the interaction energy carriers are identified in their continuum limits as light, heat and mass. At sub-microscopic scales, reference is made instead to photons, phonons and molecules/radicals, to account for the signficance of quantum mechanical effects at these scales. Excitons, free electrons and phonons are the significant energy carriers between components at scales smaller than photosynthetically active wavelengths of light. 

\begin{figure*}
\centering
\includegraphics[angle=0,width=1\textwidth]{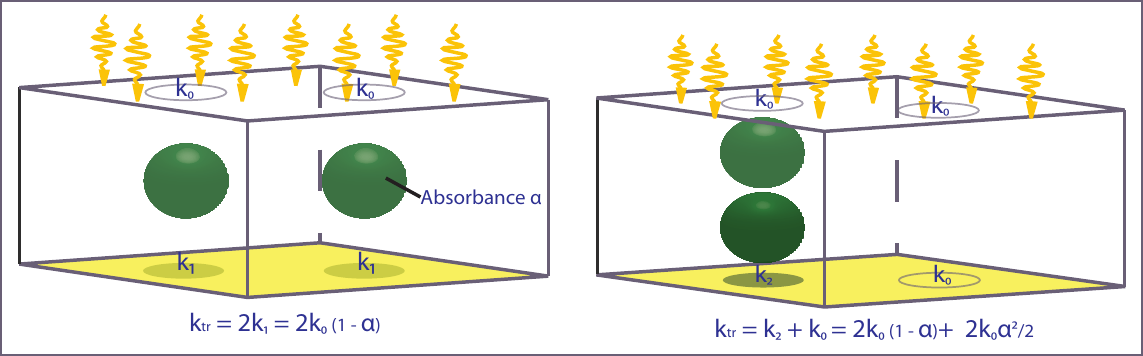}
\caption[Composition, structure and irradiance distribution: the package effect]{\textbf{Composition, structure and irradiance distribution: the package effect.} A volume in which absorbers (each of absorbance $\alpha$) are `packaged' densely to induce mutual shading transmits more irradiance $k_{tr}$ than a volume in which absorbers are more evenly dispersed though at the same overall concentration. This effect is thought to produce more equitable irradiance distribution between components at multiple scales in photosynthetic systems.}
\label{fig:package}
\end{figure*}

Distribution of energy carriers (e.g. light) between components (e.g. cells -- fig. \ref{fig:multiscale}c) within a given scale subsystem (phytoelement -- fig. \ref{fig:multiscale}b) depends on the subsystem's composition (e.g. chemical composition of the aqueous medium and pigment composition of the cells) and structure (e.g. distributions of cell size, shape, orientation and position throughout the phytoelement). Dependences on both composition and structure must be understood in order to model and optimise the energetics at each scale. For example, irradiance distribution in a microalgal cell culture is known to diverge from that of a pigment solution of equal pigment concentration\footnote{Here, the concentrations compared are volumetric averages over sample volumes large when compared with individual cells in the cell culture.} (the latter being described by the simple Beer-Lambert law: exponential decay of irradiance with depth), for two main reasons. First, by the so-called `package effect' (fig. \ref{fig:package}) the per-pigment absorptivity of a cell suspension is reduced compared with the equivalent pigment solution, due to enhanced mutual shading of tightly packed intracellular pigments (and therefore a net reduction in shading overall, per unit of cross-sectional area of the culture\footnote{Assuming the areal unit is large when compared with a cell.}) \cite{duyens1956, dubinsky1986}. Second, anisotropic scattering of light by algal cells, which depends on their shapes and sizes \cite{pottier2005}, cannot be accommodated by the Beer-Lambert law, which assumes no scattering. 

Irradiance distribution is known to be similarly affected by composition and structure across the spectrum of scales in higher-plant systems. The package effect operates in leaf canopies (fig. \ref{fig:multiscale}a,b) \cite{anisimov1993, myneni1991, ross1981} and within individual leaves (fig. \ref{fig:multiscale}c) \cite{terashima1995}, giving rise to more equitable irradiance distribution between leaves and cells respectively at these two scales, helping to increase photosynthetic productivity across the depth of the canopy/leaf compared with a homogeneous absorber of equal pigment concentration \cite{anisimov1997, terashima1995}. Melis \textit{et al} \cite{melis1987} find that in spinach chloroplasts, 65\% of the chlorophyll is concentrated within the densely stacked `grana' regions of the thylakoid (fig. \ref{fig:multiscale}d,e), and this generates an optical package effect, reducing absorption within the grana (and so in PSII) by 15--20\% compared with the stroma (PSI). Further, there is evidence that at yet smaller, sub-optical-wavelength scales, the grana (fig. \ref{fig:multiscale}e) may exploit optical interference effects to distribute irradiance more equitably between membrane layers (depending on layer thickness, spacing, pigment composition and orientation of the stack overall, relative to the propagation vector of incident light) \cite{meszena1994, paillotin1993}. However, only a small number of studies have explored this hypothesis -- which is effectively a claim that the thylakoid membrane can act as an optical metamaterial -- and these have used only relatively simple optical models; this invites further studies using more sophisticated methods.

These studies suggest that system composition and structure can be optimised for resource distribution between components at a given scale, and at multiple scales concurrently. In the case of irradiance distribution in higher-plant systems, a particular strategy -- the optical package effect -- is deployed at multiple scales existing within the same physical regime (scales larger than the wavelengths of PAR). This suggests a strategy for multiscale light distribution in engineered systems. However, it is stressed that composition and structure at each scale in a plant has evolved to support multiple energetic processes simultaneously (e.g. light, heat and mass transfer between components), as well as other complex adaptive and housekeeping processes required for life, and therefore will not be optimal for only one process. Higher plants therefore offer insights into multiscale, multiprocess optimal design of photosynthetic energy systems, although due to having different system-scale objectives and constraints, optimal energy systems should not necessarily be expected to resemble higher plants in detail.  

\begin{figure*}
\centering
\includegraphics[angle=0,width=0.9\textwidth]{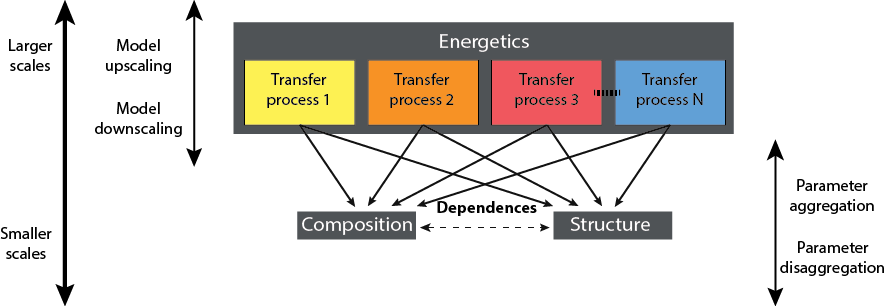}
\caption[Schematic summary of interdependences between composition, structure and energetics at each scale]{\textbf{Schematic summary of interdependences between composition, structure and energetics at each scale in the system hierarchy.} Energy transfer processes depend on system composition and structure at each scale. Composition and structure are only weakly interdependent at supracellular scales but more strongly interdependent at intracellular scales. Developing an optimisable model of the energetics at each scale requires an understanding of the interdependences summarised. Further, for collaborative, multiscale optimisation, it must also be understood how parameters and energetic models relate between scales.}
\label{fig:interdep}
\end{figure*}

Microalgal cultivation systems combine biological systems at (sub)cellular scales with artificial (sub)systems at supracellular scales (figure \ref{fig:multiscale}, `Microalgae' column). At supracellular scales, energy transfer processes can be described using relatively well-understood bulk material properties and models of continuum mechanics and radiative transfer. Moreover, composition and structure are only weakly coupled at these scales. For example, a photobioreactor with a given structure may be constructed out of different materials and contain cultures of different compositions without changing its structure. Conversely, within the cell, the protein composition of the thylakoid membrane affects its structure at multiple scales from individual photosystem supercomplexes (fig. \ref{fig:multiscale}g) to large shifts in the `ultrastructure' of the membrane at larger scales (fig. \ref{fig:multiscale}d--f) \cite{kouril2012, dekker2005}. This interdependence between protein composition and thylakoid structure, which is not yet fully understood, adds an extra level of complexity to modelling and optimisation of the thylakoid and its components (chapter \ref{chp:structure}), compared with larger-scale subsystems where composition and structure are only weakly interdependent. There also remain many open questions about interdependences between energy transfer processes (radiative transfer, heat transfer, mass transfer and EET), structure and composition at intracellular scales. Better understanding the interdependences between composition, structure and energetics at each scale in the hierarchy -- and in the thylakoid in particular, where many questions remain -- is a central challenge in multiscale, multiprocess analysis and optimisation of photosynthetic energy systems. Figure \ref{fig:interdep} provides a schematic summary of these interdependences.

Beyond modelling interactions between components at each scale, multiscale analysis requires an account of how parameters and energetic models relate between scales (section \ref{subsec:hierarchy}). Further, multiscale \textit{optimisation} requires a strategy for coordinating concurrent optimisations of weakly coupled subsystems, such as different types of processes (e.g. light and heat transfer) at a given scale, and a single type of process at different scales (e.g. light transfer through the overall photobioreactor structure and through the cell culture within each phytoelement, as well as at other scales of interest). Such a strategy is provided by the recently developed theory of decomposition-based design optimisation \cite{dewit2011, chen2010, xiong2010, li2008, allison2005, kim2003}, application of which to photosynthetic energy systems is explored in chapter \ref{chp:multiopt}. First however, chapters \ref{chp:qeet} and \ref{chp:structure} present quantitative studies utilising multiscale, hierarchical analysis to provide novel insights into interdependences between composition, structure and energy transfer in the thylakoid membrane.

\subsection{Interscale feedbacks within a dynamic hierarchy}\label{subsec:feedbacks}
The challenge of defining a hierarchical partitioning in a biological photosynthetic system is complicated by the fact that subsystems at some scales can dynamically adapt their compositions, structures and functions to the changing environment presented by upscale processes. For example, in the `energy-dependent quenching' (qE) component of NPQ, high charge-separation rates at RCs within PSII complexes induce a pH gradient ($\Delta$pH) between the lumenal and stromal fluids on opposite sides of the thylakoid membrane. This $\Delta$pH triggers a switch (the mechanical details of which remain controversial) within the LHCII antenna, activating nonphotochemical quenching sites and consequently reducing the charge-separation rate at the RCs \cite{ruban2012}. The activation of qE has frequently been observed to be associated with structural rearrangement of PSII and LHCII over the scale of an overall granal disc \cite{ruban2012}. 

It is proposed here that such a mechanism may be conceptualised within a hierarchical analysis as an interscale feedback loop: If the overall thylakoid domain transmits a supersaturating light environment to a given photosystem, it drives that photosystem's contribution (through rapid charge separation) to the $\Delta$pH, which is a larger-scale, bulk property of the fluid surrounding the thylakoid. Provided that a sufficiently large $\Delta$pH is generated (resulting from contributions of many PSIIs), this larger-scale environment then activates a switch to reduce the photochemical activity of the individual photosystem (and similarly, other photosystems). Such interscale feedbacks, which can dynamically alter the system hierarchy through structural reorganisations, would need to be accounted for in a fully detailed hierarchical analysis. Although this is beyond the scope of this thesis, it suggests a direction for future work. Chapter \ref{chp:multiopt} further discusses the concept of scale-adjacent subsystems finding mutual balance, such that the holon is configured to provide an environment optimal for its components' functioning, and the components are simultaneously configured to function optimally within the environment provided by the holon.

\section{Summary and outlook}
A key challenge in developing high-productivity photosynthetic energy systems is optimising light-harvesting efficiency under a range of light conditions. The traditional linear-process approach to analysing photosynthetic energetics shows that the dominant sources of inefficiency are nonabsorption of incident wavelengths falling outside the PAR spectrum, and dissipation of unusable electronic excitations through NPQ. These two loss mechanisms respectively dominate in light-limited and light-supersaturated irradiance conditions, which coexist in real photosynthetic systems in a typical high-light environment. Accordingly, strategies for improving light-harvesting efficiency have largely focussed on achieving more equitable irradiance distribution throughout the system. 

However, the standard linear-process analysis gives no account of how irradiance distribution depends simultaneously on parameters over a range of scales in the system, nor for how those parameters depend on each other. This is also the case for other transfer processes required for photosynthesis, such as mass and heat transfer. Consequently, strategies to improve distribution have tended to focus at a single scale (or a relatively small subset of the scales of organisation present in the system), tuning compositional and structural parameters such as LHC antenna protein composition, cell culture density, and photobioreactor structure and surface properties. The literature appears to contain no account of how to best coordinate these parameters. The prospect of developing such an account is challenged by the naturally multiscale, multiprocess complexity of photosynthetic systems; multiscale analyses of higher-plant systems have shown that processes such as light transfer can scale nonlinearly such that different mechanistic models are needed at different scales. 

Inspired by the coordinated resource-distribution strategies facilitated by multiscale, hierarchical structures in naturally evolved higher plants (such as the optical `package effect'), this chapter has introduced hierarchy theory as a system-design tool complementary to the standard linear-process approach. This theory partitions a complex dynamical system into a hierarchy of recursively nested subsystems with well-separated characteristic length and time scales in order to simplify system analysis. The multiscale analysis of the hierarchy is constructed by studying internal properties and dynamics of each scale subsystem independently, as well as couplings between scale subsystems. Hierarchy theory provides a general framework for quantitatively linking global system-scale objectives and constraints with engineerable parameters at different scales within a system, in a coordinated way.

As a basis for applying hierarchy theory, a multiscale, hierarchical system partitioning was proposed for a generic microalgal cultivation system. The system was partitioned heuristically such that each component within a given scale subsystem is a repeated functional unit that exchanges energy through multiple mechanisms with other components at the same scale and, through weaker interactions, also with components at other scales. Quantitative hierarchical analysis of the system, or a collection of scale subsystems, requires models of the energetics at each scale, and of how the models and parameters on which they depend, such as composition and structure, transform between scales. At supracellular scales, energy transfer processes can be described using relatively well-understood bulk material properties and models of continuum mechanics and radiative transfer. However, many open questions remain about interdependences between composition, structure and energetics at the nanoscales of the thylakoid membrane, where chromphores are arranged within proteins, protein complexes and supercomplexes, and overall membrane ultrastructure.

Two such questions are addressed by quantitative, hierarchical studies presented in the following two chapters, as a contribution toward the development of quantitative, hierarchical whole-system analysis. Chapter \ref{chp:qeet} analyses how the mechanisms of EET depend on structure in a chromophore network with the generic, multiscale structural and energetic features of a thylakoid membrane. Renormalisation theory, which may be thought of as a particular formulation of hierarchy theory, is used to assess the largest length scale up to which quantum dynamical effects may mediate EET. Chapter \ref{chp:structure} then asks how the multiscale structure of the thylakoid membrane depends on its protein composition. The focus is on structural changes observed in so-called `antenna-mutant' strains of microalgae, in which the size and composition of the light-harvesting antennas have been genetically modified from wild-type species. Finally, chapter \ref{chp:multiopt} extends the hierarchical \textit{analysis} framework introduced in this chapter to a framework for integrated, hierarchical system \textit{optimisation}, as a speculative basis for future work.

\chapter[Quantum-classical crossover in multiscale excitation energy transfer]{Quantum-classical crossover in multiscale photosynthetic excitation energy transfer} \label{chp:qeet}

\chapquote{Nature is just enough; but men and women must comprehend and accept her suggestions.}{A.B. Blackwell \cite{blackwell2010}}

Quantitative hierarchical analysis of a photosynthetic energy system or a collection of its scale subsystems requires models of the energetics at each scale. It must also be understood how the models and parameters on which they depend, such as composition and structure, transform between scales. At supracellular scales energy transfer processes can be described using relatively well-understood bulk material properties and models of continuum mechanics and radiative transfer. However, many open questions remain about interdependences between composition, structure and energetics in the thylakoid membrane, where chromphores are arranged within proteins, protein complexes and supercomplexes, and overall membrane ultrastructure. 

Whereas energy is distributed to photosynthetically active components through radiative transfer at larger scales, at subwavelength scales nonradiative excitation energy transfer (EET) is the dominant process. The key problem to solve at these scales is allocation of excitations to photochemical reaction centres (RCs) throughout the thylakoid at rates no greater than the RCs' maximal turnover rates. This is referred to in this chapter as the excitation-allocation problem. Successfully solving this problem ensures that excitations are not wasted through NPQ mechanisms, so the thylakoid is able to utilise energy from absorbed photons to drive primary charge separation with maximal efficiency (`Light-limited' column in fig. \ref{fig:procpart}). After many decades of study, the mechanisms of EET central to solving this problem, and their dependences on multiscale thylakoid structure, are still not fully understood. 

In this chapter the mechanisms of EET are studied in detail. The main question considered is whether and how they change with length scale in a chromophore network such as a thylakoid membrane, from individual chromophores to pigment-protein complexes (PPCs), photosystem supercomplexes and larger scales. The aim is to clarify how best to model EET at these different scales within a multiscale system optimisation. Established EET theory is briefly reviewed, including recent developments that raise intriguing questions about the role(s) of quantum dynamics in EET. A novel theoretical study is then presented, which focusses on the question of whether quantum dynamics can persist over scales that are large when compared with individual PPCs. Scaling trends in the mechanisms of EET are analysed using renormalisation theory and the critical scale of crossover from quantum to classical dynamics is assessed for different parameterisations. This study is presented in section \ref{sec:natphys} as a published journal article \cite{ringsmuth2012} together with the article's supplementary information.  

\section{Chromophores}

At the heart of every biological photosynthetic system is a network of chromophores bound within PPCs. A chromophore is a molecule or molecular subgroup in which an electronic transition may be induced by absorption of a visible photon \cite{laidler2002}.  
Chromophores take part in an array of energetic processes, facilitating energy transfer and conversion between various energy carrying particles and quasiparticles including photons, electrons, excitons and phonons. 

\begin{figure*}
\centering
\includegraphics[angle=0,width=0.875\textwidth]{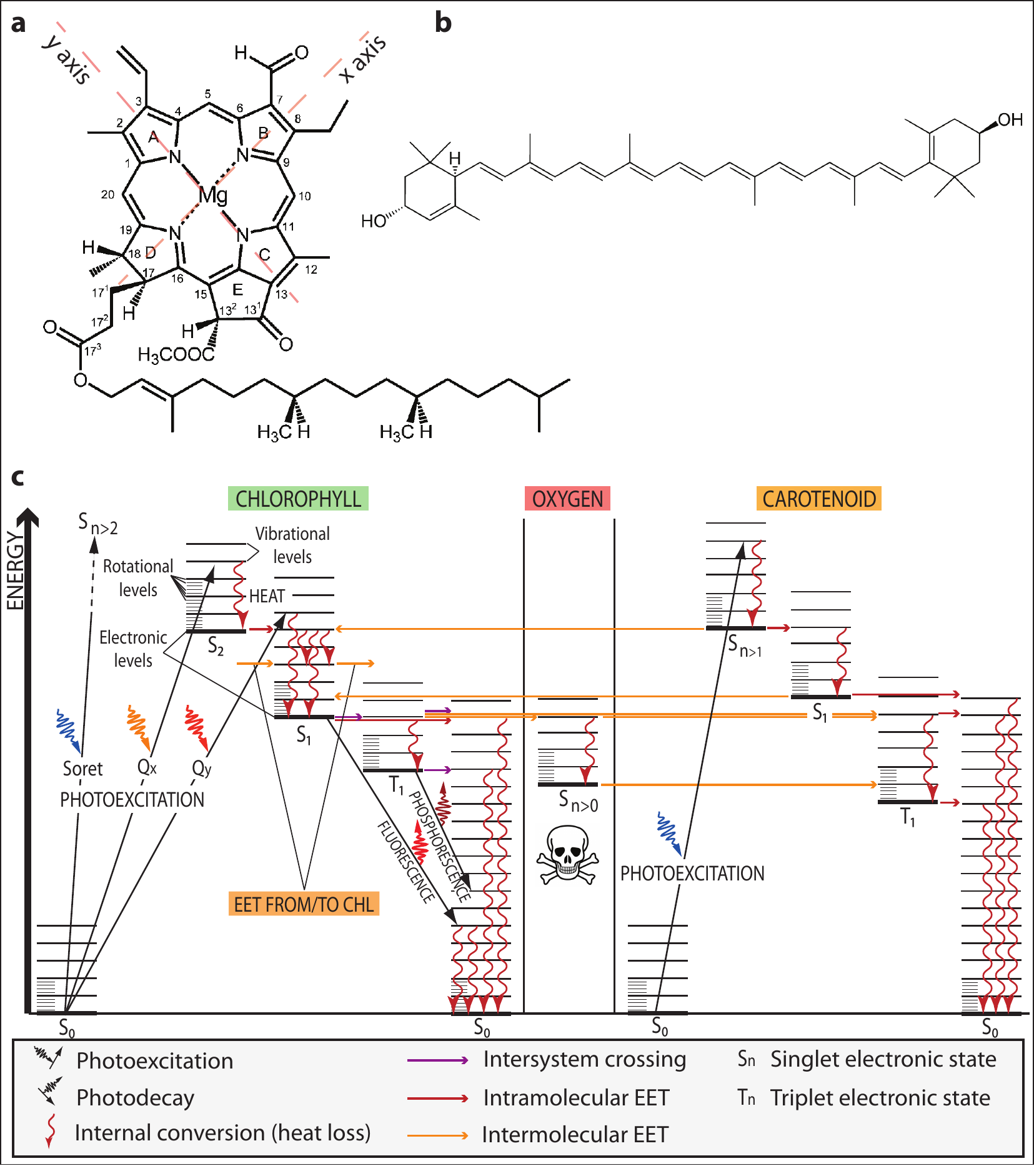}
\caption[Structure and energetics of chlorophyll and carotenoid chromophores]{\textbf{Structure and energetics of chlorophyll (Chl) and carotenoid (Car) chromophores}. \textbf{a}, Chemical structure of Chl--\textit{b}, labelled under IUPAC conventions. Chl--\textit{a} is identical except that the phytol group attached to carbon 7 is replaced with a methyl group. Image adapted from \cite{merchant2005}. \textbf{b}, Structure of (Car) lutein, the broad features of which represent most Cars found in oxygenic organisms: an extended polyene chain bounded by ring structures at both ends \cite{blankenship2002}. \textbf{c}, Jablonski diagram showing important \textit{in vivo} energy transitions within and between (generic) Chl and Car, and molecular oxygen, in the aqueous environment. Photon colours indicate approximate energy of each transition band; \textit{In vivo} absorption spectra of Chl--\textit{a} and \textit{b} are shown in figure \ref{fig:solarspectra}. The main functions of Chls are to absorb light and transfer the resulting excitations, as well as excitations received from Cars and other Chls (\textit{a} and \textit{b}), towards the RC through the Chl--\textit{a} network. Chls can also undergo intersystem crossing to a triplet state, followed by energy transfer to oxygen, which enters a dangerously oxidative singlet state. Cars absorb light and transfer energy to nearby Chls (\textit{a} and \textit{b}) for subsequent EET towards the RC through the Chl--\textit{a} network. They also serve a photoprotective role by quenching triplet chlorophyll and singlet oxygen before dissipating the energy as heat through internal conversion \cite{blankenship2002, ke2001}. Chromophore relaxation pathways are further discussed in section \ref{subsec:relax}}.
\label{fig:chromophores}
\end{figure*}

Biochemical, structural, spectroscopic and theoretical studies have generated a large literature on chromophores; references \cite{blankenship2002}, \cite{ke2001} and \cite{vanamerongen2000} give accessible introductions, and \cite{scheer2003} and references therein provide a more thorough background. The major classes of chromophores occurring in photosynthetic organisms are chlorophylls (Chls), bacteriochlorophylls, carotenoids (Cars) and bilins. The discussion here is restricted to Chl types \textit{a} and \textit{b}, and Cars, since only these occur in the photosynthetic machinery of green algae and higher plants\footnote{Bilins are ubiquitous in light-sensing phytochrome molecules in higher plants, but these do not participate directly in photosynthesis.} \cite{scheer2003}. Figure \ref{fig:chromophores} shows molecular structures representative of Chls and Cars, as well as a simplified picture of their spectroscopic properties and common energetic interactions \textit{in vivo}.

\subsection{Chlorophylls}
\paragraph{Structure}
Figure \ref{fig:chromophores}a shows the chemical structure of Chl--\textit{b}. Its two major structural regions are a square-like, planar chlorin ring of side-length $\sim 1~\textup{nm}$, centred on a magnesium atom, and a long, hydrophobic hydrocarbon (`phytol') tail. The tail functions primarily as a structual anchor into the surrounding protein matrix. The chlorin ring is the photochemically functional region of the molecule; it is the site of electron orbital rearrangements when the Chl is excited, and of unpaired electrons when it is oxidised or reduced \cite{taiz2006}. This is facilitated by an extensive delocalised $\pi$ electron system spanning most of the chlorin \cite{blankenship2002}. Surrounding the chlorin's central magnesium are four nitrogens, each part of a cyclic organic substructure derived from pyrrole; these `pyrrole rings' are labelled A--D (figure \ref{fig:chromophores}a) under the conventions of the International Union of Pure and Applied Chemistry (IUPAC\nomenclature{IUPAC}{International Union of Pure and Applied Chemistry}) and a fifth ring, E also adjoins ring C. It is conventional to define the \textit{y} molecular axis of the molecule as passing through the nitrogens of rings A and C, and the \textit{x} axis as passing through the nitrogens of rings B and D. Chl--\textit{a} is structurally identical to \textit{b} except that the formyl group attached at carbon 7 in figure \ref{fig:chromophores}a is replaced with a methyl group \cite{blankenship2002}. This change shifts the molecule's spectroscopic properties.
 
\paragraph{Spectroscopy}
\textit{In vivo} absorption spectra of  Chl--\textit{a} and Chl--\textit{b} are shown in figure \ref{fig:solarspectra}. The Chls each have two major absorption bands, one in the blue and near-UV range, and one in the red. Figure \ref{fig:chromophores}c includes a simplifed model of energetic transitions within a Chl molecule, showing electronic transitions and also their broadening by vibrational and rotational transitions. The two lowest-energy (red) transitions, from the ground (singlet\footnote{A singlet state is one in which the spin of the excited electron is anti-aligned with its ground-state partner.}) state (S$_0$) to the first (S$_1$) and second (S$_2$) excited singlet states, are called Q bands. Transitions from S$_0$ to higher excited singlet states (S$_{n>2}$) (blue) are called B or Soret bands. Each electronic transition has a transition dipole moment comprising a magnitude and a polarisation. The S$_0\rightarrow \textup{S}_1$ transition is polarized along the chlorin's \textit{y} axis and is accordingly labelled the $Q_y$ transition (figure \ref{fig:chromophores}a,c). This is the most important transition for interchromophoric EET. The S$_0\rightarrow \textup{S}_2$ transition is polarized approximately along the \textit{x} axis and is therefore the $Q_x$ transition. The Soret transitions (S$_0\rightarrow \textup{S}_{n>2}$) each have mixed polarisation \cite{renger2005, blankenship2002, scheer2003}. 

Importantly for the study of EET, these electronic transitions can be shifted and broadened by interactions with external degrees of freedom. These effects include inhomogeneous broadening (`static disorder') where individual chromophores are subject to local effects (e.g. due to the local protein environment), and homogeneous broadening and shifting where all chromophores are affected equally on average (e.g. thermal effects). The mechanisms of these effects are discussed further in section \ref{subsec:exphocoup}.

\subsection{Carotenoids}
\paragraph{Structure}
Cars are, structurally and functionally, the most diverse group of chromophores, with over 800 different types identified \cite{scheer2003}. They are long-chain, conjugated hydrocarbons, each containing a string of isoprene residues and distinguished from one another by their end groups \cite{ke2001}. An example, lutein is shown in figure \ref{fig:chromophores}b. Similarly to the chlorin rings of Chls, Cars have delocalised $\pi$ electron systems that mediate their photochemical properties \cite{blankenship2002}. 

\paragraph{Spectroscopy}
Cars have unusual spectroscopic properties. In many types the S$_0\rightarrow \textup{S}_1$ electronic transition is optically `forbidden' (allowed but improbable, according to quantum-mechanical selection rules) and the major absorption band, typically in the 400--500nm (blue) range (figure \ref{fig:solarspectra}), instead drives the S$_0\rightarrow \textup{S}_2$ transition. The transition dipole moment is oriented approximately along the molecule's major axis \cite{gruszecki1999}. Relaxation to the ground state usually proceeds though non-radiative internal conversion (section \ref{subsec:relax}) \cite{blankenship2002, scheer2003}. Cars also have unusually low-energy triplet\footnote{A triplet state is one in which the spin of the excited electron is aligned with its ground-state partner.} excited states, which are well suited to quenching low-energy excited states of molecules in their environments, such as triplet Chls and singlet oxygen (figure \ref{fig:chromophores}c). This is essential to the photoprotective role played by Cars \textit{in vivo}.

\subsection{Chromophore relaxation pathways}\label{subsec:relax}
Figure \ref{fig:chromophores}c shows various relaxation mechanisms by which a chromophore can relax to its ground electronic state. The mechanisms that kinetically compete with EET in chromophore networks are:

		\paragraph{Fluorescence}
		Radiative decay from an excited singlet state to a singlet state of lower energy, typically within 10 ns \cite{ke2001, laidler2002}. 
		
		\paragraph{Intersystem crossing}
		Non-radiative transition between electronic states of different quantum multiplicities: singlet and triplet states. Such transitions are quantum mechanically `forbidden', so triplet states are both rarer and longer lasting (ms--hr lifetimes) than singlet states resulting from photoexcitation (ns lifetimes) \cite{ke2001}. 
	
		\paragraph{Phosphorescence}
		Radiative decay of a triplet state. Due to the large range of possible lifetimes for triplet states, phosphorescence is responsible for the sustained `glowing in the dark' that is observed in some materials but not generally visible in photosynthetic tissues because photoprotective mechanisms rapidly quench excited triplet states \cite{ke2001, blankenship2002}.
			
		\paragraph{Internal conversion} 
		Intramolecular, non-radiative transitions in which an excited electron loses energy directly to internal vibrational modes of the molecule \cite{bixon1968, barber1978, ke2001}. These modes are coupled to vibrational modes of molecules in the environment, so energy is gradually dissipated among these modes; this dissipation is heat loss. Usually, a Chl or Car excited to a singlet state higher than S$_1$ will relax to that state by internal conversion within 10--100 fs \cite{ke2001, blankenship2002}. This is rapid compared with other pathways including, typically, interchromophoric EET \textit{in vivo}. The latter process is therefore dominated by S$_0\leftrightarrow \textup{S}_1$ transitions \cite{blankenship2002} and, accordingly, each chromophore may be approximated as a two-level system for the purposes of modelling EET \cite{ishizaki2010}.  
		
\section{Interchromophoric excitation energy transfer}\label{sec:inteet}
The study of energy transfer between chromophores began in the early twentieth century. Gradual progress during the intervening period, in understanding the structural and physicochemical properties of chromophores, PPCs and their networks \textit{in vivo}, preceeded a recent explosion of interest in the field. Several excellent reviews describe these developments: \cite{lambert2013}, \cite{ishizaki2012}, \cite{olaya2011}, \cite{ishizaki2010}, \cite{renger2009}, \cite{renger2005}, \cite{clegg2004} and \cite{scholes2003}.  

	\subsection{Historical summary}
	An obvious candidate mechanism for interchromophoric energy transfer is the reabsorption of fluorescence from one chromophore (the `donor') by another (the `acceptor'). This is possible, provided the donor's emission spectrum overlaps with the acceptor's absorption spectrum (there is `spectral overlap'). However, such `trivial transfer' cannot account for the energetics in thylakoid membranes. Even prior to the first formulation of EET theory it was known that `photosynthetic unit' functional domains (each comprising a RC and the minimum number of its associated antenna chromophores required to drive photosynthesis) must be small compared with photosynthetically active wavelengths of light, making it meaningless to consider photon exchange within them \cite{clegg2004}. It was also known that if the interchromophoric separations were commensurate with visible wavelengths, the probability of (isotropically emitted) fluorescence reabsorption would be insufficient to account for measured transfer efficiencies \cite{clegg2004}. 
	
	\subsubsection{F\"orster theory}	
	F\"orster published the first successful model of non-radiative interchromophoric EET in 1946 by extending earlier work of Perrin and Perrin, who proposed a mechanism utilising a Coulombic dipole-dipole interaction between chromophores \cite{forster1946, clegg2004}. In F\"orster's model, energy is transferred through an electrodynamic interaction between the chromophores' transition dipole moments. The energy of this `transfer coupling', $\Delta_{T}$ depends on the chromophores' separation, mutual orientation, spectral overlap and the dielectric of the intervening medium (section \ref{subsec:frenkel}). Importantly, in deriving his model F\"orster assumed that
	\be
	\Delta_{T}\ll \Gamma~~~~~~~~~(\text{F\"orster theory}),\label{eqn:forsterregime}
	\ee
where $\Gamma$ is the homogenous linewidth, a measure of the coupling energy between chromophore electronic excitations\footnote{Molecular electronic excitations may be treated as quasiparticles called excitons \cite{merrifield1964}. The quantum states of these quasiparticles within a chromophore network are often written in the network's electronic energy eigenbasis or site basis (these are orthogonal); site-localised and site-delocalised states are equally valid representations of excitons. However, in the literature on EET dynamics the term `exciton' is usually reserved for (site-delocalised) excitation energy eigenstates. This is largely because spectroscopic measurements usually provide information projected onto energy eigenstates and theory seeks to describe these experiments. In contrast, site-localised states are often simply called `excitations'. These conventions are followed here. Importantly, which representation is used does not affect the dynamics of EET \cite{ishizaki2009}.} and nuclear vibrations due to their thermal environment \cite{leegwater1996} (section \ref{subsec:exphocoup}). This permitted use of the dipole-dipole approximation for the transfer coupling and allowed this coupling to be treated as perturbative to the initial donor excited state. As a result, Fermi's golden rule could be used to derive the excitation transfer rate,
	\be
	k_f=\frac{{\Delta_{T}}^2}{2\pi\hbar^2}\int_{-\infty}^{\infty}d\lambda ~\varepsilon_D(\lambda)\alpha_A(\lambda),\label{eqn:forster}
	\ee
where the integral evaluates spectral overlap between the normalised donor fluorescence emission spectrum, $\varepsilon_D(\lambda)$ and the acceptor absorption (molar extinction coefficient) spectrum, $\alpha_A(\lambda)$. 

\begin{figure*}
\centering
\includegraphics[angle=0,width=0.35\textwidth]{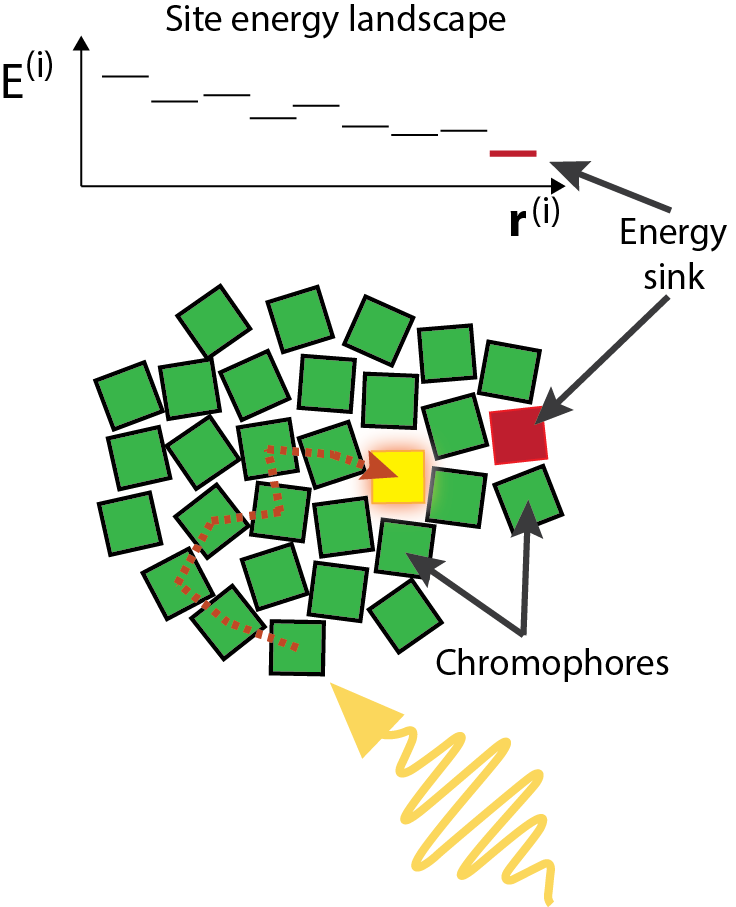}
\caption[Excitation energy transfer in the F\"orster regime]{\textbf{Excitation energy transfer in the F\"orster regime}. (Bottom) An excitation created by photon absorption is rapidly localised to a single site by large excitation-vibration coupling. This site-localised excitation then diffuses through the chromophore network in a classical random walk, biased towards the energy sink(s) by the site energy landscape (top). This landscape is typically characterised by local static and dynamic (thermal) disorder superimposed on a net decline in energy towards the sink(s). A network may have multiple such sinks and each may be an exit site from which the excitation is further transferred to another network (e.g. a neighbouring PPC), or it may be a photochemical RC or nonphotochemical quenching site. Photon wavelength shown not to scale.}
\label{fig:forster}
\end{figure*}

It follows from equation \ref{eqn:forsterregime} that EET as described by F\"orster theory is incoherent `hopping' motion of excitations localised to individual chromophores (sites) (figure \ref{fig:forster}); quantum-coherent delocalisation between sites is prevented by rapid thermal decoherence due to the strong excitation-vibration coupling \cite{ishizaki2012, ishizaki2010, leegwater1996}. Long-range EET in this regime is therefore described by classical diffusion \cite{mohseni2008}, biased by the energy landscape of site energies in the network.     
	
	\subsubsection{Beyond the F\"orster regime}
	F\"orster's model was long assumed to adequately describe EET in photosynthetic chromophore networks. However, structural and linear-spectroscopic studies of PPCs revealed interchromophoric separations small enough, and EET kinetics fast enough, to suggest that some such systems do not satisfy equation \ref{eqn:forsterregime} \cite{beljonne2009, harcourt1994, leegwater1996}. It was therefore suggested that models be developed to incorporate site-delocalised exciton states \cite{harcourt1994, leegwater1996} and quantify their influence on EET dynamics through the overall network. 
	
	In the last five years a growing body of evidence from 2D femtosecond spectroscopy performed on ensembles of solvated LHCs has supported these ideas \cite{engel2007, lee2007, calhoun2009, schlau2009, collini2010, panit2010}. The data have suggested that quantum-coherent evolution of excitons can persist over time scales significant to EET dynamics through individual PPCs, even at physiological temperatures. Questions of what mechanisms underly these effects, and whether they could be present and physiologically significant \textit{in vivo}, have been intensely examined but remain controversial \cite{lambert2013}.	
	
	Section \ref{sec:natphys} addresses the open question of whether coherent EET can persist at supra-PPC scales under physiological conditions in thylakoid membranes. A theoretical foundation for that discussion is first laid in a brief review of techniques used to model quantum-dynamical EET in individual PPCs.

	\section{System Hamiltonian}\label{sec:sysham}  
	
	\subsection{Frenkel exciton Hamiltonian}\label{subsec:frenkel}
 
Frenkel pioneered the theory of site-delocalised electronic excitations in atomic crystals \cite{frenkel1931} and this theory was later extended to molecular crystals (reviewed in \cite{silbey1976}). More recently it has been applied to chromophore networks in photosynthetic systems (reviewed in \cite{scholes2006}). 
The total electronic Hamiltonian for a photosynthetic light-harvesting system of $N$ chromophores may be written \cite{ishizaki2012, may2008, gilmore2005} in second-quantised form, in the site basis, as
\be
H^S =\sum_{i=1}^N E^{(i)} |i\rangle \langle i|+\frac{1}{2}\sum_{i\neq j}^N\Delta_T^{(ij)}(|i\rangle \langle j|+|j\rangle \langle i|),\label{eqn:hs}
\ee
where the site energies, $E^{(i)}=E^{(i)}_{excited}-E^{(i)}_{ground}$ usually correspond to S$_1\leftrightarrow \textup{S}_0$ transitions. The transfer couplings, $\Delta_T^{(ij)}$ are parameterised according to equation \ref{eqn:transcoup}.
	
	\subsubsection{Components of the electronic transfer coupling}
	Harcourt \emph{et al} modelled the `factors' responsible for the electronic transfer coupling between a pair (dimer) of equivalent molecules \cite{harcourt1994} (reviewed in \cite{olaya2011}). By considering mixtures of donor and acceptor locally excited states with bridging ionic states, these authors were able to quantify contributions to the total interchromophoric electronic coupling energy as
	\beqn
	\Delta_T=\Delta_{oo}+\Delta_{ed},\nonumber
	\eeqn	
	  where $\Delta_{oo}$ is the coupling due to electron transfer through molecular orbital overlap and $\Delta_{ed}$ is electrodynamic `inductive resonance' coupling. In modern terms the latter is an interaction between the transition densities\footnote{A molecule's transition density quantifies the way in which the oscillating electric field of incident light distorts the electron density of the molecule.}, $P^{(D)}_{S_1S_0}(\textbf{r}^{(D)})$ and $P^{(A)}_{S_0S_1}(\textbf{r}^{(A)})$, of the donor and acceptor respectively, such that the relaxation (usually S$_1\rightarrow \textup{S}_0$) of the donor resonates with the excitation (S$_0\rightarrow \textup{S}_1$) of the acceptor \cite{olaya2011}. The energy of this transfer coupling may be quantified \cite{olaya2011} as, 
	\beqn	\Delta_{ed}&=&\frac{e^2}{4\pi\epsilon_0}\int{\frac{P^{(D)}_{S_1S_0}(\textbf{r}^{(D)})P^{(A)}_{S_0S_1}(\textbf{r}^{(A)})}{|\textbf{r}^{(D)}-\textbf{r}^{(A)}|}}d\textbf{r}^{(D)}d\textbf{r}^{(A)}\nonumber\\
	      &\approx&\frac{1}{4\pi\epsilon_0}\frac{\kappa\eta^{(D)}\eta^{(A)}}{d^3}~~~~~~\text{(well-separated chromophores)}.\label{eqn:transcoup}
	\eeqn
	
	Here, $\eta^{(i)}\equiv|\boldsymbol{\eta}^{(i)}_{S_mS_n}|$ is the magnitude of the transition dipole moment between the $i$th molecule's S$_m$ and S$_n$ electronic states. $\epsilon_0$ is the vacuum permittivity (no dielectric screening is assumed) and $d$ is the centre-to-centre interchromophoric separation. The orientation factor, $\kappa=\cos\Theta-3\cos\Phi^{(D)}\Phi^{(A)}$ depends on $\Theta$, the angle between the two transition dipole moments, and $\Phi^{(i)}$, the angle between the $i$th molecule and the molecules' separation vector \cite{blankenship2002}. 
	
	Equation \ref{eqn:transcoup} is approximate because the assumption of a transition dipole-dipole interaction (the `ideal-dipole approximation' [IDA]) is strictly accurate only if the interchromophoric separation is large when compared with the chromophores themselves. Fr\"ahmcke and Walla found that for chromophores in light-harvesting complex II (LHCII) of green algae and higher plants, the IDA is reliably accurate for $d>25$ \AA~and to within an order of magnitude for smaller separations \cite{frahmcke2006} . Accurately evaluating $\Delta_{ed}$ in the latter range requires that structural details of the transition densities be accounted for and quantum chemical methods are needed \cite{olaya2011, frahmcke2006}. At extremely small separations ($d<5$ \AA), $\Delta_T$ is dominated by $\Delta_{oo}\propto \exp(-2\beta d)$, where usually $1.2\leq \beta \leq2.0~\textup{\AA}^{-1}$ \cite{olaya2011}. 
	Since interchromophoric separations in photosynthetic light-harvesting antennas generally do not enter this regime, the molecular orbital overlap interaction is not discussed here (see referencess \cite{olaya2011} and \cite{harcourt1994}). Accordingly, henceforth in this chapter, $\Delta_T\approx\Delta_{ed}\equiv\Delta$.
	
	\subsection{Excitation-phonon coupling}\label{subsec:exphocoup}
 Molecular dynamics simulations of proteins have shown that their nuclear vibrational spectra are typically dominated by harmonic modes\footnote{Vibrations with frequencies independent of their amplitudes.}. For example, 95\% of the modes in human lysozyme are harmonic, with small amplitudes and fast time scales, while only 0.5\% are anharmonic modes involving conformational changes \cite{ishizaki2010}. In PPCs the predominance of harmonic vibrations allows for a normal mode (phonon) representation of chromophore vibrations, and the vibrations' small amplitudes away from well-defined equilibrium configurations allow them to be treated perturbatively.  
 The nuclear Hamiltonian of the $j$th chromophore in its $e$th electronic state may be written \cite{ishizaki2010},
\be
H^{(j)}_e=V^{(j)}_e({\textbf{r}^{(j)}_e}_0)+\sum_i\frac{\hbar\omega_i}{2}(\textbf{p}_i^2+\textbf{q}_i^2),\nonumber 
\ee 
 where ${\textbf{r}^{(j)}_e}_0$ is the equilibrium configuration of the nuclear coordinates, and $V^{(j)}_e(\textbf{r}^{(j)})$ is the chromophore's nuclear potential energy function. $\textbf{q}_i$ are dimensionless phonon coordinates (wavevectors) with corresponding frequencies, $\omega_i$ and momenta, $\textbf{p}_i$. The Hamiltonian for the phonon bath can then be written in terms of creation and annihilation operators for the phonons:
 \be
 H^B=\hbar \sum_\textbf{q} \omega(\textbf{q}) a(\textbf{q})^\dagger a(\textbf{q}).\nonumber
 \ee
 Interaction between chromophore electronic excitations and phonons arises from the fact that $V^{(j)}_e(\textbf{r}^{(j)})$ differs in different electronic states \cite{merrifield1964, ishizaki2010, silbey1976}. This difference, for example due to an S$_0\leftrightarrow \textup{S}_1$ transition, is called the reorganisation energy. Its inverse, the reorganisation time, gives a characteristic time scale for phonon dynamics in the system \cite{ishizaki2010}. In general, the reorganisation energy will contain both linear and higher-order terms in the phonon coordinates \cite{silbey1976, mahan2000, merrifield1964}. However, for PPCs at physiological temperatures, only linear terms need be retained and these correspond to a shift in the equilibrium nuclear configuration ${\textbf{r}^{(j)}_e}_0$ of an excited chromophore \cite{ishizaki2010}. 
 
 In general, phonons will also shift electronic transfer coupling energies (by interacting with off-diagonal terms in equation \ref{eqn:hs}) \cite{silbey1976}. However, in PPCs these interactions are approximately two orders of magnitude weaker than phonon interactions with the site energies and accordingly may be neglected \cite{leegwater1996, ishizaki2010}. Consequently, the system-bath interaction Hamiltonian may be expressed as
 \be
 H^{SB}=\sum_{i,\textbf{q}} g^{(i)}(\textbf{q})|i\rangle \langle i|\left(a(\textbf{q})-{a(-\textbf{q})}^\dagger\right),\label{eqn:hsb}
 \ee 
 where $g^{(i)}(\textbf{q})$ is the coupling energy between an excitation on site $i$ and a phonon of wavevector $\textbf{q}$. 
 
 The total system Hamiltonian may then be expressed as
 \be
 H=H^S+H^B+H^{SB}.\nonumber
 \ee
 \section{Open quantum dynamics of excitation energy transfer}\label{sec:openquant}
 \subsection{Individual pigment-protein complexes}
 \subsubsection{Markovian models}
 In light of equation \ref{eqn:hsb}, the criterion for the F\"orster regime, equation \ref{eqn:forsterregime} can be reformulated as,
 \be
 \Delta^{(ij)}\ll g^{(m)}(\mathbf{q}),~~~m=i,j.\nonumber
 \ee
 Studies of photosynthetic EET outside of this regime have focussed almost exclusively at the scale of individual PPCs. The theoretical tools most commonly used have been quantum master equations, such as various formulations of the Redfield and Lindblad equations. 
  Until recently, such studies usually made three simplying assumptions concerning the system-bath coupling: 1) the Born approximation, which amounts to weak system-bath coupling ($\Delta^{(ij)}\gg g^{(m)}(\textbf{q}),~m=i,j$), allowing a pertubative treatment of the bath; 2) the Markov, or `memory-less' bath approximation (see Supplementary Information in section \ref{sec:natphys}); 3) the independent-baths approximation, which implies no correlations between baths at different sites \cite{lambert2013, ishizaki2009, palmieri2009}. 
 
 In general these equations, which describe the evolution of the reduced density matrix, $\rho_e$ for the electronic degrees of freedom\footnote{Obtained by averaging over the phonon dynamics with a partial trace of the total density matrix.}, take the form,
 \be
	\frac{\partial}{\partial t}~\rho_e(t)=-\frac{i}{\hbar}[H,\rho_e(t)]+R(\rho_e(t)).\nonumber
	\ee
	Here the first term describes unitary evolution under the Schr\"odinger equation and the second, $R(\rho)$ is a relaxation superoperator that quantifies decoherence due to system-bath interactions. This superoperator may take many forms, which can be derived from the microscopic system-bath coupling and/or invoked phenomenologically \cite{palmieri2009} to quantify processes such as pure dephasing and the various relaxation pathways described in section \ref{subsec:relax}. However, relaxation pathways are often neglected, consistent with the near-unity quantum efficiencies typical of unstressed photosynthetic light-harvesting networks. 
	
\begin{figure*}
\centering
\includegraphics[angle=0,width=1\textwidth]{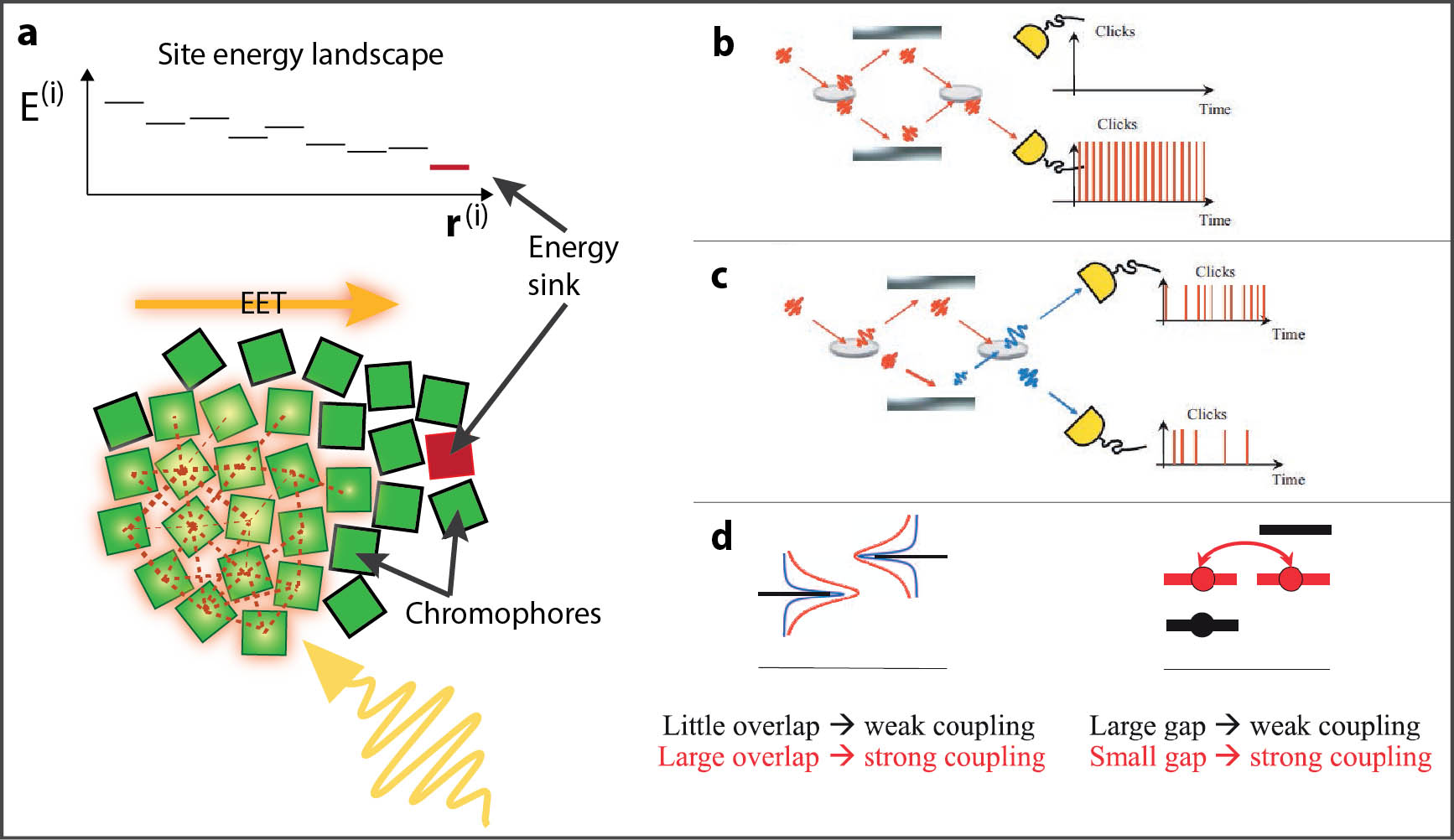}
\caption[Environment-assisted quantum excitation energy transfer]{\textbf{Environment-assisted quantum excitation energy transfer}. \textbf{a}, (Bottom) Photon absorption creates a site-delocalised exciton. The exciton migrates through the network in a wavelike manner, biased towards the energy sink(s) by the site energy landscape (top). Dashed orange links and yellow glow respectively indicate hypothetical excitation transfer couplings and instantaneous site populations. EET dynamics are subject to mechanisms shown in panels b--d (reproduced from \cite{caruso2009}). Photon wavelength shown not to scale. \textbf{b}, A noiseless Mach-Zehnder interferometer provides an analogy for multi-path EET. Quantum interference between interferometer paths closes the path to the upper detector. \textbf{c}, Introducing dephasing noise by vibrating the lower mirror opens the path to the upper detector. This is analogous to simulation results showing that thermal decoherence in photosynthetic light-harvesting networks can improve EET efficiency compared with purely coherent EET by overcoming transfer-inhibiting interference effects \cite{caruso2009}. \textbf{d}, (Left) Spectral broadening of site energies caused by thermal noise strengthens transfer couplings by overcoming localization due to static disorder.}
\label{fig:enaqt}
\end{figure*}
	
	These models have elucidated mechanisms (figure \ref{fig:enaqt}) by which excitonic coherences may improve EET performance under different measures; usually transfer efficiency or robustness to environmental perturbations.  
Thermal decoherence, which may be thought of as dynamical disorder arising from thermal fluctuations in the bath, cooperates with excitonic coherences by suppressing transfer-inhibiting interference between EET pathways and exploiting spectral broadening of site energies to overcome localization due to static disorder. The result is transfer efficiency exceeding purely coherent or purely incoherent transfer \cite{plenio2008, mohseni2008, caruso2009, ishizaki2009(1), rebentrost2009, chin2010}.     
	
	\subsubsection{Beyond Markovian models}
	The claimed relevance of these findings to photosynthetic systems has been criticised because at the scale of individual PPCs, excitation-phonon coupling energies may be comparable to excitation transfer couplings ($\Delta^{(ij)}\approx g^{(m)}(\mathbf{q}),~m=i,j$); these systems therefore fall outside the weak system-bath coupling limit and the phonon dynamics cannot simply be averaged out in order to describe the electronic dynamics \cite{ishizaki2010, ishizaki2009, palmieri2009, beljonne2009}. It has also been proposed that the Markov approximation is inadequate at the scale of PPCs. This approximation requires phonons to relax to their equilibrium states instantaneously; hence they are always in equilibrium relative to any electronic state. Conversely, EET between chromophores occurs \textit{via} nonequilibrium phonon states, according to the vertical Franck-Condon transition and this process becomes more significant when $\Delta^{(ij)}\approx g^{(m)}(\mathbf{q}),~m=i,j$ \cite{ishizaki2010, ishizaki2009}. Additionally, the independent-baths approximation has been challenged \cite{sarovar2011, lee2007}, with authors providing evidence that spatial and temporal correlations between bath fluctuations can strongly influence EET dynamics.
	
	A large body of work has emerged, focussed on verifying and understanding these proposed violations of approximations earlier used for individual PPCs. New insights have emerged but many open questions remain \cite{lambert2013}. Exploring this work is beyond the scope of this chapter and the reader is directed to the recent reviews by Lambert \textit{et al} \cite{lambert2013} and Olaya-Castro and Scholes \cite{olaya2011}.  
	The ultimate focus here is on long-range EET over scales that are large when compared with individual PPCs and strong system-bath coupling is unlikely to feature at these scales (section \ref{sec:natphys}). 
	
	\subsection{Pigment-protein aggregates: long-range energy transfer}\label{subsec:mcft}
	 
	An interesting question is whether excitonic quantum coherence can also enhance EET over scales larger than individual PPCs in aggregates such as those found in thylakoid membranes. More specifically, can it help to solve the excitation-allocation problem at photochemically active scales of the thylakoid? Furthermore, if so, is this due only to accumulated intra-PPC effects or are coherences also present over longer distances?  
	One may also ask what mechanisms ultimately limit coherence length, and how to extend it in engineered systems \cite{spano2009}.
	These questions have not been conclusively addressed.
		
	\begin{figure*}
\centering
\includegraphics[angle=0,width=0.45\textwidth]{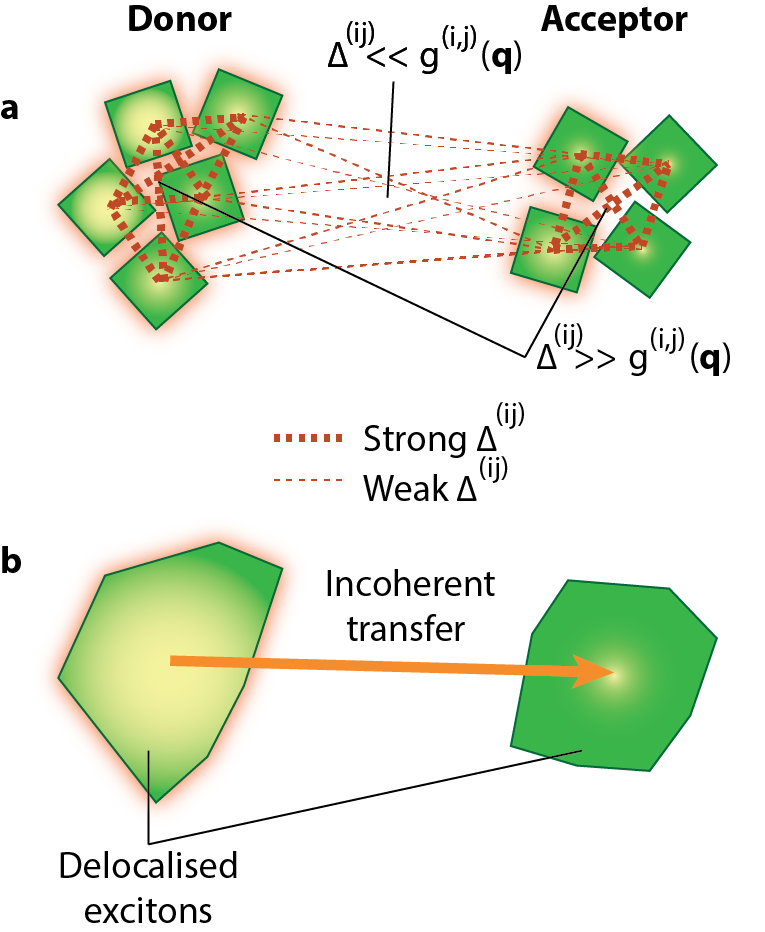}
\caption[Multichromophoric F\"orster theory]{\textbf{Multichromophoric F\"orster theory}. \textbf{a}, Schematic showing two weakly coupled clusters of strongly coupled chromophores. Yellow glow shows hypothetical instantaneous site populations; an exciton initially located on the donor cluster is transferred to the acceptor cluster through weak, pairwise intercluster chromophore couplings. \textbf{b}, The coarse-graining implicit in the MCFT formalism. Due to delocalisation of the intracluster excitons, the donor and acceptor clusters behave as singular functional units for the purposes of intercluster EET, which is assumed to be incoherent.}
\label{fig:gft}
\end{figure*}

	So-called multichromophoric F\"orster theory (MCFT\nomenclature{MCFT}{Multichromophoric F\"orster theory}) has been developed to describe EET between weakly coupled clusters of strongly coupled chromophores, such as two well-separated PPC subunits (within a single PPC) or whole PPCs \cite{banchi2013, novoderezhkin2010, beljonne2009, strumpfer2009, jang2004, scholes2000, sumi1999}. This theory accounts for intracluster delocalisation of excitons but assumes incoherent EET between clusters (no delocalisation between clusters). This is because pairwise chromophore couplings are assumed to satisfy $\Delta^{(ij)}\gg g^{(m)}(\textbf{q}),~m=i,j$ within each cluster and $\Delta^{(ij)}\ll g^{(m)}(\textbf{q}),~m=i,j$ between clusters. As shown in figure \ref{fig:gft}, an exciton delocalised over the strongly coupled donor cluster hops incoherently to an exciton on the acceptor cluster. 
	
	Depending on initial conditions, the donor's exciton may be in a nonequilibrium state which continues to evolve coherently during intercluster transfer, and this affects the intercluster transfer rate. This time-dependent rate is given \cite{jang2004} by    

	\be
	k_{mcf}(t)=\sum_{j'j''}\sum_{k'k''}\frac{\Delta^{(j'k')}\Delta^{(j''k'')}}{2\pi\hbar^2}\int_{-\infty}^{\infty}d\lambda ~\varepsilon_D^{(j''j')}(t, \lambda)\alpha_A^{(k'k'')}(\lambda),\nonumber
	\ee
	
where $\varepsilon_D^{(j''j)}(t, \lambda)$ is the time-dependent, normalised donor fluorescence emission spectrum \cite{jang2004} and other quantities are analogous to those in equation \ref{eqn:forster}, with summations over all intercluster pairings. In the limit where the donor state equilibrates rapidly compared with intercluster transfer, a stationary expression is obtained \cite{jang2004} for the latter: 
\be
	k_{mcf}=\sum_{j'j''}\sum_{k'k''}\frac{\Delta^{(j'k')}\Delta^{(j''k'')}}{2\pi\hbar^2}\int_{-\infty}^{\infty}d\lambda ~\varepsilon_D^{(j''j')}(\lambda)\alpha_A^{(k'k'')}(\lambda).\nonumber
	\ee
This stationary rate is equivalent to the rate predicted by an early formulation of MCFT, often called `generalised F\"orster theory' (GFT\nomenclature{GFT}{Generalised F\"orster theory})\cite{jang2004, sumi1999}.

Jang \textit{et al} \cite{jang2004} used MCFT to study transfer between the so-called B800 and B850 subunits (well-separated rings of chromophores) of a single LH2 light-harvesting complex from purple bacteria. They found that whereas standard F\"orster theory, which neglects intracluster coherences, underestimated the inter-ring transfer rate by an order of magnitude, MCFT predicted a faster rate in good agreement with experiments. It was inferred that LH2 exploits intra-ring excitonic coherences for enhanced EET between rings. This mechanism has also been identified as so-called `supertransfer' and suggested to underly observations of long-range EET through large artificial aggregates of LH2 over distances spanning $10^4$ complexes \cite{lloyd2010, escalante2010}.

An interesting feature of MCFT is that the distance dependence of $k_{mcf}$ can vary with static disorder and temperature in both the time-dependent and time-independent cases. This is because disorder affects the energetic and spatial features of the exciton spectra and temperature alters the relative populations of different excitons and their mutual coherences. These effects can change the effective distance(s) between the coupled donor an acceptor excitons, thereby changing the intercluster coupling and transfer rate \cite{jang2004}. MCFT therefore provides a formalism for studying how interplay between PPC-scale excitonic coherences, static disorder and excitation-phonon couplings affect EET over larger scales in the thylakoid. However, this formalism has the crucial restriction that intercluster coherences are prohibited \textit{a priori}.  
Therefore, MCFT can describe enhancement of long-range EET due to excitonic coherences only as a linear sum of enhancements at the scale of individual PPCs. In the next section, a new formalism for studying long-range EET in PPC aggregates is developed, which accommodates intercluster coherences and reveals unexpected scaling trends in the mechanisms of EET. 
	
\section{Multiscale photosynthetic and biomimetic excitation energy transfer}\label{sec:natphys}

\blankpage
\includepdf[pages={1-15}]{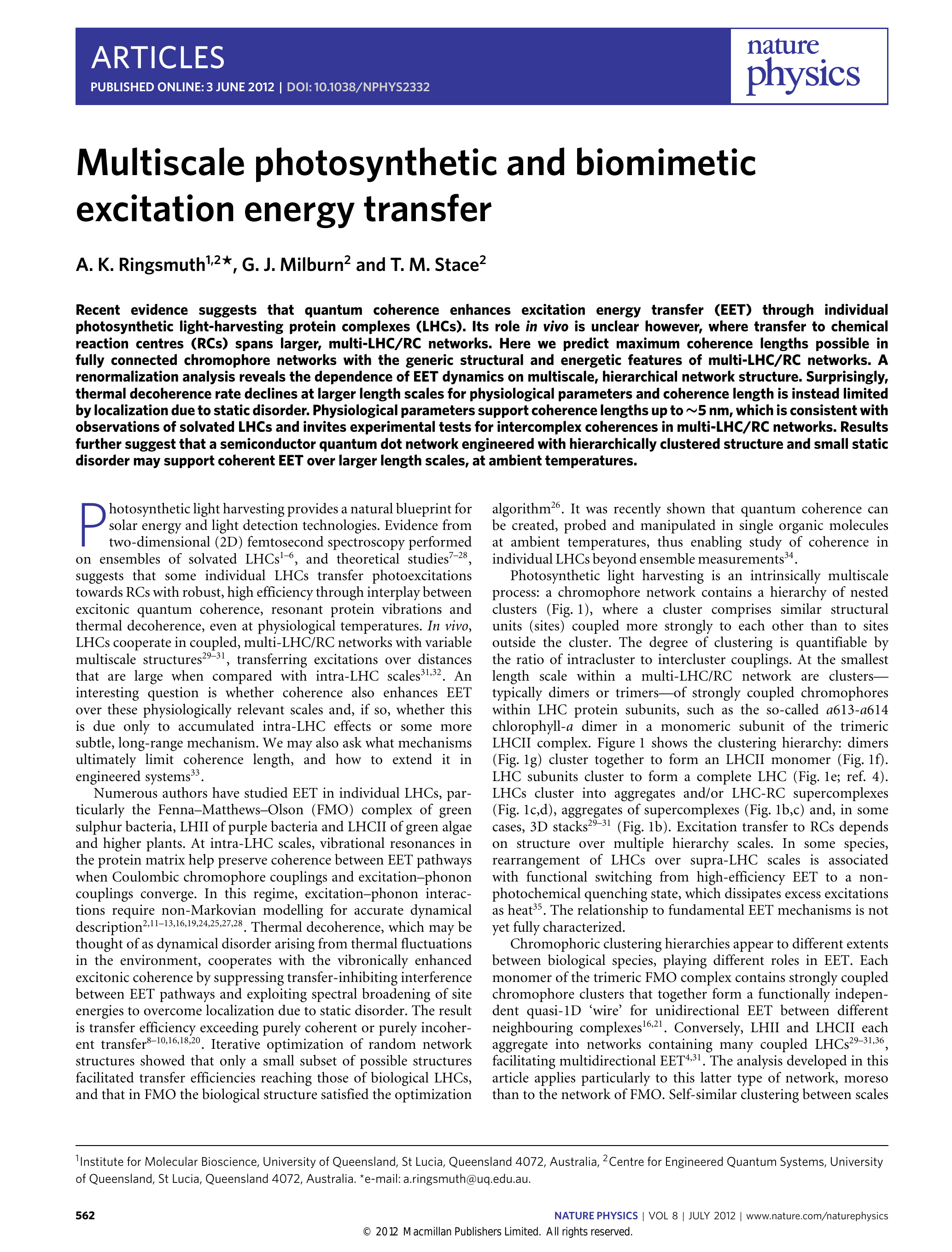}
\blankpage
\section{Conclusions and outlook}\label{sec:qeetconc}
It was concluded in section \ref{sec:natphys} that, whereas intracomplex transfer may require quantum-dynamical description, intercomplex transfer is not expected to require this (although experimental testing is required before this can be categorically ruled out). Rather, quantum dynamics is expected to be significant to intercomplex transfer in PPC aggregates only in so far as it underpins the mechanism of multichromophoric F\"orster theory (supertransfer), described in section \ref{subsec:mcft}. The results therefore suggest that MCFT will, in general, be sufficient for modelling energy transfer at scales larger than individual PPCs. This is a useful conclusion for the purpose of multiscale system modelling within the framework introduced in chapter \ref{chp:multianalysis}. This is because it suggests that accurately modelling long-range EET through the thylakoid is unlikely to require the massive computational resources that would be needed for a fully detailed quantum simulation of all chromophores in the network.    

Notwithstanding the likely absence of long-range quantum-coherent EET in thylakoid membranes under physiological conditions, the scaling principles revealed here may be of use in designing artificial light-harvesting materials that can support such long-range coherence. However, it should first be asked whether these materials would derive any functional advantage from long-range coherent EET, compared with EET by supertransfer. Some insight may be obtained by extending the model from section \ref{sec:natphys} to include dissipative sites that mimic RCs and studying dissipation rates from the network under parameterisations that support long-range coherence, and others that do not. 

In particular, it is interesting to ask whether long-range coherence in a multi-LHC/RC network could improve the network's optimal solution to the excitation-allocation problem. One potential strategy for answering this question is to use a multiscale optimisation technique such as analytical target cascading (ATC) (section \ref{sec:multiopt}) to collaboratively optimise the EET properties of the network's PPCs, photosystem supercomplexes and the arrangement of the latter across the overall network, under the network-scale functional target of optimal excitation allocation. Of interest are the possible consequences of allowing and disallowing coherent EET at supra-PPC scales in different cases. Would the optimal network energy conversion efficiencies differ significantly? Would the multiscale structures of the optimally performing networks differ much between the two cases? How would they each compare with structures seen in natural thylakoids? These questions deserve further exploration. 

However, this in turn raises the methodological question of whether existing formulations of ATC (section \ref{sec:multiopt}) can accommodate multiscale quantum mechanical models, which are based on stochastic dynamical variables. To the author's knowledge, such application of ATC has not yet appeared in the literature. It is worth noting however, that probabilitic formulations of ATC have recently emerged \cite{xiong2010, chen2010}, designed to handle stochastic uncertainties in classical dynamical variables. Perhaps these could be adapted for multiscale design of engineered quantum systems, including light-harvesting materials.

Aside from light-harvesting materials, the scaling principles revealed in section \ref{sec:natphys} may be of interest in developing other kinds of engineered quantum systems. Quantum error correction is a major field of research aimed at developing schemes for protecting quantum dynamical systems from the decohering effects of their environments \cite{williams2011}. The analysis presented in section \ref{sec:natphys} demonstrates that quantum-coherent dynamics in a network described by the Frenkel exciton Hamiltonian can be preserved over surprisingly large length and time scales, even under thermal decoherence at ambient temperature, simply by virtue of network structure. One may speculate that similar effects can be expected under a broader class of Hamiltonians, of which the multiscale EET system is merely representative. This possibility, too, deserves further exploration. 

\chapter[Light-harvesting protein composition and multiscale structure in the thylakoid membrane]{Light-harvesting protein composition and multiscale structure in the thylakoid membrane} \label{chp:structure}

\chapquote{If you want to understand function, study structure. (I was supposed to have said in my molecular biology days.)}{F. Crick \cite{crick1988}} 

\section{Introduction}\label{sec:structintro}
Energetic modelling at each scale in a photosynthetic system requires understanding how the process(es) of interest depend(s) on system parameters such as composition and structure at that scale, as well as at other scales (chapter \ref{chp:multianalysis}). In an ideal scenario, a multiscale-optimal theoretical model would guide engineering at each scale in the system, from photobioreactor design to genetic engineering of the nanoscale light-harvesting machinery in the thylakoid. The latter is complicated by the fact that the protein composition of the thylakoid membrane is known to strongly affect its structure across a range of scales from individual protein supercomplexes to overall membrane ultrastructure, and these structures in turn affect the membrane's complex, multiprocess energetics (transfer of light, electronic excitations, molecules/radicals and heat). In addition to uncertainties about structure-energetics relationships in the thylakoid, such as those addressed for excitation energy transfer (EET) in chapter \ref{chp:qeet}, composition-structure relationships also are incompletely understood.

Recent studies (reviewed in \cite{ort2011, melis2009}; see also Appendix A) have demonstrated improved photosynthetic productivity at the cell-culture scale by genetically down-regulating the expression of light-harvesting protein complexes (LHCs). The improvements have been attributed to more equitable light distribution at the culture scale, resulting in reduced nonphotochemical quenching (NPQ) in cells near to the illuminated surfaces of the culture and increased photosynthesis in previously light-limited regions \cite{ort2011, melis2009, mussgnug2007}. However, the complex interdependences between thylakoid protein composition and structure, productivity at the chloroplast scale and productivity at the overall culture scale are not yet fully understood (chapter \ref{chp:multianalysis}). Characterising these relationships is essential for targeted thylakoid engineering as a part of multiscale system engineering. 
 
 An important step in this direction is to study correlations between LHC expression levels and multiscale thylakoid structure. This chapter describes contributions to an ongoing, larger research program focussed on developing high-throughput multiscale structure assays for thylakoids harvested from microalgal species with genetically engineered LHC expression levels -- so-called `antenna-mutant' strains. Within the context of this broader project, structural techniques appropriate for studying the thylakoid and its subsystems are introduced, and results of prior studies briefly reviewed. Novel structural studies are then reported for two thylakoid membrane protein (super)complexes from the antenna-mutant green microalgal strain \textit{Chlamydomonas reinhardtii stm3}, based on the technique of single-particle electron microscopy (single-particle EM). The significance of these results within the broader multiscale-structure research program is discussed. Finally, it is considered how the structure assays under development may be integrated with complementary theoretical modelling studies to more comprehensively elucidate composition-structure relationships and, ultimately, composition-structure-energetics relationships in the thylakoid.       

\section{Thylakoid structure: spectrum of scales and structural techniques} \label{sec:thylakoidspec}
\begin{figure*}
\centering
\includegraphics[angle=0,width=0.9\textwidth]{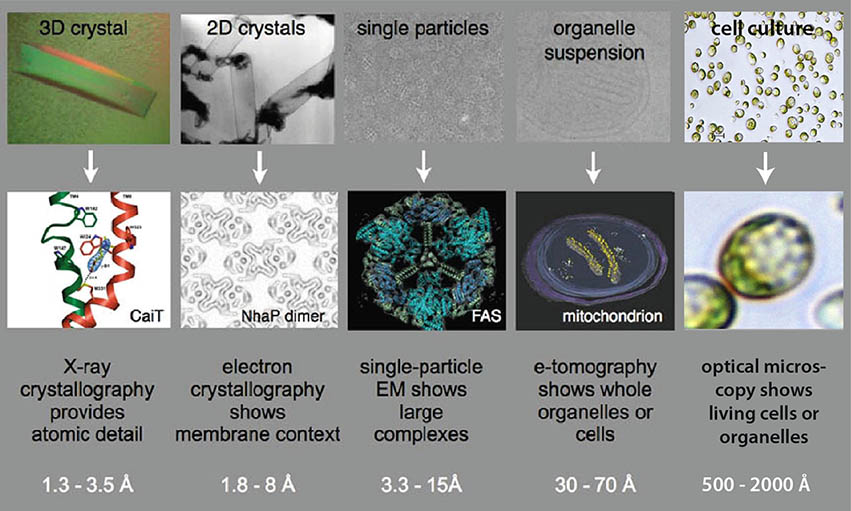}
\caption[Structural biology techniques and domains of applicability]{\textbf{Structural biology techniques and domains of applicability.} Resolution ranges (best to typical, according to \cite{kuhlbrandt2012, wang2011}) are given. Optical micrograph of cell culture courtesy of G. Jakob, used with permission. Overall figure adapted from \cite{kuhlbrandt2012}.}
\label{fig:techniques}
\end{figure*}

Resolving the 3D organisation of the thylakoid and its nested subsystems (fig. \ref{fig:multiscale}d--k) requires structural information over at least five orders of magnitude ($10^{-10}$--$10^{-6}$ m). No single technique can resolve structures across this range, so an integrated, multi-technique approach is required. Figure \ref{fig:techniques} summarises the array of applicable structural techniques. Mature techniques exist for both the coarsest and finest scales in the form of cellular optical microscopy and X-ray crystallograpy respectively, and intermediate techniques based on transmission electron microscopy (TEM\nomenclature{TEM}{Transmission electron microscope/microscopy}) are rapidly developing. Other techniques that have been applied to the analysis of thylakoid structure include nuclear magnetic resonance (NMR\nomenclature{NMR}{Nuclear magnetic resonance}) and atomic force microscopy (AFM\nomenclature{AFM}{Atomic force microscopy}), though these are not discussed here. 

\subsection{Optical microscopy and super-resolution nanoscopy}
With the resolution of conventional optical microscopy being diffraction-limited to $\sim 200$ nm in the visible spectrum, it is applicable at the scales of whole cells and organelles \cite{wang2011}. However, recently-developed super-resolution optical nanoscopes have resolving powers of up to $50$ nm \cite{wang2011} and therefore permit analysis of smaller structures as well. A rapidly growing array of such techniques \cite{kanchanawong2011, galbraith2011, wang2011, huang2010(1), schermelleh2010, toomre2010, huang2010, ntziachristos2010, huang2009, shtengel2009} can be applied to cells to observe their coarse-scale structures under varied conditions, including \textit{in vivo}. For a thorough review the reader is directed to the referenced literature.  

\subsection{X-ray crystallography}\label{subsec:xray}
 
At the opposite end of the resolution spectrum, X-ray crystallography is unsurpassed for providing atomic-scale 3D structural information, typically at resolutions of $\sim3.5~\textup{\AA}$ for membrane proteins and $\sim1.3~\textup{\AA}$ at best \cite{kuhlbrandt2012}. However, this technique requires that the structure of interest is able to form high-quality (`well-diffracting') 3D crystals for analysis by X-ray diffraction and such crystallisation can be a nontrivial challenge, particularly with intrinsic membrane proteins such as those of the thylakoid. 

The difficulty with crystallising membrane proteins arises from multiple factors.  
In order to attempt 3D crystallisation, membrane proteins must be solubilised with detergents and detergent molecules which remain bound to the proteins' hydrophobic, transmembrane regions can provide steric hindrance to the formation of crystal contacts between proteins \cite{loll2003}. Furthermore, removing proteins from their native lipid membrane environment can make them structurally unstable due to the removal of essential extrinsic or intrinsic lipids, which can subsequently make it impossible to determine the original, native structure.

\begin{figure*}
\centering
\includegraphics[angle=0,width=0.795\textwidth]{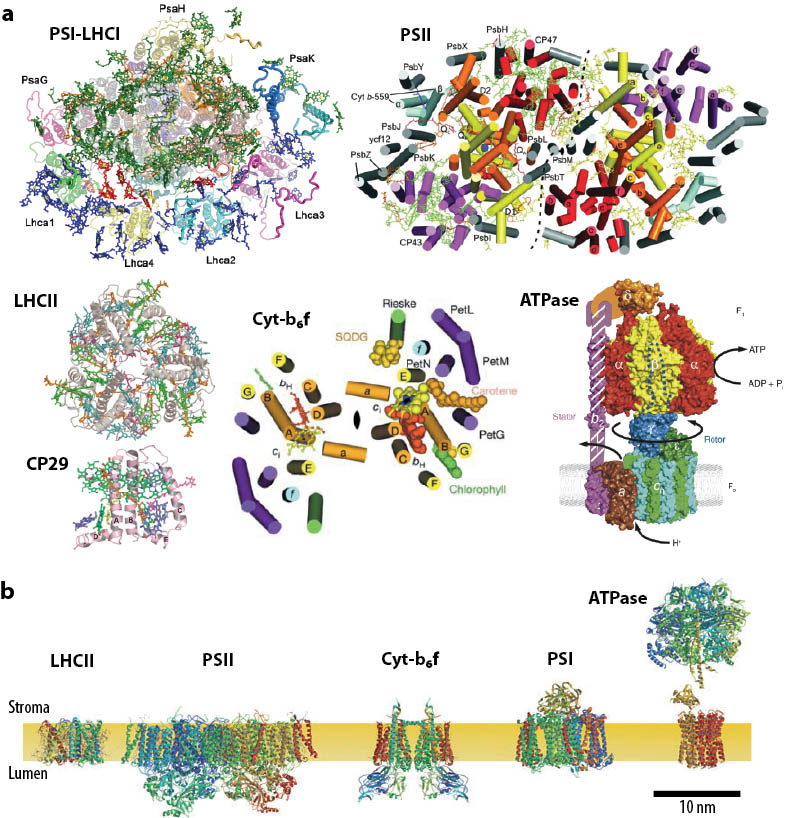}
\caption[X-ray crystal structures of thylakoid membrane proteins]{\textbf{High-resolution structures of thylakoid membrane protein complexes determined by X-ray crystallography.} \textbf{a}, \textbf{PSI-LHCI}: From plant, \textit{Pisum sativum}. View is from the stroma. Protein subunits are shown as different-coloured ribbon structures; selected, important subunits are labelled. Chlorophylls are shown in green (within the core complex), blue (in LHCs) and red (in the interstitial spaces). Carotenoids are shown in orange. Reproduced from \cite{amunts2010}. \textbf{PSII}: From cyanobacterium,  \textit{Thermosynechococcus elongatus}. Core dimer with protein subunits labelled. $\alpha$--helices are represented by cylinders and membrane-extrinsic regions are not shown.  Cofactors are shown in green (chlorophyll), orange (carotenoids) and blue (haem group). Reproduced from \cite{guskov2009}. \textbf{LHCII}: From \textit{P. sativum}. View is from the stroma. $\alpha$--helices are shown as grey ribbons, and cofactors are coloured cyan (chl--\textit{a}), green (chl--\textit{b}), orange (carotenoids) and pink (lipids). Reproduced from \cite{standfuss2005}. \textbf{CP29}: From plant, \textit{Spinacia oleracia}. View is parallel with the membrane plane. $\alpha$--helices are shown as pink ribbons, and cofactors are coloured green (chl--\textit{a}), blue (chl--\textit{b}), orange (carotenoid violaxanthin) and magenta (carotenoid neoxanthin). Reproduced from \cite{pan2011}. \textbf{Cyt}--$\boldsymbol{b_6f}$: From green microalga, \textit{C. reinhardtii}. View is from the stroma; only the stromal half of the transmembrane region is shown. Helices are represented by cylinders. Subunits are labelled and cofactors are represented by space-filling models. Reproduced from \cite{stroebel2003}. \textbf{ATPase}: From bacterium, \textit{Escherichia coli}. View is cross-sectional to the membrane plane. Each protein subunit is represented as a space-filling model and coloured differently. Extrinsic F$_1$ and membrane spanning F$_0$ regions are labelled. Reproduced from \cite{weber2007}. Structures in each case represent the most recently published structures, not to scale. \textbf{b}, X-ray structures, excluding CP29, shown to scale in thylakoid membrane environment. May be compared with fig. \ref{fig:zscheme}, which shows energetics. Adapted from \cite{eberhard2008}. Overall figure adapted from \cite{knauth2013}.}
\label{fig:xraystruct}
\end{figure*}    

\subsubsection{X-ray crystal structures of thylakoid membrane proteins}
Despite these difficulties, high-resolution structures have been determined for $\sim 300$ membrane proteins through X-ray crystallography \cite{white2013} (compared with $\sim 77,000$ X-ray structures for water-soluble proteins \cite{pdb2013}). Among these are several from the photosynthetic machinery of various species including higher plants, bacteria and algae (fig. \ref{fig:xraystruct}). These structures provide essential parameters for energetic modelling at the smallest photochemically-active scale in biological photosynthetic systems. There is a strong level of protein structural homology between species, such that homologues can to some extent be used interchangably in multiscale studies. However, the limitations of this approach should be accounted for in parameterising energetic models.

The photosystem II (PSII) core dimer isolated from cyanobacteria has been structurally characterised at resolutions of 3.5--1.9 \AA \cite{ferreira2004, guskov2009, umena2011} and crystal structures of the major light-harvesting antenna protein associated with PSII in higher-plants, LHCII have been determined at 2.7--2.5 \AA \cite{liu2004, standfuss2005}. Recently, the structure of one of the minor light-harvesting antenna proteins, CP29 was also obtained at 2.8 \AA~for higher plants \cite{pan2011}. 

X-ray crystallography has also delivered structural insights into the non-photosystem complexes of the photosynthetic light reactions (section \ref{subsec:lightreact}), cytochrome $b_6f$ (Cyt--$b_6f$) and ATP synthase (ATPase) (figs. \ref{fig:zscheme} and \ref{fig:xraystruct}). The structure of Cyt--$b_6f$ from \textit{C. reinhardtii} has been resolved at 3.1 \AA \cite{stroebel2003} and the structure from cyanobacteria has been determined at a similar resolution \cite{yamashita2007}. Although no high-resolution structure has been obtained for the complete ATPase complex, such structures have been obtained for its discrete parts \cite{gibbons2000, groth2001, weber2007} and then combined with lower-resolution structures from TEM-based techniques to propose a structure for the complete complex \cite{rubinstein2003}.  

\paragraph{Photosystem I}
The structure of the photosystem I--light-harvesting complex I (PSI--LHCI) supercomplex from higher plants (fig. \ref{fig:xraystruct}a) has been determined at 3.4 \AA~ resolution \cite{amunts2007} and recently refined to 3.3 \AA \cite{amunts2010}. This is the only photosynthetic membrane protein structure so far to include both the core complex (PSI) and dedicated light-harvesting antenna proteins (collectively LHCI; individually Lhca1--4). The structure is described here in some detail because it is significant to the novel structural studies presented in section \ref{sec:mutstruct}.

In addition to LHCI, the structure also contains protein subunits PsaA, B, C, D, E, F, G, H, I, J, K, and N (a total of 18 subunits). LHCI forms a crescent around one side of the core complex. Also assigned to the PSI-LHCI structure are 173 chlorophylls (145 revealing the orientation of the Q$_x$ and Q$_y$ transition dipole moments), 15 carotenoids, two phyloquinones and three Fe$_4$S$_4$ clusters \cite{amunts2010}. The major proteins, PsaA and B, form the core heterodimer near the centre of PSI and are associated with most of the cofactors involved in electron transport, including the special chlorophyll pair, P700 \cite{amunts2008}. There are structural and sequence similarities between these two PSI core complex proteins and the PSII core-complex proteins D1, D2, CP43 and CP47 of higher plants and cyanobacteria, indicating a strong evolutionary link between the photosystems \cite{barber1999}.

The PsaH protein, located on the opposite side of the core from LHCI, is thought to facilitate the so-called `state transitions' or `qT' mechanism of NPQ. These processes balance light absorption between photosystems I and II by shuttling major and minor LHCII proteins between the two on a time scale of tens of minutes \cite{minagawa2011, ruban2012}. The reported role of PsaH in the state transitions is to interact with LHCs that have migrated from PSII-LHCII in state 2 (fig. \ref{fig:ps1spa}e--g) \cite{kargul2005, kouril2005}.  

	\subsection{Electron tomography}\label{subsec:et}
	Three orders of magnitude separate the resolutions attainable with optical microscopy ($\sim10^3$ \AA) and X-ray crystallography ($\sim1$ \AA). This divide is rapidly being filled by TEM-based techniques: electron tomography (ET\nomenclature{ET}{Electron tomography}) at larger scales and single-particle EM at smaller scales.
	
	ET is widely applicable, being able to image up to a 3--4 $\upmu$m field of view \cite{fridman2012} and typically resolving structures at $\sim7$ nm, or 3 nm at best \cite{kuhlbrandt2012}. It is a powerful tool for resolving 3D ultrastructure of cells and organelles, and can even visualize large ($>$400 kDa), individual macromolecular assemblies. Indeed, it is currently the only technique able to visualise the macromolecular organisation of chloroplasts and their membrane systems in 3D \cite{daum2011}. 3D information is collected by recording series of tilted projections in increments of 1--5$\dg$ over a range of $\pm70\dg$. These images are then digitally reprojected to create a 3D volume \textit{in silico} \cite{daum2011}. In conventional ET, biological samples are embedded in plastic and sectioned by ultramicrotomy before imaging. This yields high-contrast images of coarse-scale cellular features such as membranes but molecular detail is lost. 
	
Higher resolution tomograms can be obtained by vitrifying the sample for imaging by electron cryo-tomography (cryo-ET\nomenclature{Cryo-ET}{Electron cryo-tomography}) \cite{daum2011, orlova2011}. Samples are prepared for cryo-ET by spreading a solution of the structure of interest (e.g. protein complexes or organelles) on an electron microscope (EM\nomenclature{EM}{Electron microscope/microscopy}) grid coated with a holey carbon film. A thin film of suspension forms within the holes and the sample is plunged into liquid ethane for flash freezing in a vitreous (near-native) state \cite{daum2011, orlova2011}. Large samples thicker than $\sim 300$ nm, such as tissues, whole cells or whole chloroplasts, are frozen under high pressure and then sectioned by cryo-ultramicrotomy \cite{daum2011, orlova2011}. Tomograms of serial sections can then be stitched together \textit{in silico} to reconstruct volumes larger than individual sections \cite{bouchet2011}. This comes at the cost of imperfect alignment between membrane fragments in adjacent sections  
due to uneven shrinkage and no overlap in image information between sections.

\begin{figure*}
\centering
\includegraphics[angle=0,width=0.9\textwidth]{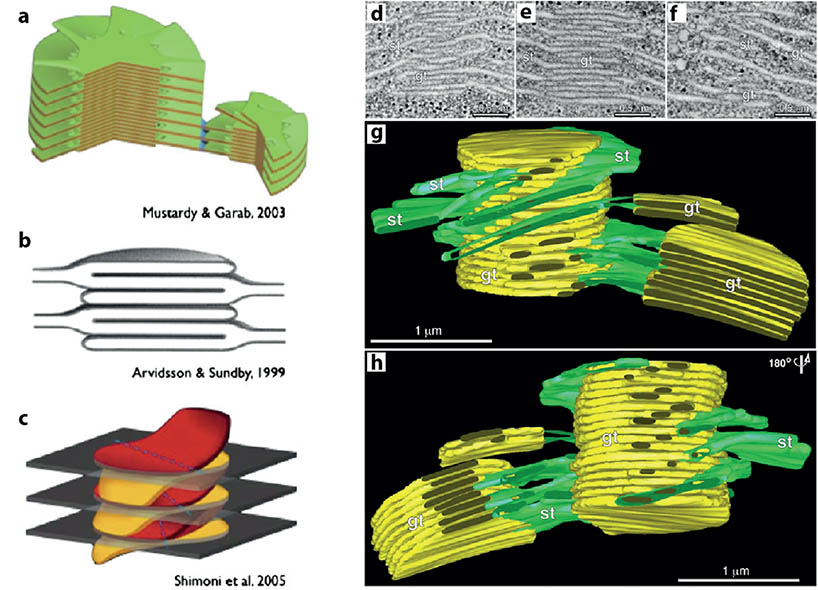}
\caption[3D models and electron-tomographic reconstructions of higher-plant thylakoids]{\textbf{3D models and electron-tomographic reconstructions of higher-plant thylakoid membranes. a,} Helix model \cite{mustardy2003} in which the stroma membranes (stroma `lamellae') wind around the granal stacks as a right-handed helix, connected to individual granal layers by narrow membrane protrusions. \textbf{b}, Simpler `fork' model \cite{arvidsson1999} suggesting that a granum consists of repetitive units, each containing three membrane layers formed by symmetrical invaginations of a thylakoid pair. \textbf{c}, The model of Shimoni \textit{et al} \cite{shimoni2005} in which the granal layers are paired units emerging from bifurcations of stroma membranes. In each pair, the upper layer bends upwards to fuse with next pair in the stack, and the lower layer bends down on the opposite side to fuse with the layer below. \textbf{d--f}, Composite tomographic slice images showing views from the front (d), middle (e), and back (f) of a granum. \textbf{g,h}, tomographic reconstructions of the granum shown in d--f. Granal membranes are shown in yellow, and stroma membranes in green. Panel h is rotated 180$\dg$ with respect to g. Panels a--c adapted from \cite{daum2011}, and panels d--h adapted from \cite{austin2011}.}
\label{fig:thyultra}
\end{figure*}

\subsubsection{Electron-tomographic studies of thylakoid ultrastructure}
Several studies have used ET to elucidate ultrastructure in higher-plant thylakoids, as recently reviewed in \cite{daum2011, kouril2012}. Earlier studies proposed contradictory models for the structures of granal stacks, stroma membanes (`lamellae') and intersections between the two (fig. \ref{fig:thyultra}a--c). However, more recent findings \cite{daum2011, austin2011, kouril2011, mustardy2008} (e.g. fig. \ref{fig:thyultra}d--h) have tended to support the so-called helical model (fig. \ref{fig:thyultra}a) originally proposed by Wehrmeyer \cite{wehrmeyer1964(1), wehrmeyer1964(2)} and advocated in particular by Mustardy and Garab \cite{mustardy2003, mustardy2008}. In this model, grana comprise stacks of flat, quasicircular membrane disks and stroma lamellae wind around each granum as a right-handed helix, connecting to individual granal layers by narrow membrane protrusions \cite{daum2011, mustardy2003, mustardy2008}. 

However, while falling broadly in support of this model, recent results suggest it to be an oversimplification of real thylakoids, in which the sizes of the junctions between stromal and granal thylakoids vary widely, from $\sim35$--$400$ nm; in some cases one stromal thylakoid forms a planar sheet with only one granal disk \cite{daum2011, austin2011}. The helical model has also been criticised because it does not present a clear mechanism for accommodating the extensive reorganisation of the thylakoid known to occur under changing light conditions \cite{daum2011, arvidsson1999, kouril2012}. An alternative `fork' model has been proposed (fig. \ref{fig:thyultra}b), involving simpler interfacing between stromal and granal domains, making it conducive to reversible unstacking of grana under changing light conditions \cite{arvidsson1999, kouril2012}. While some structural studies have found in favour of this model (as summarised in \cite{kouril2012}), on balance the current evidence is inconclusive. Further structural analyses are required in order to fully characterise thylakoid ultrastructure and its complex, dynamic variations. Cryo-ET is seen as the most promising technique by which to achieve this \cite{kouril2012}.

	 So far, no electron-tomographic studies of green microalgal thylakoids have appeared in the literature. Given the importance of thylakoid structure determination to engineering photosynthetic energy systems, cryo-ET studies of microalgal thylakoids are needed. Such studies are underway\footnote{Being conducted by other researchers within the Hankamer group at the University of Queensland (in which parts of the present research project were completed), and possibly elsewhere as well.} within a broader effort to develop high-throughput multiscale structure assays of the thylakoid \cite{knauth2013}. In order to bridge the resolution gap between ET reconstructions and published X-ray crystal structures of thylakoid protein complexes, complementary studies using single-particle EM are also underway. Since this technique is applied in the novel structural studies reported in section \ref{sec:spastm3}, it is now described in some detail.  

	\subsection{Single-particle electron microscopy}\label{subsec:spa}
		In high-resolution thylakoid electron tomograms, coarse structural features of individual protein supercomplexes can be identified \cite{daum2011, daum2010}, allowing for structure-based docking of smaller, high-resolution models. However, typical ET resolutions of 5--7 nm are unable to resolve protein subunit arrangements. X-ray crystal structures are established only for subunits of thylakoid protein complexes (except higher-plant PSI-LHCI, for which an X-ray structure exists for the whole supercomplex, as presented in section \ref{subsec:xray}). Single-particle EM reconstructions of protein supercomplexes can reveal the quaternary structure of these multi-subunit complexes, thereby bridging the gap between ET and X-ray models and facilitating multiscale structural modelling at quasi-atomic resolution \cite{rossmann2000}. The combination of these structural techniques offers a unique tool for exploring the structural configuration space\footnote{The term, `structural configuration space' is introduced here to refer to the abstract space spanned by all possible structural arrangements of the system of interest. Each point in this space maps to a complete structural configuration of the system.} of the thylakoid under different genetic and acclimation conditions.
		
\begin{figure*}
\centering
\includegraphics[angle=0,width=1\textwidth]{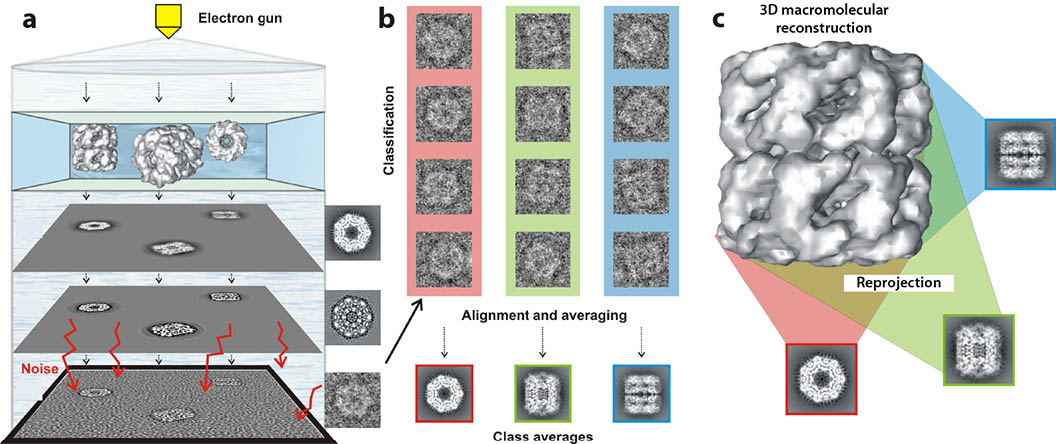}
\caption[Summary of single-particle electron microscopy (single-particle EM)]{\textbf{Summary of single-particle electron microscopy (single-particle EM). a,} TEM imaging of randomly oriented single particles yields (noisy) 2D projection images. \textbf{b}, Statistically similar images are classified, then translationally and rotationally aligned and averaged to yield high signal-to-noise ratio class averages. \textbf{c}, Class averages are projected at predefined angles to create an initial 3D model, which is then reprojected to create reference images as a basis for particle re-classification and 3D model refinement. This process is re-iterated until the 3D model structure converges (or fails to converge). Figure adapted from \cite{woolford2007}.}
\label{fig:spa}
\end{figure*}		

Single-particle EM comprises a group of EM and image processing techniques for determining structures of macromolecular complexes without requiring crystallisation (fig. \ref{fig:spa}), as recently reviewed in \cite{boekema2009, orlova2011, kuhlbrandt2012}. The complex (`particle') of interest is biochemically isolated and distributed onto a microscope grid, where single particles bind to the substrate in random orientations. The grid is then imaged using TEM to collect single-particle 2D-projection images, which are computationally post-processed into a 3D reconstruction of the complex's electron density distribution. The particles' random orientations are useful because a wide range of angular projections is required for a complete, isotropic-resolution 3D reconstruction. Electron dose must be minimised in order to avoid radiation damage to the sample, though this results in noisy images with particles barely visible. The signal-to-noise ratio (SNR\nomenclature{SNR}{Signal-to-noise ratio}) is increased by image averaging, so a large number of single-particle images is required \cite{boekema2009}; typically a dataset comprising on the order of $\sim10^4$--$10^5$ particle projection images.

\paragraph{Sample preparation and imaging}
In `negative-stain' EM, particles are stained with a heavy metal salt solution (typically uranyl acetate) prior to imaging at room temperature. The salt fills cavities in and around the complex but usually does not penetrate the hydrophobic interior. Image contrast is proportional to the sample's atomic number, so the particles' heavy-metal stain layer greatly improves contrast at the protein envelope compared with unstained particles. However, this method faces the problem that fragile complexes can collapse or disintegrate during staining and drying. The stain can also distort the structure of the complex or, if it does not cover the entire complex, can produce imaging artifacts. 

An alternative approach is cryo preparation, in which particles are suspended in vitreous ice in a manner similar to that used for cryo-ET samples. Cryo preparation avoids the problems of negative staining but results in lower image contrast, particularly for small complexes ($\lesssim100-200$ kDa). It is therefore best suited to larger, highly symmetric complexes such as viruses \cite{orlova2011, boekema2009}. Indeed, cryo-single-particle EM has achieved 3D reconstructions for such particles with quasi-atomic resolutions ($<4$ \AA) approaching those typical for X-ray crystal structures \cite{zhang2008}. Resolutions achievable with negative-stain preparations are typically lower ($\sim 10-30$ \AA) \cite{kuhlbrandt2012}, though resolutions $<10$ \AA~ have also been achieved \cite{boekema2009}. 

EM images are recorded at high magnification ($\times10^4$--$10^5$) using either film or a charge-coupled device (CCD\nomenclature{CCD}{Charge-coupled device}) camera (though these are being superceded by direct electron counters). Once raw images have been digitised, semi-automated software packages such as EMAN2 boxer \cite{tang2007} are used to select a dataset of individual particle projection images.

\paragraph{Image processing and 3D model reconstruction}
High-resolution 3D model reconstruction from low-SNR 2D projection requires both sophisticated image-processing techniques \cite{orlova2011} and considerable computing power. Numerous specialised single-particle EM reconstruction software packages are available \cite{pdbe2013}, the most popular of which include EMAN \cite{ludtke1999} (and EMAN2 \cite{tang2007}), IMAGIC \cite{vanheel1996}, Spider \cite{shaikh2008}, Xmipp \cite{sorzano2004} and Frealign \cite{grigorieff2007}. The basic reconstruction algorithm, on which these packages are based, may be summarised as follows. First, the single-particle dataset is subjected to multivariate statistical analysis to sort the 2D projections into groups or \textit{classes} of similar images, which are translationally and rotationally aligned. Each class is then averaged to create a representative `class average' (or `projection map') image with a much higher SNR than its constituent images. Class averages are systematically projected through a 3D volume at calculated angles to create an initial 3D reconstruction (electron density map). This is subsequently re-projected to create reference images for reclassification of the dataset, which begins the next round of 3D model refinement. This process is re-iterated until the 3D model structure converges (or fails to converge). The 2D projection class averages can themselves provide useful structural information, for example by indicating the arrangement of membrane protein subunits within the membrane plane.

\subsection{Single-particle analyses of thylakoid photosystem supercomplexes}\label{subsec:spathy}
Single-particle EM studies have been reported for the PSII-LHCII and PSI-LHCI membrane supercomplexes of a variety of species, under different acclimation and solubilisation conditions, revealing a range of subunit compositions and arrangements. The focus here is on green algae and higher plants.

\begin{figure*}
\centering
\includegraphics[angle=0,width=0.90\textwidth]{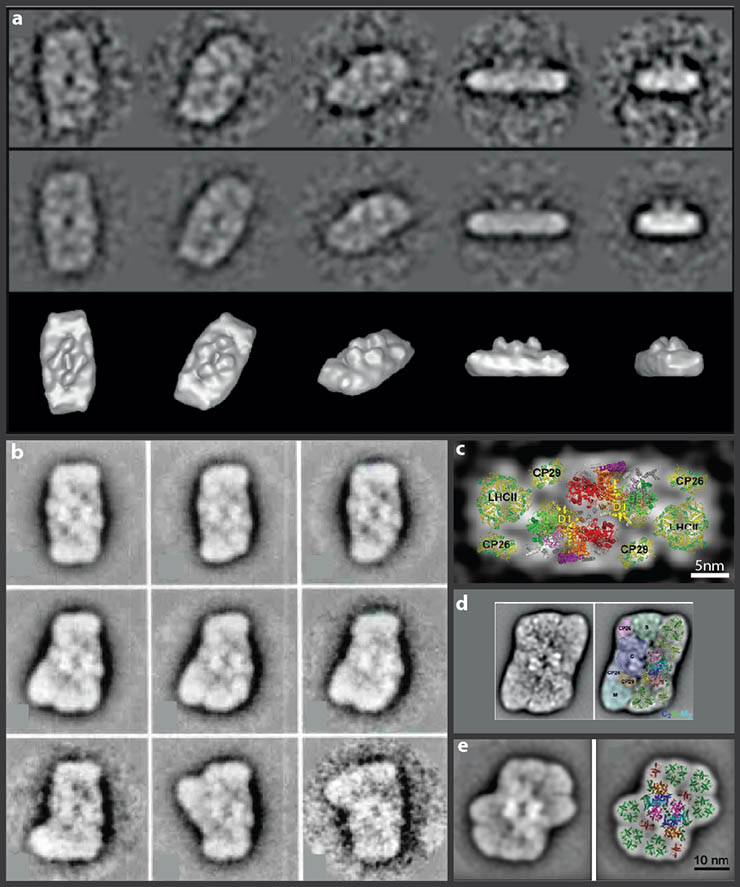}
\caption[PSII-LHCII supercomplex structures determined by single-particle analysis]{\textbf{Single-particle structures of PSII-LHCII supercomplexes from \textit{C. reinhardtii}, \textit{A. thaliana} and \textit{S. oleracia}.} Panels adapted from cited literature: \textbf{a}, (Top) Projection maps, (centre) reprojection maps, and (bottom) 3D reconstruction of C$_2$S$_2$-type supercomplex from \textit{C. reinhardtii}, at 30 \AA~resolution \cite{nield2000(1)}. \textbf{b}, Projection maps of various supercomplex configurations from partially solubilized thylakoids of \textit{S. oleracia} at $>16.5$ \AA~resolution \cite{boekema1999}. \textbf{c}, Projection map of C$_2$S$_2$-type supercomplex from \textit{S. oleracia} at 17 \AA~resolution, and assignment of subunits by fitting high-resolution crystal structures (fig. \ref{fig:xraystruct}) \cite{nield2006}. \textbf{d}, (Left) Projection map of C$_2$S$_2$M$_2$ supercomplex from \textit{A. thaliana} at 12 \AA~resolution, and (right) assignment of subunits by fitting high-resolution crystal structures \cite{caffarri2009, kouril2012}. \textbf{e}, (Left) Projection map of C$_2$S$_2$M$_2$L$_2$-type supercomplex from \textit{C. reinhardtii} at 16.8 \AA~resolution, and (right) assignment of subunits by fitting high-resolution crystal structures \cite{tokutsu2012}.}
\label{fig:ps2spa}
\end{figure*} 

\paragraph{PSII--LHCII}
Nield and coworkers provided the first 3D reconstruction of PSII--LHCII from \textit{C. reinhardtii} \cite{nield2000(1)}, showing the C$_2$S$_2$-type\footnote{This nomenclature specifies the numbers of core complexes (C), strongly-bound (S) LHCII complexes, moderately-bound (M) LHCIIs, and weakly-bound (W) LHCIIs in a PSII-LHCII supercomplex.} supercomplex (fig. \ref{fig:ps2spa}a) previously identified in higher plants \cite{boekema1995}. Another 3D reconstruction of the same structure from higher plants was also reported by Nield \textit{et al} in the same year \cite{nield2000}. This contrasted with a variety of PSII-LHCII types, and thus subunit arrangements, reported by Boekema \textit{et al} for higher plants \cite{boekema1999}, including several that contain extra LHCII trimers (fig. \ref{fig:ps2spa}b). In follow-up work, Nield \textit{et al} fitted high-resolution X-ray structures within a 17 \AA-resolution single-particle EM projection map to obtain a tentative, quasi-atomic structure for the C$_2$S$_2$ supercomplex (fig. \ref{fig:ps2spa}c). A C$_2$S$_2$M$_2$-type supercomplex was subsequently purified from \textit{A. thaliana} \cite{caffarri2009} (fig. \ref{fig:ps2spa}d) and more recently a C$_2$S$_2$M$_2$L$_2$ type was observed in \textit{C. reinhardtii} \cite{tokutsu2012} (fig. \ref{fig:ps2spa}e). 

Neither of the latter two configurations have yet been observed in other species \cite{tokutsu2012}. However, existing evidence does not dictate whether the supercomplex configurations are possible in different species because their isolation involves different membrane solubilisation procedures and artifacts resulting from these cannot be ruled out \cite{tokutsu2012}. It thus remains an open question whether or not the particular configurations observed are of special functional importance \textit{in vivo} or rather, are mere structural fragments selectively preserved during solubilisation of larger core complex--LHC interspersions across the membrane, comprising a variety of structural arrangements. Indeed, additional to the family of known PSII-LHCII supercomplexes, numerous larger arrangements of PSII and its major and minor associated LHCII complexes have frequently been observed in EM and AFM studies of stacked thylakoids \cite{dekker2005}. These include so-called `megacomplexes', each made up of a bound pair of PSII-LHCII supercomplexes; Dekker and Boekema reported five different known structural types of megacomplexes \cite{dekker2005}. Semicrystalline arrays of PSII-LHCII supercomplexes have also been observed, spanning distances of up to hundreds of nanometres across the granal regions of higher-plant thylakoids, where PSII and LHCII are predominantly localised (fig. \ref{fig:multiscale}e--g) \cite{kouril2012, dekker2005, hankamer1997}. Relatively disordered `fluid' interspersions of PSII and LHCII have also been routinely observed \cite{kouril2012}. Various functional roles have been proposed for these different structural configurations, typically related to exciton distribution between photochemical reaction centres (RCs) and/or nonphotochemical quenching sites, or protein and electron-carrier mobility \cite{schneider2013}. However, no quantitatively predictive models of the proposed structure-energetics relationships have yet been established \cite{schneider2013}. 

\begin{figure*}
\centering
\includegraphics[angle=0,width=0.95\textwidth]{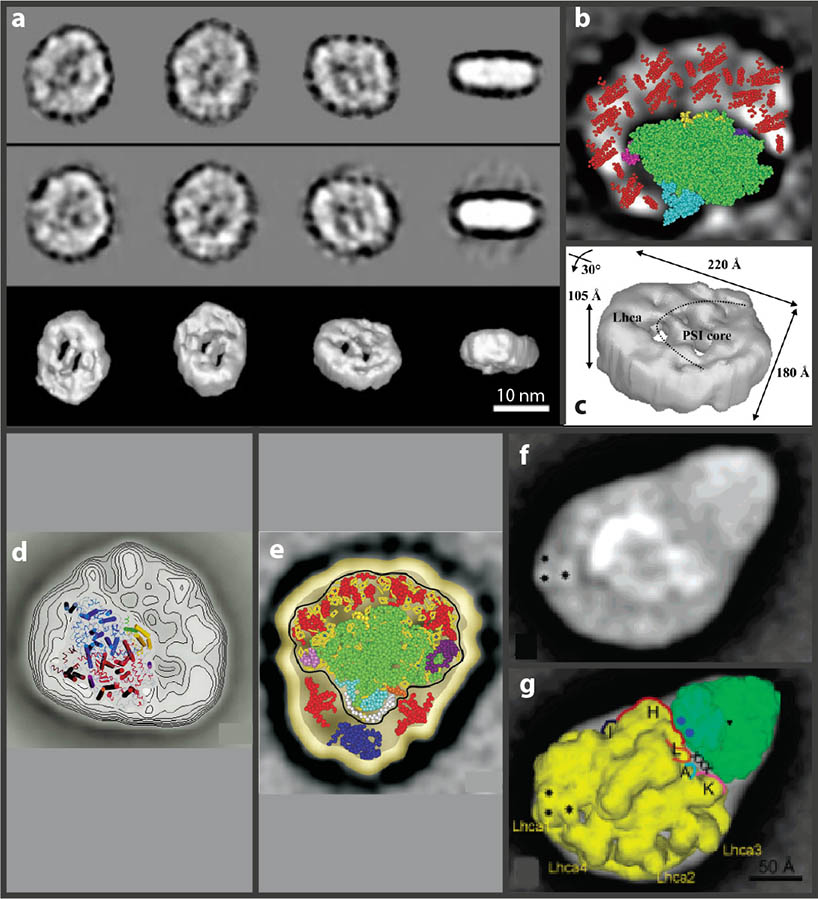}
\caption[PSI-LHCI supercomplex structures determined by single-particle analysis]{\textbf{Single-particle structures of PSI-LHCI supercomplexes from \textit{C. reinhardtii} and \textit{A. thaliana}.} Panels adapted from cited literature: \textbf{a}, (Top) Projection maps, (centre) reprojection maps, and (bottom and panel \textbf{c}) 3D reconstruction of supercomplex from \textit{C. reinhardtii}, at 30 \AA~resolution \cite{kargul2003}. Scale bar represents 10 nm. \textbf{b}, Modelling study using a top view (from the stroma) difference projection map from the same investigation \cite{kargul2003}; the PSI core density has been subtracted to reveal the Lhca densities, and X-ray crystal structures (fig. \ref{fig:xraystruct}) tentatively fitted for PSI (green, cyan, magenta, yellow, purple) and LHCI antenna (red). \textbf{d}, Projection map of supercomplex from \textit{C. reinhardtii} at 14 \AA~resolution with electron density contours shown, and fitting of the PSI core X-ray structure \cite{germano2002}. \textbf{e}, Projection map of supercomplex from \textit{C. reinhardtii} with fitting of PSI (green), LHCI (red) and CP29 (dark blue) suggested \cite{kargul2005}. \textbf{f}, Projection map of supercomplex from \textit{A. thaliana} at 15 \AA~resolution, suggesting association between PSI and a large additional density \cite{kouril2005}. \textbf{g}, Structural assignment of PSI-LHCI and LHCII to the large supercomplex in panel f \cite{kouril2005}.}
\label{fig:ps1spa}
\end{figure*} 

\paragraph{PSI-LHCI}
The high-resolution X-ray crystal structure of higher-plant PSI-LHCI, which contains four Lhca subunits (fig. \ref{fig:xraystruct}) forms the basic building block for all known configurations of PSI-LHCI in higher plants and green algae \cite{kargul2012}. However, structural, proteomic and biochemical studies have identified variations containing an additional 2--7 Lhca subunits, as well as associations with LHCII and minor PSII-associated LHCs during state 2 of the state transitions in both algae and higher plants \cite{kargul2012}. 

Kargul \textit{et al} \cite{kargul2003} used single-particle EM to reconstruct the first 3D model of PSI-LHCI from \textit{C. reinhardtii} at 30 \AA~resolution (fig. \ref{fig:ps1spa}a,c). Fitting high-resolution X-ray structures (fig. \ref{fig:ps1spa}b) suggested an extra band of Lhca subunits addtional to the Lhca1--4 in higher-plant PSI-LHCI. This was consistent with an earlier finding by Germano and coworkers \cite{germano2002}, who studied higher-resolution (14 \AA) 2D projection maps of PSI-LHCI from \textit{C. reinhardtii} (fig. \ref{fig:ps1spa}d). A more recent single-particle EM study \cite{drop2011} further supported these findings, identifying at 15 \AA~resolution a PSI-LHCI configuration in \textit{C. reinhardtii} with nine Lhca subunits, all located on one side of the core complex (fig. \ref{fig:ps1spa}e) (fig. \ref{fig:xraystruct}). 

In addition to the various PSI-LHCI supercomplexes so far identified under state 1 conditions, Kargul and coworkers have more recently presented evidence \cite{kargul2005} for an association between phosphorylated CP29 and PSI under state 2 conditions (fig. \ref{fig:ps1spa}f). A similar association with LHCII in state 2 was also identified \cite{kouril2005} (fig. \ref{fig:ps1spa}g,h). 

For brevity, non-EM findings are not detailed here. However, of particular note is proteomic work reported by Stauber \textit{et al} \cite{stauber2009}, who studied the stoichiometry of Lhca subunits with the PSI core in thylakoids extracted from \textit{C. reinhardtii}. Results revealed that the thylakoid LHCI population was heterogeneously composed (the antenna-to-core stoichiometry varied across the membrane) and on average LHCI contained $\sim7.5\pm1.4$ Lhca subunits. This suggests a need for studying supercomplex structures within the context of larger membrane structures, to elucidate their significance within the overall membrane configuration space. Multiscale structural studies combining results from single-particle EM and ET reconstructions offer a unique tool towards this end. 

\section{Thylakoid structure in antenna mutants of \textit{C. reinhardtii}} \label{sec:mutstruct}
The structural studies so far reviewed in this chapter have all dealt with wild-type organisms. However, as discussed in section \ref{sec:structintro} and chapter \ref{chp:multianalysis}, genetic engineering of thylakoid membrane protein expression (typically, down-regulation of LHC expression levels) is increasingly used to improve productivity in microalgal cultivation systems. The complex, multiscale effects of engineered LHC expression are known to include changes in thylakoid structure \cite{mussgnug2007}. However, the details of these changes, the composition-structure relationships underlying them, and their effects on multiscale system energetics are not yet well characterised.

\begin{figure*}
\centering
\includegraphics[angle=0,width=0.6\textwidth]{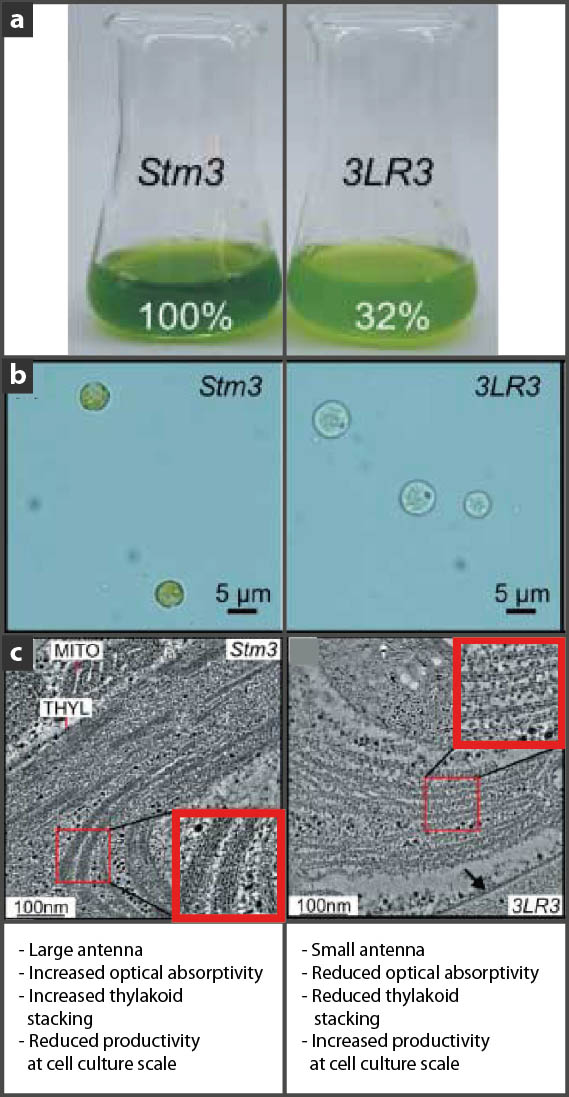}
\caption[Comparison of antenna-mutant strains \textit{stm3} and \textit{stm3LR3}]{\textbf{Comparison of antenna-mutant strains \textit{stm3} and \textit{stm3LR3}.} \textbf{a}, \textit{stm3LR3} (\textit{`3LR3'}) cell culture chlorophyll concentration was 32\% of \textit{stm3} at the same cell density, and visibly less absorptive. \textbf{b}, Individual \textit{3LR3} cells are also visibly less absorptive than \textit{stm3} cells. \textbf{c}, Transmission electron micrographs showing resin-embedded cell sections of 80 nm thickness and stained with uranyl acetate/lead citrate, imaged at 300 keV. Thylakoid stacking is increased in \textit{stm3} and reduced in \textit{3LR3}. Overall figure adapted from \cite{mussgnug2007}.}
\label{fig:mutantthy}
\end{figure*}

Mussgnug \textit{et al} created an insertional mutant of \textit{C. reinhardtii}, \textit{stm3} (state transition mutant 3), which was shown to have increased levels of LHCs associated with PSII compared with the wild type \cite{mussgnug2005}. Subsequently, \textit{stm3LR3} (\textit{stm3} LHC reduction mutant 3) was created using RNAi technology, with \textit{stm3} as the parent strain \cite{mussgnug2007}. In \textit{stm3LR3}, the expression of all 20 genes encoding for LHCI, LHCII, CP26 and CP29 were simultaneously and strongly down-regulated, to levels between 0.1\% and 26\% of that in \textit{stm3}.

Cell cultures of \textit{stm3LR3} (fig. \ref{fig:mutantthy}a,b) show higher cell growth and replication rates, and lower photoinhibition (measured as reduced O$_2$ evolution at the culture scale) than \textit{stm3} under high-light conditions (1000 $\upmu$E m$^{-2}$ s$^{-1}$). This is attributed to reduced levels of NPQ in the thylakoid and more equitable light distribution at the culture scale \cite{mussgnug2007}. 

The work reported below in section \ref{sec:spastm3} forms part of a larger program  
which aims to better elucidate interdependences between LHC composition, thylakoid structure and photosynthetic performance at the culture scale. As part of this work, thylakoids from the two `antenna mutants', \textit{stm3} and \textit{stm3LR3} (large and small antennae respectively), along with the wild-type parent strain, have been imaged in 2D using TEM to investigate differences in ultrastructure between strains (fig. \ref{fig:mutantthy}c). Preliminary investigations have revealed a greater level of thylakoid stacking in \textit{stm3}, including formation of `pseudograna' (fig. \ref{fig:multiscale}d,e [right], fig. \ref{fig:mutantthy}c [left]), compared with \textit{stm3LR3}, which showed only long stretches of parallel double bilayers (fig. \ref{fig:mutantthy}c [right]). This is consistent with the established role of LHCII in facilitating attraction between layers \cite{chow2005, dekker2005, schneider2013}. However, detailed structural changes at the scale of photosystem supercomplexes have not been determined, and how these relate to ultrastructural and functional differences at larger scales is not yet well understood.

In the next section this work is furthered through a structural survey of pigment-protein (super)complexes isolated from \textit{stm3}. The method of choice has been negative-stain single-particle EM because it combines relatively straightforward sample preparation with resolving power ample to facilitate structure-based docking of smaller, high-resolution structures. Moreover, the electron density maps resulting from the single-particle reconstructions may be used as structural `building blocks' for docking into larger-scale ET structures. The ultimate aim is for the resulting multiscale 3D structural map to be compared with similar maps from \textit{stm3LR3} and wild-type strains.   

\section[Single-particle EM analysis of two thylakoid protein complexes from \textit{stm3}]{Single-particle analysis of two thylakoid protein complexes from large-antenna mutant, \textit{stm3}} \label{sec:spastm3}

\subsection{Experimental procedures}\label{subsec:methods}
\subsubsection{Thylakoid membrane purification and solubilisation}
A 300 mL culture of \textit{stm3} cells was innoculated at an OD$_{750}$ of 0.2 and grown to mid-log phase at 100 $\upmu$E m$^{-2}$ s$^{-1}$ irradiance, then centrifuged (10 min, $2200 g$, 4 $\dg$C) and washed by re-suspending in 30 mL of buffer A (25 mM HEPES pH 7.5, 1 mM MgCl$_2$, 0.3 M sucrose). The washed cells were then pelleted by centrifugation (10 min, $2200 g$, 4 $\dg$C) before being re-suspended in 8 mL of buffer A and sheared open by two passes through a French press (2000 psi, 4 $\dg$C). The sample was then diluted to a final volume of 30 mL using buffer A and the thylakoid membranes were precipitated by centrifugation (45 min, $20,000 g$, 4 $\dg$C). Thylakoid membranes were washed with 30 mL buffer B (5 mM HEPES, pH 7.5, 10 mM EDTA, 0.3 M sucrose) and pelleted (45 min, 4 $\dg$C, $48,000 g$). The thylakoid membranes were then re-suspended in 2.65 mL of buffer B and 1 mL buffer C (5 mM HEPES, pH 7.5, 10 mM EDTA, 2.2 M sucrose) to adjust the final sucrose concentration to 1.82 mM (not including volume of the pellet). This suspension was then dispensed into a Beckman SW32Ti centrifugation tube before being overlaid with 1.75 M sucrose solution (half of the tube volume [15.3 mL]) and a further layer (5 mL) of buffer D (5 mM HEPES, pH 7.5, 0.5 M sucrose) before centrifuging (60 min, $100,000 g$, 4 $\dg$C). The centrifuged thylakoid membrane sample formed a dense green band at the position of the sucrose cushion step. Upon harvesting, the sample was diluted with five volumes of buffer E (20 mM MES, pH 6.3, 5 mM MgCl$_2$, 15 mM NaCl, 10\% (v/v) glycerol) to facilitate pelleting of the purified thylakoid membranes on centrifugation (20 min, $40,000 g$, 4 $\dg$C). The thylakoid membranes were then re-suspended in a minimal volume ($\sim$1 mL) of MMNB buffer (25 mM MES, pH 6.0, 5 mM MgCl$_2$, 10 mM NaCl, 2 M betaine). 

Membrane protein complexes were isolated from the thylakoid membranes (0.5 mg mL$^{-1}$ Chl) by solubilizing the samples in loading buffer (25mM MES pH5.5, 10mM NaCl, 5mM CaCl$_2$, 33 mM $\beta$-dodecyl maltoside, 500 $\upmu$L final volume) and resolving them on a sucrose density gradient (16 h, $\sim250 ,000 g$, 4 $\dg$C) (fig. \ref{fig:sucromicro}a). Sucrose gradients were formed by freeze-thawing centrifuge tubes filled with 25 mM MES pH 5.5, 500 mM sucrose, 10 mM NaCl, 5 mM CaCl$_2$, 10 mM NaHCO$_3$, and 0.03\% $\beta$-dodecyl maltoside according to \cite{hankamer1997}.

\subsubsection{Electron microscopy}

\begin{figure*}
\centering
\includegraphics[angle=0,width=1\textwidth]{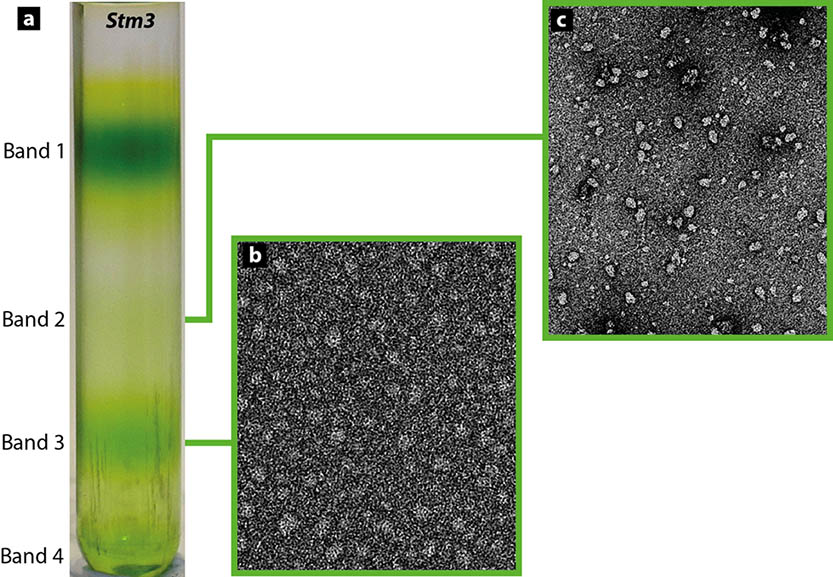}
\caption[Sucrose gradient and electron micrographs of \textit{stm3} thylakoid fractions]{\textbf{Sucrose gradient and electron micrographs of \textit{stm3} thylakoid fractions. a,} Green bands contain pigment-protein (super)complexes of increasing mass (from top to bottom). More optically dense green bands indicate higher chlorophyll content, suggestive of higher LHC content. Samples harvested from bands 3 and 2 were examined using negative-stain EM; resulting micrographs are shown in \textbf{b} and \textbf{c} respectively, at 67,000$\times$ nominal magnification. Protein densities are white.  
Higher optical density and larger particle size in band 3 compared with band 2 are consistent with higher LHC content, suggesting photosystem supercomplexes and core complexes in these bands respectively.}
\label{fig:sucromicro}
\end{figure*}

\paragraph{Band 3}
Sample harvested from the band-3 sucrose gradient fraction (fig. \ref{fig:sucromicro}a) was diluted to $1:25$ concentration using dilution buffer (10 mM MES, pH 5.5, 5 mM CaCl$_2$, 5 mM NaCl), applied to glow-discharged carbon-coated copper grids, allowed to bind for 30 s and then negatively stained for 20 s with 1\% uranyl acetate. Electron micrographs were recorded on a Tecnai T12 electron microscope (FEI) at 100 kV, 67,000$\times$ nominal magnification and an underfocus value of $\sim0.7$ $\upmu$m. Images free from astigmatism and drift were scanned from film at 4000 dots per inch (dpi) using a Nikon Super Coolscan 8000, resulting in a pixel size of 0.95 \AA. 

\paragraph{Band 2}
The same method was followed as for the band-3 fraction, with the exceptions that freshly prepared 0.7\% uranyl formate stain was used and EM images were captured on an FEI Eagle CCD camera (4096$\times$4096 pixels) at a pixel size of 1.7 \AA.

\paragraph{Bands 1 and 4} Electron microscopy and single-particle analyses of bands 1 and 4 were conducted in a parallel study \cite{knauth2013}. Detailed results are not reported here but are summarised briefly at the end of this chapter. 

\subsubsection{Single-particle analysis}
\begin{figure*}
\centering
\includegraphics[angle=0,width=0.95\textwidth]{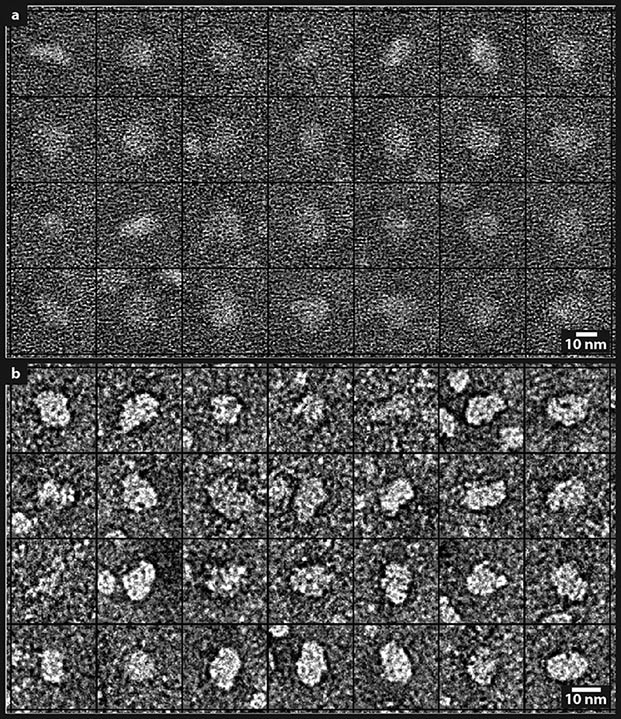}
\caption[Sample images from particle projection datasets for bands 3 and 2]{\textbf{Sample images from particle projection datasets for bands 3 (a) and 2 (b).} Band 3 particles were band-pass filtered with a high-frequency cutoff of 17 \AA~and a low-frequency cutoff of 230 \AA, and a circular mask of 40-pixel radius applied. Band 2 particles were band-pass filtered to 190 and 9 \AA. A mask was deemed unnecessary due to the higher level of contrast visible in the band 2 data.}
\label{fig:particles}
\end{figure*}

\paragraph{Band 3}
Electron micrographs were coarsened to 3.8 \AA/px using proc2d software from the EMAN package. From 110 micrographs, 9,548 particles were extracted in a 108$\times$108-pixel area with SWARMps software and band-pass filtered with a high-frequency cutoff of 17 \AA~and a low-frequency cutoff of 230 \AA~using proc2d. Subsequently a circular mask with 40-pixel radius was applied to each image. The resulting particle projection dataset was analysed using the EMAN, IMAGIC and Xmipp packages. 3D reconstructions were refined using EMAN and visualised using UCSF Chimera \cite{pettersen2004}, as described in section \ref{subsec:resanddis}.  

\paragraph{Band 2}
Electron micrographs were coarsened to 3.3 \AA/px with proc2d. From 180 micrographs, 13,958 particle projections were extracted in a 90$\times$90-pixel area with e2boxer software from the EMAN2 package \cite{tang2007} and band-pass filtered with a high-frequency cutoff of 9 \AA~and a low-frequency cutoff of 190 \AA~using proc2d. The resulting particle projection dataset was analysed using the EMAN, IMAGIC and Xmipp packages. 3D reconstructions were refined using EMAN and visualised using UCSF Chimera, as described in section \ref{subsec:resanddis}.    
	
\subsection{Results and discussion}\label{subsec:resanddis}

\subsubsection{Thylakoid membrane purification and solubilisation, and electron microscopy}
Four protein bands were resolved in the sucrose gradient after centrifugation (fig. \ref{fig:sucromicro}a). Transmission electron microscopy of sample harvested from band 3 generated micrographs such as that shown in figure \ref{fig:sucromicro}b. A representative micrograph of sample harvested from band 2 is shown in figure \ref{fig:sucromicro}c. Single-particle projection images extracted from band 3 and band 2 micrographs are shown in figure \ref{fig:particles}. These were band-pass filtered and masked for use in 3D model reconstruction. 

\subsubsection{Single-particle analysis}

\paragraph{Band 3}
\begin{figure*}
\centering
\includegraphics[angle=0,width=1\textwidth]{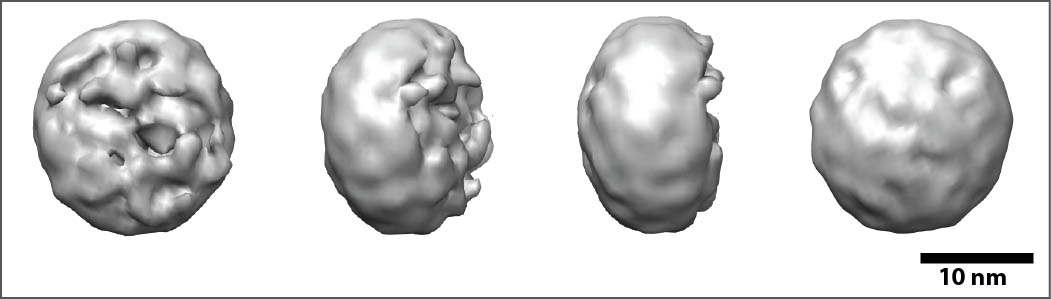}
\caption[Initial 3D single-particle reconstruction from band 3 data (model 3.1)]{\textbf{Initial 3D single-particle reconstruction from band 3 data (model 3.1) at 31 \AA~resolution.} Top, tilt, side and bottom views are shown (left to right).}
\label{fig:model31}
\end{figure*}

\begin{figure*}
\centering
\includegraphics[angle=0,width=0.8\textwidth]{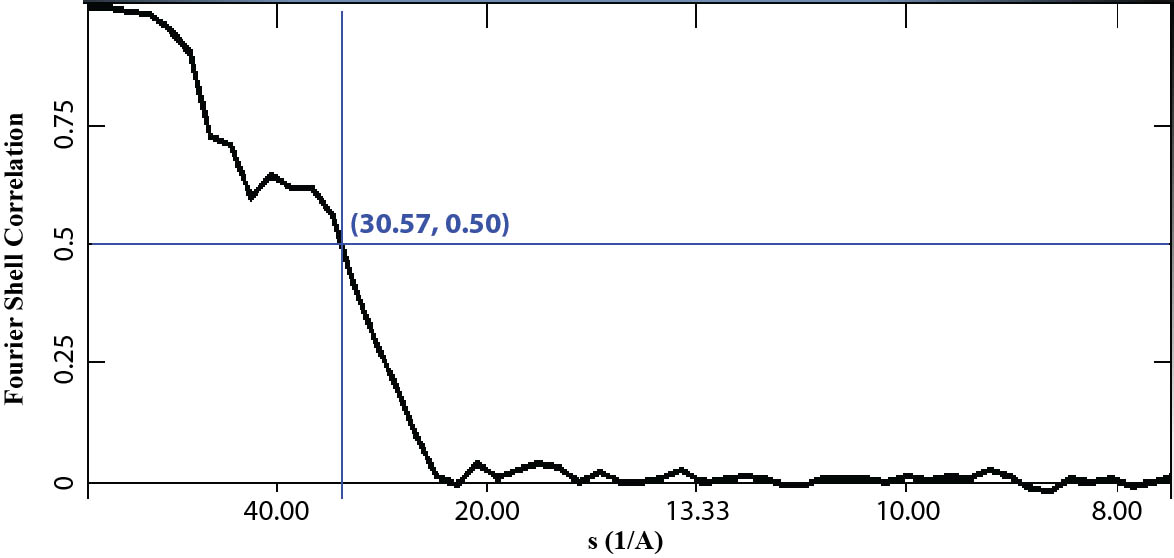}
\caption[Resolution assessments for model 3.1 by even-odd test]{\textbf{Resolution assessment for model 3.1 by standard even-odd test.}}
\label{fig:model31fsc}
\end{figure*}

Initial image analysis and 3D model refinement were carried out using refine software from EMAN over 30 iterations, starting from a Gaussian-sphere initial model. Following \cite{landsberg2009}, iterations were carried out at progressively finer angular sampling, with each step iterated to pseudoconvergence as judged by Fourier shell correlation \cite{ludtke2001}. Within each angular sampling step, the number of rounds of `iterative class averaging' was also decreased in a stepwise manner, to minimise model bias in early rounds of refinement and blurring of model features in later rounds \cite{ludtke2004}.  
The resolution of the resulting 3D model was estimated to be 30.6 \AA~using a standard even-odd test; Fourier shell correlations between models generated from even-numbered and odd-numbered particle images are shown in figure \ref{fig:model31fsc}. The model was therefore low-pass filtered to 31 \AA~using proc3d software from EMAN and the result (model 3.1) is shown in figure \ref{fig:model31}.

Overall size and shape of model 3.1 were broadly consistent with previous results for the PSI-LHCI supercomplex from \textit{C. reinhardtii} (fig. \ref{fig:ps1spa}). (This was consistent with expectations because \textit{stm3} had previously been shown only to have increased levels of LHCII compared with the wild type; no modification of LHCI levels had been shown, though also had not been ruled out \cite{mussgnug2005}.) However, the quarternary structure of model 3.1 was not sufficiently well resolved for subunits to be assigned. Side and bottom views of the model revealed a continuous, broadly convex surface on one side of the membrane plane, contrasting with the opposite side which was more planar overall, though discontinuous and convoluted over shorter scales. These large qualitative structural differences between opposite surfaces of the complex were inconsistent with previous findings for PSI-LHCI. Moreover, the ratio of the model's thickness in side view ($13$ nm) to its largest width in top view ($19$ nm), 0.68, was $\sim$40\% larger for model 3.1  than for the result previously reported by Kargul \textit{et al} \cite{kargul2003} ($10.5~\text{nm}/22~\text{nm}\sim$0.48). It was postulated that these results suggest underrepresentation of side views in the dataset and consequent biasing of 3D model refinement by the Gaussian-sphere initial model and alignment of top and tilt views with model side view reprojections. 

\begin{figure*}
\centering
\includegraphics[angle=0,width=1\textwidth]{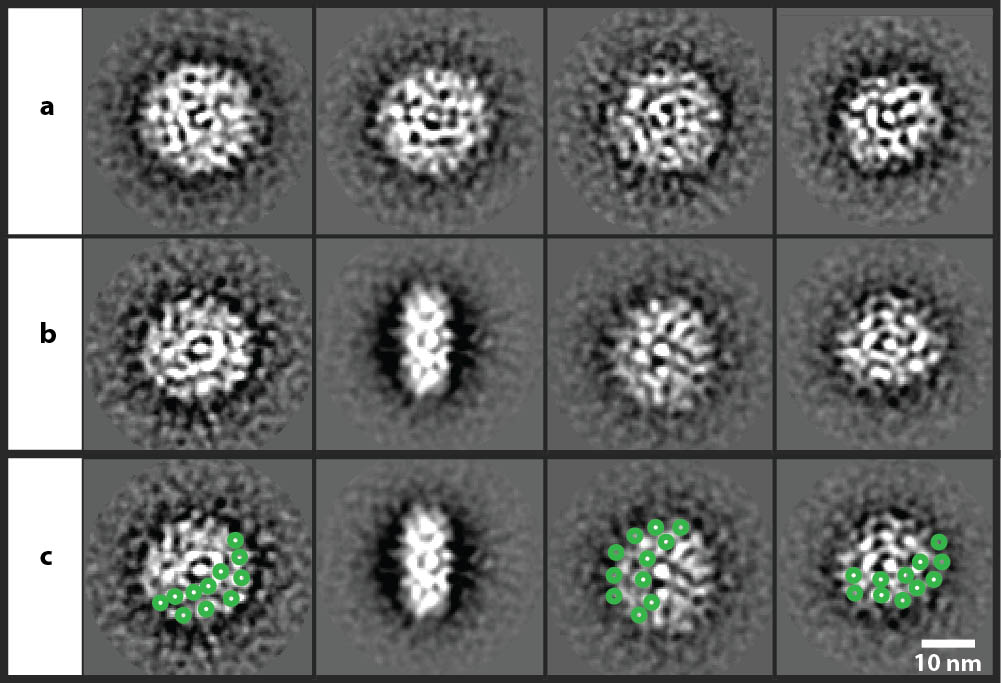}
\caption[Reference-free class averages for band 3 particle data]{\textbf{Reference-free class averages for band 3 particle data.} Eight representative classes are shown from a total of 100 generated from the 8990-particle dataset. \textbf{a}, Many such as these showed limited structural ordering. \textbf{b}, Some such as these showed repeated densities suggesting bands of light harvesting antenna proteins adjacent to a region suggestive of the PSI core complex. \textbf{c}, Reproduction of class averages shown in row b, marked to suggest possible regions of LHCs (green circles). Green circles label clear densities in each putative LHC band and do not necessarily each represent a single LHC. Side-view class average suggests particle thickness of $\sim$10.5 nm, 23\% less than model 3.1 and consistent with the finding by Kargul \textit{et al} \cite{kargul2003}. The pronounced, central density visible in all class averages shown was suggested to be artifactual and therefore likely not of structural significance.}
\label{fig:band3classes}
\end{figure*}

Subsequently, lower-quality top-view and tilt-view images were manually deleted from the dataset, leaving 8990 images for further analysis. The Xmipp package was then used to generate 100 reference-free class averages from the data. Eight representative class averages are shown in figure \ref{fig:band3classes}a,b. Particle side views were represented by only one class (shown in figure \ref{fig:band3classes}b) out of 100, supporting their postulated underrepresentation in the data. Top views were represented by most class averages, many showing apparently disordered quarternary structure (fig. \ref{fig:band3classes}a) and a small number (b,c) showing repeated densities suggesting bands of light-harvesting antenna proteins adjacent to a region suggestive of the PSI core complex. 

Most class averages clearly showed a central density, approximately 1--2 nm wide and visible in all classes shown in figure \ref{fig:band3classes}. This suggested the presence of a dense structural feature near the geometric centre of the particle, surrounded by a region of low density. The findings of Kargul \textit{et al} \cite{kargul2003} included a similar feature (fig. \ref{fig:ps1spa}a), though the `central' density was less pronounced and more varied in appearance and position between class averages of different orientations.  
Other structural models previously obtained for PSI-LHCI from green algae and higher plants (fig. \ref{fig:ps1spa}) have not clearly indicated the presence of such a prominent, central feature. The consistently central location of this density in the class averages for band 3 data,  irrespective of particle orientation (fig. \ref{fig:band3classes}), may suggest it to have been at least partly artifactual, possibly resulting from alignment between artifactual densities in the relatively noisy band 3 particle data (fig. \ref{fig:particles}). The prominence of this putative artifact across the set of reference-free class averages may herald a source of alignment bias also affecting 3D model reconstruction.   

\begin{figure*}
\centering
\includegraphics[angle=0,width=0.75\textwidth]{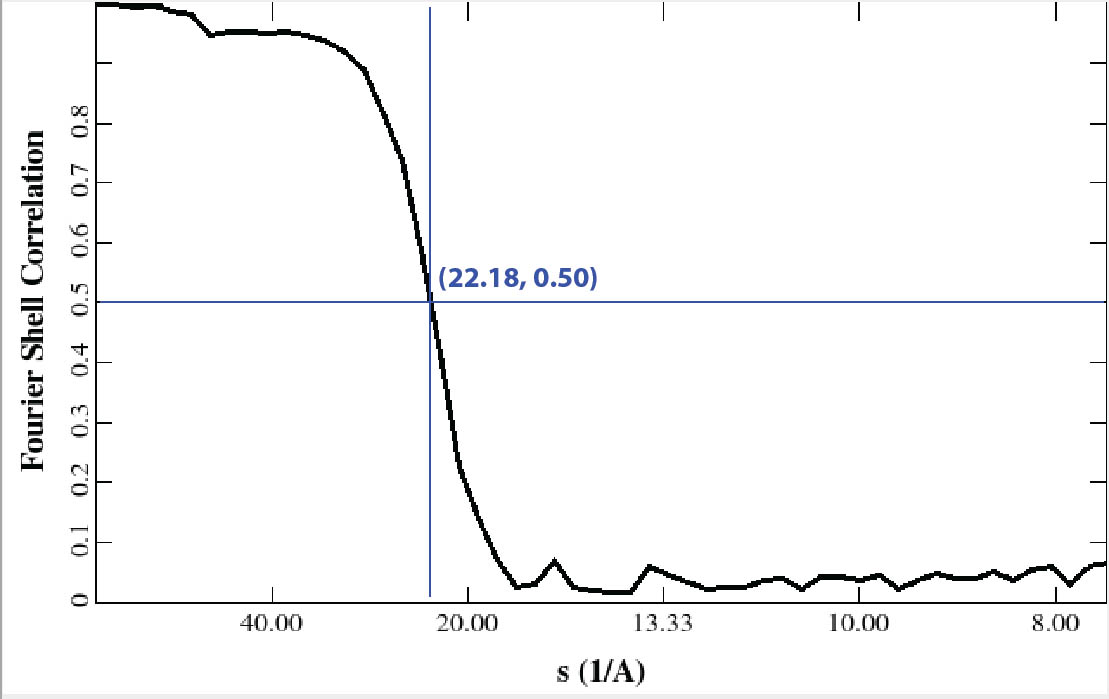}
\caption[Resolution assessments for model 3.2 by even-odd test]{\textbf{Resolution assessment for model 3.2 by standard even-odd test.}}
\label{fig:model32fsc}
\end{figure*} 

\begin{figure*}
\centering
\includegraphics[angle=0,width=1\textwidth]{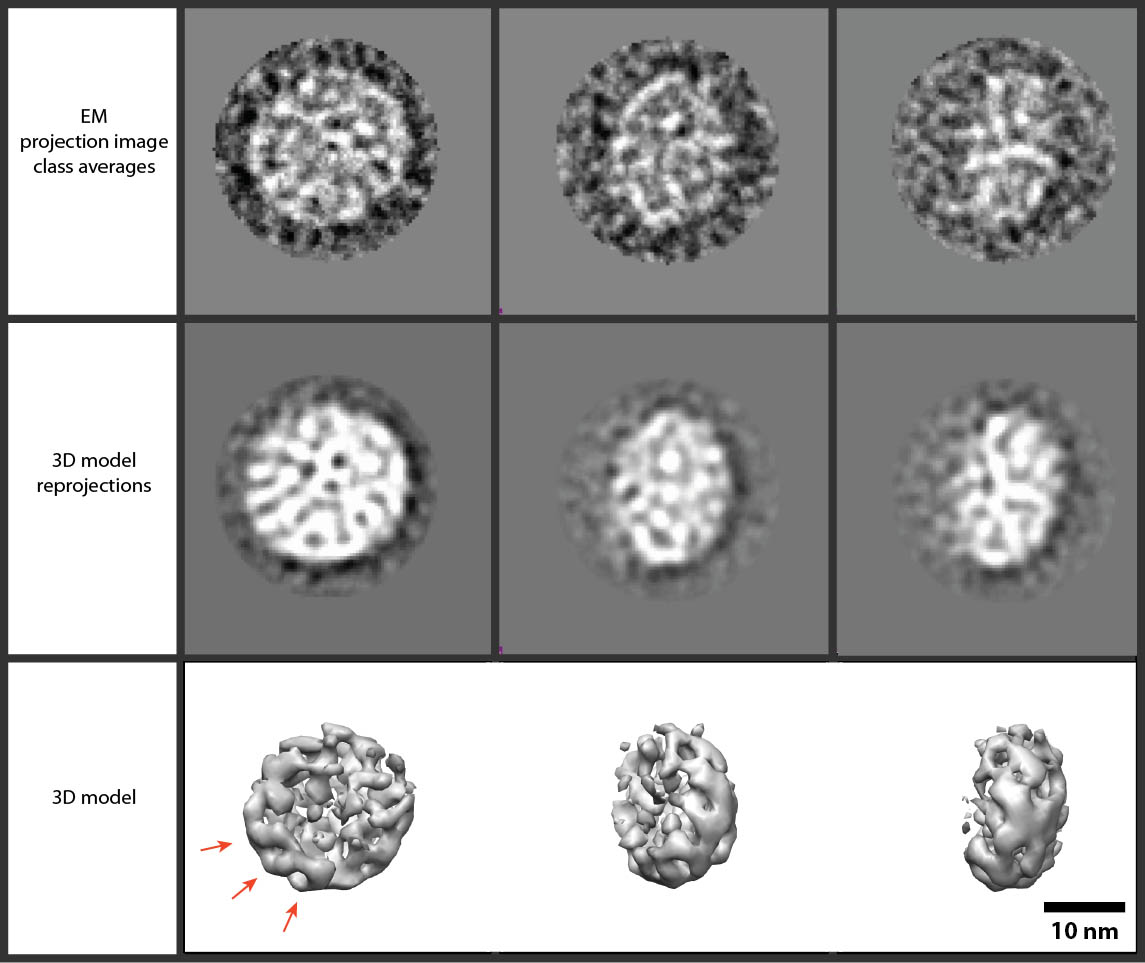}
\caption[Final single-particle reconstruction from band 3 (model 3.2)]{\textbf{Final single-particle reconstruction from band 3 at 24\AA~resolution (model 3.2).} Top, tilt and side views are shown from left to right. Size and shape are broadly consistent with the PSI-LHCI supercomplex (compare figure \ref{fig:ps1spa}). Arrows indicate regularly spaced densities suggestive of Lhca antenna complexes. Class averages are free of the well-defined central density seen in figure \ref{fig:band3classes}. Top view model reprojection shows an approximately central feature in which a clear density is surrounded by a `triangle' of vacancies.}
\label{fig:model32}
\end{figure*}

An attempt was made to generate a 3D model with better-resolved quarternary structure than that in model 3.1. The IMAGIC package was used to create a preliminary 3D model by reprojecting reference-free class averages representing top, tilt and side views of the complex. This model was used as an initial model for 3D refinement using refine software from EMAN. Forty-one iterations of model refinement were completed using the refinement procedure described above for model 3.1, resulting in model 3.2. A standard even-odd test estimated the model resolution to be 22.2 \AA~(fig. \ref{fig:model32fsc}). Model 3.2 was accordingly low-pass filtered to 22.2 \AA and then further to 24.0 \AA~to remove noise still apparent at 22.2 \AA. The resulting model is shown in figure \ref{fig:model32} together with class averages and model reprojections from the final round of refinement. 

Compared with model 3.1, model 3.2 displayed overall size and shape more consistent with structures previously determined for PSI-LHCI. Particularly notable were its reduced thickness in side view ($\sim11$ nm) and increased largest width in top view ($\sim20$ nm; ratio 0.55) which compared more closely with the $\sim10.5$ nm and $\sim22$ nm (ratio 0.48), respectively, found by Kargul \textit{et al} \cite{kargul2003}. However, model 3.2 retained the anomalous convexity on one side parallel to the membrane plane, albeit to a lesser extent than model 3.1. In top view model 3.2 revealed features consistent with the earlier result from Kargul \textit{et al} \cite{kargul2003}. Close to the centre of the complex is a region of lower average density, surrounded by the higher-density regions near the periphery. On one side a crescent of regularly spaced densities was visible (indicated in figure \ref{fig:model32}), which when compared with earlier structural models suggested a band of LHCs. Accordingly, it was inferred that the high-density region near the opposite edge of the model was likely to be the core complex.

\begin{figure*}
\centering
\includegraphics[angle=0,width=0.9\textwidth]{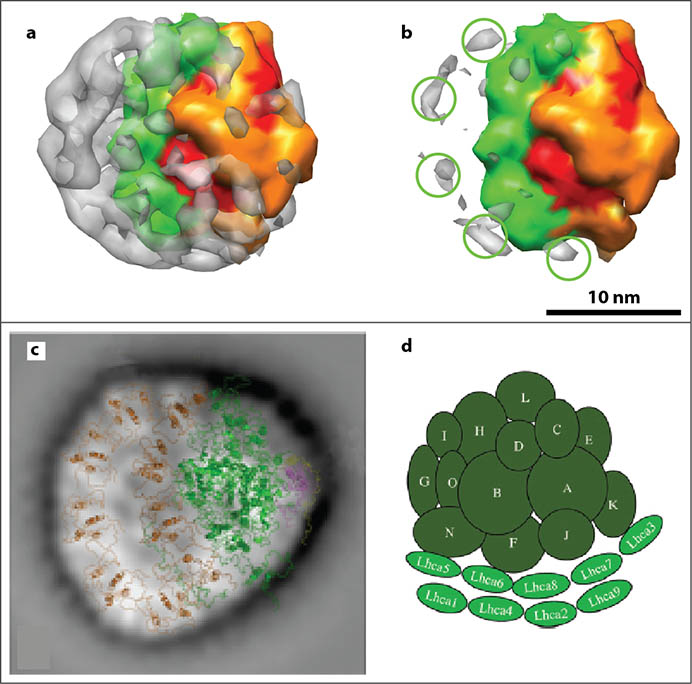}
\caption[Higher-plant PSI-LHCI X-ray structure optimally fitted to model 3.2]{\textbf{Higher-plant PSI-LHCI X-ray structure optimally fitted to model 3.2.} X-ray structure (PDB code 2WSC \cite{amunts2010}) from \textit{P. sativum}. PsaN and PsaR subunits have been deleted (see text). Coloured electron density map was simulated at 24 \AA~resolution from 2WSC using UCSF Chimera. Coloured subunits are Lhca antenna complexes (green), PsaA and PsaB main core heterodimer (red), and remaining subunits of the core complex (orange). \textbf{a}, Simulated 2WSC EM map optimally fitted to model 3.2 (grey). \textbf{b}, Volume of model 3.2 has been reduced using volume viewer software in UCSF Chimera, to reveal regions of highest electron density. Region of model 3.2 external to the 2WSC EM map is shown to comprise five strong densities (green circles) suggestive of additional Lhca antenna complexes. \textbf{c}, Projection map of PSI-LHCI supercomplex from \textit{C. reinhardtii} at 15 \AA~resolution with 2WSC and additional Lhca subunits overlaid; adapted from \cite{drop2011}. \textbf{d}, Schematic representation of PSI-LHCI model in \textit{C. reinhardtii} based on structural and proteomic modelling; reproduced from \cite{yadavalli2011}.}
\label{fig:model32xrayfits}
\end{figure*}

Evidence in support of these inferences was provided by optimally fitting the X-ray structure of PSI-LHCI from \textit{P. sativum} (fig. \ref{fig:xraystruct}a, Protein Data Bank (PDB\nomenclature{PDB}{Protein Data Bank}) code 2WSC) to model 3.2 using fit-model-in-map software\footnote{This software performs a rigid rotation and translation of the X-ray structure to maximise the sum of the EM map values at the atom positions in the X-ray structure \cite{goddard2008}.} from UCSF Chimera (fig. \ref{fig:model32xrayfits}). The 2WSC structure was modified by deleting the PsaN and PsaR subunits; structural and proteomic studies have suggested that PsaN occupies a different position in \textit{C. reinhardtii} compared with higher plants \cite{yadavalli2011}, while PsaR is a protein of unknown identity, loosely associated with the core complex \cite{amunts2010}. 

Initial results (fig. \ref{fig:model32xrayfits}a,b) suggested the presence of an additional band of subunits, postulated to be Lhca subunits, located on one side of the core complex and separated from it by the LHCI band present in 2WSC. An attempt was made to ascertain the number and arrangement of putative addtional Lhca subunits by optimally fitting Lhca subunits from 2WSC to model 3.2. This failed to achieve unambiguous results. However, in order to more clearly determine the number of densities present in the additional LHCI band, volume viewer software in Chimera was used to reduce the volume of model 3.2, revealing the regions of highest electron density. The band of model 3.2 external to 2WSC was shown to comprise five strong densities (indicated by green circles in figure \ref{fig:model32xrayfits}c,d), suggesting a total of nine Lhca antenna complexes in the \textit{stm3} PSI-LHCI supercomplex. This was consistent with results for \textit{C. reinhardtii} from two independent studies published while the present study was in progress (fig. \ref{fig:model32xrayfits}e,f). However, whereas Drop \textit{et al} \cite{drop2011} described a band of five additional Lhca complexes all located to one side of the 2WSC LHCI band (e), results here were consistent with the finding by Yadavalli \textit{et al} \cite{yadavalli2011} suggesting one of the additional Lhca complexes to be located adjacent to the PsaK subunit (f).

\paragraph{Band 2}

\begin{figure*}
\centering
\includegraphics[angle=0,width=0.75\textwidth]{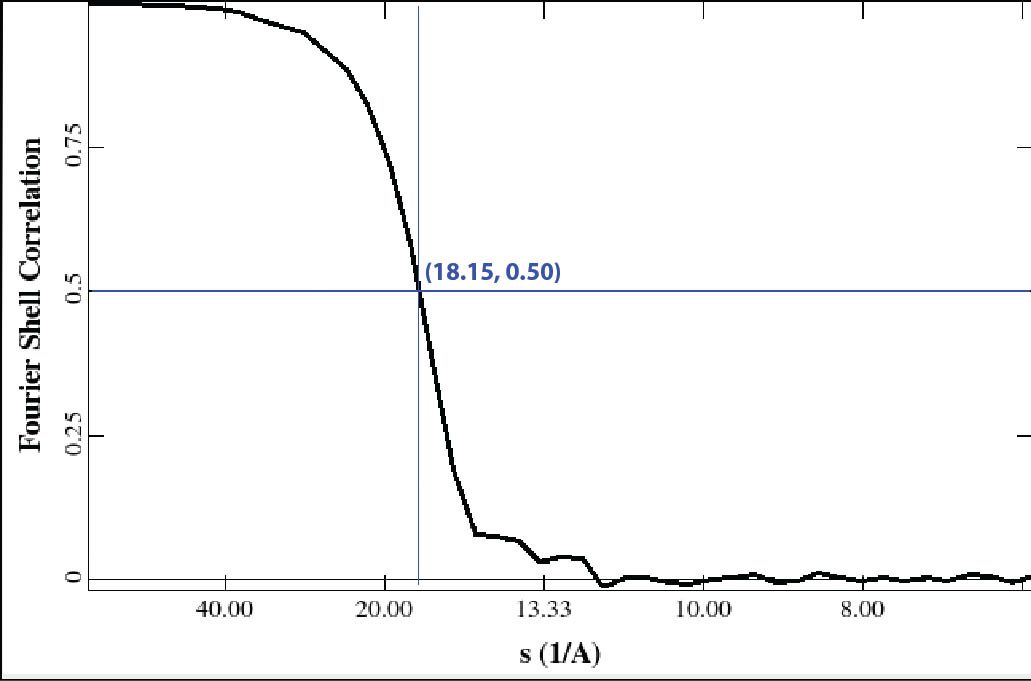}
\caption[Resolution assessment for models 2.1 by even-odd test]{\textbf{Resolution assessment for model 2.1 by standard even-odd test.}}
\label{fig:model21fsc}
\end{figure*}

\begin{figure*}
\centering
\includegraphics[angle=0,width=1\textwidth]{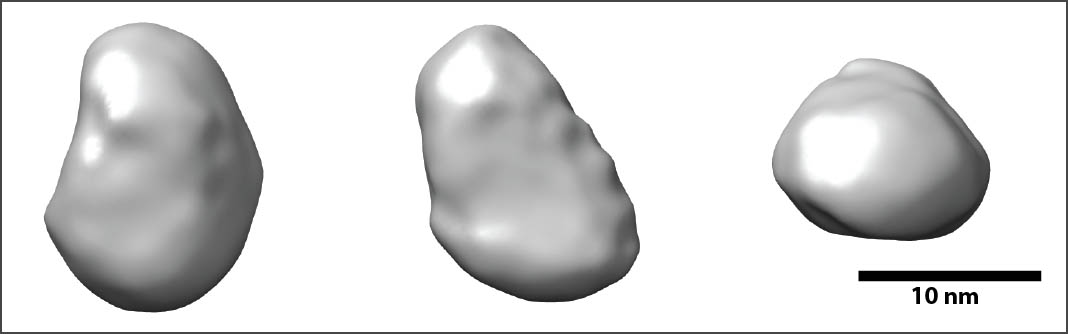}
\caption[Initial single-particle reconstruction from band 2 (model 2.1)]{\textbf{Initial single-particle reconstruction from band 2 at 18.2 \AA~resolution (model 2.1).} Top, side and end views are shown (left to right). The model's smallest dimension in end view was $\sim10$ nm.}
\label{fig:model21}
\end{figure*}

Initial image analysis and 3D model refinement were carried out using refine software from EMAN over 35 iterations using the same procedure as for models 3.1 and 3.2, starting from a Gaussian-sphere initial model. The resolution of the resulting model was estimated to be 18.2 \AA~ using a standard even-odd test (fig. \ref{fig:model21fsc}). Consequently the model was low-pass filtered to 18.2 \AA~using proc3d software from EMAN. The result, model 2.1, is shown in figure \ref{fig:model21}.

\begin{figure*}
\centering
\includegraphics[angle=0,width=0.9\textwidth]{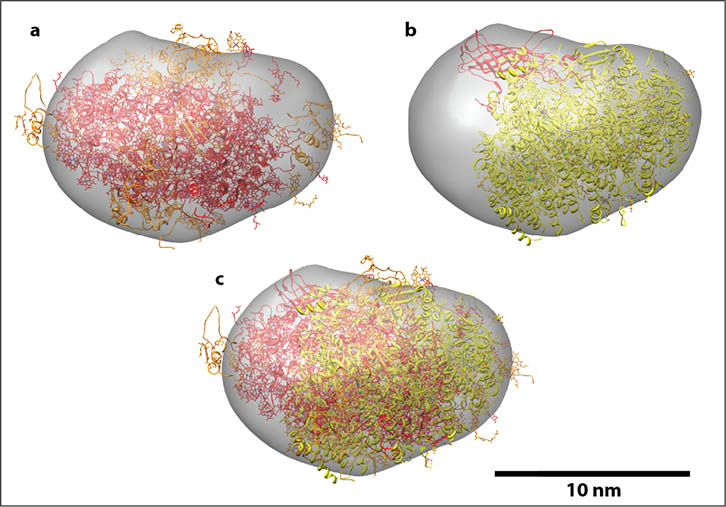}
\caption[X-ray structures of PSI and PSII core complexes optimally fitted to model 2.1]{\textbf{X-ray structures of PSI (a,c) and PSII (b,c) core complexes optimally fitted to model 2.1.} \textbf{a}, Top view of model 2.1 fitted to PSI core complex X-ray structure from \textit{P. sativum} (PDB code 2WSC). Core heterodimer comprising PsaA and PsaB is shown in red with remaining subunits shown in orange. Subunits PsaN and PsaR have been removed (see text). \textbf{b}, Top view of model 2.1 fitted to monomeric PSII core complex X-ray structure from \textit{T. elongatus} (PDB code 3KZI). Oxygen-evolving complex is shown in red with remaining subunits shown in yellow. \textbf{c}, Both the PSI and PSII core complexes simultaneously fitted to model 2.1. Fitting was done using fit-model-in-map software in the UCSF Chimera package.}
\label{fig:model21fits}
\end{figure*}

X-ray structures of core complexes from PSI in \textit{P. sativum} (PDB code 2WSC) and (monomeric) PSII in \textit{T. elongatus} (3KZI) were optimally fitted to model 2.1 using fit-model-in-map software in UCSF Chimera (fig. \ref{fig:model21fits}). Visual assessment revealed that although PSI fit more closely than PSII, neither case achieved a close fit, with prominent regions of the EM map and atomic models remaining unmatched. However, the fits appeared to be `complimentary' such that optimally fitting both X-ray models simultaneously to model 2.1 (fig. \ref{fig:model21fits}c) achieved a closer overall match to the EM map, assessed visually. It was therefore hypothesised that the band 2 particle dataset may be heterogeneous, with both the PSI and PSII core complexes significantly represented.  
The presence of additional particle species (e.g. subunits of Cyt--$b_6f$ and ATPase) also could not be ruled out, and was supported by visual inspection of the complete dataset (not presented).

\begin{figure*}
\centering
\includegraphics[angle=0,width=1\textwidth]{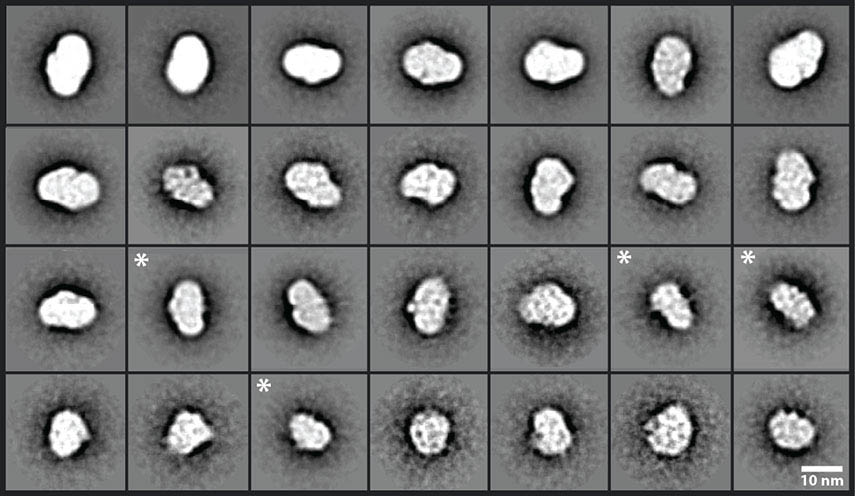}
\caption[Reference-free class averages for band 2 particle data]{\textbf{Reference-free class averages for band 2 particle data.} Twenty-eight representative classes are shown from a total of 200 generated from the 13,958-particle dataset. Classes marked with a white star have at least one 1D projection of length less than 10 nm and therefore smaller than the shortest 1D projection of model 2.1.}
\label{fig:band2classes}
\end{figure*}

Reference-free class averages were generated for the band-2 particle dataset using Xmipp software. A set of 28 representative class averages, from 200 in total, is shown in figure \ref{fig:band2classes}. Visual inspection revealed consistency between many class averages and projections of model 2.1 (figs. \ref{fig:model21}, \ref{fig:model21fits} and \ref{fig:model2comp}). However, some class averages (labelled) had at least one 1D projection of length less than 10 nm and therefore smaller than the shortest 1D projection of model 2.1. This supported the hypothesis of heterogeneity in the band 2 dataset.

\begin{figure*}
\centering
\includegraphics[angle=0,width=0.75\textwidth]{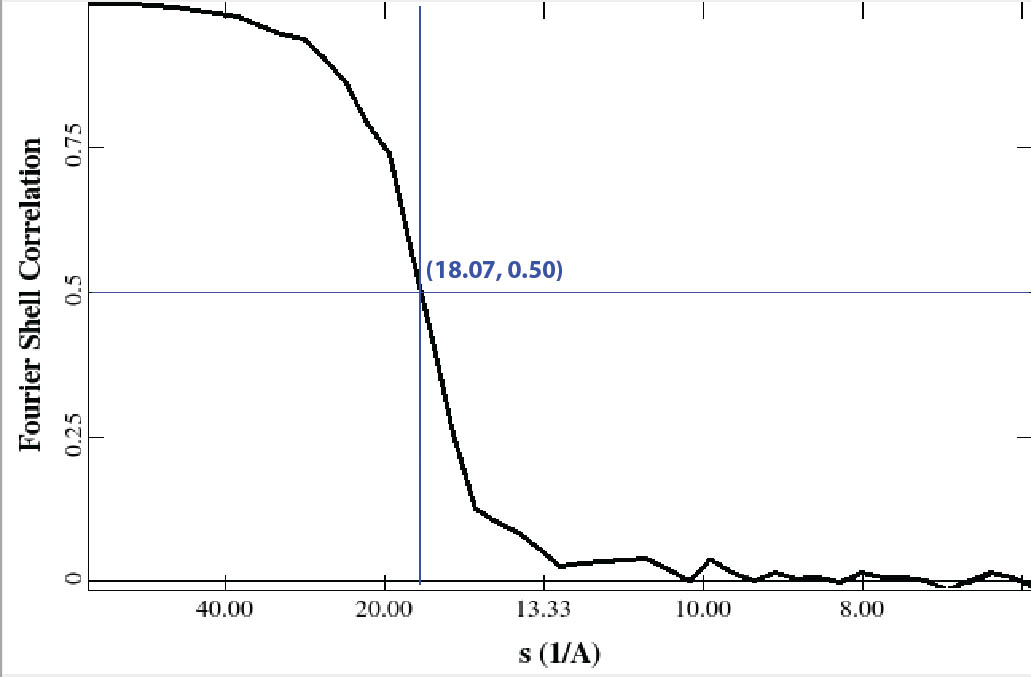}
\caption[Resolution assessment for models 2.2 by even-odd test]{\textbf{Resolution assessment for model 2.2 by standard even-odd test.}}
\label{fig:model22fsc}
\end{figure*}

With the aim to `purify' the dataset \textit{in silico}, multirefine software from EMAN was used to reconstruct two 3D models simultaneously from different initial models\footnote{This software separates a particle projection image dataset into multiple, not necessarily equal subsets, based on which model they most closely correlate with.}. Initial models were created by coarsening X-ray structures of the core complexes from PSI in \textit{P. sativum} and (monomeric) PSII in \textit{T. elongatus} to 15 \AA~resolution using proc3d. The PDB models used were respectively 2O01 \cite{amunts2007} and 1S5L \cite{ferreira2004}, which were current at the time but have since been superceded by the core complex from 2WSC and 3KZI \cite{broser2010}.

\begin{figure*}
\centering
\includegraphics[angle=0,width=1\textwidth]{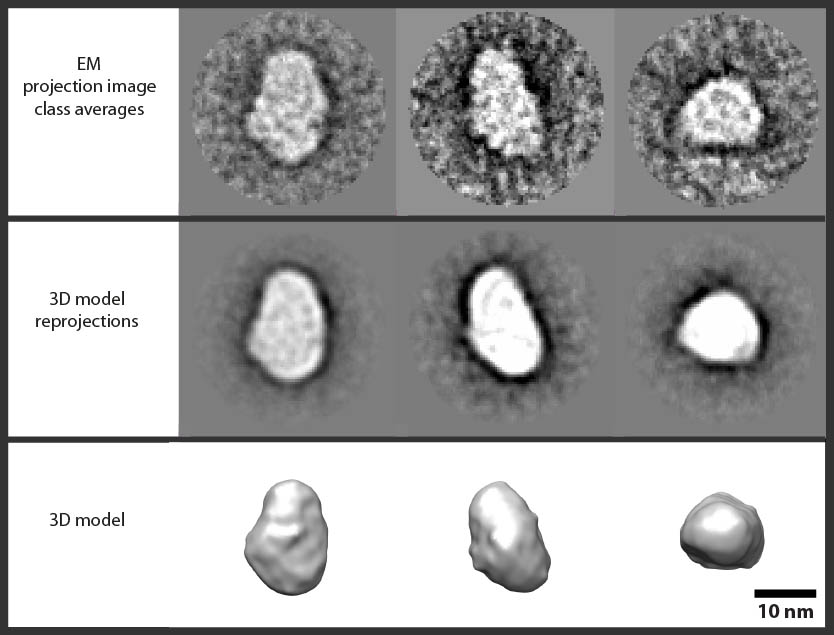}
\caption[Final single-particle reconstruction from band 2 (model 2.2)]{\textbf{Final single-particle reconstruction from band 2 at 18.1 \AA~resolution (model 2.2): putative PSI core complex.}}
\label{fig:model22panel}
\end{figure*} 

Forty-one iterations of multi-refinement were completed with invariant angular sampling and `iterative class averaging'. The usefilt option in EMAN's refine software permitted two analogous data sets to be used: one band-pass filtered to 9 and 190 \AA~for the classification and alignment routines, and a second unfiltered data set for the 3D reconstruction. The model initiated from the higher-plant PSI core complex converged to a stable structure, while the cyanobacterial PSII-based model failed to converge. This finding suggested that the dataset was dominated by PSI particles despite likely being heterogeneous, consistent with the observation that model 2.1 more closely resembled PSI than PSII. 

\begin{figure*}
\centering
\includegraphics[angle=0,width=0.9\textwidth]{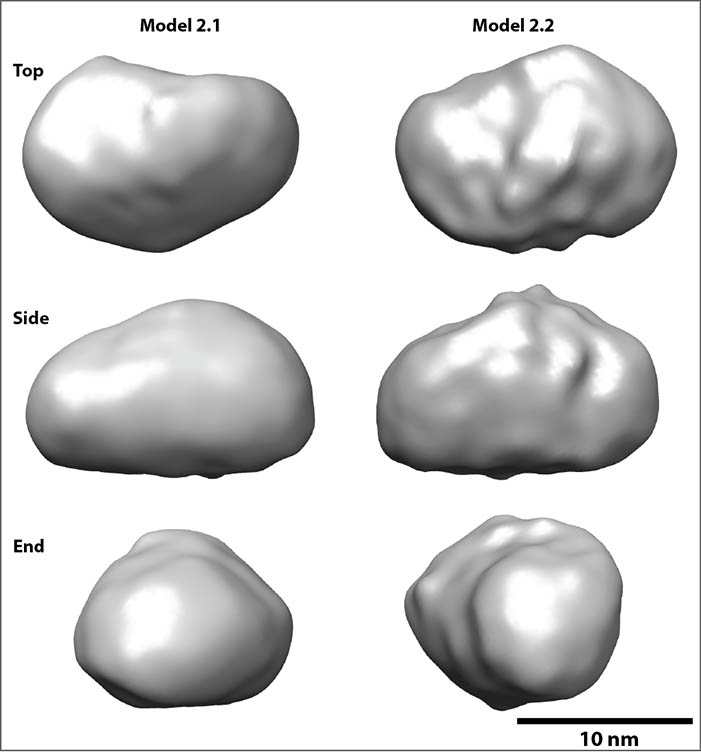}
\caption[Comparison of models 2.1 and 2.2]{\textbf{Comparison of models 2.1 and 2.2.} Top, side and end views are shown.}
\label{fig:model2comp}
\end{figure*}

The resolution of the successfully refined model (model 2.2) was estimated by even-odd test to be 18.1 \AA~(fig. \ref{fig:model22fsc}). The result after low-pass filtering to 18.1 \AA~is shown in figure \ref{fig:model22panel} together with class averages and model reprojections from the final round of refinement. Model 2.2 was compared with model 2.1 (fig. \ref{fig:model2comp}). The two showed clear similarities, although appeared to be of different handedness. Given that model 2.2 was refined from the higher-plant X-ray crystal structure, which itself is of correct handedness, it is likely that model 2.2 reflected the correct handedness and model 2.1 was inverted.

\begin{figure*}
\centering
\includegraphics[angle=0,width=0.9\textwidth]{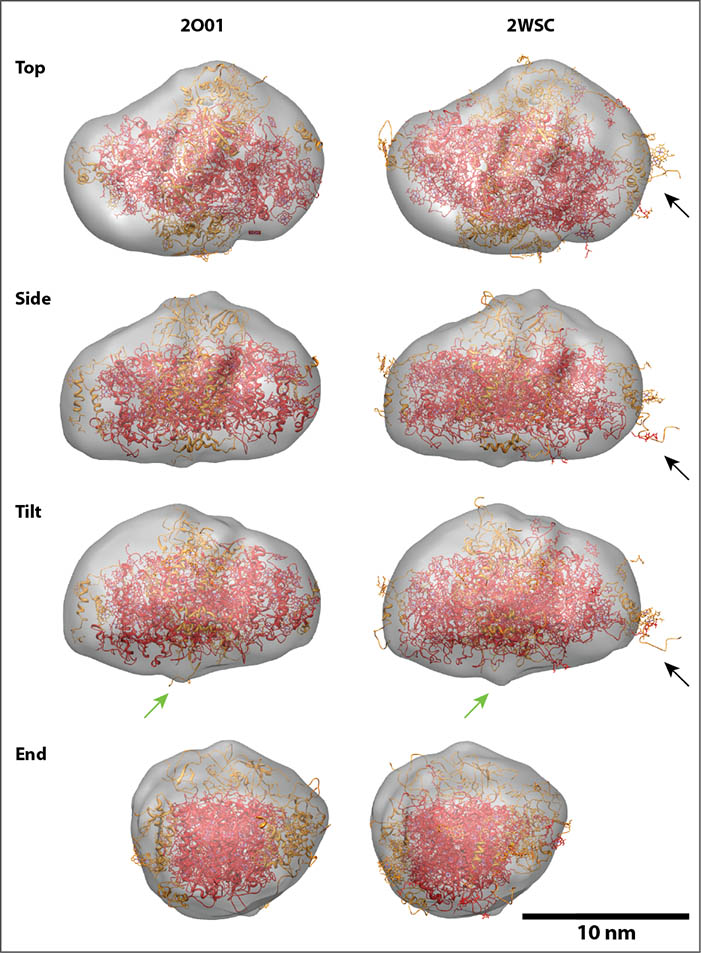}
\caption[Comparison of optimal fits of PSI core X-ray structures to model 2.2]{\textbf{optimal fits of PSI core X-ray structures to model 2.2: comparing 2O01 with 2WSC.} Subunits PsaN and PsaR have been removed (see text). Black arrows indicate subunit PsaK, which was characterised with higher precision in 2WSC and found to be larger than in the earlier 2O01 structure \cite{amunts2010}. Green arrows indicate a protrustive density in the EM map which closely fits a protein chain extending from the PsaH subunit structure in 2O01 but which is not closely fit by any part of the 2WSC structure.}
\label{fig:model22xray}
\end{figure*}

Figure \ref{fig:model22xray} shows fits of PSI core X-ray models to model 2.2, comparing the 2O01 structure, from which model 2.2 was refined, with the more recent 2WSC structure. Both cases achieved a closer fit, visually assessed, than either achieved for model 2.1 in figure \ref{fig:model21fits}. This suggested that refinement of model 2.2 successfully extracted PSI core particle projections from the dataset to reconstruct a model of the PSI core complex for \textit{stm3}. Whereas the 2O01 structure was largely contained within the model 2.2 EM map, multiple subunits of the 2WSC structure were seen to protrude substantially from the EM map. In particular, the PsaK subunit, the structure of which was well resolved for the first time in the 2WSC structure \cite{amunts2010} and shown to be larger than previously accounted for in 2O01, did not closely fit the model 2.2 EM map. As shown by the top, side and tilt views in figure \ref{fig:model22xray}, the PsaA and PsaB core heterodimer was shifted laterally in the 2WSC fit compared with the 2O01 fit. Correspondingly, the PsaG subunit at the opposite end of the heterodimer from PsaK protruded from the EM map in the 2WSC fit while being well contained within the EM map in the 2O01 fit. These shifts may have resulted from the need for the optimal fitting algorithm to accommodate the larger PsaK subunit in 2WSC. However, other regions of the EM map were shown to be more closely matched by small peripheral subunits in 2WSC that were not resolved in 2O01 (compare top, side and tilt views in figure \ref{fig:model22xray}). 

One notable exception was a protrustive density in the EM map (indicated by green arrows in tilt view in figure \ref{fig:model22xray}) which closely fit a protein chain extending from the PsaH subunit in 2O01 but was not closely fit by any part of the 2WSC structure. This mismatch between the 2O01 and 2WSC fits in particular raises the question of initial-model bias; whether utilising 2O01 coarsened to 15 \AA~as an initial model for the refinement of model 2.2 introduced structural biases that persisted even after the model refinement process. 
Notwithstanding these relatively small differences between the two X-ray model fits, both showed clear correspondences between the X-ray models and the model 2.2 EM map in prominent structural features such as the membrane-extrinsic subunits (PsaC, PsaD and PsaE) seen at top in side and end views. In both cases the model fits were unambiguous and strongly suggested that model 2.2 described the PSI core complex in \textit{stm3}. The small differences between the higher-plant X-ray structures and model 2.2, may of course have resulted from organismal differences.  
 
\paragraph{Combined band 3 and band 2 results}

\begin{figure*}
\centering
\includegraphics[angle=0,width=0.95\textwidth]{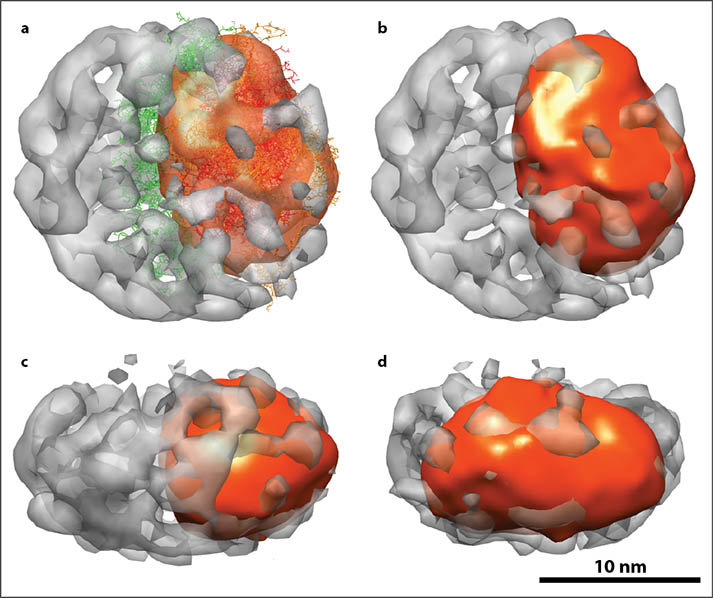}
\caption[Final single-particle EM model for \textit{stm3} PSI-LHCI supercomplex]{\textbf{Combination of models 3.2 and 2.2: \textit{stm3} PSI-LHCI supercomplex}. \textbf{a}, EM maps were combined through simultaneous fitting to the 2WSC higher-plant X-ray structure (with PsaN and PsaR subunits deleted - see text). The resulting combined EM map is shown independently of the X-ray structure in top (b), side (c) and front views (d).}
\label{fig:triplefit}
\end{figure*}

Quarternary structure was not sufficiently clearly resolved within the region of model 3.2 attributed to the PSI core complex (fig. \ref{fig:model32xrayfits}) for unambiguous assignment of the PSI core in the absence of LHCI. Conversely, model 2.2 achieved a close fit to X-ray structures of a higher-plant PSI core complex and was resolved at 18.1 \AA~resolution compared with 24 \AA~for model 3.2. Therefore it was proposed that model 2.2 may provide a useful `augmentation' for the less-well resolved putative PSI core region of model 3.2. As shown in figure \ref{fig:triplefit}a, the two models were combined by first optimally fitting the higher-plant PSI supercomplex (2WSC) to model 3.2 and then manually fitting model 2.2 to 2WSC consistent with the optimal fit shown in figure \ref{fig:model22xray} (fit-model-in-map software in Chimera did not support simultaneous optimal fitting of the X-ray structure to both EM maps). The combined EM map of models 2.2 and 2.3 is shown in figure \ref{fig:triplefit}b,c,d.   

\subsection{Conclusions and outlook}\label{subsec:concandfut4}
This study has provided the first 3D structural models of the PSI-LHCI supercomplex and PSI core complex from the antenna-mutant algal strain \textit{stm3}, by negative-stain single-particle EM, at 24 \AA~and 18.1 \AA~resolution respectively. To the author's knowledge, the model of PSI obtained here is the first 3D reconstruction of PSI from any green algal species. Since the genetic mutation which differentiates \textit{stm3} from wild-type \textit{C. reinhardii} is not known to affect PSI or LHCI, it is likely that the structural models obtained here for these (super)complexes in \textit{stm3} are also applicable to wild-type \textit{C. reinhardtii}. 

By fitting the high-resolution X-ray structure of PSI-LHCI from \textit{P. sativum} to the EM map here determined for PSI-LHCI in \textit{stm3} it was inferred that the latter is likely to contain five additional Lhca subunits (nine in total) in positions consistent with the recent findings of Yadavalli \textit{et al} \cite{yadavalli2011} for wild-type \textit{C. reinhardtii}. However, quarternary structure in the PSI-LHCI supercomplex was less well-resolved than in the PSI core complex and it was determined that this was likely to be partly attributable to underrepresentation of side-view particle projection images in the PSI-LHCI dataset. The dataset also showed lower contrast compared with the PSI dataset, which was stained and imaged using different methods. Accordingly, it is suggested that improved results may be obtained through a repeat study of the \textit{stm3} PSI-LHCI supercomplex using 0.7\% uranyl formate stain and electron-microscopic imaging using a high-resolution CCD camera (4096$\times$4096 pixels or greater). Sample preparation methods that encourage greater acquisition of side-view particle projection images are also recommended. For example, extending the glow-discharging time for the EM grids and/or embedding particles in a thick stain layer by rapidly drying grids after staining. 

The model obtained here for the PSI core complex showed a high level of consistency with X-ray structures previously determined for \textit{P. sativum}. However, inconsistencies between fits of the EM structure to two different X-ray structures, the earlier of which provided the initial model for EM model refinement, suggested that initial-model bias may have been present in the final EM model. Moreover, the multirefine algorithm used to generate the PSI EM model failed to concurrently produce a stable structure from an initial model based on a higher-plant PSII core monomer. To what extent particle heterogeneity was present in the dataset therefore remains an open question. It is suggested that to address the question of initial-model bias in the final PSI EM model, the multirefine procedure should be repeated using the most recent X-ray structures for PSI and PSII core monomers (suitably coarsened) as initial models. Addressing the possibility of dataset heterogeneity may be less straightforward, requiring proteomic characterisation and more stringent biochemical purification of the band 2 sucrose gradient fraction.

\section[Towards predictive modelling of composition-structure relationships in the thylakoid]{Towards predictive modelling of composition-structure relationships in microalgal thylakoids}\label{sec:towpred}

\subsection{Future directions in experimental multiscale structure determination}
\subsubsection{Completing the set of thylakoid membrane protein structural models}
The structural studies reported in section \ref{sec:spastm3} were completed in parallel with single-particle EM studies \cite{knauth2013} of protein (super)complexes harvested from bands 1 and 4 of the sucrose gradient shown in figure \ref{fig:sucromicro}. These studies produced a 3D model of LHCII at 21.5 \AA~resolution and a model of the C$_2$S$_2$-type PSII-LHCII supercomplex at 23 \AA~resolution. Together with the results reported here for PSI and PSI-LHCI, these models provide a set of `building blocks' (table \ref{tab:blocks}) with which to explore the structural configuration space of the thylakoid through docking into larger structural maps derived by ET. 

\begin{table*}
\centering
\includegraphics[angle=0,width=1\textwidth]{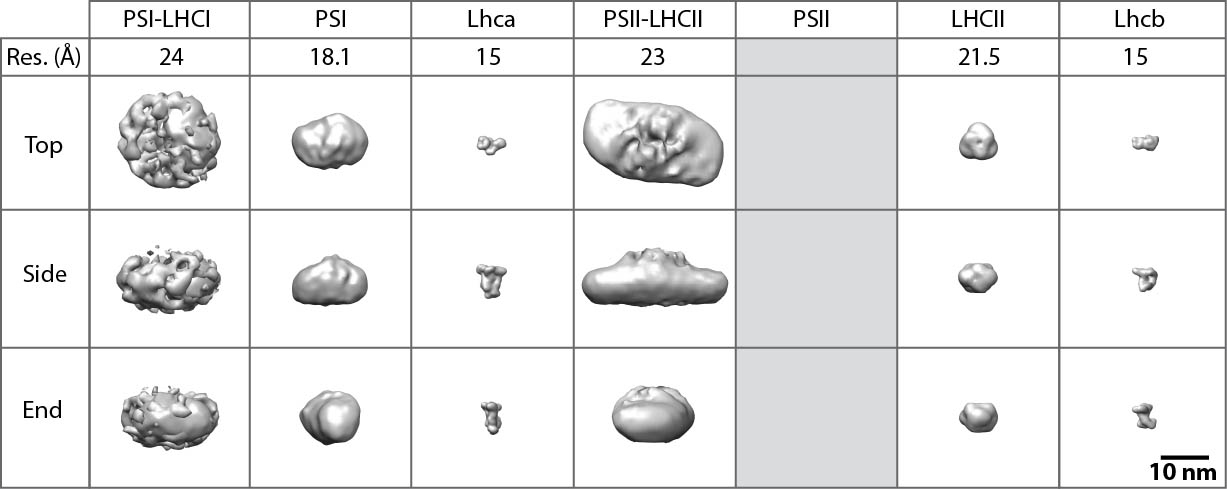}
\caption[Set of thylakoid membrane pigment-protein (super)complex 3D models]{\textbf{Set of thylakoid membrane pigment-protein (super)complex 3D models}, for use in modelling thylakoid ultrastructure. PSI-LHCI and PSI models were created as described in section \ref{sec:spastm3}. Lhca model was generated by coarsening the Lhca1 subunit of the higher-plant PSI-LHCI supercomplex X-ray structure (2WSC) to 15 \AA~resolution using proc3d software in EMAN. Lhcb model was generated in the same way as Lhca, from the X-ray structure of CP29 (PDB code 3PL9). PSII-LHCII and LHCII (from \textit{stm3}) models were created by Knauth as described in \cite{knauth2013} (images used with permission). Reconstruction of a 3D model for monomeric and/or dimeric PSII from \textit{stm3} remains a subject for future work.} 
\label{tab:blocks}
\end{table*}

Example structures of Lhca and Lhcb antenna proteins coarsened from higher-plant X-ray structures have been added to the set of larger pigment-protein (super)complexes in table \ref{tab:blocks} for completeness. Single-particle EM studies of homologous structures from \textit{stm3} have not been attempted because the dimensions of these particles are roughly commensurate with the current resolution limit of ET (section \ref{subsec:et}). It is therefore unclear whether they may be reliably identifiable in tomograms and, furthermore, it is unlikely that structural differences between particle types (e.g. different Lhcas within or between species) could be resolved in tomograms. 

The absence of a 3D model for the PSII core complex from table \ref{tab:blocks} provides a subject for future work. Obtaining such a model would complete the set of pigment-protein (super)complexes for \textit{stm3}. Further addition of models for Cyt--$b_6f$ and ATPase would achieve a complete set of thylakoid membrane proteins. The former has previously been modelled at high resolution for \textit{C. reinhardtii} (section \ref{subsec:xray} \cite{stroebel2003}), while the latter has so far been only partially structurally characterised and only in higher plants and bacteria (section \ref{subsec:xray}). Therefore, while a suitable structure for Cyt-$b_6f$ may be straightforwardly obtained by coarsening the X-ray structure, \textit{de novo} study of ATPase is required.

\subsubsection{Analysis of thylakoid tomograms for (super)complex model docking}

\begin{figure*}
\centering
\includegraphics[angle=0,width=0.86\textwidth]{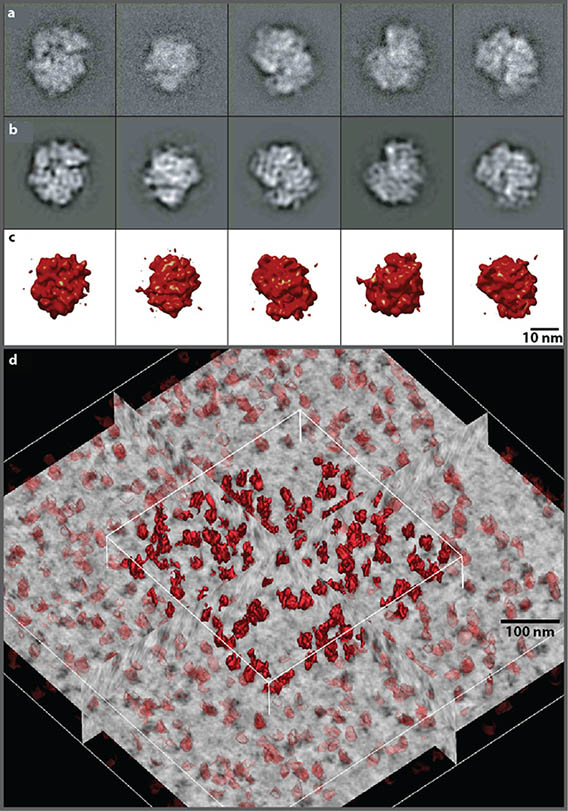}
\caption[Towards automated single-particle docking into electron tomograms]{\textbf{Towards automated single-particle docking into electron tomograms: ribosome example}. \textbf{a}, \textbf{b} and \textbf{c}, adapted from \cite{manuell2005}, respectively show class averages, model reprojections and 3D model views of 80S cytosolic ribosomes from \textit{C. reinhardtii}, reconstructed by cryo-EM at 25 \AA~resolution. \textbf{d}, Electron tomogram of a region within a \textit{C. reinhardtii} chloroplast in which ribosomes have been automatically segmented using the 3D bilateral edge filter \cite{ali2012}. Image courtesy of R. Ali, used with permission.}
\label{fig:ribosomes}
\end{figure*}

Complementary to the single-particle EM studies presented in section \ref{sec:spastm3}, electron-tomographic analysis of \textit{stm3} chloroplasts is underway by other researchers in the Hankamer group at the University of Queensland \cite{knauth2013, ali2012}. To facilitate high-throughput structural docking of single-particle models into electron tomograms, computational tools for automated 3D tomogram image analysis is also under development. 

Traditional methods for identifying \textit{in situ} macromolecular structures within images of crowded cellular landscapes such as thylakoid membranes are labor intensive and time consuming. However, Ali \textit{et al} recently developed the 3D bilateral edge (3D BLE) filter for parameter-free segmentation of macromolecular structures from electron tomograms \cite{ali2012}. The performance of this filter was comparable or superior to that of established 3D image filters and was used to successfully segment single ribosomes \textit{in situ} within a tomogram of a chloroplast in \textit{C. reinhardtii} \cite{ali2012} (fig. \ref{fig:ribosomes}d). As shown in figure \ref{fig:ribosomes}a--c, these ribosomes are comparable in size to thylakoid membrane protein supercomplexes (section \ref{sec:spastm3}). This result therefore provided proof of principle towards high-throughput segmentation of pigment-protein (super)complexes \textit{in situ} in the thylakoid membrane. Software is currently being developed for fitting high-resolution (relative to tomograms) single-particle models of these (super)complexes to the automatically detected single particle contours in tomograms.  

\subsection{Potential for integration with theoretical modelling of composition-structure relationships} \label{subsec:potentinteg}
Experimental determination of a multiscale, 3D structural map of the thylakoid membrane at quasi-atomic resolution would provide an unprecedented platform for modelling structure-energetics relationships in the photosynthetic light harvesting machinery. However, a single such map can capture only a single configuration of a highly structually variable system. Even accounting for ongoing technical development towards high-throughput structural characterisation, the complete construction of a single multiscale map is likely to remain a labor-intensive and time-consuming undertaking in coming years. Efforts to characterise the complex interrelationships between thylakoid protein composition, multiscale structure and energetics may therefore benefit from an approach which integrates experimental and theoretical structural studies.    

\begin{figure*}
\centering
\includegraphics[angle=0,width=0.5\textwidth]{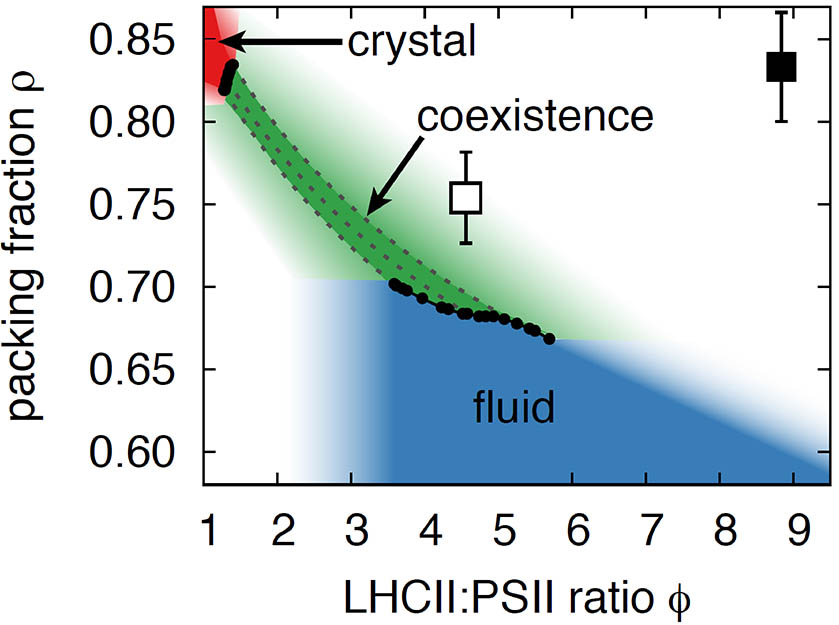}
\caption[Phase diagram of thylakoid structure in simulated, stacked PSII-LHCII membranes]{\textbf{Phase diagram of thylakoid structure in simulated, stacked PSII-LHCII membranes analogous to higher-plant grana layers.} Structural phase is parameterised by membrane packing fraction and LHCII:PSII ratio. Structure transitions between a crystalline phase, a fluid (disordered) phase, and a phase in which the crystalline and fluid phases coexist. Reproduced from Schneider and Geissler \cite{schneider2013}, used with permission.}
\label{fig:psiiphase}
\end{figure*}

Schneider and Geissler recently reported a Monte Carlo simulation-based study relating thylakoid protein composition to structure in a stacked pair of membranes simulating granal layers in a higher-plant thylakoid \cite{schneider2013}. The membranes were populated with structurally simplified C$_2$S$_2$-type PSII-LHCII complexes and LHCII complexes, and were allowed to diffuse through the membrane, interacting \emph{via} simple lateral (intramembrane) and transverse (intermembrane) interaction potentials. This simple model was able to reconstruct a diverse range of experimentally observed structural configurations, including different types of supercomplexes, megacomplexes and larger semicrystalline arrays. Moreover, membrane-scale ordering of proteins was shown to depend strongly on membrane packing fraction and LHCII:PSII ratio (fig. \ref{fig:psiiphase}). Consistent with experimental findings, varying either or both of these parameters induced transitions between a crystalline phase, a fluid (disordered) phase and a phase in which the crystalline and fluid phases coexisted \cite{schneider2013}. This elegant result suggests that in addition to cataloguing solubilised supercomplex configurations observed in structural studies, it is important to study the interactions underlying those configurations. This helps to reveal their significance within the overall membrane configuration space.

A model of the type presented by Schneider and Geissler, generalised to accommodate the more heterogeneous membrane structures of microalgae, may provide a useful complement to the multiscale structural determination program described in this chapter. It may be speculated that such a model might make it possible to predict membrane structural features in antenna mutant strains based on proteomic assessments of membrane protein composition. These predictions could then be tested using the experimental structure assays described here. Ultimately it may be possible to couple energetic models to these structural models with the aim to eludicate composition-structure-energetics relationships in the thylakoid to help inform targeted engineering.
\chapter[Towards a multiscale optimal design framework for photosynthetic energy systems]{Towards a multiscale optimal design framework for light harvesting in photosynthetic energy systems} \label{chp:multiopt}

\chapquote{You see, we cannot draw lines and compartments and refuse to budge beyond them... In the end, it's all a question of balance.}{R. Mistry \cite{mistry2010}}

Chapter \ref{chp:multianalysis} introduced hierarchy theory as a basis for multiscale, multiprocess analysis of photosynthetic energy systems. This theory partitions a complex dynamical system into a hierarchy of recursively nested, quasi-separable subsystems with well-separated characteristic length and time scales in order to simplify system analysis. The multiscale analysis of the hierarchy is constructed by studying internal properties and dynamics of each scale subsystem independently, as well as couplings between scale subsystems. Hierarchy theory provides a general framework for linking system-scale objectives and constraints with engineerable parameters at smaller scales within a system, in a coordinated way. In section \ref{subsec:suggestedhierarchy} this framework was heuristically mapped onto a generic photosynthetic energy system and the studies presented in chapters \ref{chp:qeet} and \ref{chp:structure} contributed new insights into composition-structure-energetics relationships at particular scales in the system. This chapter presents a speculative theoretical study exploring how the multiscale \textit{analysis} framework provided by hierarchy theory may be extended for multiscale system \textit{optimisation}. It is stressed that this work is preliminary and is intended as a foundation for future development.

\section{Metamodel for optimising productivity at each characteristic scale}\label{subsec:generic}

\begin{figure*}
\centering
\includegraphics[angle=0,width=0.745\textwidth]{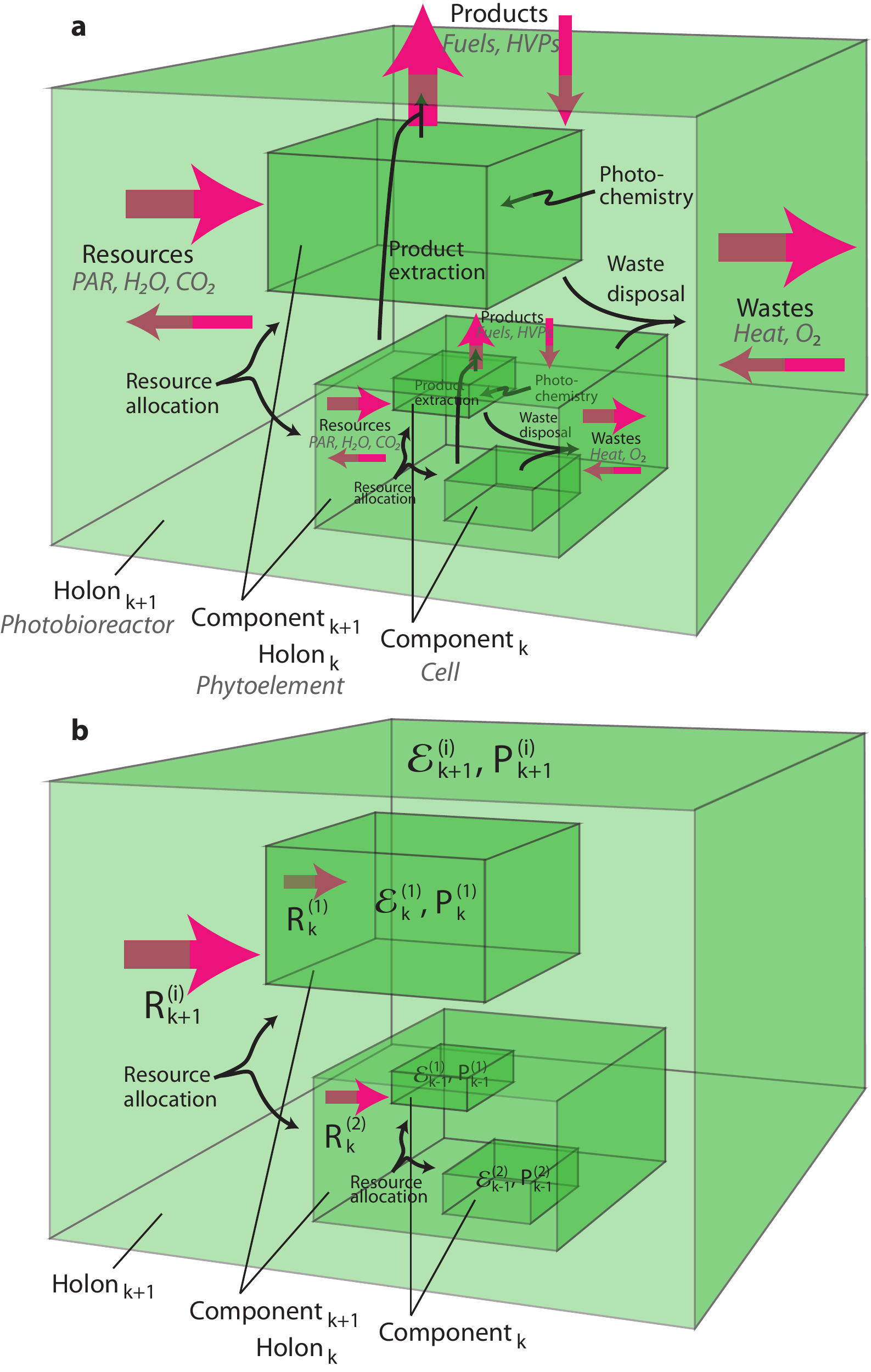}
\caption[Generic scale-subsystem productivity metamodel]{\textbf{Generic scale-subsystem productivity metamodel. (a)} Qualitative metamodel for all scales larger than, and including, the photosynthetic unit (PSU), which is the smallest photochemically-active component in the system. Grey italic text describes a particular realisation of the generic metamodel for photobioreactor, phytoelement and cell scales (see fig. \ref{fig:multiscale}). \textbf{(b)} Simple quantification of the metamodel, linking resource allocation and productivity between scales. The resource ($R$) considered here is assumed to be rate-limiting for the productivity ($P$). It is assumed that this metamodel may be applied recursively over multiple scales; see text.}
\label{fig:generic}
\end{figure*}

\subsection{Conceptual formulation}
Figure \ref{fig:generic}a presents a `toy' metamodel for a subsystem at some scale within an idealised photosynthetic system, focussing on its interactions with the adjacent upscale and downscale subsystems. Fluxes of `resources' (e.g. photosynthetically active radiation (PAR), H$_2$O and CO$_2$), `wastes' (e.g. heat and O$_2$) and photosynthetic products (e.g. fuels and/or high-value products [HVPs]) are passed between the holon\footnote{Nomenclature from the presentation of hierarchy theory in section \ref{subsec:hierarchy} is adopted in this chapter.} and its photosynthetically active components. The system is assumed to be self-similar between scales. However, in order to obtain a closed description at the focal scale, each component is treated phenomenologically (as a `black box' with unspecified internal mechanics). It is assumed to convert resources into products with efficiencies determined by its (unspecified) internal configuration and also by the rates at which resources are incident upon it (the latter dependence accounts for saturation effects such as nonphotochemical quenching [NPQ]). The fraction of each resource converted to wastes varies inversely with the component's production efficiency. 

This metamodel characterises a fundamental problem, which must be solved at each scale in a photosynthetic energy system: if the functional target for the holon is maximal productivity, its optimal internal configuration must include a transport network which allocates incident resources to, and disposes of wastes from, the components at rates corresponding to each component's maximum achievable productivity. (At some scales in some real systems, products are also extracted from the components immediately [e.g. hydrogen gas being excreted by a microalgal cell in a photobioreactor], and these transport processes must also be accommodated by the optimal configuration.) Since the total resource fluxes are limited by upscale processes (e.g. PAR irradiance), maximal subsystem \textit{productivity} requires maximal \textit{efficiency} in the subsystem's resource allocation processes (e.g. maximally efficient allocation of PAR to the components).   
   
The grey text labels in figure \ref{fig:generic}a map the metamodel specifically to the nested scales of the photobioreactor, phytoelement and cell. An \textit{ideal} photobioreactor will allocate PAR, H$_2$O and CO$_2$ to, and dispose of heat and O$_2$ from, \textit{all} of its constituent phytoelements at rates commensurate with each phytoelement's maximum achievable productivity, with minimal resource losses during allocation. Whether or not photochemical products are also extracted from the phytoelements continuously during production or at a later time depends on the particulars of a given system (e.g. continuous H$_2$ excretion vs. delayed biomass harvest). 

Similarly applying the metamodel at the nanoscale, an ideal photosynthetic unit\footnote{A photosynthetic unit comprises a single photochemical reaction centre and the minimum number of antenna chromophores it requires to drive photosynthesis. It is the smallest photosynthetically-active component in the photosynthetic system.} (PSU) will allocate excitons and H$_2$O (resources) to, and dispose of heat and O$_2$ (wastes) from, its reaction centre (RC) at rates commensurate with the RC's maximum achievable productivity (its maximal turnover rate multiplied by its charge-separation energy), with theoretically minimal resource losses during allocation. It will also extract energised electrons and protons (products) from the PSU for transmission to the energetically-downstream components of the thylakoid domain (fig. \ref{fig:multiscale}e), at rates which ensure that PSU productivity is not inhibited by saturation effects such as NPQ.   
Optimising each scale subsystem involves determining an internal configuration which approaches such an ideal, subject to constraints at the focal scale, and also constraints imposed by upscale and downscale processes. 

\subsection{Simple quantitative formulation}
The general question of how to maximise the productivity of a photosynthetic (sub)system at some scale can be placed on a quantitative footing, guided by a simplified formulation of the generic metamodel, shown in figure \ref{fig:generic}b. Consider the $i$th subsystem at hierarchy level $k+1$ (holon$_{k+1}$), which contains $N$ photosynthetically active components at level $k+1$ (components$_{k+1}$/holons$_k$). Holon$_{k+1}$ receives multiple resources from upscale processes, one of which is assumed to be rate limiting (e.g. PAR in a low-light environment, or CO$_2$ in a high-light environment, under normal atmospheric conditions) and therefore of primary importance in the optimisation. Let the net flux of this resource incident at the holon surface be $R_{k+1}^{(i)}$. The holon converts the resource into product at rate (productivity),
\be
P_{k+1}^{(i)}=\sum_{j=1}^N P_k^{(j)}=\varepsilon_{k+1}^{(i)}(R_{k+1}^{(i)})~R_{k+1}^{(i)},\label{compprod}    
\ee
where $P_k^{(j)}$ is the productivity of the $j$th component at scale $k+1$ (holon$_k$) and $\varepsilon_{k+1}^{(i)}(R_{k+1}^{(i)})$ is the production efficiency of holon$_{k+1}$ as a function of the incident resource flux. 

Due to the quasi-separability between the dynamics in scale-adjacent subsystems, $R_{k+1}^{(i)}$ is assumed to be time-independent over time scales significant to intra-holon$_{k+1}$ dynamics. Therefore, $P_{k+1}^{(i)}$ will be maximised if $\varepsilon_{k+1}^{(i)}(R_{k+1}^{(i)})$ is maximised for the given constant $R_{k+1}^{(i)}$. Assuming that the metamodel is also valid for each (the $j$th) holon$_k$ (i.e. it applies equally well, one scale down), we can express $\varepsilon_{k+1}^{(i)}(R_{k+1}^{(i)})$ as a functional of the distributions of resource fluxes ($R_{k}^{(j)}$) incident on holons$_k$ and their production efficiencies ($\varepsilon_{k}^{(j)}(R_{k}^{(j)})$):    
\be
\varepsilon_{k+1}^{(i)}(\varepsilon_{k}^{(j)}(R_{k}^{(j)}), R_{k}^{(j)})=\frac{1}{R_{k+1}^{(i)}} ~\sum_{j=1}^N\varepsilon_{k}^{(j)}(R_{k}^{(j)})~R_{k}^{(j)}.\label{holonef}
\ee 
The subsystem optimisation problem at level $k+1$ is therefore a \textit{variational} problem, the goal of which is to find particular (though not necessarily unique) distributions for $R_{k}^{(j)}$ and $\varepsilon_{k}^{(j)}(R_{k}^{(j)})$ such that $\varepsilon_{k+1}^{(i)}(\varepsilon_{k}^{(j)}(R_{k}^{(j)}), R_{k}^{(j)})$ is maximised. This can be stated as a \textit{variational principle}: the photosynthetic productivity of a subsystem at a given scale will be maximised when resource fluxes (ultimately limited by upscale processes) incident on the photosynthetically active components are matched as closely as possible to the components' photosynthetic production efficiencies (ultimately limited by downscale processes), \textit{and} when those photosynthetic production efficiencies are simultaneously matched as closely as possible to the incident resource fluxes. The goal at each scale is \emph{balance} between upscale and downscale processes. 

The root cause of the need for this balance is the dependence of $\varepsilon_{k}^{(j)}$ on $R_{k}^{(j)}$; without this, maximal productivity for holon$_{k+1}$ would be ensured simply by allocating all resources to the most efficient component$_{k+1}$. However, this is not the case; components can become saturated by incident resources such as irradiance and the only way for excess resource to then be used productively within the holon is for it to be allocated to other, sub-saturated components. The vehicle for executing the required allocation at a given scale is the compositional and structural configuration of the subsystem at that scale. The possibility of a configuration that achieves or optimally approaches this balance depends on the mechanisms and parameters underlying resource allocation and production efficiency at the focal scale. For example, the production efficiency of a phytoelement may be estimated by modelling radiation transfer through its cell culture and coupling the radiant field to a productivity model for each individual cell \cite{murphy2011}. However, these models (both for radiative transfer and cell productivity), and the parameters underlying them, in turn depend on models and parameters at other scales, in accordance with the principles of hierarchy theory. Therefore, for whole-system optimisation, `\textit{negotiation}' is required between subsystem optimisations at different scales. Its result may be that locally-suboptimal subsystem configurations are required in order to optimise the system overall. The multiscale optimisation algorithms described in section \ref{sec:multiopt} provide a basis for exploring these complex negotiations numerically. 

The metamodel described in this section generalises the concept of \textit{acclimation}, which appears in the microalgal-cultivation and higher-plant-ecology literature. It has been emphasised that a central principle of photobioreactor design is to provide an environment for the cell culture that is optimal for cell productivity \cite{dillschneider2013, posten2009}, and also that a natural leaf canopy tends to maintain an internal environment which supports leaf growth \cite{terashima1995} by supplying PAR, CO$_2$ and H$_2$O to cells/leaves in balance with their needs. Reciprocally, individual microalgal cells should be optimised for productivity within the environment supplied by the encompassing photobioreactor \cite{mitra2008}. Similarly, leaves are known to acclimate to their canopy environment (e.g. by reorienting or wilting to change their absorption cross-sections), and individual plant cells similarly acclimate to their environment within the leaf (e.g. by reorienting and relocating chloroplasts to alter the cell's absorption cross section) \cite{terashima1995, ogren1993}. (Though at both scales the acclimation often fails to achieve theoretically optimal subsystem production efficiency) \cite{terashima1995, ogren1993}. Here it is suggested that abstracting the concept of \textit{multiscale balance between resource allocations and component production efficiencies} may provide a useful, general principle for engineering photosynthetic systems. Rather than applying this principle only at obvious scale partitions in pre-existing system structures (e.g. cells within a pre-defined phytoelement geometry such as a tubular or flat-plate bioreactor), it may be applied at each partition within the system scaling hierarchy, defined either heuristically (e.g. figure \ref{fig:multiscale}) or \emph{via} a quantitative algorithm \cite{ahn2010, clauset2008, sales2007, allison2009}. In principle, resource allocations may then be balanced with component production efficiencies at each scale in the hierarchy, using a multiscale, collborative design optimisation algorithm such as those described in the next section.

\section{Framework for multiscale, multiprocess optimisation}\label{sec:multiopt}

The theory of decomposition-based optimal system design allows for co-ordinated optimisation of coupled subsystems under system-scale functional targets and constraints, as well as targets and constraints localised to individual subsystems \cite{dewit2011, xiong2010, chen2010, allison2009, li2008, luo2007, allison2005, li2004, kim2003}. Targets are quantified by so-called `objective' functions, and the overall system objective is equal to the sum of the subsystem objectives. Subsystem coupling can be hierarchical (when one or more subsystems dictate the output of other subsystems) and/or non-hierarchical (when subsystems do not prescribe each other's outputs). Subsystem analysis models (e.g. describing radiation transfer within a holon) are largely separable, excepting a small number of linking variables, which are exchanged between subsystems as targets and responses; a coordination algorithm drives these to mutual consistency as the system optimisation proceeds. The design variables (DVs) over which subsystems are optimised may be localised to individual subsystems or shared between them. Various algorithms have been developed for coordinating subsystem optimisations in order to drive the system to convergence, some able to account for propagation of stochastic uncertainty between subsystems within and/or between scales \cite{xiong2010, chen2010}. Here the algorithms for Analytical Target Cascading (ATC) and Collaborative Optimisation (CO) \cite{li2008, allison2005, kim2003} are briefly described because they are well suited to multiscale, multiprocess optimisation of photosynthetic systems. Detailed conceptual and mathematical formulations of these and other algorithms can be found in the cited literature.

\begin{FPfigure}
\centering
\includegraphics[angle=0,width=1\textwidth]{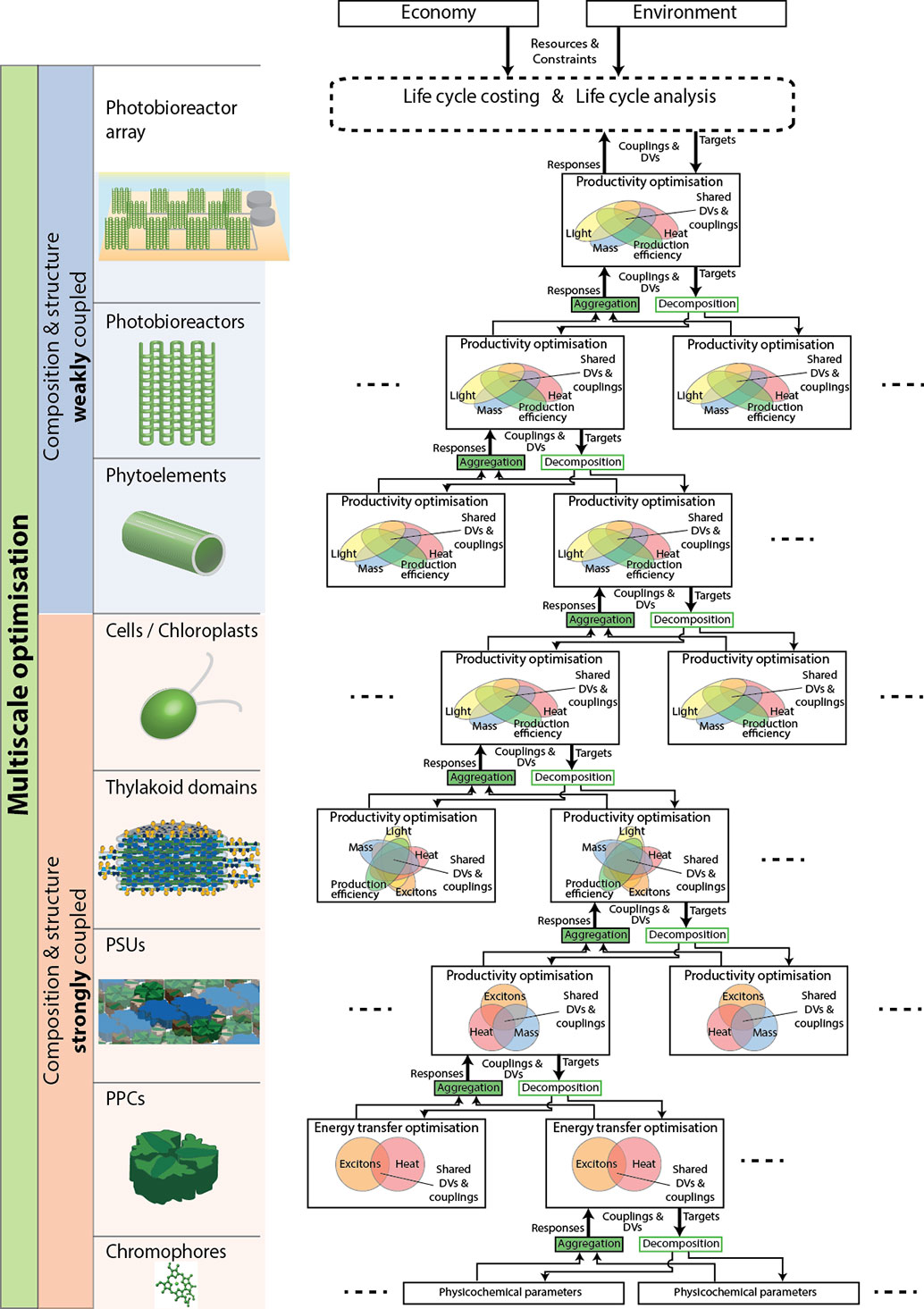}
\caption[Multiscale, multiprocess optimisation of a microalgal fuel production system]{\textbf{Scheme for multiscale, multiprocess optimisation of a microalgal fuel production system.} The system is partitioned into subsystems by scale and energy transfer process according to figure \ref{fig:multiscale}. It is proposed that subsystem optimisations may be coordinated between scales using an Analytical Target Cascading (ATC) algorithm, and between processes at each scale using a Collaborative Optimisation (CO) algorithm (such that CO is nested within ATC). Venn diagrams are used to represent energy transfer processes at each scale, with overlaps indicating that multiple processes may depend on the same `design variables' (DVs -- such as compositional and structural parameters). It may also be the case that dynamical models of different transfer processes are not entirely separable; coupling variables must then be passed between process optimisations by the CO algorithm. The ATC algorithm proceeds as follows: functional targets at the scale of the photobioreactor array are cascaded down the scalar hierarchy to the smallest scale that is assumed to include engineerable DVs. DVs at that scale are then collaboratively optimised for multiple energy transfer processes and the resulting configuration is passed upscale as a response, incorporating suitable parameter aggregations. The upscale subsystem is then similarly optimised for multiple energy transfer processes and the resulting configuration is used to update targets for the next iteration of optimisation downscale. The `negotiation' between scale-adjacent subsystems, and processes within each, is iterated until convergence. Responses are then passed to the next upscale subsystem, which `enters negotiations', and so on until convergence of the overall system configuration. See text for further details.} 
\label{fig:atcco}
\end{FPfigure}

ATC applies to hierarchically coupled subsystems and is summarised \cite{allison2006, li2008, kim2003} as follows: System targets are cascaded down to subsystems (holons) within the system hierarchy. A specific optimisation problem is formulated for each subsystem, under the target received from upscale subsystems, which coordinates the interactions between components (downscale subsystems) in order to ensure optimality and feasibility of the holon design. The solution to this problem produces targets for component optimisation problems. The objective for components is then to match these targets as closely as possible, while in some cases also seeking to satisfy local component objectives. Results of component optimisations are then communicated upscale as responses to the holon, which adjusts component targets appropriately. This negotiation process is repeated until responses match targets, resulting in a consistent and optimal system design. It can be shown that the ATC algorithm is guaranteed to converge to a stable solution for the system \cite{li2008, michelena2003}, and recent developments have significantly reduced the computing time required for convergence \cite{li2008}. 

By contrast, CO involves non-hierarchically coupled optimisation problems executed concurrently over a common subsystem \cite{allison2005}. It is intended for integration of optimisations using analysis models from different fields or disciplines (e.g. concurrent optical and thermal optimisation of a phytoelement, assuming these two processes to be independent). CO can be nested within ATC, such that each subsystem optimisation executed within the ATC algorithm constitutes an execution of CO \cite{allison2005}. 

In figure \ref{fig:atcco} a framework is presented for applying a nested ATC-CO algorithm to a microalgal cultivation system with multiscale, multiprocess partitioning. Subsystem optimisation at each scale larger than and including the PSU constitutes a variant of the generic optimisation problem described in section \ref{subsec:generic}. At yet smaller scales, the optimisation concerns only energy transfer since the subsystems do not contain photosynthetically active components.  

It is suggested that under the proposed framework, system optimisation would ideally proceed as follows; concerns of computational practicality are temporarily suspended, to be discussed below. Economic and environmental targets and constraints at the scale of the overall photobioreactor array dictate productivity targets for the system and these are cascaded downscale as far as the smallest-scale subsystem to be included in the optimisation. This may be the PSUs if, for example, LHC expression levels are taken as the smallest-scale engineerable parameters. Or, if internal parameters of the PPCs themselves (e.g. chromophore properties and arrangements) are considered engineerable, they may also be included in the optimisation; this may be of greater interest in artificial photosynthetic systems, although genetic manipulation allows such engineering to an extent in biological systems as well. Parameters at smaller scales are taken to be fixed (e.g. internal parameters of individual chromophores), and system optimisation begins at the smallest scale that includes engineerable parameters.   

In each PSU, for example, composition (expression levels) and arrangements of PPCs is collaboratively optimised for light absorption, EET, and transfer of mass and heat, subject to targets and boundary conditions cascaded from upscale subsystems, as well as PSU-scale constraints (e.g. maximal turnover rate and charge-separation efficiency at the RC, and limitations on the possible arrangements of PPCs) and any responses or fixed parameters from chromophores downscale. Following the completion of the multiprocess PSU optimisation, DVs (parameters) are aggregated over the PPCs into bulk parameters for the PSU (e.g. shape, optical absorptivity, thermal conductivity, etc.). These PSU parameters, together with the responses for any dynamical linking variables (e.g. irradiance, temperature, chemical concentrations) are passed upscale to be used in optimising the thylakoid domain. Upon convergence between PSUs and thylakoid domain, a response is passed to the next, larger-scale subsystem --- the cell/chloroplast --- which then `enters negotiations'. This bottom-up progression continues until responses are passed to the top-most system scale, which then adjusts targets for re-cascading to the bottom of the hierarchy, where the next iteration of bottom-up optimisation begins. The overall process is repeated until system convergence.

Although the decomposition-based optimisation framework formulated here offers a large reduction in complexity compared with an all-in-one system optimisation, practical implementation under realistic assumptions still poses formidable computational cost. Recent developments have helped to reduce the cost of ATC by enabling parallelisation between subsystem optimisations \cite{li2008}. The complexity of the individual subsystem optimisations can also be reduced through judicious selection of analysis models that balance simplicity with required levels of accuracy.   
Moreover, the scaling hierarchy itself can be reduced by omitting scales considered unimportant to the goal of a given system optimisation. For example, if the goal were to optimise a photobioreactor for a fixed microalgal strain, the scaling hierarchy could be truncated to exclude intracellular scales. Upscale subsystems could then be optmised hierarchically, with productivity at each scale reducible only to an empirical model for production efficiency at the cellular scale; arguably, this approach is a logical first step beyond current modelling approaches (section \ref{sec:hierarchy}). In another case, a designer might choose to neglect some intermediate-scale subsystem(s) and, instead, re-formulate parameter aggregations and dynamical links to map directly between previously non-adjacent scales. For example, a phytoelement could be optimised considering only its chromophore content, ignoring the effects of intermediate levels of organisation within the cell and cell culture. 

Ultimately, the key challenges of implementing \textit{in silico} multiscale optimisation in photosynthetic systems can be properly identified and addressed only in practice, beginning with simple systems and gradually increasing complexity towards that of realistic photosynthetic energy systems. 

\section{Conclusions and outlook}
This chapter has presented a preliminary theoretical investigation into optimising idealised, multiscale photosynthetic systems. This provides a testing ground for the tools required to optimise real systems. Beginning from a hierarchically structured toy model relating flows of resources, products and wastes between nested scale subsystems, a variational principle was proposed for optimising photosynthetic productivity at a given scale in the system: the productivity of a subsystem at a given scale will be maximised when resource fluxes (ultimately limited by upscale processes) incident on the photosynthetically active components are matched as closely as possible to the components' photosynthetic production efficiencies (ultimately limited by downscale processes), \textit{and} when those photosynthetic production efficiencies are simultaneously matched as closely as possible to the incident resource fluxes. It was suggested that this principle may underlie the acclimation processes observed at multiple scales in natural photosynthetic systems, and that abstracting the concept of \textit{multiscale balance between resource allocations and component production efficiencies} may provide a useful, general principle for engineering photosynthetic systems.

Overall system optimisation is complicated by the fact that processes at a given scale often depend on parameters at multiple scales simultaneously (explored in detail in chapter \ref{chp:multianalysis}), so optimisation of these parameters requires \textit{negotiation} between optimisations for the processes at different scales. The result may be that locally-suboptimal subsystem configurations are required in order to optimise the system overall. The recently developed theory of decomposition-based optimal design allows for co-ordinated optimisation of coupled subsystems under system-scale functional targets and constraints, as well as targets and constraints localised to individual subsystems. It was speculated in this chapter that this theory may be applied to a hierarchically partitioned photosynthetic system to facilitate quantitative multiscale, multiprocess optimisation under system-scale functional targets. This proposal invites a future program of research beginning with modelling and optimising simple multiscale systems and gradually increasing complexity towards that of realistic photosynthetic energy systems.

\chapter{Conclusions and future directions} \label{chp:conc}

Photosynthesis provides the energetic bedrock for biological and economic activity on Earth. Due to concerns over the unsustainability of fossil-fuelled energy systems, recent years have seen increasing interest in the potential for photosynthesis to help meet humanity's energy needs sustainably, through harnessed photosynthetic energy systems such as higher plants and microalgal cultivation systems. Three billion years of evolution has engineered biological photosynthetic systems for solar-powered survival and reproduction across diverse ecosystems. This thesis has asked whether and how such systems and/or biomimetic artificial photosynthetic systems can be (re-)engineered as solar energy harvesting technologies able to sustainably power the human economy on a globally significant scale. This question links the needs of global-scale environmental and economic systems with the properties and functions of photosynthetic systems and their component subsystems, which span length scales from landscape to chromophore, and time scales from annual climatic cycles to femtosecond energy-transfer processes at the molecular level. Accordingly, this thesis has taken a multiscale, multidisciplinary approach, combining reductionistic studies of specific system components (chapters 3 and 4, and appendix A) with holistic studies (chapters 1, 2 and 5) of how those components work together in service of system-scale objectives. Both types of study have generated novel insights and opened new avenues for future work.

Chapter \ref{chp:global} presented a detailed literature review, quantifying limitations of current global energy systems and propects for a global-scale transition to sustainable energy systems powered by photosynthesis. Chapter \ref{chp:multianalysis} introduced hierarchy theory to the study of photosynthetic energy systems, as a novel system-design tool complementary to the linear-process analysis usually applied to such systems, and better able to accommodate their multiscale, multiprocess complexity. Based on this hierarchical approach, chapter \ref{chp:qeet} pioneered the use of renormalisation theory to quantitatively analyse how the mechanisms of excitation energy transfer (EET) depend on structure and scale in a chromophore network with the generic, multiscale structural and energetic features of a thylakoid membrane. Chapter \ref{chp:structure} contributed to research elucidating how thylakoid membrane protein composition determines multiscale thylakoid structure in antenna-mutant strains of green microalgae. Novel structural studies were reported for two membrane protein (super)complexes from the strain, \textit{Chlamydomonas reinhardtii stm3}, based on single-particle electron microscopy. Finally, chapter \ref{chp:multiopt} proposed a novel variational principle for theoretically maximising productivity at each scale in an idealised photosynthetic system, and it was speculated that this may in future be used together with decomposition-based design optimisation theory to construct multiscale-optimal, idealised photosynthetic systems \textit{in silico}. The main findings of each chapter, and resulting new avenues for future work, are summarised in this final chapter.

\section{Photosynthesis in global energy systems: realities and prospects}\label{sec:globalconc}
\subsubsection{Summary and conclusions}
Geophysical, ecological and economic limitations of global energy systems dictate the functional requirements of photosynthetic energy systems at smaller scales. A detailed understanding of these limitations is therefore important in engineering photosynthetic energy systems. 

Solar energy is available far in excess of global economic demand but is incident at power densities that are low compared with incumbent thermal power generating systems, which therefore challenges its economic utility. Earth's current biota photosynthetically store only $\sim0.1\%$ of the solar energy annually available, though the resulting net primary production (NPP) is still nine times larger than the human economy's current total primary energy supply (TPES). This NPP also fixes ten times more atmospheric carbon than total annual anthropogenic CO$_2$ emissions contain but currently, $\sim96\%$ returns to the atmosphere on the same time scale. Simultaneously, the economy appropriates $\sim25\%$ of current NPP and this level is thought to be significantly above the sustainable limit. When sustainability criteria are considered, agro-biofuels can be expected to supply only a small fraction of future global energy demand. 


Carbon-based fuels directly supply the majority of global final energy consumption, and the transition to electricity is gradual compared with the speed required for climate-change and peak-oil mitigation. Rapid development of fungible, sustainable fuels is therefore critically important. Time scales of previous global energy transitions indicate that an unprecedented level of sociopolitical cooperation is required to accelerate the transition from fossil to sustainable fuels. Sustainable, increased reliance on NPP from non-fossil sources, to supply low-carbon fuels on globally significant scales, may be possible. This will likely require an increase in Earth's total potential net primary productivity (NPP$_0$) through deployment of bioengineered and/or artificial photosynthetic systems on otherwise unproductive or low-productivity lands. 

Microalgal cultivation systems are rapidly emerging as photosynthetic energy systems with potential to offer increased NPP$_0$ simultaneously with other economic and environmental services. However, comprehensive assessments indicate that technical innovation is still required to achieve economic and environmental feasibility. Artificial photosynthetic systems are promising technologies for the longer term and stand to benefit from ongoing advances in understanding biological photosynthetic systems. A key challenge in developing high-productivity photosynthetic energy systems is optimising light-harvesting efficiency under a range of light environments including high-irradiance conditions well suited to solar fuel production. Addressing this challenge through a multiscale approach formed the main subject of this thesis.  

\subsubsection{Future directions} 
In addition to bioengineered and artificial photosynthetic systems, a broader class of technologies may help to meet the urgent need for sustainable, carbon-based fuels. In principle, energy from any low-carbon\footnote{When compared with fossil fuels.} source, such as hydro, wind, nuclear, geothermal and other renewables, can be used to drive chemical fuel production. However, techno-enviro-economic comparisons of the various possible pairings between energy sources and fuel production cycles are largely absent from the literature. 

A notable recent exception by Blankenship \textit{et al} \cite{blankenship2011} compared photosynthetic and photovoltaically-powered fuel production technologies, although the metric for comparison was only first-order energy conversion efficiency. A comprehensive comparison of technologies' economic and environmental feasibilities requires second-order life-cycle analyses (LCAs). Broader, systematic application of assessment and comparison frameworks based on second-order LCA, such as that recently developed by Beal \textit{et al} \cite{beal2012} for microalgal cultivation systems, is in order. 
   
\section{Multiscale systems analysis of light harvesting in photosynthetic energy systems}\label{sec:multiscaleconc} 
	\subsubsection{Summary and conclusions}
Guided by traditional linear-process analysis of photosynthetic light-harvesting efficiency, strategies for improvement have largely focussed on achieving more equitable irradiance distribution throughout the photosynthetic system. However, standard linear-process analysis gives no account of how irradiance distribution (and distribution of other resources such as CO$_2$ and H$_2$O) depends simultaneously on parameters over a wide range of scales in the system, nor for how those parameters depend on each other. Consequently, strategies to improve distribution have tended to focus at a single scale (or narrow range), tuning parameters such as LHC antenna protein composition, cell culture density, and photobioreactor structure and surface properties. The literature appears to contain no account of how to best coordinate these parameters. The prospect of developing such an account is challenged by the naturally multiscale, multiprocess complexity of photosynthetic systems. 

Inspired by the coordinated resource-distribution strategies facilitated by multiscale, hierarchical structures in higher plants (such as the optical `package effect'), chapter \ref{chp:multianalysis} introduced hierarchy theory as a system-design tool complementary to the standard linear-process approach. This theory partitions a complex dynamical system into a hierarchy of recursively nested subsystems with well-separated characteristic length and time scales in order to simplify system analysis. The multiscale analysis of the hierarchy is constructed by studying internal properties and dynamics of each scale subsystem independently, as well as couplings between scale subsystems. Hierarchy theory provides a general framework for quantitatively linking global system-scale objectives and constraints with engineerable parameters at different scales within a system, in a coordinated way.

As a basis for applying hierarchy theory, a multiscale, hierarchical system partitioning was proposed for a generic microalgal cultivation system. The system was partitioned heuristically such that each component within a given scale subsystem was a repeated functional unit that exchanged energy through multiple mechanisms (e.g. light, mass and heat transfer) with other components at the same scale and, through weaker interactions, also with components at other scales. 

Quantitative hierarchical analysis of the system, or a collection of scale subsystems, requires models of the energetics at each scale, and understanding of how the models and parameters on which they depend, such as composition and structure, transform between scales. At supracellular scales, energy transfer processes can be described using relatively well-understood bulk material properties and models of continuum mechanics and radiative transfer. However, many open questions remain about interdependences between composition, structure and energetics at the nanoscales of the thylakoid membrane, where chromphores are arranged within proteins, protein complexes and supercomplexes, and overall membrane ultrastructure. Two such questions were addressed by quantitative, hierarchical studies in chapters \ref{chp:qeet} and \ref{chp:structure}, as a contribution toward the development of quantitative, hierarchical whole-system analysis. Other questions remain to be addressed in future work.

  \subsubsection{Future directions}
Further developing the hierarchical framework for quantitative systems analysis and optimisation will require work on both the overall framework and on scale-focussed models relating composition, structure and energetics. Future work related to optimisation is discussed in section \ref{sec:concmultiopt}; this section focusses on analysis. 

\paragraph{Quantitative network partitioning for photosynthetic systems}
The multiscale, hierarchical system partitioning proposed in chapter \ref{chp:multianalysis} (fig. \ref{fig:multiscale}) was based on intuition about component interaction strengths within and between levels. In principle this can be tested using recently developed methods for quantifying structure in complex networks \cite{ahn2010, clauset2008, sales2007}, assuming that enough is known about component interaction mechanisms at each scale to parameterise network topology (e.g. In the hierarchical study of EET mechanisms in chapter \ref{chp:qeet}, network topology was parameterised using transition dipole-dipole couplings between chromophores to quantify the partitioning between scales). Quantifying network structure is important because accurate hierarchical modelling requires a well-formed hierarchy in which interaction strengths between components at different levels are scale-separated. Where no natural scale paritioning can be found or well approximated, multiscale modelling techniques more sophisticated than hierarchy theory will be required.

\paragraph{Radiative transfer in the chloroplast}
Section \ref{subsec:suggestedhierarchy} raised the possibility that stacked thylakoid layers in higher-plant grana may exploit optical interference effects at subwavelength scales to distribute irradiance more equitably between membrane layers; effectively a claim that the thylakoid membrane can act as an optical metamaterial. While this is an intriguing posibility with implications for both bioengieered photosynthetic systems and artificial light-harvesting materials, radiative transfer models previously used to support these claims have typically relied on basic ray optics and highly simplified models of thylakoid composition and structure. More sophisticated models are needed. 

One approach may be to use a detailed thylakoid structural model such as those discussed in chapter \ref{chp:structure}, and consider its optical response as a sum over the linear responses of its constituent pigment-protein complexes (PPCs). Radiative transfer could be modelled numerically, using a finite-element-based solution to Maxwell's equations \cite{margalit2010} or possibly a T-matrix calculation \cite{loke2009}. The optical effects of structurally rearranging PPCs within the thylakoid, and also overall thylakoid structure, could be studied using such an approach. Recent studies have attempted to understand the dependence of quantum EET dynamics on the state of the driving optical field \cite{brumer2012, fassioli2012, kassal2012}. However, these studies have not considered potential effects due to thylakoid ultrastructure mediating interactions between incident radiation and PPCs. This possibility deserves investigation. It should be noted, however, that the nonlinear optical responses of PPCs, which heralded the presence of quantum dynamical effects in EET (chapter \ref{chp:qeet}) would not be accounted for by models based only on linear optical responses of PPCs; such models may need to be viewed as a stepping stone to a more sophisticated approach, which bridges the optical properties of thylakoid ultrastructure with the nonlinear optical reponse of PPCs. 

\paragraph{Heat transfer in the chloroplast}
There is an abundance of literature describing the `safe dissipation as heat' of excess excitation energy through nonphotochemical quenching (NPQ) mechanisms. However, the fate of that heat and how it relates to thylakoid structure have rarely been discussed. It is currently unknown whether there is a mechanistic relationship between the heating and thylakoid structural rearrangements associated with NPQ. Modelling heat transfer through thylakoids and their subsystems may lend insight into these questions. Molecular dynamics simulations of cooling within the protein and lipid phases of the thylakoid (and their surrounding aqueous solvent) could be used to infer bulk thermal conductivities at the scale of PPCs and the lipid spaces between them. These may in turn be used to study membrane conductivity at larger scales, in different structural configurations. Of particular interest is how thermoregulation of the chloroplast interacts with NPQ, and what role multiscale thylakoid structure plays in this.

\section{Quantum-classical crossover in multiscale photosynthetic excitation energy transfer}\label{sec:qeetconc}
	\subsubsection{Summary and conclusions}
	At the scale of the thylakoid membrane, energy is allocated to photosynthetically active components first through radiative transfer and then through EET after light absorption (though the interactions between these two processes are the subject of ongoing research). The key problem to solve at this scale is allocation of excitations to photochemical reaction centres (RCs) throughout the thylakoid in balance with their maximum-achievable turnover rates. After many decades of study, the mechanisms of EET central to solving this problem, and their dependences on multiscale thylakoid structure, are still not fully understood. 
	
	It was long assumed that EET in photosynthetic systems is adequately described by F\"orster theory, which assumes excitations to be localised on individual chromophores such that EET is a classical, diffusive process. However, in recent years, structural, spectroscopic and theoretical studies have strongly suggested that individual PPCs may exploit quantum coherence between excitation states to improve the efficiency and robustness of EET compared with F\"orster transfer. Questions of what mechanisms underly these effects, and whether they could be present and physiologically significant \textit{in vivo}, have been intensely examined but remain controversial.
	
	Detailed studies of quantum-dynamical EET mechanisms to date have focussed almost exclusively within individual PPCs. Multichromophoric F\"orster theory (MCFT) was developed to account for the effect of intra-PPC quantum coherences on inter-PPC transfer rate, which is known as `supertransfer'. However, the theory assumes \textit{a priori} that inter-PPC transfer is incoherent. Therefore, MCFT can describe coherence-enhanced long-range EET in PPC aggregates only as a linear sum of enhancements at the scale of individual PPCs. 
	
	In section \ref{sec:natphys}, a new formalism for studying long-range EET in PPC aggregates was developed, which accommodates intercluster coherences and reveals unexpected scaling trends in the mechanisms of EET. A renormalisation analysis revealed the dependence of EET dynamics on multiscale, hierarchical network structure. Surprisingly, thermal decoherence rate was found to decline at larger length scales for physiological parameters and coherence length was instead limited by localization due to static disorder. Physiological parameters supported coherence lengths up to $\sim5$ nm, which was consistent with observations of solvated light-harvesting complexes (LHCs) and invites experimental tests for intercomplex coherences in multi-LHC/RC networks. Results further suggested that a semiconductor quantum dot network engineered with hierarchically clustered structure and small static disorder may support coherent EET over larger length scales, at ambient temperatures. 
This surprising result may also hint at a more general principle for networks of electronically coupled quantum systems in condensed matter: hierarchically clustered network structures carefully designed to complement the dielectric and mechanical properties of their surroundings may aid in sustaining long-range coherent interactions at high temperatures. This possibility warrants exploration and may be of interest for application in quantum information systems and/or other quantum technologies. 
	
  \subsubsection{Future directions}
The results suggest that MCFT will, in general, be sufficient for modelling energy transfer at scales larger than individual PPCs in a biological system, within the multiscale system analysis framework developed in chapter \ref{chp:multianalysis}. The development of that framework has therefore been moved one step forward and the next logical step may be to focus on a different energetic process at the thylakoid scale, such as radiation transfer or heat transfer. Nonetheless, important questions regarding EET also remain.

Notwithstanding the likely absence of long-range quantum-coherent EET in thylakoid membranes under physiological conditions, the scaling principles revealed here may be of use in designing artificial light-harvesting materials that can support such long-range coherence. However, it should first be asked whether these materials would derive any functional advantage from long-range coherent EET, compared with EET by supertransfer. Some insight may be obtained by extending the model from section \ref{sec:natphys} to include dissipative sites that mimic RCs, and studying dissipation rates from the network under parameterisations that support long-range coherence, and others that do not. 

In particular, it is interesting to ask whether long-range coherence in a multi-LHC/RC network could improve the network's optimal solution to the excitation-allocation problem. One potential strategy for answering this question is to use a multiscale optimisation technique such as analytical target cascading (ATC) to collaboratively optimise EET properties of scale-separated subsystems within the network (e.g. chromophore clusters, PPC subunits, PPCs, PPC aggregates/supercomplexes) under the network-scale functional target of optimal excitation allocation. Of interest are the possible consequences of allowing and disallowing coherent EET at supra-PPC scales in different cases. Would the optimal network energy conversion efficiencies differ significantly? Would the multiscale structures of the optimally performing networks differ markedly between the two cases, and how would they each compare with structures seen in natural thylakoids? 

However, this in turn raises the methodological question of whether existing formulations of ATC can accommodate multiscale quantum mechanical models, which are based on stochastic dynamical variables. To the author's knowledge, such application of ATC has not yet appeared in the literature. It is worth noting, however, that probabilitic formulations of ATC have recently emerged \cite{xiong2010, chen2010}, designed to handle stochastic uncertainties in classical dynamical variables; perhaps these could be adapted for multiscale design of engineered quantum systems, including light-harvesting materials.

Aside from light-harvesting materials, the scaling principles revealed in section \ref{sec:natphys} may be of interest for developing other kinds of engineered quantum systems. Quantum error correction is a major field of research aimed at developing schemes for protecting quantum dynamical systems from the decohering effects of their environments. The analysis in section \ref{sec:natphys} demonstrates that quantum-coherent dynamics in a network described by the Frenkel exciton Hamiltonian can be preserved over surprisingly large length and time scales, even under thermal decoherence at ambient temperature, simply by virtue of network structure. One may speculate that similar effects can be expected under a broader class of Hamiltonians, of which the multiscale EET system is merely representative. This possibility, too, deserves further exploration.

\section[Protein composition and multiscale structure in the thylakoid membrane]{Light-harvesting protein composition and multiscale structure in the thylakoid membrane}\label{sec:structconc}
	\subsubsection{Summary and conclusions}
	Energetic modelling at each scale in a photosynthetic system such as a microalgal cultivation system requires understanding how the process(es) of interest depend(s) on system parameters such as composition and structure at that scale, as well as at other scales. Recent studies have demonstrated improved photosynthetic productivity at the cell-culture scale by genetically down-regulating the expression of light-harvesting protein complexes (LHCs). The improvements have been attributed to more equitable light distribution at the culture scale, resulting in reduced NPQ in cells near to the illuminated surfaces of the culture and increased photosynthesis in previously light-limited regions. However, the complex interdependences between thylakoid protein composition and structure, productivity at the chloroplast scale and productivity at the overall culture scale are not yet fully understood. Characterising these relationships is essential for targeted thylakoid engineering as a part of multiscale system engineering.
	
	Chapter \ref{chp:structure} described contributions to an ongoing, larger research program focussed on developing high-throughput multiscale structure assays for thylakoids harvested from microalgal `antenna-mutant' strains. Novel structural studies were reported for two thylakoid membrane protein (super)complexes from the antenna-mutant green microalgal strain, \textit{Chlamydomonas reinhardtii stm3}, based on negative-stain single-particle electron microscopy (single-particle EM). These studies provided the first 3D structural models of the PSI-LHCI supercomplex and PSI core complex from \textit{stm3}, at 24 \AA~and \\18.1 \AA~resolution respectively. To the author's knowledge, the model of PSI obtained here is the first 3D reconstruction of PSI from any green algal species. By fitting the high-resolution X-ray structure of PSI-LHCI from \textit{P. sativum} to the EM map determined here for PSI-LHCI in \textit{stm3}, it was inferred that the latter is likely to contain five additional Lhca subunits (nine in total) in positions consistent with the recent findings of Yadavalli \textit{et al} \cite{yadavalli2011} for wild-type \textit{C. reinhardtii}. Since the genetic mutation which differentiates \textit{stm3} from wild-type \textit{C. reinhardii} is not known to affect PSI or LHCI, it is likely that the structural models obtained here for these (super)complexes in \textit{stm3} are also applicable to wild-type \textit{C. reinhardtii}. 

\subsubsection{Future directions}
Quarternary structure in the PSI-LHCI supercomplex was less well-resolved than in the PSI core complex and this was likely to be partly attributable to underrepresentation of side-view particle projection images in the PSI-LHCI dataset. The dataset also showed lower contrast compared with the PSI dataset, which was stained and imaged using different methods. Accordingly, it is suggested that improved results may be obtained through a repeat study of the \textit{stm3} PSI-LHCI supercomplex using 0.7\% uranyl formate stain and electron-microscopic imaging using a high-resolution CCD camera (4096$\times$4096 pixels or greater). Sample preparation methods that encourage greater acquisition of side-view particle projection images are also recommended. For example, extending the glow-discharging time for the EM grids and/or embedding particles in a thick stain layer by rapidly drying grids after staining. 

The model obtained here for the PSI core complex showed a high level of consistency with X-ray structures previously determined for \textit{P. sativum}. However, inconsistencies between fits of the EM structure to two different X-ray structures, the earlier of which provided the initial model for EM model refinement, suggested that initial-model bias may have been present in the final EM model. Moreover, the multirefine algorithm used to generate the PSI EM model failed to concurrently produce a stable structure from an initial model based on a higher-plant PSII core monomer. To what extent particle heterogeneity was present in the dataset therefore remains an open question. It is suggested that to address the question of initial-model bias in the final PSI EM model, the multirefine procedure should be repeated using the most recent X-ray structures for PSI and PSII core monomers (suitably coarsened) as initial models. Addressing the possibility of dataset heterogeneity may be less straightforward, requiring proteomic characterisation and more stringent biochemical purification of the sucrose gradient fraction. 

These structural studies of PSI-LHCI and PSI were completed in parallel with other similar studies of PSII-LHCII and LHCII from \textit{stm3}, by Knauth \cite{knauth2013}. Together, the resulting 3D models provide a set of `building blocks' with which to explore the structural configuration space of the thylakoid, through docking into larger structural maps derived by electron tomography. This exploration will be assisted by software currently under development by Ali \cite{ali2012} for fitting single-particle models to automatically detected single-particle contours in tomograms.

Efforts to characterise the complex interrelationships between thylakoid protein composition, multiscale structure and energetics may benefit from an approach which integrates experimental and theoretical structural studies. A theoretical model of the type recently presented by Schneider and Geissler \cite{schneider2013}, generalised to accommodate the more heterogeneous membrane structures of microalgae, may provide a useful complement to the multiscale structural determination program described here. It may be speculated that such a model might make it possible to predict membrane structural features in antenna-mutant strains based on proteomic assessments of membrane protein composition. These predictions could then be tested using the experimental structure assays described here. Ultimately it may be possible to couple energetic models to these structural models with the aim to eludicate composition-structure-energetics relationships in the thylakoid to help inform targeted engineering.
	
\section[Towards multiscale optimal design of light harvesting in photosynthetic systems]{Towards a multiscale optimal design framework for light harvesting in photosynthetic energy systems}\label{sec:concmultiopt}

\subsubsection{Summary and conclusions}
Chapter \ref{chp:multiopt} presented a speculative, preliminary theoretical investigation into optimising idealised, multiscale photosynthetic systems. This provides a testing ground for the tools required to optimise real systems. Beginning from a hierarchically structured toy model relating flows of resources, products and wastes between nested scale subsystems, a variational principle was proposed for optimising photosynthetic productivity at a given scale in the system: the productivity of a subsystem at a given scale will be maximised when resource fluxes (ultimately limited by upscale processes) incident on the photosynthetically active components are matched as closely as possible to the components' photosynthetic production efficiencies (ultimately limited by downscale processes), \textit{and} when those photosynthetic production efficiencies are simultaneously matched as closely as possible to the incident resource fluxes. It was suggested that this principle may underlie the acclimation processes observed at multiple scales in natural photosynthetic systems, and that abstracting the concept of \textit{multiscale balance between resource allocations and component production efficiencies} may provide a useful, general principle for engineering photosynthetic systems.

Overall system optimisation is complicated by the fact that processes at a given scale often depend on parameters at multiple scales simultaneously (explored in detail in chapter \ref{chp:multianalysis}), so optimisation of these parameters requires \textit{negotiation} between optimisations for the processes at different scales. The result may be that locally-suboptimal subsystem configurations are required in order to optimise the system overall. The recently developed theory of decomposition-based optimal design allows for co-ordinated optimisation of coupled subsystems under system-scale functional targets and constraints, as well as targets and constraints localised to individual subsystems. It was speculated that this theory may be applied to a hierarchically partitioned photosynthetic system to facilitate quantitative multiscale, multiprocess optimisation under system-scale functional targets. This proposal invites a future program of research beginning with modelling and optimising simple multiscale systems and gradually increasing complexity towards that of realistic photosynthetic energy systems.

\subsubsection{Future directions}

\paragraph{Numerical optimisation of photobioreactor configuration}
At the macroscopic scales of phytoelements, photobioreactors and the overall photobioreactor array, material composition and structure are only weakly coupled, and energy transfer processes are relatively well understood. This contrasts with a more complex and less-well understood situation inside the cell. A first step towards multiscale system optimisation (beyond toy models) may therefore be to treat intracellular processes phenomenologically (e.g. using a standard light-response curve) and optimise supracellular subsystems collaboratively. It is anticipated that this may result in photobioreactor configurations beyond standard geometries (e.g. flat-plate or tubular) and potentially beyond the reach of intuition.

%

 \section{Closing comments}
This thesis has reported a multidisciplinary investigation into optimising light harvesting in photosynthetic systems for economically scalable, sustainable energy production. The research reported has delivered novel insights into the complex interrelationships between composition, structure and energetics at the scale of the thylakoid membrane, as well as holistic insights into how components at different scales in a photosynthetic system work together under system-scale objectives. This opens new avenues for future research and constitutes progress toward the urgent goal of developing photosynthetic energy systems able to sustainably power the human economy on a globally significant scale.


\cleardoublepage
\addcontentsline{toc}{chapter}{List of references}
\bibliography{bib}       
\cleardoublepage
\markboth{Appendices}{}
\chapter*{Appendices}\label{chp:apps}
\addcontentsline{toc}{chapter}{Appendices}

\section*{Appendix A: RNAi knock-down of LHCBM1, 2 and 3 increases photosynthetic hydrogen production efficiency of the green alga \textit{Chlamydomonas reinhardtii}}\label{sec:appa}
\addcontentsline{toc}{section}{App. A: Knock-down of LHCII increases hydrogen production in \textit{C. reinhardtii}}

The following publication reports work completed during the final year of PhD candidature. This work demonstrates that engineering reduced LHCII antenna systems in the microlalga, \textit{Chlamydomonas reinhardtii} can significantly improve the light-to-biomass and light-to-H$_2$ conversion efficiency at the laboratory scale. It is therefore consistent with the exposition of this thesis. However, it is included here as an appendix because the author's contributions were limited to co-authorship of the introduction and creation of figure 6.

\blankpage
\includepdf[pages={1-12}]{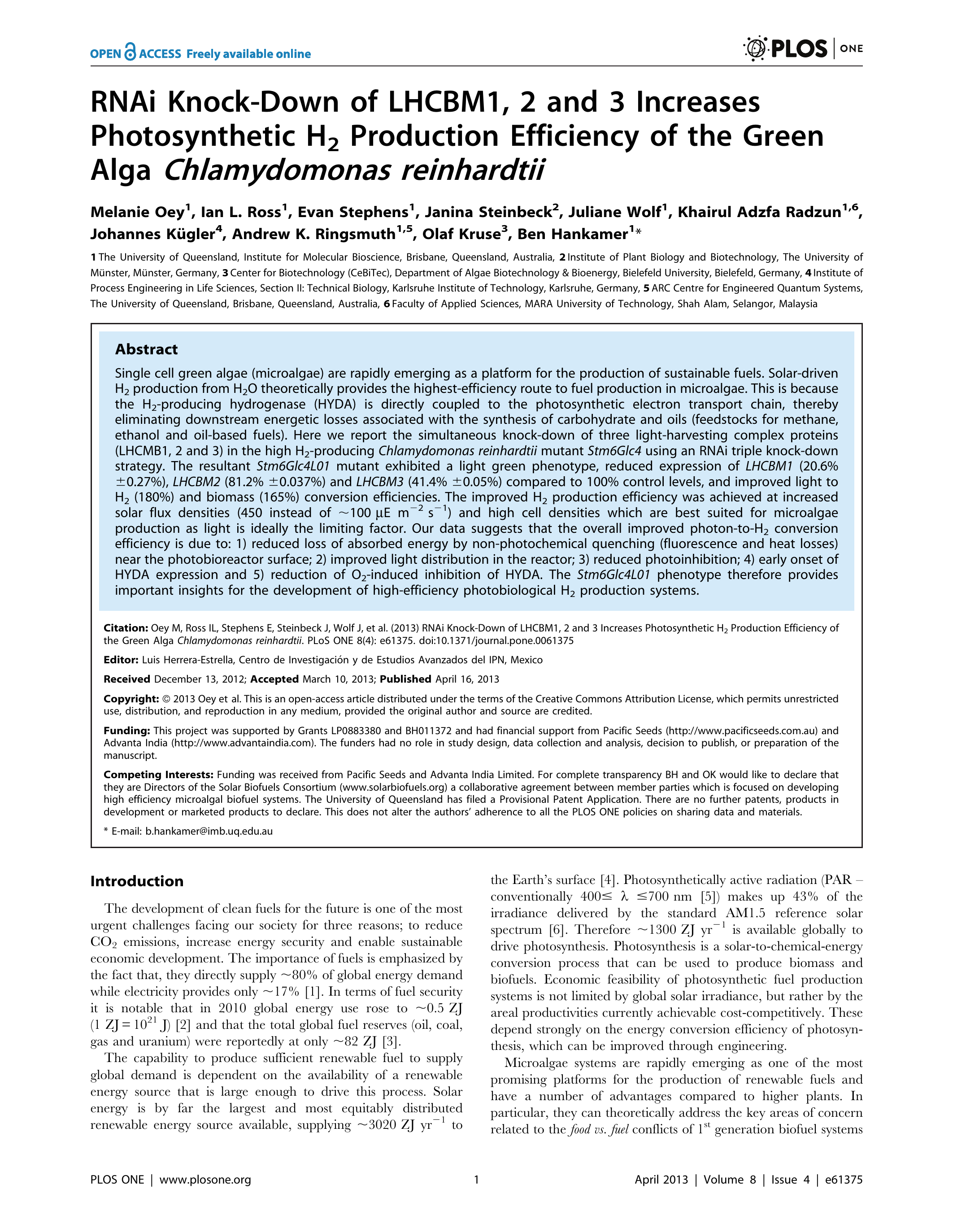}

\section*{Appendix B: Entangling a nanomechanical resonator with a microwave field}\label{sec:appb}
\addcontentsline{toc}{section}{App. B: Entangling a nanomechanical resonator with a microwave field}

The following publication reports work completed during the first year of PhD candidature, before the project was shifted to focus on photosynthetic energy systems. It is included here as an appendix because the techniques which were learned and applied in this work for modelling open quantum systems provided an important foundation for techniques later applied to modelling photosynthetic excitation energy transfer, presented in chapter \ref{chp:qeet}.

\blankpage
\includepdf[pages={1-14}]{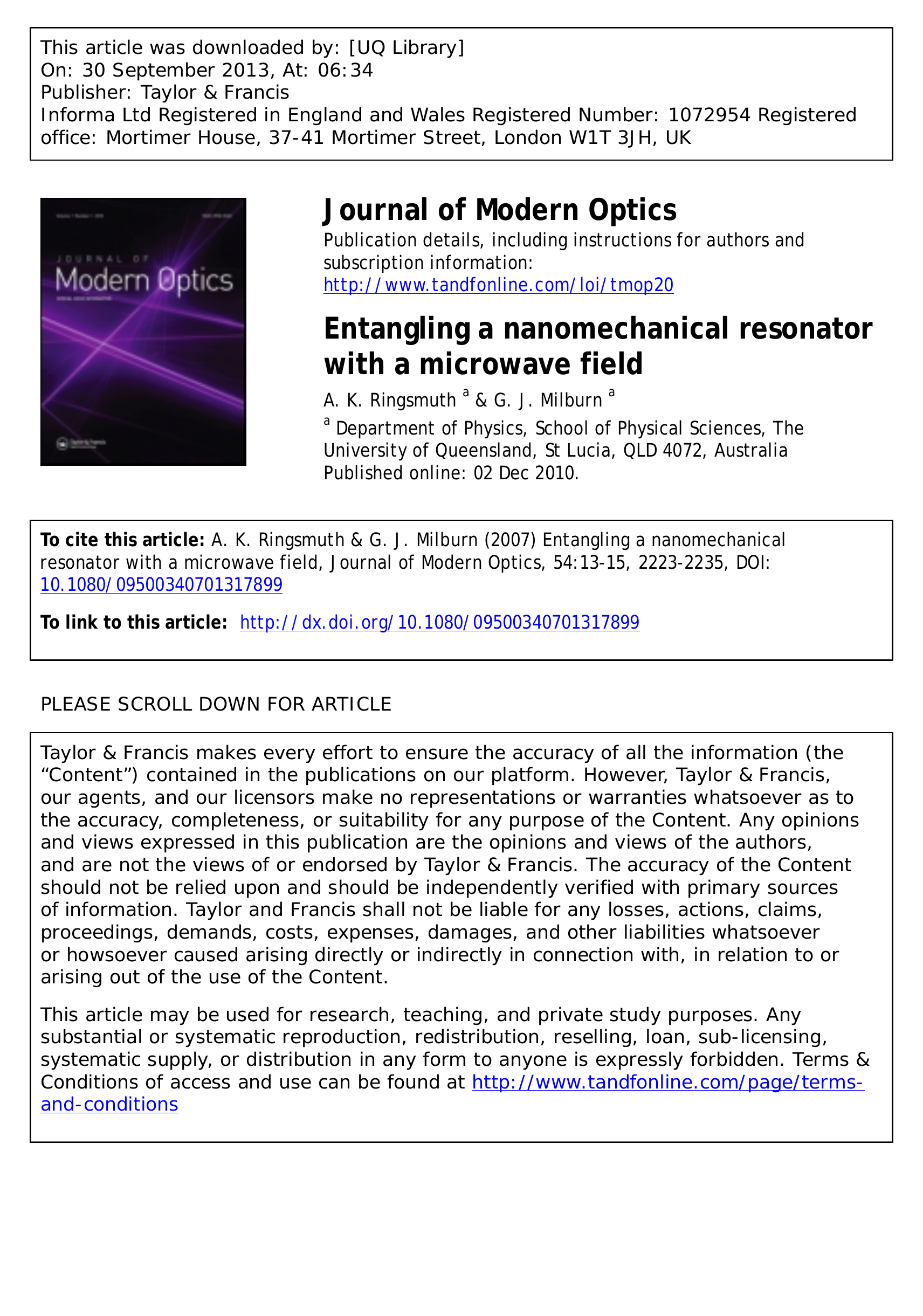}


\end{document}